\documentclass[a4paper, 12pt]{book}
\usepackage[dutch,english]{babel}
\usepackage{latexsym}
\usepackage{amsbsy}
\usepackage{fancyhdr}
\usepackage{sectsty}
\usepackage{amssymb}
\usepackage{amstext}
\usepackage{amsmath}
\usepackage{latexsym,amssymb,amstext}

\newcommand{\be}{\begin{equation}}
\newcommand{\ee}{\end{equation}}
\newcommand{\bea}{\begin{eqnarray}}
\newcommand{\eea}{\end{eqnarray}}

\newcommand{\bay}{\begin{array}}
\newcommand{\eay}{\end{array}}
\newcommand{\lb}{\label}

\newcommand{\m}{\mu}
\newcommand{\n}{\nu}

\def\ep{\epsilon}
\def\vep{\varepsilon}

\def\si{\sigma}
\def\al{\alpha}
\def\be{\beta}
\def\ga{\gamma}
\def\ze{\zeta}
\def\de{\delta}

\def\Om{\Omega}

\def\bX{\bar{X}}
\def\bF{\bar{F}}

\def\bep{\bar{\epsilon}}

\def\bal{\bar{\alpha}}
\def\bbe{\bar{\beta}}
\def\bga{\bar{\gamma}}
\def\bze{\bar{\zeta}}
\def\bps{\bar{\psi}}

\def\bOm{\bar{\Omega}}

\renewcommand{\L}{\mathcal{L}}

\renewcommand{\S}{\mathcal{S}}
\renewcommand{\H}{\mathcal{H}}
\newcommand{\G}{\mathcal{G}}

\newcommand{\V}{\mathcal{V}}
\newcommand{\F}{\mathcal{F}}
\newcommand{\D}{\mathcal{D}}
\newcommand{\M}{\mathcal{M}}
\newcommand{\N}{\mathcal{N}}
\newcommand{\Y}{\mathcal{Y}}
\newcommand{\Z}{\mathcal{Z}}

\def\bV{\bar{V}}
\def\bX{\bar{X}}
\def\bY{\bar{Y}}

\def\tA{{\tilde A}}
\def\tB{{\tilde B}}

\def\tF{{\tilde F}}
\def\tG{{\tilde G}}
\def\tL{{\tilde \L}}
\def\tS{{\tilde S}}
\def\tX{{\tilde X}}
\def\tY{{\tilde Y}}
\def\tN{{\tilde N}}
\def\tZ{{\tilde Z}}
\def\tH{{\tilde H}}

\def\hH{\hat{H}}

\def\tOm{{\tilde \Omega}}
\def\tch{{\tilde \chi}}

\def\bmN{\bar{\mathcal{N}}}

\def\btOm{\bar{\tilde{\Omega}}}

\def\ra{\sf{a}}
\def\rb{\sf{b}}

\def\Im{\,{\rm Im}\, }
\def\Re{\,{\rm Re}\, }

\def\({\left(}
\def\){\right)}
\def\[{\left[}
\def\]{\right]}
\def\p{\partial}

\newcommand{\half}{\frac{1}{2}}
\newcommand{\vier}{\frac{1}{4}}
\newcommand{\acht}{\frac{1}{8}}
\newcommand{\nn}{\nonumber}

\def\slash#1{\rlap{\hbox{$\mskip 1 mu /$}}#1}      
\def\Slash#1{\rlap{\hbox{$\mskip 3 mu /$}}#1}      

\newcommand{\spa}{\slash{\partial}}
\newcommand{\SD}{\Slash{D}}
\newcommand{\SDm}{\Slash{\D}}
\newcommand{\SH}{\Slash{\H}}

\DeclareSymbolFont{AMSa}{U}{msa}{m}{n}
\DeclareSymbolFont{AMSb}{U}{msb}{m}{n}
\DeclareMathSymbol{\fieldR}{\mathalpha}{AMSb}{"52}

\usepackage{type1cm}

\subsubsectionfont{\large}
\subsectionfont{\large}

\renewcommand{\chaptermark}[1]{\markboth{ #1}{}}

\renewcommand{\headrulewidth}{0pt}
\renewcommand{\footrulewidth}{0pt}
\begin{document}

\frontmatter \pagestyle{empty}
\thispagestyle{empty}
\vspace*{-2.2cm}

\begin{flushright} \small
 ITP--UU--07/42 \\ SPIN--07/30
\end{flushright}
\smallskip

\begin{center}
 {\LARGE\bfseries N=2 Supersymmetric Theories, \\[3mm]
  Dyonic Charges and Instantons}
\\[10mm]
Mathijs de Vroome \\[5mm]
 {\small\slshape
 Institute for Theoretical Physics \emph{and} Spinoza Institute \\
 Utrecht University, 3508 TD Utrecht, The Netherlands \\[2mm]
 {\upshape\ttfamily M.T.deVroome@phys.uu.nl} \\[3mm]
 }
\vspace{12mm}

\begin{minipage}[]{11.5cm}
\begin{center}{\bfseries Abstract}\end{center}
{\small This paper contains the results of our investigations of
BPS instantons and of our work on $N=2$ supersymmetric gauge
theories. The BPS instantons we study appear in type II string
theory compactifications on Calabi-Yau threefolds. In the
corresponding four-dimensional effective supergravity actions the
BPS instantons arise as finite action solutions to the Euclidean
equations of motion. For $N=2$ supersymmetric gauge theories we
construct general Lagrangians involving gauge groups with
(non-abelian) electric and magnetic (dyonic) charges. In this work
a coupling to hypermultiplets is included.}
\end{minipage}\\
\end{center}
\newpage
\thispagestyle{empty}
\clearpage{\pagestyle{empty}\cleardoublepage}

\makeatletter
\def\thickhrulefill{\leavevmode \leaders \hrule height 1ex \hfill \kern \z@}
\def\@makechapterhead#1{%
  \vspace*{10\p@}%
  {\parindent \z@ \centering \reset@font
        \Huge \bfseries #1\par\nobreak
        \par
        \vspace*{10\p@}%
    \vskip 60\p@
  }}
\def\@makeschapterhead#1{%
  \vspace*{10\p@}%
  {\parindent \z@ \centering \reset@font
        \Huge \bfseries #1\par\nobreak
        \par
        \vspace*{10\p@}%
    \vskip 60\p@
  }}

\pagestyle{fancy}

\tableofcontents \thispagestyle{plain}

{\fancyhead{} \fancyhead[LE, RO]{\thepage}
\fancyhead[CO]{\slshape\leftmark}
\fancyhead[CE]{\slshape\leftmark}

\chapter{Preface}
The following text is based on my Ph.D. thesis, which was defended
on the eighteenth of June $2007$. The research on which this
thesis is based was performed at the Spinoza Institute at the
University of Utrecht, as part of the research programme of the
'Stichting voor Fundamenteel Onderzoek der Materie (FOM)', which
is financially supported by the 'Nederlandse Organisatie voor
Wetenschappelijk Onderzoek (NWO)'. For their great support in
finishing my Ph.D. in a successful way, I thank my supervisors
Stefan Vandoren and Bernard de Wit.

\begin{flushright}
Mathijs de Vroome, August $2007$
\end{flushright}

 \clearpage{\pagestyle{empty}\cleardoublepage}

\chapter{Introduction}

Three of the four fundamental forces of nature are described by
the Standard Model. This is a quantum field theory, which
presupposes all elementary particles to be point-like objects. The
fourth force of nature is gravity. On large (classical) length
scales it behaves according to the laws of General Relativity. On
small enough length scales, or high enough energy scales, quantum
effects become important and the classical theory should be
modified. The typical energy scale at which these modifications
are expected to be necessary is the Planck scale, $M_P c^2 =
\sqrt{\hbar c^5 / G_N} \sim 1,2 \times 10^{19} GeV$, where $G_N$
is the gravitational coupling constant. It is important to realize
that this scale is far beyond experimental reach. Present day
particle accelerators can produce collisions in which energies up
to order $1 \, TeV$ are involved, which is a factor $10^{16}$ away
from the Planck scale. Quantum gravity therefore is a theoretical
problem whose solution needs to be found without much help from
experimental side.

The obvious first guess for a quantum gravity theory is a quantum
field theory of General Relativity, set up along the same lines as
the quantum theory of the other forces. This theory gives rise to
infinities \footnote{In this respect we need to mention that
recently discussion arose about a possible finiteness of $N=8$
supergravity in four spacetime dimensions (see e.g.
\cite{BDR,GRV}), which is the maximally supersymmetric extension
of General Relativity.}. By itself these do not need to be
disastrous; also the Standard Model contains them. However,
whereas in the latter case all infinities can be absorbed in the
parameters of the theory and sensible physical predictions can be
extracted, this turns out to be impossible for the quantized
version of General Relativity. In other words, quantum General
Relativity is \emph{non-renormalizable}. Therefore a more drastic
modification of the classical theory is called for. This is
provided by \emph{string theory}, which is no longer based on
point-like particles, but on one-dimensional extended objects,
called strings.

Strings come in two varieties. There are strings with endpoints
(open strings) and strings without (closed strings). Both are
described by a two-dimensional action, with the coordinates
parameterizing a surface, called the worldsheet. This worldsheet
is the surface swept out by the string in spacetime. Its embedding
coordinates are functions of the two-dimensional worldsheet
coordinates.

Quantizing such a - bosonic and relativistic - string yields
interesting states: First of all, the closed spring spectrum
contains a massless spin-two state, which can be identified with
the graviton, the massless particle that mediates the
gravitational force. One can therefore say that string theory not
only describes quantum gravity, it even predicts it! Moreover, the
spectrum of the open string incorporates massless spin-one states,
which could play the role of vector gauge particles mediating the
Standard Model forces. This makes string theory a candidate for
being a unified description of the forces of nature. Bosonic
string theory also contains tachyons, which are spinless objects
with negative mass squared. They imply that the theory is
unstable. We come back to this shortly.

String theory has two parameters. One of them is $\alpha'$, which
has dimension $l^2$ and sets the scale of the string length. The
length of a string is taken about $10^{-35} \, m$, which is around
the Planck scale (although alternative scenarios do exist, for
instance based on \cite{ADD}). The other parameter is $g_s$, the
string coupling constant. It arises as the vacuum expectation
value of the dilaton, which is another (scalar) field in the
(closed) string spectrum. Perturbatively it controls the number of
loops in a stringy Feynman diagram.

To obtain particles with half-integer spin, anticommuting fields
are included on the worldsheet. The resulting theory is invariant
under local Weyl transformations, diffeomorphisms and
supersymmetry transformations. Furthermore, demanding spacetime
supersymmetry (a symmetry interchanging bosonic and fermionic
states) makes the theory free of tachyons. This leads to five
possible superstring theories, which all live in ten spacetime
dimensions.

As first conjectured by Witten, these five string theories are
related \cite{W4}. They all arise as appropriate descriptions in
special limits of one big theory, named M-theory. The different
descriptions, i.e. the different string theories, are then
connected by \emph{duality} transformations.

Duality transformations by definition relate different
descriptions of the same physical system. Three different types of
them can be distinguished. First of all, there are dualities
between descriptions that are based on different theories.
Secondly, dualities may relate two descriptions that are based on
the same theory, but with different values of the parameters
involved. These are called selfdualities. Finally, there are
dualities relating two descriptions that are based on the same
theory, including equal values of the parameters. These latter
dualities are invariances of the theories under consideration.

Electric/magnetic duality is an early discovered example of a
(self)duality. In short, electric/magnetic duality comes with a
rotation of elementary (electric) and solitonic (magnetic
monopole) states, as well as an inversion of the coupling
constant. Many aspects of the duality web connecting the five
string theories are directly or indirectly related to dualities of
this type.

A specific example of a duality in string theory is
\emph{T-duality}, which shows up as a selfduality in its purely
bosonic (closed string) version. When there is a compact
dimension, states in bosonic string theory are labelled by the
(quantized) internal momenta and the number of winding modes
(around the compact dimension). T-duality implies that there is an
alternative description in which the momentum (winding) modes of
the original formulation are described as (winding) momentum
modes. This duality involves an inversion of the radius of the
compact dimension; it relates bosonic string theory in a spacetime
background with a compact dimension with radius $R$ to the same
theory in a background with a compact dimension with radius
$\frac{\al'}{R}$ (note that when the radius is taken to be $R =
\sqrt{\alpha'}$ this T-duality is an invariance of the theory).

In the context of superstrings, there is a T-duality involving
\emph{type IIA} and \emph{type IIB} string theory (which differ in
the massless sector of their ten-dimensional spectra). Starting
from a description based on one of these theories - in a
background with a compact dimension - T-duality again implies the
existence of an alternative description, in which the momentum
modes are described as winding modes and vice versa. As before,
this alternative description is based on a theory in which the
radius of the compact dimension is the inverse of the radius of
the compact dimension in the original theory. But, contrary to the
bosonic case, this dual theory is \emph{different} than the
original one; T-duality relates type IIA string theory in a
background with a compact dimension with radius $R$ to type IIB
string theory in a background with a compact dimension with radius
$\frac{\al'}{R}$.

The T-duality between type IIA and type IIB string theory is an
example of a perturbative duality (in $g_s$), which means that the
perturbative region of one theory (i.e. the region of its
parameter space where perturbation theory is applicable) is mapped
to the perturbative region of another. This implies that (for
finite $R$) type IIA and type IIB  give useful descriptions of the
same corner of M-theory. Some other dualities in the duality web
are non-perturbative, in the sense that perturbative regions are
mapped to non-perturbative regions (like the electric/magnetic
dualities mentioned above). In these cases the domains of
applicability of the two theories on both sides of the duality are
different.

Coming back to T-duality, on open strings it affects the boundary
conditions on the endpoints. Strings that have their endpoints
fixed on a $p$-dimensional hypersurface are mapped to strings
whose endpoints live on a $(p+1)$/$(p-1)$-dimensional (depending
on whether the compact dimension involved is part of the
hypersurface or not) space.

The hypersurfaces appearing in the context of open strings are
called \emph{D-branes}. Polchinski discovered that
non-perturbatively these D-branes become dynamical \cite{Po}. This
can be understood best from the fact that their mass is
proportional to the inverse of $g_s$. The type IIA and type IIB
string theories (which are the ones mainly important for us)
contain D-branes of even and odd dimensionality respectively (note
the consistency with T-duality). These D-branes take over the role
of strings as fundamental objects in the non-perturbative regime.
The latter is consistent with the duality web, which involves for
instance a class of $SL(2, \mathbb{Z})$ selfdualities of type IIB
string theory, rotating the fundamental string and the D1-brane (a
one-dimensional D-brane, the `D-string') as fundamental objects
\cite{Sch}.

This important work on the duality web and the role of D-branes
has greatly increased the understanding of string theory and
certainly of its non-perturbative aspects. However, there remain
areas to be explored. The central question in this respect is:
What is M-theory? We know that in certain limits of the parameter
space of this theory string theories appear as the appropriate
descriptions. Away from these limits there is less clarity. What
is known, is that the heart of M-theory can be reached from type
IIA string theory by increasing its coupling constant to a value
much bigger than one. As the string coupling of IIA string theory
at low energies can be related to the radius of an extra
dimension, it indicates that the appropriate theory should not be
ten- but eleven-dimensional. Nowadays much research concentrates
on this sector and on the full non-perturbative completion of
string theory in general.

\vspace{3mm}

Another important area of research is concerned with the
construction of phenomenologically interesting models out of
string theory. As string theory necessarily lives in ten (or
eleven) dimensions this involves first of all the assumption that
six (or seven) of these parameterize a compact space. At energies
low compared to the scale associated with the size of the compact
space, four-dimensional theories then arise as appropriate
effective descriptions.

What four-dimensional model is obtained (and how much
supersymmetry is preserved) depends on the scheme chosen,
involving e.g. the type of string theory and the properties of the
six-dimensional compact space. As there are many options to choose
from, many different four-dimensional models are obtainable.
However, to get a four-dimensional model that can be related to
our universe turns out to be difficult.

We point out two of the characteristics of our universe that are
non-trivial to reproduce from string theory in particular. First
of all there is the fact that no massless scalar fields are
observed in nature, while string theory typically gives rise to
massless moduli, associated with the geometry of the compact
space. Secondly, the positive cosmological constant that our
universe seems to have is not easy to obtain from string theory
compactifications \cite{dWSH,MN}.

The need to generate masses for the moduli of the internal
manifold is usually referred to as the problem of \emph{moduli
stabilization}. It requires a potential for these fields in the
four-dimensional theory. Such a potential can be provided by
background \emph{fluxes} (non-vanishing background values of the
ten-dimensional fields) in the internal manifold.

A positive cosmological constant, it seems, can only be found in
metastable string theory vacua (which are vacua that are unstable
under tunneling effects, but with lifetimes large on cosmological
scales) and requires the inclusion of non-perturbative effects in
the string coupling constant $g_s$. The latter provides another
motivation for investigations of non-perturbative $g_s$ physics.

`KKLT' sketches a type IIB scenario in which it would be possible
to stabilize all the four-dimensional moduli in a - metastable -
de Sitter vacuum \cite{KKLT}. This scenario involves as
ingredients background fluxes, three-dimensional (anti-)D-branes
transversal to the internal manifold and so-called instanton
effects (to which we come back later on in this introduction).
These days a lot of research focusses on realizing scenarios of
this type in practice, either in type IIB or another type of
string theory.

\vspace{3mm}

Above we gave a brief introduction to string theory and treated
some relevant issues in the research concerning it. To summarize
the latter, two important (and not unrelated) categories are
investigations focussing on its non-perturbative (in $g_s$)
completion and work on finding phenomenologically interesting
solutions with all moduli stabilized and a positive cosmological
constant. Let us now come a bit more to the point: What are the
important issues in the material presented in this thesis and how
do they fit in the framework outlined above?

One of our main results is a spacetime description of instanton
solutions that can be related to D-branes and other
non-perturbative string theory objects. The other main topic
involves a construction of supersymmetric gauge theories with both
electric and magnetic (\emph{dyonic}) charges, which can be
related to string theory through compactifications with background
fluxes turned on. As we will explain in more detail later on, our
study thus fits in both categories just mentioned.

The context of our work is formed by four-dimensional $N=2$
supersymmetric models. All necessary properties and details of
models of the latter type can be found in chapter \ref{ch2}. For
now it is just important to know that these models are field
theories (with or without a gravitational coupling), symmetric
under the action of two independent supersymmetry generators. They
contain models that descend from string theory at low energies.
Shortly we will see how, but before going to that we first
elaborate a bit on why they are interesting in the first place.

An important aspect of supersymmetry is the control it gives over
a model. From a calculational point of view it is therefore
preferable to consider systems which have (some) supersymmetry.
$N=2$ is an interesting amount of supersymmetry to have in this
respect as it is just enough to have good control over
calculations, while the amount is sufficiently low to allow for
non-trivial (quantum) effects. Requiring ($N=2$) supersymmetry has
for instance proven to be useful in handling gauge theories. Of
course these are theoretical rather than phenomenological
arguments; they help to extract physics from a model, but the
model and corresponding physics are not necessarily relevant for a
description of the real world.

Nevertheless, regardless of the calculational ease it offers,
there are also bottom-up motivations for considering
supersymmetric models. One of these is the following. The energies
associated with the heaviest Standard Model particles are of the
order of $100 \, GeV$. However, the natural cutoff of the Standard
Model - the energy scale where it might lose its applicability -
is the Planck scale \footnote{Or the nearby grand unification
scale ($\sim 10^{16} GeV$) where the running couplings of the
Standard Model seem to meet.}. A priori one would expect the
masses involved to be of the latter order, but as we just saw,
their typical scale is many orders of magnitude lower. This
difference in scales is referred to as the \emph{hierarchy
problem}. It is indicative of new physics just on or above levels
reachable by present day particle accelerators (which is more or
less the mass scale of the Standard Model particles). This new
physics may correspond to a supersymmetric theory of (initially)
massless fields. Supersymmetry then needs to be spontaneously
broken at the appropriate scale such that the masses of the
Standard Model particles arise accordingly. Whether this is
realized in practice may be determined by upcoming LHC experiments
\footnote{LHC stands for \emph{Large Hadron Collider}. Currently
under construction (in Geneva), it is supposed to become the
world's highest energy particle accelerator.}. To have a scenario
like described above, $N=1$ supersymmetry suffices. In fact, since
$N>1$ models cannot accommodate chiral fermions, they are hard to
relate to realistic models. $N=2$ models are therefore not of
direct value for phenomenology. However, either they could be
useful in the exceptional scenario where $N=2$ does give rise to a
realistic non-supersymmetric theory or it might be hoped that they
are relevant toy models for $N=1$.

Besides the arguments given above, the arguably most intriguing
aspect of a supersymmetric field theory is that its local version
(with transformation parameters being free in their spacetime
dependence) automatically includes gravity. Models exhibiting
local $N=2$ supersymmetry are therefore called $N=2$ supergravity
systems.

$N=2$ supergravity theories arise as low energy effective actions
from string theory. This can be understood as follows. At energies
low compared to the Planck scale explicit knowledge of the
behavior of the massive string states - which have Planck scale
masses - is not important. So they can be integrated out, i.e. an
effective (field theory) description in terms of the massless
states suffices. When string theory is taken in a background with
ten non-compact dimensions, supersymmetry completely fixes the
form of the action at two-derivative level. In case of type II
strings this leads to the maximally supersymmetric type IIA and
type IIB supergravity. However, we want to end up with only four
non-compact dimensions. As said above, this can be achieved by
considering the remaining six dimensions to be a (small) compact
space. The type of six-dimensional compact space then determines
how many supersymmetry is left in four dimensions. In case of type
II strings, to get a four-dimensional $N=2$ supergravity model the
compact space should be a so-called (compact) \emph{Calabi-Yau}
(CY) manifold \footnote{The same class of manifolds is found when
demanding \emph{heterotic} strings to descend to
(phenomenologically interesting) four-dimensional models with
$N=1$ supersymmetry. $N=1$ models can also be obtained from
compactifying type II strings on Calabi-Yau orientifolds. These
might be considered as arguments for hoping that low energy $N=2$
type II theories are indeed relevant toy models.}.

As we will see in chapter \ref{ch2}, $N=2$ supersymmetry has two
important representations: The vector and the hypermultiplet. The
bosonic part of a vector multiplet consists of a vector gauge
field and a complex scalar, while its hypermultiplet counterpart
has four real scalars. The scalars of the theory parameterize a
$(2n + 4(m+1))$-dimensional manifold (where $n$ and $m+1$ are the
numbers of vector and hypermultiplets). This manifold is of a
quite restricted type, due to the constraints of supersymmetry. To
be precise, the $2n$-dimensional manifold of the vector multiplet
scalars should be \emph{special K\"{a}hler} (SK)
\cite{deWit:1984pk}, while the $4(m+1)$-dimensional space
parameterized by the scalars of the hypermultiplet sector
necessarily is \emph{quaternionic-K\"{a}hler} (QK) \cite{BW}. For
more details we refer to chapter \ref{ch2}.

The number of multiplets emerging after compactifying type II
strings on a CY is determined by the topological properties of the
latter. Details of this can be found in chapter \ref{ch3}. For now
the only relevant issue is the relation between the numbers of
multiplets following from IIA and IIB compactifications. In case
IIA gives $n$ vector multiplets and $m+1$ hypermultiplets, IIB
compactified on the same CY yields $n'=m$ vector and $m'+1=n+1$
hypermultiplets.

To obtain the precise form of the low energy effective $N=2$
supergravity theory corresponding to type II string theory
compactified on a CY \footnote{When the numbers of vector and
hypermultiplets are equal, the four-dimensional effective actions
of type IIA and type IIB are the same. This is the result of
\emph{mirror symmetry} between the corresponding CY's.} would in
principle require a full string theory calculation, which at
present seems to be far too complicated. Fortunately one can get
quite far without performing such a calculation, by doing a
supergravity analysis.

Let us consider this supergravity analysis in more detail. First
recall that the low energy effective action of type II
superstrings in ten uncompact dimensions (with maximal
supersymmetry) is known. This allows a compactification to be
performed at supergravity level. It involves an expansion of the
ten-dimensional fields in eigenfunctions of the CY wave operator.
Keeping only the zero-modes, a classical four-dimensional $N=2$
effective action in terms of massless fields is obtained (see for
instance \cite{BCF}, where this supergravity compactification was
explicitly performed in the context of type IIA strings).
Importantly, the dilaton (whose vacuum expectation value, we
recall, is the string coupling constant $g_s$) one always find
back in a hypermultiplet.

All scalar fields together make up a $SK \times QK$ manifold in
the four-dimensional Lagrangian. The precise form of the scalar
geometry fixes the rest of the action as well. There only is an
issue concerning isometries, which may be present in the geometry
of the scalar manifolds. In compactifications where the background
values of the fields other than the ten-dimensional metric are put
to zero, they correspond to rigid invariances of the total action.
However, when appropriate background fluxes are turned on, these
invariances are local, i.e. the isometries are gauged. Gauged
isometries in a $N=2$ supersymmetric theory imply the presence of
a scalar potential, which, we recall, is important for moduli
stabilization.

The classical version of the scalar geometry as obtained from the
CY compactification considered above receives corrections. These
are of two types, ($\al'$) corrections associated with quantum
effects on the worldsheet and corrections corresponding to
spacetime quantum behavior. Here we only focus on the latter and
just mention that the former are under some - and in a few cases
complete - control \cite{COGP,HKSY}.

The spacetime quantum effects appear as $g_s$ corrections (recall
that perturbatively $g_s$ controls the number of loops in a
stringy Feynman diagram). As the dilaton lives in a
hypermultiplet, the quaternion-K\"{a}hler manifold of the type II
theories is modified in this way. The perturbative $g_s$
corrections to the quaternion-K\"{a}hler space are fixed using
some general knowledge about their properties and the constraints
imposed by supersymmetry (see \cite{RSV} and references therein).
The non-perturbative $g_s$ corrections, however, are not yet
found, although some partial results were obtained in
\cite{ASV,RRSTV}.

Microscopically these non-perturbative $g_s$ effects correspond to
Euclidean $p$-branes wrapping $(p+1)$-dimensional cycles in the
CY. From a four-dimensional perspective these branes are points in
Euclidean space and in the classical supergravity action they
appear as \emph{instanton} solutions.

Instantons are by definition solutions to Euclidean equations of
motion with finite action. They are for example known from
Yang-Mills theory where they are associated with tunneling effects
between different classical vacua. The value of the action
evaluated on instantons is typically of the form $S_{inst} =
|q|/g$ (or with higher negative powers of $g$), where $g$ is the
coupling constant of the theory and $q$ is some charge. Hence they
give rise to non-perturbative $(e^{ - |q|/g})$ contributions to
the path integral.

\vspace{5mm}

In chapter \ref{ch3} of this thesis instanton solutions are
described that correspond to CY wrapping branes. More precisely,
two classes of such solutions are determined in the general
hypermultiplet model arising from type II strings on a CY with
background fluxes turned to zero.

The first class is derived from known black hole solutions in the
vector multiplet sector. In doing this the \emph{c-map}
\cite{CFG,FS} is exploited, which involves a dimensional reduction
of the ($n$) ungauged vector multiplet sector of type II CY
compactifications. The resulting three-dimensional action can then
be uplifted to the ($n+1$) hypermultiplet sector of an $N=2$
supergravity theory of the same type. Note from what we said
earlier that this basically is a map from type IIA to type IIB or
vice versa. In fact, the underlying mechanism is the T-duality we
described before, which says that type IIA compactified on a CY
times a circle with radius $R$ is equivalent to type IIB
compactified on the same CY times a circle with radius
$\frac{\alpha'}{R}$. The solutions found using the c-map on the
black hole solutions are the D-brane instantons, which correspond
to D-branes wrapping cycles in the CY.

The second class of instantons are obtained using a
``Bogomol'nyi-bound-like" method, similar to the one described in
\cite{TV1}. These solutions arise from so-called
\emph{NS-fivebranes} (other higher-dimensional objects in string
theory) wrapping the entire CY.

For both classes of instantons the value of the action is
determined. This is important with respect to the corresponding
deformation of the scalar manifold metric, since these corrections
involve exponential factors with (minus) the instanton action
appearing in the exponent.

Obviously this work concentrates on obtaining a better picture of
non-perturbative string theory. Furthermore, as we already
mentioned in the context of the KKLT scenario, understanding
instantons and their effects is also important in relation to the
construction of models with phenomenologically interesting
aspects, such as a stabilization of moduli and a positive
cosmological constant.

The latter can be understood from the fact that the scalar
potential in a gauged $N=2$ supersymmetric model depends on the
geometrical properties of the scalar sigma manifold. With respect
to the CY compactifications of type II strings it therefore
matters if we take the sigma manifold with or without quantum
corrections. \cite{DSTV} considered the case of one (the
universal) hypermultiplet and the associated quaternion-K\"{a}hler
manifold with $g_s$ corrections corresponding to (membrane)
instantons included \footnote{A model with just one hypermultiplet
can be realized in a geometric compactification of type IIA
only.}. As it turns out, the corresponding scalar potential
allows, contrary to its analog without $g_s$ corrections,
metastable de Sitter vacua with the moduli in the universal
hypermultiplet stabilized. As said above, we analyzed the theory
and corresponding instanton solutions that type II strings on an
arbitrary CY give rise to. It would be interesting to see what
scalar potential and corresponding vacua are generated from the
associated higher-dimensional quaternion-K\"{a}hler manifold with
the more general instanton corrections taken into account.

\vspace{5mm}

Let us then focus on the vector multiplets. An important feature
of ungauged $N=2$ supersymmetric actions based on $n$ vector
supermultiplets is the existence of the $Sp(2n, \mathbb{R})$ group
of electric/magnetic duality transformations. Under these duality
transformations the Lagrangian changes. Different Lagrangians
related by a duality transformation belong to the same equivalence
class, meaning that their sets of equations formed by equations of
motion and Bianchi identities are equivalent. It may happen that
the Lagrangian does not change under such a duality transformation
(possibly up to redefinitions of the other fields), in which case
one is dealing with an invariance of the theory. To appreciate
this, it is important to note that the Lagrangian does not
transform as a function under the duality transformation (although
this may be the case for a restricted subgroup of the full
invariance group). For a detailed treatment of this we refer to
chapters \ref{ch1} and \ref{ch2}.

Electric/magnetic duality transformations are realized by a
constant rotation of the electric and magnetic field strengths.
The new field strengths can then be solved in terms of new (dual)
vector fields, which are not locally related to the original
vector fields by a local field redefinition. The fact that
electric/magnetic duality acts on the field strengths rather than
the gauge fields is the reason why charges have to be absent when
applying electric/magnetic duality, because they couple to the
gauge fields. However, this does not preclude the possibility that
one can describe a gauge theory with electric charges from a dual
point of view.

One may wonder what is gained by using such a description. We
mention two important advantages. First of all, when one is
interested in gauging a certain subgroup of the rigid invariance
group, the standard procedure is to first convert the theory to a
suitable electric/magnetic duality frame, in which all the
potential charges will appear as electric.  This is a cumbersome
procedure in general and it would be convenient if it could be
avoided. Secondly, in the context of string theory the charges
correspond to turning on fluxes in the internal manifold that
emerges in a compactification to four spacetime dimensions. These
fluxes are associated to background values of antisymmetric tensor
fields on non-trivial cycles of the internal manifold, to
background quantities associated with the geometry of the manifold
itself or even to quantities associated with manifolds without a
definite geometry (see \cite{Tr} and references therein). In all
known cases, the fluxes give rise to parameters in the
four-dimensional theory that correspond to gauge charges
\footnote{For gauged supergravity to be a reliable low
  energy description, the fluxes should be chosen such that the
  backreaction on the internal geometry is negligible.}.  When
appropriate fluxes in the internal manifold are turned on, both
electric and magnetic charges show up in four dimensions, giving
rise to supergravity theories that are not of the canonical type
(see for instance \cite{LM}). Moreover, the fluxes are subject to
certain transformations defined in the internal manifold, which
manifest themselves as electric/magnetic duality transformations
in the four-dimensional theory. It is obviously advantageous to
keep such symmetry aspects manifest where possible.

Recently, in a general - non-supersymmetric - context, a formalism
was developed, which indeed allows the introduction of both
electric and magnetic charges \cite{dWST} \footnote{The charges
involved are mutually local, which means that an electric/magnetic
frame exists in which they are all of the electric type.}. It
involves an extra set of magnetic gauge fields which couple to the
magnetic charges, accompanied by a set of antisymmetric tensor
fields. These new fields come with additional gauge
transformations and, as a consequence, the total number of
physical degrees of freedom remains unaltered. The electric and
magnetic charges are contained in a so-called \emph{embedding
tensor}. This embedding tensor is treated as a spurionic quantity,
which implies that it transforms non-trivially under the
electric/magnetic dualities.  In this way gauge theories are
obtained that still contain the duality structure of the ungauged
theories.

In the second part of this thesis we apply this formalism to $N=2$
supersymmetric theories. We derive the supersymmetric Lagrangian
and transformation rules for gaugings that involve both electric
and magnetic charges (which we call dyonic gaugings). On the
scalar fields the gauge symmetries are generated by isometries of
the scalar sigma manifold of both the vector and the
hypermultiplet sector. As one of the results, one finds a scalar
potential that is independent of the electric/magnetic duality
frame. In particular, in a subclass of our models, the potentials
of \cite{LM} and \cite{Mi} are reproduced.

Our work provides a unified description of whole classes of string
theory models. It should facilitate the construction of
phenomenologically interesting models, which may, for instance,
lead to groundstates with a positive cosmological constant and to
stabilization of the moduli.

\vspace{5mm}

This thesis is organized as follows. The first two chapters are
meant as an introduction to the later chapters. Chapter \ref{ch1}
deals with electric/magnetic duality, studied in a wider context
than $N=2$ supersymmetric theories. In this chapter we also derive
a new result concerning symmetries of Lagrangians that are subject
to electric/magnetic duality transformations. The last section of
the chapter is about related duality transformations between
scalars and tensors. In chapter \ref{ch2} we treat the relevant
aspects of $N=2$ supersymmetry. We introduce the so-called
superconformal method, which is useful for the construction of the
supergravity theory. The different multiplets (Weyl, vector and
hyper) are considered and the c-map mentioned above is performed
explicitly. We also give some new insights associated with
electric/magnetic duality.

Then in chapter \ref{ch3} we derive the instantons present in the
hypermultiplet sector of $N=2$ supergravity. First we treat the
relatively simple case of the universal hypermultiplet, after
which the D-brane and NS-fivebrane instantons of the general
hypermultiplet theory - together with their action - are analyzed.
We work in a formulation of the hypermultiplets where a set of
scalars is dualized to tensors; the tensor multiplet formulation.

In chapter \ref{ch4} we construct $N=2$ supersymmetric gauge
theories with both electric and magnetic charges. These gaugings
are performed in the vector as well as the hypermultiplet sector
of the theory. We derive the supersymmetric action and the
supersymmetry transformation rules.

Several technical details of our work can be found in one of the
Appendices. We refer to these when necessary.

\newpage

\thispagestyle{empty}
}

\makeatletter
\def\thickhrulefill{\leavevmode \leaders \hrule height 1ex \hfill \kern \z@}
\def\@makechapterhead#1{%
  \vspace*{10\p@}%
  {\parindent \z@ \centering \reset@font
        {\fontsize{35}{15.6pt}\selectfont \bfseries \thechapter \fontsize{13}{15.6pt}\selectfont}
        \par\nobreak
        \vspace*{15\p@}%
        \interlinepenalty\@M
        \vspace*{10\p@}%
        \Huge \bfseries #1\par\nobreak
        \par
        \vspace*{10\p@}%
    \vskip 60\p@
  }}
\def\@makeschapterhead#1{%
  \vspace*{10\p@}%
  {\parindent \z@ \centering \reset@font
        {\fontsize{35}{15.6pt}\selectfont \bfseries \vphantom{\thechapter} \fontsize{13}{15.6pt}\selectfont}
        \par\nobreak
        \vspace*{15\p@}%
        \interlinepenalty\@M
        \vspace*{10\p@}%
        \Huge \bfseries #1\par\nobreak
        \par
        \vspace*{10\p@}%
    \vskip 60\p@
  }}

\renewcommand{\chaptermark}[1]{\markboth{\thechapter\ #1}{}}
\fancyhf{} \fancyhead[LE, RO]{\thepage}
\fancyhead[CO]{\slshape\rightmark}
\fancyhead[CE]{\slshape\leftmark}
\renewcommand{\headrulewidth}{0pt}
\renewcommand{\footrulewidth}{0pt}
\addtolength{\headheight}{0pt} \mainmatter
\chapter{Electric/magnetic duality}\lb{ch1}

As is well-known, the eight equations that form the basis of all
electromagnetic phenomena we observe in our universe are
\begin{eqnarray}\label{max}
\vec{\nabla} \cdot \vec{B} = 0\ , \quad - \vec{\nabla} \times
\vec{E} = \frac{\partial
\vec{B}}{\partial t}\ ,\nn\\
\vec{\nabla} \cdot \vec{D} =  \rho_e \ , \quad \vec{\nabla} \times
\vec{H} = \frac{\partial \vec{D}}{\partial t} +  \vec{J}_e\ ,
\end{eqnarray}
as derived by J.C. Maxwell in 1864 \cite{M}. Here $\vec{E}$ and
$\vec{H}$ are the electric and magnetic fields, $\vec{D}$ is the
electric displacement and $\vec{B}$ is the magnetic induction.
$\vec{D}$ and $\vec{B}$ are related to $\vec{E}$ and $\vec{H}$
through the polarization $\vec{P}$ and the magnetization $\vec{M}$
of a material medium, via
\begin{eqnarray}\label{maxt}
\vec{D} = \vec{E} +  \vec{P}\ ,\quad  \vec{B} = \vec{H} +
\vec{M}\ .
\end{eqnarray}
$\rho_e$ and $\vec{J}_e$ are the electric charge and current
density. We employ Heaviside-Lorentz units and put $c = 1$ (as we
will do later on with $\hbar$ as well).

Note the striking similarity between the way the magnetic
($\vec{B}$ and $\vec{H}$) and electric ($\vec{D}$ and $\vec{E}$)
fields enter these equations. The only difference lies in the
absence of sources in the equations of the first line, which
corresponds to the fact that we do not observe magnetic monopoles
in nature. Ignoring the latter fact and including a magnetic
charge and current density nonetheless, we see that (\ref{max})
remains equivalent under the transformations
\begin{eqnarray}\label{emdu1}
\( \begin{array}{c} \vec{E}\\ \vec{H} \end{array} \)
\longrightarrow \(
\begin{array}{c} \tilde{\vec{E}}\\ \tilde{\vec{H}} \end{array} \) = \(
\begin{array}{cc} \cos \alpha & \sin \alpha\\ - \sin \alpha & \cos \alpha \end{array}
\) \( \begin{array}{c} \vec{E}\\ \vec{H} \end{array} \)\ ,\nn\\
\end{eqnarray}
accompanied by
\begin{eqnarray}\label{emdu1b}
\( \begin{array}{c} \vec{P}\\ \vec{M} \end{array} \)
\longrightarrow \(
\begin{array}{c} \tilde{\vec{P}}\\ \tilde{\vec{M}} \end{array} \) = \(
\begin{array}{cc} \cos \alpha & \sin \alpha\\ - \sin \alpha & \cos \alpha \end{array}
\) \( \begin{array}{c} \vec{P}\\ \vec{M} \end{array} \)\ ,
\end{eqnarray}
and \footnote{Note that we do not take charge quantization into
account. We come back to this point in section \ref{monol}.}
\begin{eqnarray}\label{emdus}
\( \begin{array}{c} \rho_e\\ \rho_m \end{array} \) \longrightarrow
\( \begin{array}{c} \tilde{\rho}_e\\ \tilde{\rho}_m \end{array} \)
= \( \begin{array}{cc} \cos \alpha & \sin \alpha\\ - \sin \alpha &
\cos \alpha \end{array} \) \( \begin{array}{c} \rho_e\\ \rho_m \end{array} \)\ ,\nn\\
\( \begin{array}{c} \vec{J}_e\\ \vec{J}_m \end{array} \)
\longrightarrow \(
\begin{array}{c} \tilde{\vec{J}}_e\\ \tilde{\vec{J}}_m \end{array} \) = \(
\begin{array}{cc} \cos \alpha & \sin \alpha\\ - \sin \alpha & \cos \alpha \end{array}
\) \( \begin{array}{c} \vec{J}_e\\ \vec{J}_m \end{array} \)\ .
\end{eqnarray}
That (\ref{emdu1}) - (\ref{emdus}) are equivalence transformations
of (\ref{max}) and (\ref{maxt}) (with magnetic sources included)
means by definition that they leave its space of solutions
invariant. However, although, using $\tilde{\vec{D}} =
\tilde{\vec{E}} +  \tilde{\vec{P}}$ and $\tilde{\vec{B}} =
\tilde{\vec{H}} +  \tilde{\vec{M}}$, the new set of equations can
be written in the same form as (\ref{max}) and (\ref{maxt}),
(\ref{emdu1}) - (\ref{emdus}) are generically \emph{not}
invariances of (\ref{max}) and (\ref{maxt}). This is due to the
fact that there is input needed, contained in $(\vec{P},
\vec{M})$, $(\rho_e, \rho_m)$ and $(J_e, J_m)$, to solve these
equations. This input is of fixed value, but forced to transform
non-trivially via (\ref{emdu1b}) and (\ref{emdus}). Only when
$(\rho_e, \rho_m) = 0$ and $(J_e, J_m) =0 $ and the polarization
and magnetization of the medium are of the form $(\vec{P},
\vec{M}) \propto (\vec{E}, \vec{H})$ - which implies that the
appropriate transformation of $(\vec{P}, \vec{M})$ is induced by
the transformation of $(\vec{E}, \vec{H})$ - the transformations
above are invariances of (\ref{max}) and (\ref{maxt}).

\vspace{5mm}

Notwithstanding the latter fact, the remarkable equivalence of
Maxwell's equations under (\ref{emdu1}) - (\ref{emdus}) seems to
indicate that not only electric and magnetic phenomena are
intimately related, but that a distinction between phenomena in
this way has no intrinsic meaning. Whether a physical phenomenon
is electric or magnetic just depends on which description is used.
The transformation between the different descriptions is what is
called electric/magnetic duality.

In the rest of this chapter we explore the status of
electric/magnetic duality at a more fundamental level and in a
more general context. We have to stress that as electric/magnetic
duality is so big and diverse a subject, on which so much work is
done, it is by far not possible to cover all aspects and certainly
not in as much detail as they deserve. We mainly focus on those
issues that are important for us in later chapters. They will be
treated in some detail, embedded in a fairly qualitative
discussion of the subject as a whole.

\section{Maxwell theory}\label{maxwellt}

Let us consider the relativistically covariant version of
(\ref{max}). The electric and magnetic fields then combine into
the covariant tensor $F^{\mu \nu}$, which takes the form
\begin{eqnarray}\label{fineenb}
F^{\mu \nu} & = & \( \begin{array}{cccc} 0 & -E_x & -E_y & -E_z\\
E_x & 0 & -B_z & B_y\\ E_y & B_z & 0 & -B_x\\ E_z & -B_y & B_x & 0
\end{array} \)\ .
\end{eqnarray}
The first line of (source-free) Maxwell equations (\ref{max})
becomes $\p_{[\mu} F_{\nu \rho]} =0$, which is a Bianchi identity
when we demand (locally) $F_{\mu \nu} = 2 \p_{[\mu} A_{\nu]}$.
$A_{\mu}$ is called the gauge potential. This allows the second
line of Maxwell's equations to be derived from the action
\footnote{Despite the factor $i$ appearing in the prefactor, the
second term in (\ref{ac2}) is real. See Appendix \ref{notcon} for
our conventions.}
\begin{equation}\label{ac2}
\S = - \frac{\pi}{g^2} \int d^4x \; F_{\mu \nu} F^{\mu \nu} - i
\frac{\theta}{16 \pi} \int d^4x \; \vep^{\mu \nu \rho \sigma}
F_{\mu \nu} F_{\rho \sigma}\ ,
\end{equation}
using the variational principle. $g$ is the coupling constant of
the theory and $\theta$ is called the theta-angle. The second line
of Maxwell's equations thus becomes a field equation. It can be
formulated as $\p_{[\mu} G_{\nu \rho]} = 0$, where we have defined
\begin{equation}\label{defg2}
G_{\mu \nu} = i \vep_{\mu \nu \rho \sigma} \frac{\delta \L}{\delta
F_{\rho \sigma}}\ .
\end{equation}
Note that for these classical considerations there does not seem
to be a reason to include the second term in (\ref{ac2}), as it is
a total divergence. Furthermore, the coupling constant $g$ can be
scaled away. However, shortly we will see why it is useful to
include them both.

It is convenient to rewrite (\ref{ac2}) as
\begin{equation}\label{ac2s}
\S =  - \vier i \int d^4x \; \left[ \bar{\tau} F_{\mu \nu}^+ F^{+
\mu \nu} - \tau F_{\mu \nu}^- F^{- \mu \nu} \right]\ ,
\end{equation}
where the complex parameter $\tau$ is given by
\begin{equation}\label{taum}
\tau = \frac{4 \pi i}{g^2} + \frac{\theta}{2 \pi}\ .
\end{equation}
$F^{\pm}_{\mu \nu}$ is the (anti-)selfdual field strength, defined
as $F^{\pm}_{\mu \nu} \equiv \half (F_{\mu \nu} \pm \half
\vep_{\mu \nu \rho \sigma} F^{\rho \sigma})$ (see Appendix
\ref{notcon}).

In terms of the (anti-)selfdual parts of $F_{\mu \nu}$ and $G_{\mu
\nu}$, the set of source-free Maxwell equations takes the form
\begin{eqnarray}\label{eq1}
\p_{\mu} \( \begin{array}{c} F^{+ \mu \nu} - F^{- \mu \nu}\\
G^{+ \mu \nu} - G^{- \mu \nu}
\end{array} \) & = & 0 \ .
\end{eqnarray}
Obviously this set of equations remains equivalent under
\begin{eqnarray}\label{emdu3}
\( \begin{array}{c} F_{\mu \nu}^{\pm}\\ G_{\mu \nu}^{\pm}
\end{array} \)
\longrightarrow \( \begin{array}{c} \tilde{F}_{\mu \nu}^{\pm}\\
\tilde{G}_{\mu \nu}^{\pm}
\end{array} \) = \( \begin{array}{cc} U & Z\\ W & V \end{array} \)
\( \begin{array}{c} F_{\mu \nu}^{\pm}\\ G_{\mu \nu}^{\pm}
\end{array} \)\ ,
\end{eqnarray}
where $U$, $Z$, $W$ and $V$ are real numbers satisfying $UV - WZ
=1$, i.e. form a matrix with determinant equal to one (this last
condition excludes transformations that act as constant rescalings
of the field strength and Lagrangian). Two-by-two matrices with
determinant equal to one are elements of $Sp(2, \mathbb{R}) \sim
SL(2, \mathbb{R})$. Note that (\ref{emdu1}) is included in
(\ref{emdu3}) as the  maximal compact subgroup of $Sp(2,
\mathbb{R})$ ($U=V=\cos \al$ and $Z=-W=\sin \al$).

The effect of (\ref{emdu3}) is a rotation of the Bianchi identity
of $F_{\mu \nu}$ and the field equation of $A_\mu{}$, as following
from (\ref{ac2}). In other words, the equation $\p_{\mu} (\tF^{+
\mu \nu} - \tF^{- \mu \nu}) =0$ is interpreted as the new Bianchi
identity, while $\p_{\mu} (\tG^{+ \mu \nu} - \tG^{- \mu \nu})  =0$
is the new field equation. From $\p_{\mu} (\tF^{+ \mu \nu} -
\tF^{- \mu \nu}) =0$ being a Bianchi identity it follows that
(locally) $\tF_{\mu \nu} = 2 \p_{[\mu} \tA_{\nu]}$. $\tA_{\mu}$ is
the new gauge potential, which is \emph{not} locally related to
the old gauge potential $A_{\mu}$.

Since the transformed system of equations includes a new Bianchi
identity and a new field equation that are not separately
equivalent to their untransformed versions, the Lagrangian
associated with the new system is non-trivially related to the
original one. The expression for the transformed Lagrangian,
$\tL$, follows from $\tilde{G}_{\mu \nu} = i \vep_{\mu \nu \rho
\sigma} \frac{\delta \tilde{\L}}{\delta \tilde{F}_{\rho \sigma}}$.
It can be written in the same form as (\ref{ac2s}),
\begin{equation}\label{ac2sb}
\tS =  - \vier i \int d^4x \; \bar{\tilde{\tau}} \tF^+_{\mu \nu}
\tF^{+ \mu \nu} + \mathrm{h.c.}\ ,
\end{equation}
when we transform $\tau$ as
\begin{eqnarray}\label{trtau}
\tau \longrightarrow \tilde{\tau} = \frac{W +V \tau}{U+ Z \tau}\ .
\end{eqnarray}
Observe that the transformations (\ref{trtau}) contain an
inversion of the coupling constant as a special case.

Despite the fact that we have written (\ref{ac2sb}) in the same
form as (\ref{ac2s}), (\ref{emdu3}) is \emph{not} a symmetry of
the action. A symmetry requires the Lagrangian to transform as an
invariant function $\tL (\tF (F)) = \L (F) = \L (\tF (F))$.
However under (\ref{emdu3}) the Lagrangian does not transform as a
function, $\tL (\tF) \neq \L (F)$, nor is it invariant, $\tL (\tF)
\neq \L (\tF)$. The latter refers to the fact that the value of
the input parameter $\tau$, which plays a similar role as
$\vec{P}$ and $\vec{M}$ in (\ref{max}) and (\ref{maxt}), is
different in (\ref{ac2sb}) as compared to (\ref{ac2s}).

Note that in case $\tau = \pm i$, which is the analog of $\vec{P}
= \vec{M} = 0$ in (\ref{max}) and (\ref{maxt}), the Lagrangian is
invariant under the maximal compact subgroup of $Sp(2,
\mathbb{R})$. This is consistent with the fact that (\ref{emdu1})
is an invariance of (\ref{max}) and (\ref{maxt}) in case $\vec{P}
= \vec{M} = 0$ (and $(\rho_e, \rho_m) = (J_e, J_m) =0 $).

 \vspace{3mm}

Next we include sources in the equations (\ref{eq1}),
\begin{eqnarray}\label{eq1so}
\p_{\mu} \( \begin{array}{c} F^{+ \mu \nu} - F^{- \mu \nu}\\
G^{+ \mu \nu} - G^{- \mu \nu}
\end{array} \) & = &  \( \begin{array}{c} J_e^\nu\\
J_m^\nu \end{array} \)  \ ,
\end{eqnarray}
where $(J_e^\nu, J_m^\nu)$ is the vector formed by the covariant
electric and magnetic currents,
\begin{equation}
J_e^\nu = (\rho_e, \vec{J}_e)\ ,\quad J_m^\nu = (\rho_m,
\vec{J}_m)\ .
\end{equation}
Obviously, to preserve electric/magnetic duality, $(J_e^\nu,
J_m^\nu)$ should transform as a symplectic vector.


\vspace{3mm}

So electric/magnetic equivalences are found of the classical set
of Bianchi identities and equations of motion (\ref{eq1so}). This
raises the question: What about full quantum theories, do they
exhibit electric/magnetic duality as well? In the next section we
briefly turn to this issue, basing ourselves on \cite{O} and
\cite{dV}. For a more extended review we refer to these papers.

\section{The Montonen-Olive conjecture}\label{monol}

The electric charges of (\ref{eq1so}) would appear dynamically
when, in addition to the $U(1)$ gauge field, there are charged
elementary fields contained in the model. Magnetic charges, on the
other hand, correspond to (magnetically charged) \emph{solitons}.

Solitons are classical, localized, finite energy solutions, which
typically travel undistorted in space with a uniform velocity and
can therefore be seen as (classical) particles. They correspond to
local minima of the potential. Often solitons are characterized by
a topological index, which implies that they are stable.

Magnetically charged solitons - magnetic monopoles - were first
found by 't Hooft and Polyakov in the Georgi-Glashow model. The
latter is an $SU(2)$ gauge theory, broken to a $U(1)$ subgroup by
a Higgs mechanism \cite{H,P}. The two massive vector fields
emerging due to the spontaneous symmetry breaking are charged
under the unbroken $U(1)$, with charges $\pm q_0$.

Later on, the Georgi-Glashow model turned out to contain dyons as
well \cite{JZ}.

The mass of a gauge particle of the Georgi-Glashow model is
\begin{equation}\label{mel}
M (q_0, 0) = a |q_0|\ ,
\end{equation}
where $a$ is the vacuum expectation value of the (triplet of)
Higgs scalar fields and $q_0$ is the (electric) charge of the
gauge particle. On the other hand, the mass of a (classical) dyon
is bounded from below by
\begin{equation}\label{bound}
M (q, p) \geq a \sqrt{q^2 + p^2}\ ,
\end{equation}
with $q$ and $p$ the electric and magnetic charge of the dyon.
(\ref{bound}) is called the ``Bogomol'nyi bound" \cite{B}. So, for
dyons saturating the Bogomol'nyi bound, which is realized in the
so-called BPS - Bogomol'nyi, Prasad-Sommerfield - limit \cite{PS},
(\ref{bound}) becomes
\begin{equation}\label{bounds}
M (q, p) = a \sqrt{q^2 + p^2}\ .
\end{equation}
Remarkably, the mass formula (\ref{bounds}) is universal as it
also applies to the gauge particles. Therefore the way a particle
emerges (as the excitation of an elementary field or as a soliton)
is irrelevant when computing its mass from its charges.

Let us now consider the corresponding quantum theory. First of
all, we note that a semi-classical quantization around a local
minimum of the potential associated with a stable soliton, can be
performed in the same way as around the absolute minimum of the
theory. The lowest lying state in the spectrum this gives rise to
can be identified with the ground state of the soliton. For a
review on this subject we refer to \cite{R}. The charges of the
quantum states then make up an integer lattice in the plane formed
by electric and magnetic charges. Ignoring dyons (to which we come
back shortly), the single particle states are associated to five
points of the lattice. The Higgs field and the $U(1)$ gauge field
are chargeless and so correspond to the origin, $(0,0)$. The gauge
particles have electric charges $\pm q_0$. They correspond to the
points $(1,0)$ and $(-1,0)$. The soliton states come with magnetic
charges $\pm p_0$ and so are associated to $(0,1)$ and $(0,-1)$
(all in units of $q_0$ and $p_0$). Rotations over an angle of
$\frac{\pi}{2}$ just rearrange these points. Furthermore, assuming
the Bogomol'nyi bound is valid for quantum states as well, the
masses involved in the spectrum remain the same.

The above led Montonen and Olive (1977) to the conjecture that
there exists a dual or magnetic formulation of the theory under
consideration, which is of the same form, but in which the
elementary field excitations of the electric formulation should
appear as solitons and vice versa. Furthermore, in view of the
Dirac condition \cite{D}
\begin{eqnarray}\label{Dirac}
q^1 p^2 = 2 \pi n\ ,\quad n \in \mathbb{Z}\ ,
\end{eqnarray}
(where $q^1$ is the electric charge of a purely electrically
charged particle and $p^2$ is the magnetic charge of a magnetic
monopole) they suggested that the coupling constant of the dual
theory is the inverse of the electric coupling constant
\footnote{To be precise, (\ref{Dirac}) implies that the ``magnetic
fine structure constant" should be $\frac{p_0^2 c}{4 \pi \hbar} =
\frac{n_0^2}{4 \al}$, where $\alpha = \frac{q_0^2}{4 \pi \hbar c}$
is the electric fine structure constant and $n_0$ is an integer
depending on the theory under consideration \cite{J} (here we have
temporarily reinstalled $\hbar$ and $c$).}.

However, this raises some questions. First of all, can the
assumption that the Bogomol'nyi bound is preserved after
quantization be justified? Secondly, the electrically charged
states have unit spin. For the duality to hold this should be the
same for the soliton states. How is this realized? And thirdly,
the conjecture does not take dyons into account. How do the
corresponding quantum states fit in the duality scheme?

As we explain below, these questions can all be answered in a
satisfactory way when the Georgi-Glashow model is embedded in
$N=4$ supersymmetric Yang-Mills theory (SYM) with $SU(2)$ gauge
group. Moreover, with the dyons taken into account the duality
group becomes $SL(2, \mathbb{Z}) \sim Sp(2, \mathbb{Z})$.

In $N=4$ SYM with $SU(2)$ gauge group the Bogomol'nyi bound is a
consequence of the supersymmetry algebra and is therefore
presumably quantum exact \cite{WO,Os}. When $M(q,p) = a \sqrt{q^2
+p^2}$ the structure of the supersymmetry algebra is such that the
corresponding states fill out the massive version of a so-called
short multiplet. As there is only one such multiplet in $N=4$ SYM,
the multiplets filled out by the electrically charged states
corresponding to elementary field excitations and the multiplets
associated with soliton states are necessarily isomorphic  (this
was made explicit in \cite{Os}). In particular, both have spin-one
states as the states with highest spin. In addition to this, $N=4$
SYM is quantum conformally invariant. Amongst other things this
implies that the coupling constant does not renormalize, which in
turn makes the question less pressing whether the Dirac
quantization condition should be applied to the bare or
renormalized coupling constant (as both are the same) \cite{Ro}.

When dyons are taken into account, we need to know what the
allowed values of their charges are. This follows from the
Schwinger-Zwanziger quantization condition \cite{Sc,Z}, which is
the generalization of the Dirac condition (\ref{Dirac}) and reads
as
\begin{equation}\lb{dsz}
q^1 p^2 - p^1 q^2 = 2 \pi n\ ,\quad n \in \mathbb{Z}\ .
\end{equation}
Here $(q^1, p^1)$ and $(q^2, p^2)$ are the electric and magnetic
charges of two dyons. Furthermore, we assume that the charges are
conserved and that the TCP-theorem is valid, which implies that
the set of allowed values of the charges must be closed under both
addition and reversal of sign. Then the allowed values of the
charges span a lattice. More precisely, they satisfy
\begin{eqnarray}\label{charges}
q+ip = q_0(m \tau +n),\quad m,n \in \mathbb{Z}\ ,
\end{eqnarray}
where \footnote{We have used $n_0=2$ (see footnote below
(\ref{Dirac})), which holds for $N=4$ SYM \cite{O}.}
\begin{eqnarray}\label{tau}
\tau = \frac{4 \pi i}{q_0^2} + \frac{\theta}{2 \pi}\ .
\end{eqnarray}
Here $\theta$ is a parameter of the theory under consideration. In
the case of $N=4$ SYM this parameter can be identified with the
Yang-Mills theta-angle, which is the analog of the Maxwell
theta-angle of the last section \cite{W3}. Note that this makes
(\ref{tau}) the analog of the complex parameter (\ref{taum}) of
Maxwell theory.

In a quantum theory, stable single particle states should
correspond to the primitive vectors of the charge lattice
(assuming single particle states obey (\ref{bounds})) \footnote{A
point of the lattice corresponds to a primitive vector when the
line connecting it with the origin contains no other point of the
lattice.}. As shown by Sen, this is indeed the case for $N=4$ SYM
with an $SU(2)$ gauge group broken to $U(1)$ \cite{Sen}. The
transformations that act as a rearrangement of the primitive
vectors form the group $SL(2, \mathbb{Z})$.

In view of the above, $N=4$ SYM with (spontaneously broken)
$SU(2)$ gauge group is conjectured to have exact $SL(2,
\mathbb{Z})$ electric/magnetic duality. Its set of quantum states,
corresponding to both (electrically charged) gauge particles and
(dyonic) solitons, rotates under the action of this group.
Furthermore, the coupling constant and theta-angle of the theory
transform as their Maxwell analogs,
\begin{eqnarray}\label{trtaum}
\tau \longrightarrow \tilde{\tau} = \frac{W +V \tau}{U+ Z \tau}\ .
\end{eqnarray}
As a non-trivial test of this conjecture, the partition function
of a ``twisted", topological version of $N=4$ SYM has been
evaluated and indeed found to exhibit an $SL(2, \mathbb{Z})$
symmetry \cite{VW}.

\vspace{3mm}

The $SL(2, \mathbb{Z})$ dualities of classical Maxwell theory
(when charge quantization is taken into account) and of the full
$N=4$ SYM with $SU(2)$ gauge group are in fact related, since the
former is (the gauge field part of) the low energy effective
action of the latter.

Contrary to $N=4$ SYM, $N=2$ SYM with $SU(2)$ gauge group does not
have exact electric/magnetic duality. However, in the Wilsonian
effective abelian theory \footnote{The effective Wilsonian action
is based on integrating out the massive degrees of freedom. It
describes the correct physics for energies between appropriately
chosen infrared and ultraviolet cutoffs.} arising from it at low
energies electric/magnetic equivalences of the Maxwell type do
show up. Making use of the latter fact, Seiberg and Witten managed
to solve this low energy theory completely \cite{SW}. This can be
understood as follows. In these effective models the vacuum
expectation values of the scalar fields parameterize a
target-space manifold. Singularities in the moduli space
parameterization signify the breakdown of the effective
description due to additional degrees of freedom becoming
massless. These additional degrees of freedom can be identified
with magnetic monopoles or dyons of the non-abelian theory.
Seiberg and Witten realized that a description of the exact low
energy theory requires the use of electric/magnetic dual frames -
and so dual parameterizations - in different regions of the moduli
space. In this way the entire moduli space can be covered.
Moreover, by determining the monodromies around the coordinate
singularities and patching them together, the theory can be fixed
completely.

\vspace{3mm}

The models of the next section we also treat as effective actions.
They have an arbitrary number of gauge fields, come with an
abelian gauge group and are coupled to gravity. These models
contain the $N=2$ supersymmetric systems we consider in the next
chapter.

\section{Effective actions and e/m duality}\label{efac}

The actions we study in this section are of the form
\begin{eqnarray}\label{acsie}
\S & = & \int  d^4x \, e \; \Big[ (- \vier i \bar{\tau}_{\Lambda
\Sigma}
F^+_{\mu \nu}{}^\Lambda F^{+ \mu \nu \Sigma} - \half i F^+_{\mu \nu}{}^\Lambda O^{+ \mu \nu}{}_{\Lambda}\nn\\
& & + \frac{1}{8} ((Im \tau)^{-1})^{\Lambda \Sigma} O^+_{\mu \nu
\Lambda} O^{+ \mu \nu}{}_{\Sigma} + \mathrm{h.c.}) + \L' \Big] \ .
\end{eqnarray}
$F_{\mu \nu}{}^\Lambda$ are the field strengths of the vector
gauge fields $A_{\mu}{}^\Lambda$ ($\Lambda = 1,...,n$). The
couplings of these field strengths are encoded in the complex
matrix $\tau_{\Lambda \Sigma}$, which can be field-dependent
(typically it depends on scalar fields). Furthermore, we allow for
a linear coupling of the field strengths to field dependent
tensors $O_{\mu \nu \Lambda}$ (which are usually bilinear in
spinor fields). $\L'$ is arbitrary but independent of the vector
gauge fields. Note that we could have absorbed the $O^2$ term in
$\L'$. However, as we will later see, it is useful to include it
explicitly.

The set of Bianchi identities and equations of motion of the gauge
fields take the same form as in source-free Maxwell theory
\begin{eqnarray}\label{eqse}
D_{\mu} \( \begin{array}{c} F^{+ \mu \nu \Lambda} - F^{- \mu \nu \Lambda} \\
G^{+ \mu \nu}{}_\Lambda - G^{- \mu \nu}{}_\Lambda \end{array} \) =
0\ ,
\end{eqnarray}
where the derivatives $D_{\mu}$ are covariantized with respect to
general coordinate transformations. The dual field strengths
$G_{\mu \nu \Lambda} \equiv i e \vep_{\mu \nu \rho \sigma}
\frac{\delta \L}{\delta F_{\rho \sigma}{}^\Lambda}$ are explicitly
given by
\begin{eqnarray}
G_{\mu \nu \Lambda}^+ = \bar{\tau}_{\Lambda \Sigma} F^+_{\mu
\nu}{}^\Sigma + O^+_{\mu \nu \Lambda}\ .
\end{eqnarray}

Obviously the set of equations (\ref{eqse}) is invariant under
\begin{eqnarray}\label{emduu}
\( \begin{array}{c} F_{\mu \nu}{}^{\Lambda}\\ G_{\mu \nu \Lambda}
\end{array} \)
\longrightarrow \( \begin{array}{c} \tilde{F}_{\mu \nu}{}^{\Lambda}\\
\tilde{G}_{\mu \nu \Lambda}
\end{array} \) = \( \begin{array}{cc} U^\Lambda{}_\Sigma & Z^{\Lambda \Sigma}\\
W_{\Lambda \Sigma} & V_\Lambda{}^\Sigma
\end{array} \) \( \begin{array}{c} F_{\mu \nu}{}^{\Sigma}\\ G_{\mu \nu \Sigma}
\end{array} \)\ ,
\end{eqnarray}
where $U$, $Z$, $V$ and $W$ are real-valued matrices. Similar to
the Maxwell-case we interpret the equations $D_{\mu} ( \tF^{+ \mu
\nu \Lambda} - \tF^{- \mu \nu \Lambda}) =0$ as the new Bianchi
identities, while the equations $D_{\mu} (\tG^{+ \mu
\nu}{}_\Lambda - \tG^{- \mu \nu}{}_\Lambda) =0$ are the new field
equations.

So we find again equivalence transformations of the set of
equations formed by Bianchi identities and field equations of the
gauge potentials. However, recall that the field strengths
appearing in (\ref{acsie}) are coupled to matter and gravity.
These fields therefore take part in (\ref{emduu}) as they appear
in $G_{\mu \nu \Lambda}$. For this reason we also have to worry
about their field equations as the transformations (\ref{emduu})
are real duality transformations only when these field equations
remain equivalent.

To determine what happens with the field equations of the matter
and gravitational fields we have to turn to the Lagrangian and see
how it transforms under (\ref{emduu}). Doing this shows another
feature of duality in more general models that was not present in
Maxwell theory: A dual Lagrangian can only be obtained for a
special class of matrices $U$, $Z$, $W$ and $V$.

This can be understood as follows. A new Lagrangian $\tL (\tF, \tG
(\tF))$ is implicitly given by $\tilde{G}_{\mu \nu \Lambda} = i e
\vep_{\mu \nu \rho \sigma} \frac{\delta \tilde{\L}}{\delta
\tilde{F}_{\rho \sigma}{}^\Lambda}$, which can be rewritten as
\begin{eqnarray}\label{intc}
e \frac{\delta \tilde{\L}}{\delta F_{\eta \lambda}{}^\Sigma} & = &
- \frac{1}{4} i \vep^{\rho \sigma \mu \nu} \frac{\delta
\tilde{F}_{\rho \sigma}{}^{\Lambda}}{\delta F_{\eta
\lambda}{}^\Sigma} \tilde{G}_{\mu \nu \Lambda}\nn\\
& = & - \frac{1}{4} i \vep^{\eta \lambda \mu \nu} \bigl(
(U^TW)_{\Sigma \Gamma} F_{ \mu\nu}{}^{\Gamma} +
(U^TV)_\Sigma{}^\Gamma G_{\mu \nu \Gamma} \bigr)\nn\\
& & - \frac{1}{4} i \vep^{\rho \sigma \mu \nu} (Z^TV)^{\Omega
\Gamma} \frac{\delta G_{\rho \sigma \Omega}}{\delta
F_{\eta \lambda}{}^{\Sigma}} G_{\mu \nu \Gamma}\nn\\
& & - \frac{1}{4} i \vep^{\rho \sigma \mu \nu}
(W^TZ)_\Gamma{}^\Omega \frac{\delta G_{\rho \sigma \Omega}}{\delta
F_{\eta \lambda}{}^{\Sigma}} F_{\mu \nu}{}^{\Gamma}\ .
\end{eqnarray}
To obtain $\tL$, (\ref{intc}) should be integrated over $F_{ \eta
\lambda}{}^\Sigma$. In the case of Maxwell theory, when the $U$,
$Z$, $W$ and $V$ are just numbers, this can always be done. In the
general case, however, $U$, $Z$, $W$ and $V$ are $n \times
n$-matrices and having an integrable expression on the right-hand
side of (\ref{intc}) is not automatically guaranteed.

Nevertheless, when we demand
\begin{eqnarray}\label{con1}
U^TV - W^TZ = \alpha \mathbb{I}\ ,\nn\\
U^TW = W^TU,\quad Z^TV = V^TZ\ ,
\end{eqnarray}
(\ref{intc}) becomes
\begin{eqnarray}\lb{intem}
e \frac{\delta \tilde{\L}}{\delta F_{\eta \lambda}{}^\Sigma} & = &
- \frac{1}{8} i \vep^{\rho \sigma \mu \nu} (U^T W)_{\Lambda
\Gamma} \frac{\delta}{\delta F_{\eta \lambda}{}^\Sigma} ( F_{\rho
\sigma}{}^\Lambda F_{\mu \nu}{}^\Gamma)\nn\\
& & - \frac{1}{8} i \vep^{\rho \sigma \mu \nu} (Z^T V)^{\Lambda
\Gamma} \frac{\delta}{\delta F_{\eta \lambda}{}^\Sigma} (G_{\rho
\sigma \Lambda} G_{\mu \nu \Gamma})\nn\\
& & - \vier i \vep^{\rho \sigma \mu \nu} (W^TZ)_\Lambda{}^\Gamma
\frac{\delta}{\delta F_{\eta \lambda}{}^\Sigma} (F_{\rho
\sigma}{}^\Lambda G_{\mu \nu \Gamma})\nn\\
& & - \vier i \vep^{\eta \lambda \mu \nu} \al G_{\mu \nu \Sigma}\
,
\end{eqnarray}
which is obviously integrable. Putting $\alpha$ to $1$ (again
leaving out total rescalings of the Lagrangian and field
strengths), the conditions (\ref{con1}) can be written as
\begin{eqnarray}\lb{conn1}
M^T \Omega M = \Omega\ ,
\end{eqnarray}
with
\begin{eqnarray}
M = \( \begin{array}{cc} U^\Lambda{}_\Sigma & Z^{\Lambda \Sigma}\\
W_{\Lambda \Sigma} & V_\Lambda{}^\Sigma
\end{array} \)\ ,\quad \Omega = \( \begin{array}{cc} 0 & \delta^{\Lambda \Sigma}\\ -\delta_{\Lambda \Sigma} & 0 \end{array} \)\ .
\end{eqnarray}
So the matrices in (\ref{emduu}) should be chosen such that they
leave the skewsymmetric matrix $\Om$ invariant. This means - by
definition - that they are elements of the \emph{symplectic} group
$Sp(2n, \mathbb{R})$ (the connection between electric/magnetic
duality and the symplectic group was first observed in \cite{GZ}).
We will call objects $(\al^{\Lambda}, \al_{\Lambda})$ that
transform as in (\ref{emduu}) (so for example $(F_{\mu
\nu}{}^\Lambda, G_{\mu \nu \Lambda})$) \emph{symplectic vectors}
from now on.

Restricting ourselves to transformations involving $Sp(2n,
\mathbb{R})$ matrices we can integrate (\ref{intem}) to obtain
\begin{eqnarray}\label{ts}
\tS = \int d^4x \; e \;  \Big[ \L  - \vier i(\tF^+_{\mu
\nu}{}^\Lambda \tG^{+ \mu \nu}{}_{\Lambda} - F^+_{\mu
\nu}{}^\Lambda G^{+ \mu \nu}{}_{\Lambda} - \mathrm{h.c.}) +
\hat{\L} \Big] \ .
\end{eqnarray}
Note that as $\tS$ is a functional of the new field strengths
$\tF_{\mu \nu}{}^\Lambda$, the tensors $F_{\mu \nu}{}^\Lambda$,
$G_{\mu \nu \Lambda}$ and $\tG_{\mu \nu \Lambda}$ should be
understood as functions of the former. $\hat{\L}$ is an
integration constant, i.e. an arbitrary functional of the matter
fields.

Now that we have found the new Lagrangian we can determine what
has happened with the field equations of the matter fields. We
first vary (\ref{ts}) with respect to all the fields. This way we
obtain
\begin{eqnarray}\label{vartL}
\delta \tL & = & \delta \L - i \delta \tA_{\mu}{}^\Lambda D_{\nu}
(\tG^{+ \mu \nu}{}_\Lambda - \tG^{- \mu \nu}{}_\Lambda) + i \delta
A_{\mu}{}^\Lambda D_{\nu} (G^{+ \mu \nu}{}_\Lambda - G^{- \mu
\nu}{}_\Lambda )\nn\\
& & + \delta \hat{\L}\ ,
\end{eqnarray}
where $\tA_{\mu}{}^\Lambda$ are the new gauge potentials,
satisfying $\tF_{\mu \nu}{}^\Lambda = 2 \p_{[\mu}
\tA_{\nu]}{}^\Lambda$. In deriving (\ref{vartL}) we have used that
$\tF^+_{\mu \nu}{}^\Lambda \delta \tG^{+ \mu \nu}{}_\Lambda -
F^+_{\mu \nu}{}^\Lambda \delta G^{+ \mu \nu}{}_{\Lambda} = \delta
\tF^+_{\mu \nu}{}^\Lambda \tG^{+ \mu \nu}{}_\Lambda - \delta
F^+_{\mu \nu}{}^\Lambda G^{+ \mu \nu}{}_{\Lambda}$. Also we have
thrown away a total divergence. From (\ref{vartL}) we see that a
variation of $\tL$ with respect to the fields other than
$\tA_\mu{}^\Lambda$ only differs from a variation of $\L$ with
respect to the fields other than $A_\mu{}^\Lambda$ by the term
$\delta \hat{\L}$. This implies that the field equations of the
matter and gravitational fields remain equivalent when $\hat{\L}$
vanishes. In other words, when we take $\hat{\L}=0$ the set of
transformations (\ref{emduu}) are proper duality transformations.

Note that this whole derivation, involving the transformation of
the Lagrangian, the appearance of the symplectic group and the
equivalence of the equations of motion of the matter and
gravitational fields does \emph{not} depend on the explicit form
of the original Lagrangian. All we need is that the gauge
potentials only appear in the Lagrangian through their field
strengths. This implies that the results (including the appearance
of $Sp(2n, \mathbb{R})$) also hold for higher derivative theories
in which higher powers of field strengths are involved. An example
of such a theory is the Born-Infeld Lagrangian of non-linear
electrodynamics.

Returning to the class of models given by (\ref{acsie}), it turns
out that the new action ({\ref{ts}) (with $\hat{\L}=0$) can be
written back in the form (\ref{acsie}),
\begin{eqnarray}
\tilde{\S} & = & \int d^4x \; e \; \Big[ (- \vier i
\tilde{\bar{\tau}}_{\Lambda \Sigma} \tF^+_{\mu \nu}{}^\Lambda
\tF^{+ \mu \nu \Sigma} -
\half i \tF^+_{\mu \nu}{}^\Lambda \tilde{O}^{+ \mu \nu}{}_{\Lambda}\nn\\
& & + \frac{1}{8} ((Im \tilde{\tau})^{-1})^{\Lambda \Sigma}
\tilde{O}^+_{\mu \nu \Lambda} \tilde{O}^{+ \mu \nu}{}_{\Sigma} +
\mathrm{h.c.}) + \L' \Big] \ .
\end{eqnarray}
i.e. the transformation of the Lagrangian is induced by a
transformation of the objects appearing in (\ref{acsie}), which is
of the form
\begin{eqnarray}\label{tranfob}
F^+_{\mu \nu}{}^\Lambda & \longrightarrow & \tF^+_{\mu \nu}{}^\Lambda\ ,\nn\\
\bar{\tau}_{\Lambda \Sigma} & \longrightarrow &
\bar{\tilde{\tau}}_{\Lambda \Sigma} \equiv ((W+V \bar{\tau})(U+ Z
\bar{\tau})^{-1})_{\Lambda \Sigma}\
,\nn\\
O^+_{\mu \nu \Lambda} & \longrightarrow & \tilde{O}^+_{\mu \nu
\Lambda} \equiv ((U + Z
\bar{\tau})^{-1})^\Sigma{}_\Lambda O^+_{\mu \nu \Sigma}\ ,\nn\\
\L' & \longrightarrow & \L'\ .
\end{eqnarray}
We now see why it was useful to start with the explicit $O^2$ term
in (\ref{acsie}).

Just as is the case in Maxwell theory, we stress that although the
new action is of the same form as the original one, the
electric/magnetic duality transformation involved is not an
ordinary symmetry (we recall that a symmetry requires $\tL (\tF) =
\L (F) = \L (\tF)$). From (\ref{ts}) it immediately follows that
$\tL (\tF) \neq \L (F)$. Generically we also have $\tL (\tF) \neq
\L (\tF)$, implying that the electric/magnetic duality is a
duality equivalence.

Only when the transformations of the objects in (\ref{tranfob})
are induced by transformations of fields on which they depend, the
duality is a duality invariance,
\begin{equation}\label{inva}
\tL (\tF, \tilde{\phi}) = \L (\tF, \tilde{\phi})\ .
\end{equation}
Here the matter fields are denoted by $\phi$. In chapter
\ref{ch4}, in the context of gaugings in $N=2$ supersymmetric
models, we are precisely interested in this subclass of
electric/magnetic dualities.

The condition for having a duality invariance comes down to the
requirement that the transformation of $\tau_{\Lambda \Sigma}$ and
$O_{\mu \nu \Lambda}$ is induced by a transformation of the matter
fields, combined with the demand that this transformation of the
matter fields is a symmetry of $\L'$,
\begin{eqnarray}\lb{invar}
\tilde{\tau}_{\Lambda \Sigma} (\tilde{\phi}) & = & \tau_{\Lambda
\Sigma} (\tilde{\phi})\ ,\nn\\
\tilde{O}_{\mu \nu \Lambda} (\tilde{\phi}) & = & O_{\mu \nu
\Lambda} (\tilde{\phi})\
,\nn\\
\L' (\tilde{\phi}) & = & \L' (\phi)\ .
\end{eqnarray}
We stress that these relations do \emph{not} imply that
$\tau_{\Lambda \Sigma}$ and $O_{\mu\nu \Lambda}$ should transform
as functions (which would mean $\tilde{\tau}_{\Lambda \Sigma}
(\tilde{\phi}) = \tau_{\Lambda \Sigma} (\phi)$ and $\tilde{O}_{\mu
\nu\Lambda} (\tilde{\phi}) = O_{\mu \nu \Lambda} (\phi)$). For
continuous duality invariances (\ref{invar}) gives rise to the
following identity
\begin{eqnarray}\lb{idvar}
C_{\Lambda \Sigma} (F^+_{\mu \nu}{}^\Lambda F^{+ \mu \nu \Sigma} -
F^-_{\mu \nu}{}^\Lambda F^{- \mu \nu \Sigma}) + D^{\Lambda \Sigma}
(G_{\mu \nu \Lambda}^+ G^{+ \mu \nu}{}_\Sigma - G_{\mu \nu \Lambda}^+ G^{+ \mu \nu}{}_\Sigma)\nn\\
- 2 B^\Sigma{}_\Lambda (F^+_{\mu \nu}{}^\Lambda G^{+ \mu
\nu}{}_\Sigma - F^-_{\mu \nu}{}^\Lambda G^{- \mu \nu}{}_\Sigma)  =
2 i \frac{\delta \L}{\delta \phi} \delta{\phi}\ ,
\end{eqnarray}
where the matrices $B$, $C$, $D$ are defined by an expansion of a
symplectic matrix around $\mathbb{I}$,
\begin{eqnarray}\label{expm}
\( \begin{array}{cc} U & Z\\
W & V \end{array} \) \approx \mathbb{I} + \( \begin{array}{cc} B & -D\\
C & -B^T \end{array} \)\ .
\end{eqnarray}
Here $C_{\Lambda \Sigma}$ and $D^{\Lambda \Sigma}$ are symmetric.
(\ref{idvar}) can be viewed as an equation determining (if
possible) the transformation rules for the matter fields to obtain
an invariance. Alternatively, when there are natural
transformation rules for these matter fields - as will be the case
in $N=2$ supersymmetric systems - (\ref{idvar}) is a condition on
the matrices $B$, $C$ and $D$.

\vspace{5mm}

 \textbf{Symmetries versus electric/magnetic duality}

\vspace{5mm}

In the next chapter we will consider $N=2$ supersymmetric systems.
The vector multiplet sector of these models is of the form
(\ref{acsie}), so electric/magnetic duality is realized. A
relevant question in this respect is: Do electric/magnetic duality
transformations preserve the symmetries of the Lagrangian (and in
particular supersymmetry)? Below we answer this question. Although
our main interest is in in $N=2$ supersymmetric models with at
most two derivatives, our treatment is also valid for general
models with arbitrary powers of field strengths appearing.

We start our exposition by considering a Lagrangian $\L$ and an
electric/magnetic dual of it, $\tL$. We assume that both $\L$ and
$\tilde{\L}$ are invariant under a certain set of symmetry
transformations, up to a total divergence and upon using the
Bianchi identities of the field strengths. Leaving out the total
divergences, the variations of the Lagrangians under the symmetry
transformations can be denoted as
\begin{eqnarray}\label{var2}
\delta \L & = & - i \alpha_{\mu \Lambda} D_{\nu} (F^{+ \mu \nu
\Lambda} - F^{- \mu \nu \Lambda})\ ,\nn\\
\delta \tilde{\L} & = & - i \tilde{\alpha}_{\mu \Lambda} D_{\nu}
(\tF^{+ \mu \nu \Lambda} - \tF^{- \mu \nu \Lambda})\ .
\end{eqnarray}
The objects $\alpha_{\mu \Lambda}$ and $\tilde{\al}_{\mu \Lambda}$
are field dependent quantities. Note that they are defined up to
total divergences.

In case the transformations of the fields other than the vector
fields are the same for $\L$ and $\tilde{\L}$, we also have
\begin{eqnarray}\label{vartLL}
\delta \tL + i \delta \tA_{\mu}{}^\Lambda D_{\nu} (\tG^{+ \mu
\nu}{}_\Lambda - \tG^{- \mu \nu}{}_\Lambda) = \delta \L + i \delta
A_{\mu}{}^\Lambda D_{\nu} (G^{+ \mu \nu}{}_\Lambda - G^{- \mu
\nu}{}_\Lambda)\ .
\end{eqnarray}
as follows directly from (\ref{vartL}) (with $\hat{L} = 0$).

Combining (\ref{vartLL}) and (\ref{var2}) gives
\begin{eqnarray}\label{con4}
 - i \tilde{\alpha}_{\mu \Lambda} D_{\nu}
(\tF^{+ \mu \nu \Lambda} - \tF^{- \mu \nu \Lambda}) + i \delta
\tA_{\mu}{}^\Lambda D_{\nu} (\tG^{+ \mu \nu}{}_\Lambda -
\tG^{- \mu \nu}{}_\Lambda)\nn\\
= - i \alpha_{\mu \Lambda} D_{\nu} (F^{+ \mu \nu \Lambda} - F^{-
\mu \nu \Lambda}) + i \delta A_{\mu}{}^\Lambda D_{\nu} (G^{+ \mu
\nu}{}_\Lambda - G^{- \mu \nu}{}_\Lambda)\ .
\end{eqnarray}

We saw above that $(F_{\mu \nu}{}^\Lambda, G_{\mu \nu \Lambda}$)
transforms as a symplectic vector. It then follows that we can
solve (\ref{con4}) by demanding $\delta A_{\mu}{}^\Lambda$ and
$\alpha_{\mu \Lambda}$ to form a symplectic vector as well
(observe that both $\de A_\mu{}^\Lambda$ and $\al_{\mu \Lambda}$
are defined up to a total divergence) \footnote{Note that this
does imply that neither $\delta A_{\mu}{}^\Lambda$ nor
$\alpha_{\mu \Lambda}$ is allowed to depend explicitly on
$A_{\mu}{}^\Lambda$.}. In that case we see the appearance of a
symplectic inner product on both sides of (\ref{con4}). This
equation is then nothing more than the statement that a symplectic
inner product transforms as a scalar under $Sp (2n, \mathbb{R})$.

To summarize, we have found the following: When we start with a
Lagrangian that is invariant under a certain set of
transformations, the electric/magnetic dual Lagrangian is
symmetric under a dual set of transformations. This set of
transformations is the same for the fields other than the vector
fields. The transformation rule for the dual vector field is found
by a transformation of the symplectic vector formed by the
symmetry transformation of the old gauge field and the prefactor
of the Bianchi-identity-term to which the symmetry variation of
the old Lagrangian gives rise. A large class of symmetries -
amongst which is supersymmetry - is therefore indeed necessarily
preserved under electric/magnetic duality transformations.

\section{Scalar-tensor duality}\lb{sctd}

In this last section of the first chapter we treat dualities
involving scalar and tensor gauge fields rather than vector gauge
fields. They also appear in four-dimensional field theories and
low energy effective actions and are directly related to the
electric/magnetic duality transformations described above.

To start with, we consider the simple model \footnote{We take a
flat background for notational convenience, but our analysis
straightforwardly generalizes to arbitrary backgrounds.}
\begin{eqnarray}\lb{tym}
\S^e = \int d^4x \; \Big[ \frac{1}{6} H_{\mu \nu \rho} H^{\mu \nu
\rho} + \frac{1}{3} i \lambda \vep^{\mu \nu \rho \sigma} \p_{\mu}
H_{\nu \rho \sigma} \Big] \ ,
\end{eqnarray}
where $\lambda$ is a scalar and $H_{\mu \nu \rho}$ is a tensor
field of rank three. For future convenience we take a Euclidean
setting.

From the action (\ref{tym}) two dual effective theories can be
obtained. One by eliminating $\lambda$,
\begin{eqnarray}\lb{tyma}
\S^e = \int d^4x \; \frac{3}{2} \p_{[\mu} B_{\nu \rho]} \p^{[\mu}
B^{\nu \rho]}\ ,
\end{eqnarray}
(where $H_{\mu \nu \rho} = 3 \p_{[\mu} B_{\nu \rho]}$), the other
by eliminating $H_{\mu \nu \rho}$,
\begin{eqnarray}\lb{tymb}
\S^e = \int d^4x \; \p_{\mu} \lambda \p^{\mu} \lambda\ .
\end{eqnarray}

Depending on whether the theory to start with is viewed as
fundamental or as an effective action, the elimination of the
field should be done by solving its equation of motion or by
integrating over it. How to do the latter is for example explained
in \cite{C}. Eliminating a field in (\ref{tym}) by solving its
equation of motion either yields the Bianchi identity of $H_{\mu
\nu \rho}$ (to give (\ref{tyma})) or the relation $i \vep^{\mu \nu
\rho \sigma} \p_\mu \lambda = H^{\nu \rho \sigma}$ (to yield
(\ref{tymb})).

This type of duality straightforwardly generalizes to more
complicated actions depending either on tensor gauge fields or
scalar fields. Also it can be realized in any dimension $D$, where
it relates effective theories of rank $p$ gauge potentials to
effective theories with gauge potentials of rank $D-p-2$.

(\ref{tyma}) and (\ref{tymb}) come with both a Bianchi identity
and a field equation
\begin{eqnarray}
\p_{\mu} \( \begin{array}{c} \vep^{\mu \nu \rho \sigma} H_{\nu
\rho \sigma}\\ - H^{\mu \nu \rho}
\end{array} \) = 0\ ,\quad \p_{\mu} \( \begin{array}{c} - \vep^{\mu \nu \rho \sigma} \p_{\sigma} \lambda\\
6 \, \p^{\mu} \lambda
\end{array} \) = 0\ .
\end{eqnarray}
Via (\ref{tym}) the Bianchi identity (field equation) of one
theory is related to the field equation (Bianchi identity) of the
other. This sounds very much like electric/magnetic duality.
Indeed its (Minkowskian) version with $D=4$ and $p=1$,
\begin{eqnarray}
\S = \int d^4x \;  \Big[ \L(F) + \half i \vep^{\mu \nu \rho
\sigma} \p_{\mu} \tA_{\nu} F_{\rho \sigma} \Big] \ ,
\end{eqnarray}
can be shown to effectuate the electric/magnetic duality rotation
with rotation matrix $M$,
\begin{eqnarray}\lb{maq}
M = \( \begin{array}{cc} 0 & 1\\ -1 & 0 \end{array} \)\ .
\end{eqnarray}

Put differently, the duality discussed here is the scalar-tensor
analog of the special electric/magnetic duality transformation
with matrix (\ref{maq}). This naturally brings up the question:
What about electric/magnetic duality transformations with general
matrix $M$, do they also have their scalar-tensor analogs?

We answer this question in a Euclidean context. One of the reasons
to do so is that in chapter \ref{ch3} we work with the Euclidean
version of a ($N=2$ supergravity) scalar-tensor model. There is
also another reason, to which we come shortly.

Consider the Euclidean action of a model with both a scalar and a
tensor gauge field.
\begin{eqnarray}\lb{stl}
\S^e = \frac{2 \pi} {g^2}  \int d^4x \; \Big[ \p_{\mu} \chi
\p^{\mu} \chi + \frac{1}{6} H_{\mu \nu \rho} H^{\mu \nu \rho}
\Big] - i \frac{\theta}{12 \pi} \int d^4x \; \vep^{\mu \nu \rho
\sigma} \p_{\mu} \chi H_{\nu \rho \sigma}\ .
\end{eqnarray}
Its set of equations of motion and Bianchi identities we formulate
as
\begin{eqnarray}\lb{eqbi}
\vep^{\mu \nu \rho \sigma} \p_{\mu} \( \begin{array}{c} H_{\nu \rho \sigma}\\
G^{\chi}_{\nu \rho \sigma} \end{array} \) = 0\ ,\quad \vep^{\mu
\nu \rho \sigma} \p_{\rho} \( \begin{array}{c} F^{\chi}_{\sigma}\\
G^{H}_{\sigma} \end{array} \) = 0\ ,
\end{eqnarray}
where we have used
\begin{eqnarray}
F^{\chi}_{\mu} & \equiv & - i \p_{\mu} \chi\ ,\nn\\
G^{\chi}_{\mu \nu \rho} & \equiv & - \vep_{\mu \nu \rho \sigma}
\frac{\de \L^e}{\de F^{\chi}_{\sigma}}\ ,\nn\\
G^{H}_{\mu} & \equiv & \vep_{\mu \nu \rho \sigma} \frac{\de
\L^e}{\de H_{\nu \rho \sigma}}\ .
\end{eqnarray}
A source term in the first set of equations in (\ref{eqbi}) would
involve a scalar charge density. This is different in a set of
equations associated with vector gauge fields. Then the source
term comes with a current vector, which is the natural covariant
object describing the charge and current of a point particle.
Likewise a current tensor of rank $p$ naturally corresponds to a
$p-1$-dimensional source. According to this, when the charge
density is a scalar field $(p=0)$ (as here) the associated objects
need to be a localized in \emph{all} directions. This naturally
forces one to take a Euclidean setting, in which objects of the
latter type make sense.

In both sets of (\ref{eqbi}) the first equation is a Bianchi
identity, while the second one is an equation of motion. These
sets therefore are very similar to the set of equations of Maxwell
theory (\ref{eq1}). Also just as in Maxwell theory we can perform
the rotation
\begin{eqnarray}\label{emdust}
\( \begin{array}{c} H_{\mu \nu \rho}\\ G^{\chi}_{\mu \nu \rho}
\end{array} \)
\longrightarrow \( \begin{array}{c} \tH_{\mu \nu \rho}\\
\tG^{\chi}_{\mu \nu \rho}
\end{array} \) = \( \begin{array}{cc} U & Z\\ W & V \end{array} \)
\( \begin{array}{c} H_{\mu \nu \rho}\\ G^{\chi}_{\mu \nu \rho}
\end{array} \)\ ,\nn\\
\( \begin{array}{c} F^{\chi}_{\mu}\\ G^{H}_{\mu}
\end{array} \)
\longrightarrow \( \begin{array}{c} \tF^{\chi}_{\mu}\\
\tG^{H}_{\mu}
\end{array} \) = \( \begin{array}{cc} U & Z\\ W & V \end{array} \)
\( \begin{array}{c} F^{\chi}_{\mu}\\ G^{H}_{\mu}
\end{array} \)\ .
\end{eqnarray}
Again we take $UV - WZ = 1$ to leave out total rescalings of the
Lagrangian and field strengths, i.e. the matrix in (\ref{emdust})
belongs to $Sp(2, \mathbb{R})$. We then interpret the first line
of the sets of equations as the new Bianchi identities. So
$\tH_{\mu \nu \rho} = 3 \p_{[\mu} \tB_{\nu \rho]}$ and
$\tF^{\chi}_{\mu} = - i \p_{\mu} \tilde{\chi}$ where $\tB_{\mu
\nu}$ and $\tilde{\chi}$ are not locally related to $B_{\mu \nu}$
and $\chi$. What we have effectively done is rotating the scalar
(tensor) Bianchi identity and the tensor (scalar) equation of
motion.

A dual Lagrangian is (consistently) defined by
\begin{eqnarray}
\tG^{\chi}_{\mu \nu \rho} =  - \vep_{\mu \nu \rho \sigma}
\frac{\de \tL^e}{\de \tF^{\chi}_{\sigma}}\ ,\quad  \tG^{H}_{\mu} =
\vep_{\mu \nu \rho \sigma} \frac{\de \tL^e}{\de \tH_{\nu \rho
\sigma}}\ .
\end{eqnarray}
It takes the same form as (\ref{stl}),
\begin{eqnarray}\lb{stld}
\tS^e = \frac{2 \pi}{\tilde{g}^2}  \int d^4x \; \Big[ \p_{\mu}
\tch \p^{\mu} \tch + \frac{1}{6} \tH_{\mu \nu \rho} \tH^{\mu \nu
\rho} \Big] - i \frac{\tilde{\theta}}{12 \pi} \int d^4x \;
\vep^{\mu \nu \rho \sigma} \p_{\mu} \tch \tH_{\nu \rho \sigma}\ ,
\end{eqnarray}
where we used, just as in Maxwell theory,
\begin{eqnarray}
\tau = \frac{4 \pi i}{g^2} + \frac{\theta}{2 \pi}\ ,\quad \tau
\longrightarrow \tilde{\tau} = \frac{W +V \tau}{U+ Z \tau}\ .
\end{eqnarray}
We find that generically $\tL (\tF^{\chi}, \tH) \neq \L
(\tF^{\chi}, \tH) \neq \L (F^{\chi}, H)$, so (\ref{emdust}) are
duality equivalences. They can be viewed as the analogs of the
electric/magnetic duality transformations of Maxwell theory with
general matrix $M$. The question posed above is therefore found to
have a positive answer. Note that the special transformation with
matrix $M$ given by (\ref{maq}) does not give back the duality of
(\ref{tym}), but rather a doubled version of it.

The electric/magnetic duality of Maxwell theory and the duality of
(\ref{stl}) are related. Namely, it is not difficult to show that
the Maxwell action (\ref{ac2}) and the action (\ref{stl}) yield
the same three-dimensional theory after a dimensional reduction
(for Maxwell theory this reduction should either be over time or
be accompanied by a Wick rotation). This involves the
identification of the dimensionally reduced version of the
symplectic vectors $(F_{\mu \nu}, G_{\mu \nu})$ and the ones
appearing in (\ref{emdust}).

Like the electric/magnetic duality of Maxwell theory generalizes
to the duality relations hidden in the (effective) action
(\ref{acsie}), the duality of (\ref{stl}) has a straightforward
generalization to more complicated actions (with more scalar and
tensor gauge fields and field dependent coupling matrices). In
fact, these theories and their corresponding duality relations can
be found via a dimensional reduction of the vector theory and an
uplifting of the resulting three-dimensional theory to a
scalar-tensor theory. In chapter \ref{ch2} we will explicitly
perform this map in the context of $N=2$ supergravity, where it
goes by the name \emph{c-map}.

What is the relevance of these `general $M$' scalar-tensor
dualities? In case the scalar-tensor theories under consideration
are viewed as fundamental theories, which should be quantized, the
`dualities' we found in fact are only classical equivalences. To
find out if they can be promoted to honest dualities the full
quantum theories should be considered. When we have to do with
effective actions, we can properly use the term dualities for the
equivalence relations we found. However, then these dualities can
be translated to the dual (in the sense of (\ref{tym})) all-scalar
theory, where they turn out to correspond to (ordinary) coordinate
transformations on the sigma model manifold. So it is not so clear
(yet) what the relevance of the scalar-tensor dualities is.
Nevertheless, they are interesting equivalence relations between
different scalar-tensor actions. Especially in the context of
string theory, where quite often one naturally gets scalar-tensor
theories instead of their all-scalar counterparts, they deserve
some further study. For instance, it may be that in string theory
four-dimensional scalar-tensor dualities are the low energy
effective realizations of some string dualities, i.e. that they
relate different string compactifications in the same way as the
(generalized) electric/magnetic dualities in $N=2$ supergravity do
that we describe in chapter \ref{ch4}.

\newpage

\thispagestyle{empty}

\chapter{N=2 supersymmetry}\lb{ch2}

In the introduction of this thesis we have motivated our use of
$N=2$ supersymmetric models. In short, the main reason for our
interest is the realization of $N=2$ supergravities as low energy
effective actions of string theory, and in particular the
controllable yet rich environment they offer for both doing
(quasi-) string phenomenology and studying non-perturbative string
effects.

In the present chapter our aim is to introduce and explain the
aspects of $N=2$ supersymmetric systems that are relevant in the
later chapters. For a more extended review of the subject we refer
for example to \cite{Kl}. Our setup means that a lot of notation
is introduced and therefore this chapter is a little on the
technical side.

Supersymmetry transformations transform bosons into fermions and
vice versa. Supersymmetric theories are invariant under
transformations of this type. This can be summarized schematically
as
\begin{eqnarray}
B & \longrightarrow & \tilde{B} \sim  F\nn\\
F & \longrightarrow & \tilde{F} \sim  B\nn\\
T(B, F) & \longrightarrow & T (\tilde{B}, \tilde{F}) \sim T (B,
F)\ ,
\end{eqnarray}
where $B$/$F$ denote the bosons/fermions of the theory $T (B, F)$.
As bosons have integer spin and fermions half-integer spin it
automatically follows that supersymmetry generators ($Q$) are
fermions. Naturally they have spin one-half and therefore should
transform in the spinor representation of the Lorentz group. In
four dimensions the smallest spinor has four real components. It
can be obtained from a Majorana constraint on a Dirac spinor,
which has eight real components (our notation and conventions can
be found in Appendix \ref{notcon}). The number of four-component
supersymmetry generators is denoted by $N$.

According to \cite{HLS}, supersymmetry generators have to obey the
following anticommutation relation
\begin{eqnarray}\lb{sal}
\{ Q_{\al}, \bar{Q}_{\be} \} = 2 (\ga_{\mu})_{\al \be} P^{\mu}\ ,
\end{eqnarray}
where $P^{\mu}$ are the generators of translations, which are part
of the Poincar\'{e} group. The group formed by supersymmetry and
Poincar\'{e} generators is called the Poinca\-r\'{e} supergroup.

Taking a field-theoretic setting, the Poincar\'{e} superalgebra
should be represented on the fields. One can choose to either fix
the spacetime dependence of the supersymmetry parameters or leave
it unfixed. Supersymmetries of the former type are called `global'
or `rigid'. The latter type are the `local' supersymmetries.
(\ref{sal}) then implies that the translational symmetry is also
local, which means that the theory is invariant under general
coordinate transformations and therefore automatically includes
gravity. Despite cancellations between Feynman diagrams induced by
supersymmetry, $N=2$ supergravity is not finite. We should
therefore view it as an effective action of an underlying
fundamental theory. As already explained, in our setup this
fundamental theory is string theory.

An important tool in the construction of $N=2$ supergravity is the
so-called `superconformal method' \cite{dWvHvP2}. It involves an
extension of the symmetry group from super Poincar\'{e} to the
superconformal group. The $N=2$ superconformal group contains,
besides the symmetries of the super Poincar\'{e} group, additional
symmetries, which include for instance dilatations. The
corresponding superconformally invariant theories are \emph{gauge
equivalent} to Poincar\'{e} supergravity. This means that models
of the latter type, which do not necessarily exhibit the extra
superconformal symmetries, can be obtained by gauge fixing these
symmetries or by writing the theory in terms of gauge invariant
quantities. The fields disappearing in the process are called
\emph{compensating fields}. Let us clarify this using a model of
pure gravity. The Lagrangian
\begin{equation}
  \label{eq:scale-inv-R}
  \sqrt{g} \mathcal{L}\propto \sqrt{g} \left[
  g^{\mu\nu} \partial_\mu\phi\, \partial_\nu\phi - \frac{1}{6} R\, \phi^2
  \right]\ ,
\end{equation}
is invariant under local dilatations with parameter $\Lambda(x)$:
$\delta\phi= \Lambda \,\phi$, $\delta g_{\mu\nu} = -2 \Lambda\,
g_{\mu\nu}$. It is gauge equivalent to the Einstein-Hilbert
Lagrangian, which itself does not exhibit local scale invariance.
To make the equivalence manifest one either rewrites
(\ref{eq:scale-inv-R}) in terms of a scale invariant metric
$\phi^2 g_{\mu\nu}$, or one simply imposes the gauge condition and
sets $\phi$ equal to a constant ($\phi$ therefore is a
compensating field). For a further explanation of the principle of
gauge equivalence we refer to \cite{dWC} and references therein.
In the context of supersymmetric systems, keeping the
superconformal invariance manifest, one realizes a higher degree
of symmetry, which facilitates the construction of supergravity
Lagrangians and clarifies the geometrical features of the
resulting theories.

The $N=2$ superconformal algebra has three irreducible field
representations that are important for us. These are the Weyl
multiplet \cite{dWvHvP,FV} (containing for instance the graviton),
the vector multiplet (containing a vector gauge field) and the
hypermultiplet (with a bosonic sector consisting of scalar fields
only) (the last two have also - non-conformal - rigidly
supersymmetric versions). The first three sections of this chapter
are devoted to the introduction of these three multiplets and the
rigidly supersymmetric and superconformally invariant actions they
give rise to. In the fourth section we describe the procedure to
go from a conformal model to Poincar\'{e} supergravity.

In this chapter, the gauge group associated with the gauge fields
of the vector multiplets is abelian. Furthermore, we do not
consider couplings of the gauge fields to the hypermultiplets (for
both we refer to chapter \ref{ch4}). As a result the vector fields
only appear through their field strengths. In turn it follows
that, as already alluded to several times in the last chapter, the
vector multiplet sector exhibits electric/magnetic duality
\footnote{Some early references to electric/magnetic duality in
supersymmetric theories are
\cite{CFG,FSZ,CSF,dW79,CJ,GZ,dWN,dR}.}. Involving these
electric/magnetic dualities, we produce a new result concerning
the behavior of the auxiliary field $Y_{ij}$.

As we found in chapter \ref{ch1}, between certain theories
consisting of scalars and tensors there exist duality relations
that are similar to electric-magnetic duality. In a subsector of
the tensor formulation of the hypermultiplet part of $N=2$
supergravity this scalar-tensor duality is realized. The duality
structure in these tensor multiplet models can be understood from
the existence of the c-map, which is a map from the vector
multiplet sector to the hypermultiplet (or tensor multiplet)
sector. In view of the use we will make of it in the next chapter,
we treat this map in detail in section \ref{twvijf}.

\section{The N=2 superconformal group and the Weyl
multiplet}\label{cswe}

Let us start by considering the different generators present in
the $N=2$ superconformal group. First of all there are the
generators of the conformal group itself. These are the
translations ($P^a$) and the Lorentz transformations ($M^{ab}$) of
the Poincar\'{e} group together with the scale ($D$) and special
conformal ($K^a$) transformations. Furthermore, the superconformal
group contains two supersymmetry generators $Q^i$. The algebra of
this set of generators does not close. As it turns out, two more
fermionic generators ($S^i$, `special supersymmetries') and the
generators of $U(2)_R$ (to which we come back shortly) should be
included. The algebra may furthermore contain a central charge,
which would appear in the anticommutator of two different
supersymmetry generators. However on the representations we
consider it vanishes \footnote{Notwithstanding this fact, a
central charge is generated \emph{dynamically} for field
configurations that are electrically and/or magnetically charged
\cite{WO}. The same holds for $N=4$ supersymmetry, and it leads to
the realization of the Bogomol'nyi bound at the level of the
algebra mentioned in section \ref{monol}.} and therefore we
neglect it from now on.

With the latter taken into account, the full set of
anticommutation relations between the supersymmetry generators
reads as \footnote{The complete set of (anti-)commutation
relations of the superconformal algebra can for instance be found
in \cite{Kl}.}
\begin{eqnarray}\lb{salx} \{
Q^i, \bar{Q}^j \} = 2 \ga^{\mu} P_{\mu} \de^{ij}\ .
\end{eqnarray}
Observe that this expression is invariant under $U(2)$
transformations. In fact, these $U(2)$ transformations are an
invariance of the whole $N=2$ superconformal algebra and are
therefore by definition part of its automorphism group. More
precisely, they form the part of the automorphism group that
commutes with the Lorentz group and are as such denoted by
$U(2)_R$. They are included in the $N=2$ superconformal group as
we saw above. Note that for the $U(2)$ transformations to be
compatible with the Majorana condition they  should act in a
chiral way: The chiral projections $Q^i_R = \half (\mathbb{I} -
\ga_5) Q^i$ and $Q^i_L = \half (\mathbb{I} + \ga_5) Q^i$ transform
in conjugate representations $\mathbf{2}$ and $\mathbf{\bar{2}}$
of $U(2)_R$ respectively. It is useful to introduce a notation for
the chiral projections of the supercharges that also infers the
representation of $U(2)_R$: $Q^i = Q^i_R$ and $Q_i = Q^i_L$ with
the upper index transforming in the $\mathbf{2}$ and the lower
index in the $\bar{\mathbf{2}}$ representation. Similarly, the
chiral projections of other Majorana fermions are assigned upper
or lower indices (see the tables in Appendix \ref{supcon} for our
conventions).

The gauge fields corresponding to each of the generators above
form a representation of the $N=2$ superconformal algebra.
However, as such this is not yet a multiplet we are interested in.
To have a theory including Einstein gravity, the local
translations generated by $P^a$ should be related to general
coordinate transformations. This yields a set of constraints,
which for instance makes the gauge field of local Lorentz
transformations ($\omega_{\mu}{}^{a b}$), dependent on the other
fields. Furthermore, also the gauge fields for special conformal (
$f_{\mu}{}^{a}$), and special supersymmetry transformations
transformations ($\phi_{\mu}{}^{i}$), turn out to be dependent
quantities. Their expressions in terms of independent fields can
be found in Appendix \ref{supcon}. In addition to this, the
constraints imply the inclusion of some new, auxiliary fields.

The representation arising in this way is called the Weyl
multiplet. It contains the gauge fields of local translations
($e_{\mu}{}^{a}$), $Q$-supersymmetry ($\psi_{\mu}{}^{i}$),
dilatations ($b_{\mu}$), and the gauge fields of the $U(1)_R$ and
$SU(2)_R$-part ($W_{\mu}$ and $\V_\mu{}^i{}_j$) of $U(2)_R$. They
are accompanied by the auxiliary fields $T_{ab}^{ij}$
(antiselfdual, $ij$-antisymmetric) and $D$, which are both
bosonic, and the fermionic $\chi^i$. The superconformal
transformation rules for all these fields can be found in Appendix
\ref{supcon}.

The algebra on the Weyl multiplet closes \emph{off-shell}, which
means that no field equations are needed to make it a genuine
representation of the $N=2$ superconformal algebra. As an
off-shell representation of a supersymmetry algebra it necessarily
has the same $(24+24)$ number of off-shell bosonic and fermionic
degrees of freedom.

The Weyl multiplet provides the gauge fields needed to promote an
action invariant under global $N=2$ superconformal transformations
to a locally $N=2$ superconformally invariant action. The fact
that the algebra closes off-shell on the Weyl multiplet makes it
relatively easy to construct such a coupling. In section
\ref{poin}, after we have considered the other relevant
representations of the superconformal algebra and their rigidly
superconformal actions, we explicitly show how this is done.

\section{Vector multiplets}\label{vmul}

First we discuss the vector multiplet as a representation of rigid
$N=2$ supersymmetry. Later on we treat its superconformal version.

\vspace{5mm}

\textbf{Rigid vector multiplets}

\vspace{5mm}

The field content and rigid supersymmetry transformations of the
vector multiplet are given by \cite{dWvHvP2}
\begin{eqnarray}\label{susyr}
\delta X & = & \bar{\epsilon}^i \Omega_i\
,\nonumber\\
\delta A_{\mu}& = & \varepsilon^{ij} \bar{\epsilon}_i \gamma_{\mu}
\Omega_j + \varepsilon_{ij}
\bar{\epsilon}^i \gamma_{\mu} \Omega^j\ ,\nonumber\\
\delta \Omega_i& = & 2 \spa X \epsilon_i + \half \gamma_{\mu \nu}
F^{- \mu \nu} \varepsilon_{ij} \epsilon^j + Y_{ij}
\epsilon^j\ ,\nonumber\\
\delta Y_{ij} & = & 2 \bar{\epsilon}_{(i} \spa \Omega_{j)} + 2
\varepsilon_{ik} \varepsilon_{jl} \bar{\epsilon}^{(k} \spa
\Omega^{l)}\ .
\end{eqnarray}
Here $X$ is complex scalar, $A_{\mu}$ a real vector and $Y_{ij}$
(symmetric in $i,j$) a triplet of scalar fields, subject to the
reality condition
\begin{eqnarray} \lb{ry}
Y_{ij} = \vep_{ik} \vep_{jl} Y^{kl}\ .
\end{eqnarray}
$\Omega_i$ are (the chiral projections of) two Majorana fermions.
Observe that the transformation rules (\ref{susyr}) are manifestly
covariant under (global) $SU(2)_R$ transformations (acting as
$\Om_i \rightarrow \tOm_i = S_i{}^j \Om_j$ and $Y_{ij} \rightarrow
\tY_{ij} = S_i{}^k S_j{}^l Y_{kl}$, where $S_i{}^j$ is the
generator of $SU(2)_R$ in the fundamental representation).

Also note that under supersymmetry the complex structure of the
scalar sector and the chiral structure of the fermionic sector are
related. This can be understood from the fact that the vector
multiplet is a reduced $N=2$ chiral multiplet. The Bianchi
identity $\p_{[\mu} F_{\nu \rho]} = 0$, implying $F_{\mu \nu} = 2
\p_{[\mu} A_{\nu]}$, and the reality condition on $Y_{ij}$ both
arise in the reduction (see for instance \cite{Kl}). This turns
out to be relevant later on.

Like the Weyl multiplet, the vector multiplet, consisting of the
fields in (\ref{susyr}), forms an off-shell representation of the
algebra. This is consistent with the fact that the number of
bosonic and fermionic off-shell degrees of freedom is the same
($8+8$).

The rigidly $N=2$ supersymmetric action of $n$ of these abelian
multiplets reads \cite{deWit:1984px},
\begin{eqnarray}\label{vecsh}
  \L &=&
  \Big[ i \partial_\mu F_\Lambda\,\partial^\mu \bar X^\Lambda
  + \half i F_{\Lambda\Sigma} \, \bar\Omega_i{}^\Lambda
  \slash{\partial} \Omega^{i\Sigma}
  + \vier i F_{\Lambda\Sigma}\, F^-_{\mu\nu}{}^\Lambda\,
  F^{-\Sigma\,\mu\nu}\nn\\
  & & - \frac{1}{8} i F_{\Lambda\Sigma}\, Y_{ij}{}^\Lambda \,
  Y^{ij\Sigma}\nn\\
  & & -\frac{1}{16} i F_{\Lambda\Sigma\Gamma}\,
  \bar\Omega_i{}^\Lambda\gamma^{\mu\nu} \Omega _j^\Gamma\,
  \varepsilon^{ij} \,F_{\mu\nu}^-{}^{\Sigma}
  + \frac{1}{8} i F_{\Lambda\Sigma\Gamma}\,
  Y^{ij\Lambda}\, \bar{\Omega}_{i}^{\Sigma} \Omega_{j}^{\Gamma}\,
  \nonumber\\
   &&
   - \frac{1}{48} i \varepsilon^{ij} \varepsilon^{kl} \,
    F_{\Lambda\Sigma\Gamma\Xi}\, \bar{\Omega}_i^{\Lambda}
   \Omega_k^{\Sigma} \; \bar{\Omega}_j^{\Gamma} \Omega_l^{\Xi}\, + \mathrm{h.c.} \Big]\ ,
\end{eqnarray}
where $\Lambda, \Sigma, \cdots$ run from $1$ to $n$. Observe that
$\L$ has a manifest global $SU(2)_R$ invariance.

We have used
\begin{eqnarray}
F_{\Lambda_1 ... \Lambda_q} \equiv \frac{\delta}{\delta
X^{\Lambda_1}} ... \frac{\delta}{\delta X^{\Lambda_q}} F(X)\ ,
\end{eqnarray}
and
\begin{eqnarray}
N_{\Lambda \Sigma} \equiv -i (F_{\Lambda \Sigma} - \bF_{\Lambda
\Sigma})\ ,\quad N^{\Lambda \Sigma} \equiv (N^{-1})^{\Lambda
\Sigma}\ .
\end{eqnarray}
$F(X)$ is a holomorphic function of the scalar fields
$X^{\Lambda}$. The precise form of this function fully fixes
(\ref{abel}), i.e. \emph{a choice of Lagrangian is equivalent to a
choice of a holomorphic function $F(X)$} \cite{ST,dWLPSvP}.

The sigma model contained in (\ref{abelm1}) exhibits an
interesting geometry. The complex scalars $X^{\Lambda}$
parameterize an $n$-dimensional target-space with metric
$g_{\Lambda \bar{\Sigma}} = N_{\Lambda \Sigma}$. This is a
K\"{a}hler space: Its metric equals
\begin{eqnarray}
g_{\Lambda \bar{\Sigma}} = \frac{\delta}{\de X^\Lambda}
\frac{\de}{\de \bX^\Sigma} K(X, \bX)\ ,
\end{eqnarray}
with K\"{a}hler potential
\begin{eqnarray}
K(X, \bX) = i X^{\Lambda} \bF_{\Lambda} (\bX) - i \bX^{\Lambda}
F_{\Lambda} (X)\ .
\end{eqnarray}
The resulting geometry is known as \emph{(rigid) special
geometry}.

Note that the fields $Y_{ij}{}^{\Lambda}$ only appear without
derivatives acting on them. This implies that they are auxiliary
fields, which can be eliminated from the Lagrangian by solving
their field equations. The resulting vector multiplets, consisting
of $X^{\Lambda}$, $A_{\mu}{}^\Lambda$ and $\Omega^{\Lambda}$ are
\emph{on-shell} representations of the supersymmetry algebra,
which means that the algebra only closes when the field equations
are satisfied. Correspondingly, only the number of on-shell
degrees of freedom of the bosonic and fermion sector should be
equal. This is indeed the case $(4+4)$ as the use of their field
equations reduces the number of degrees of freedom of the vectors
$A_{\mu}{}^\Lambda$ from three to two and of the Majorana fermions
from four to two.

As the gauge fields only appear through their field strengths,
(\ref{vecsh}) obviously realizes electric/magnetic duality. Before
considering this in detail, it is convenient to rewrite
(\ref{vecsh}) as
\begin{equation}\lb{abel}
\L = \L_{\mathrm{vector}} + \L_{\mathrm{matter}} +
\L_{\mathrm{\Om^4}} + \L_{\mathrm{Y}}\ ,
\end{equation}
where the different parts are given by
\begin{eqnarray}\label{abelf}
\L_{\mathrm{vector}} & = & \Big[ - \frac{1}{4} i \bF_{\Lambda
\Sigma} F^+_{\mu \nu}{}^\Sigma F^{+ \mu \nu \Lambda} +
\frac{1}{16} i \bF_{\Lambda \Sigma \Gamma} \bar{\Omega}^{ i
\Lambda} \gamma^{\mu \nu} \Omega^{j \Sigma} \vep_{ij} F_{\mu
\nu}^{+ \Gamma}\nn\\
& & + \frac{1}{256} i N^{\Delta \Omega} (\bF_{\Delta \Lambda
\Sigma} \bOm^{i \Lambda} \ga_{\mu \nu} \Om^{j \Sigma} \vep_{ij})
(\bF_{\Gamma \Xi \Omega} \bOm^{k \Gamma} \ga^{\mu \nu} \Om^{l \Xi}
\vep_{kl})\nn\\ & & + \mathrm{h.c.} \Big]\ ,
\end{eqnarray}\lb{abelm1}
\begin{eqnarray}
\L_{\mathrm{matter}} & = & - N_{\Lambda \Sigma} \p_{\mu}
X^{\Lambda} \p^{\mu} \bX^{\Sigma} - \frac{1}{4} N_{\Lambda \Sigma}
(\bOm^{i \Lambda} \spa
\Om_i{}^{\Sigma} + \bOm_i{}^{\Lambda} \spa \bOm^{i \Sigma})\nn\\
& & - \frac{1}{4} i (\bOm_i{}^{\Lambda} \spa F_{\Lambda \Sigma}
\Om^{i \Sigma} - \bOm^{i \Lambda} \spa \bF_{\Lambda \Sigma}
\Om_i{}^{\Sigma})\ ,
\end{eqnarray}
\begin{eqnarray}\lb{abelm2}
\L_{\mathrm{\Om^4}} & = & - \frac{1}{384} i (\bF_{\Lambda \Sigma
\Gamma \Xi} - 3i N^{\Delta \Omega} \bF_{\Delta (\Lambda \Gamma}
\bF_{\Sigma \Xi) \Omega}) \bOm^{i \Lambda} \ga_{\mu \nu} \Om^{j
\Sigma} \vep_{ij}
\bOm^{k \Gamma} \ga^{\mu \nu} \Om^{l \Xi} \vep_{kl}\nn\\
& & + \mathrm{h.c.} - \frac{1}{16} N^{\Delta \Omega} F_{\Delta
\Lambda \Sigma} \bF_{\Gamma \Xi \Omega} \bOm^{i \Gamma} \Om^{j
\Xi} \bOm_i{}^{\Lambda} \Om_j{}^{\Sigma}\ ,
\end{eqnarray}
\begin{eqnarray}\lb{abely}
\L_{\mathrm{Y}} & = & \frac{1}{8} i \varepsilon_{ik}
\varepsilon_{jl} \bF_{\Lambda \Sigma} Y^{ij \Lambda} Y^{kl \Sigma}
- \frac{1}{8}i \vep^{ik}
\vep^{jl} F_{\Lambda \Sigma} Y_{ij}{}^{\Lambda} Y_{kl}{}^{\Sigma}\nonumber\\
& & - \frac{1}{8} i \varepsilon_{ik} \varepsilon_{jl} \bF_{\Lambda
\Sigma \Gamma} Y^{kl \Lambda} \bar{\Omega}^{i \Sigma} \Omega^{j
\Gamma} + \frac{1}{8} i \varepsilon^{ik} \varepsilon^{jl}
F_{\Lambda \Sigma \Gamma} Y_{kl}{}^{\Lambda}
\bar{\Omega}_i{}^{\Sigma}
\Omega_j{}^{\Gamma}\nn\\
& & - \frac{1}{32} \vep_{ik} \vep_{jl} N^{\Lambda \Sigma}
\bF_{\Lambda \Omega \Delta} \bF_{\Sigma \Gamma \Xi} \bOm^{i
\Omega} \Om^{j \Delta} \bOm^{k \Gamma} \Om^{l \Xi} + \mathrm{h.c.}\nn\\
& & + \frac{1}{16} N^{\Lambda \Sigma} \bF_{\Lambda \Omega \Delta}
F_{\Sigma  \Gamma \Xi} \bOm^{i \Omega} \Om^{j \Delta}
\bOm_i{}^{\Gamma} \Om_j{}^{\Xi}\ .
\end{eqnarray}

At first we restrict ourselves to the on-shell vector multiplets,
so with $Y_{ij}{}^\Lambda$ integrated out and $\L_{\mathrm{Y}} =
0$, as the electric/magnetic duality relations contained in
(\ref{vecsh}) are most clear for this version of the action. Later
on we include the $Y_{ij}{}^\Lambda$ in the electric/magnetic
duality framework. Due to the reality condition (\ref{ry}) this is
a non-trivial exercise. Nevertheless, as a new result, we show
that the $Y_{ij}{}^\Lambda$ can be incorporated in a natural way.

Observe that the Lagrangian $\L = \L_F + \L_{\mathrm{matter}} +
\L_{\mathrm{\Om^4}}$ is of the form (\ref{acsie}) (except for the,
in this respect irrelevant, lack of a gravitational coupling),
with
\begin{eqnarray}\lb{nistem}
\tau_{\Lambda \Sigma} & = & F_{\Lambda \Sigma}\ ,\nn\\
O_{\mu \nu \Lambda}^+ & = & - \frac{1}{8} \bF_{\Lambda \Sigma
\Gamma} \bar{\Omega}^{i \Sigma} \gamma_{\mu \nu} \Omega^{j
\Gamma} \vep_{ij}\ ,\nn\\
\L' & = & \L_{\mathrm{matter}} + \L_{\mathrm{\Om^4}}\ ,
\end{eqnarray}
and so $G_{\mu \nu \Lambda}^+$ is given by
\begin{eqnarray}\label{ggedef}
G_{\mu \nu \Lambda}^+ = i \vep_{\mu \nu \rho \sigma} \frac{\de
\L_{\mathrm{vector}}}{F^+_{\rho \sigma}{}^\Lambda} = \bF_{\Lambda
\Sigma} F^+_{\mu \nu}{}^\Sigma - \frac{1}{8} \bF_{\Lambda \Sigma
\Gamma} \bOm^{i \Sigma} \ga_{\mu \nu} \Om^{j \Gamma} \vep_{ij}\ .
\end{eqnarray}

Recall that an electric/magnetic duality transformation of a
Lagrangian of this type is induced by the transformations
(\ref{tranfob}) on the objects in (\ref{nistem}). Leaving the
matter fields $X^{\Lambda}$ and $\Omega^{\Lambda}$ as they are
then gives a Lagrangian in which the matrix-components $U$, $Z$,
$W$, $V$ of the symplectic transformation explicitly appear. Since
electric/magnetic duality preserves supersymmetry (as we saw in
chapter \ref{ch1}), it should nevertheless be possible to rewrite
the dual Lagrangian back in the form (\ref{abel}), so without the
components of the symplectic matrix appearing. Note that the fact
that we can write the dual Lagrangian in the form (\ref{abel}),
implies the existence of a dual holomorphic function $\tF$, which
may be a different one then the one started from.

To obtain the dual Lagrangian in the formulation (\ref{abel}), a
suitable field redefinition of the matter fields $X^{\Lambda}$ and
$\Omega^{\Lambda}$ is required. To find out what these field
redefinitions are, and which new function corresponds to the dual
Lagrangian, we consider the transformed version of $F_{\Lambda
\Sigma}$,
\begin{eqnarray}\lb{fls}
\tF_{\Lambda \Sigma} (\tX) & = & [W_{\Lambda \Gamma} +
V_\Lambda{}^\Xi F_{\Xi \Gamma}] [\S^{-1}]^\Gamma{}_\Sigma\ ,
\end{eqnarray}
where $\S^\Lambda{}_\Sigma = U^\Lambda{}_\Sigma + Z^{\Lambda
\Gamma} F_{\Gamma \Sigma}$. It is easily seen that (\ref{fls}) is
satisfied when $\tF_{\Lambda \Sigma} (\tX)$ is taken to be the
derivative of a new $\tF_{\Lambda}$ with respect to a new
$\tX^{\Sigma}$, with $\tF_{\Lambda}$ and $\tX^{\Sigma}$ given by
\begin{eqnarray}\label{xenf}
\( \begin{array}{c} \tX^{\Lambda}\\
\tF_{\Lambda}
\end{array} \) = \( \begin{array}{cc} U^\Lambda{}_\Sigma & Z^{\Lambda \Sigma}\\
W_{\Lambda \Sigma} & V_\Lambda{}^\Sigma
\end{array} \) \( \begin{array}{c} X^{\Sigma}\\ F_{\Sigma}
\end{array} \)\ ,
\end{eqnarray}
i.e. the scalars $X^{\Lambda}$ and the derivatives of the function
$F(X)$, $F_{\Lambda} (X)$, transform as a symplectic vector
$(X^{\Lambda}, F_{\Lambda})$. Using the $X^{\Lambda}$-dependence
of $F_{\Lambda}$, the appropriate field redefinition of the
scalars can be read of from the first line of (\ref{xenf}). The
second line of (\ref{xenf}) contains the information about the new
function $\tF (\tX)$. This new function is defined (up to an
irrelevant constant) by requiring $\tF_{\Lambda}$ to be its
derivative (with respect to $\tX^{\Lambda}$).

It is easily shown that (\ref{xenf}) implies that
\begin{eqnarray}
\tF_{\Lambda \Sigma \Gamma} (\tX) & = & F_{\Xi \Delta
\Omega}[\S^{-1}]^\Xi{}_\Lambda [\S^{-1}]^\Delta{}_\Sigma
[\S^{-1}]^\Omega{}_\Gamma\ ,\nn\\
\tN_{\Lambda \Sigma} (\tX) & = & N_{\Xi \Delta}
[\S^{-1}]^\Xi{}_\Lambda [\bar{\S}^{-1}]^\Delta{}_\Sigma\ .
\end{eqnarray}

The correct field redefinition of $\Omega^{\Lambda}$ follows from
a supersymmetry transformation of (\ref{xenf}). We obtain
\begin{eqnarray}\lb{ome}
\( \begin{array}{c} \tOm_i{}^{\Lambda}\\
\tF_{\Lambda \Gamma} \tOm_i{}^{\Gamma}
\end{array} \) = \( \begin{array}{cc} U^\Lambda{}_\Sigma & Z^{\Lambda \Sigma}\\
W_{\Lambda \Sigma} & V_\Lambda{}^\Sigma
\end{array} \) \( \begin{array}{c} \Om_i{}^{\Sigma}\\ F_{\Sigma \Xi}
\Om_i{}^{\Xi}
\end{array} \)\ .
\end{eqnarray}
Note that all information is already contained in the first line;
the second line is just consistent with the first (using
(\ref{xenf})).

To convince oneself that (\ref{xenf}) and (\ref{ome}) give indeed
the right field redefinitions and the correct new function, one
should take $\L_F + \L_{\mathrm{matter}} + \L_{\mathrm{\Om^4}}$,
perform an electric/magnetic duality transformation and write the
result in terms of $\tF_{\mu \nu}^{\Lambda}$,
$\tOm_i{}^{\Lambda}$, $\tX^\Lambda$, $\tF_{\Lambda \Sigma}$ and
further derivatives of $\tF (\tX)$. The `tilded' version of $\L_F
+ \L_{\mathrm{matter}} + \L_{\mathrm{\Om^4}}$ is then obtained.

Starting from a Lagrangian described by a holomorphic function
$F(X)$ we thus find a new Lagrangian determined by a new function
$\tF(\tX)$, as defined by (\ref{xenf}). The new and original
function are related in the following way
\begin{eqnarray}\lb{fx}
F(X) \longrightarrow \tF (\tX) & = & F(X) - \half F_{\Lambda} (X)
X^{\Lambda} + \half \tF_{\Lambda}
(\tX) \tX^{\Lambda} \nn\\
& = & F(X) - \half F_{\Lambda} (X) X^{\Lambda} + \half
(U^TW)_{\Lambda \Sigma} X^\Lambda X^\Sigma\nn\\
& & + \half(U^TV + W^TZ)_\Lambda{}^\Sigma X^\Lambda F_\Sigma (X)\nn\\
& & + \half (Z^TV)^{\Lambda \Sigma} F_\Lambda (X) F_\Sigma (X)\ ,
\end{eqnarray}
up to a (irrelevant) constant. Note the similarity between the
symplectic behavior of the function $F(X)$ and the transformation
property of the Lagrangian (\ref{ts}); The fact that the
Lagrangian does not transform as a scalar $\tL (\tF) \neq \L (F)$
translates into the statement that the function $F(X)$ does not do
so, $\tF (\tX) \neq F (X)$.

Also the new function may be a different one than the original
function, in the sense that $\tF (\tX) \neq F (\tX)$. So different
functions $F(X)$ and therefore different Lagrangians belong to the
same equivalence class. The electric/magnetic duality
transformations relating these different functions and Lagrangians
are duality equivalences rather than duality invariances. However
duality invariances ($\tF (\tX) = F (\tX)$) do arise as special
cases. In fact, these are of particular importance in chapter
\ref{ch4}.


 \vspace{5mm}

\textbf{Electric/magnetic duality and} $\mathbf{Y_{ij}}$

\vspace{5mm}

So far we have considered the on-shell version of the vector
multiplets, i.e. with the auxiliary fields $Y_{ij}{}^\Lambda$
eliminated from the action by their field equations. Now we turn
to the electric/magnetic duality behavior of  off-shell vector
multiplets, so in the presence of $Y_{ij}{}^\Lambda$.

Like we determined the symplectic behavior of $\Om_i{}^\Lambda$
from a supersymmetry transformation on $(X^\Lambda, F_\Lambda)$,
we can extract the appropriate transformation of
$Y_{ij}{}^\Lambda$ under symplectic transformations from a
supersymmetry transformation on (\ref{ome}). The result is
\begin{eqnarray}\label{yenz}
\( \begin{array}{c} Y_{ij}{}^{\Lambda}\\
Z_{ij \Lambda} \end{array} \) \longrightarrow \( \begin{array}{c} \tY_{ij}{}^\Lambda\\
\tZ_{ij \Lambda}
\end{array} \) = \( \begin{array}{cc} U^\Lambda{}_\Sigma & Z^{\Lambda \Sigma}\\
W_{\Lambda \Sigma} & V_\Lambda{}^\Sigma
\end{array} \) \( \begin{array}{c} Y_{ij}{}^{\Sigma}\\ Z_{ij \Sigma}
\end{array} \)\ ,
\end{eqnarray}
where
\begin{eqnarray}\lb{defiz}
Z_{ij \Lambda} = F_{\Lambda \Sigma} Y_{ij}{}^{\Sigma} - \half
F_{\Lambda \Sigma \Gamma} \bOm_i{}^{\Sigma} \Om_j{}^{\Gamma}\ .
\end{eqnarray}
We note that
\begin{eqnarray}
Z_{ij \Lambda} =  4 i \vep_{ik} \vep_{jl} \frac{\delta
\L_{\mathrm{Y}}}{\delta Y_{kl}{}^{\Lambda}}\ ,
\end{eqnarray}
(with $Y_{ij}{}^\Lambda$ and $Y^{ij \Lambda}$ treated as
independent fields) which becomes important short\-ly. One can
check that the rotation (\ref{yenz}) is consistent with a
transformation of $Z_{ij \Lambda}$ as expressed in terms of the
matter fields on which it depend, i.e. $\tZ_{ij \Lambda} =
\tF_{\Lambda \Sigma} \tY_{ij}{}^{\Sigma} - \half \tF_{\Lambda
\Sigma \Gamma} \btOm_i^{}{\Sigma} \tOm_j{}^{\Gamma}$.

However, taking the complex conjugate of (\ref{yenz}) and
(\ref{defiz}) gives
\begin{eqnarray}\label{yenzc}
\( \begin{array}{c} Y^{ij \Lambda}\\
Z^{ij}{}_\Lambda \end{array} \) \longrightarrow \( \begin{array}{c} \tY^{ij \Lambda}\\
\tZ^{ij}{}_\Lambda
\end{array} \) = \( \begin{array}{cc} U^\Lambda{}_\Sigma & Z^{\Lambda \Sigma}\\
W_{\Lambda \Sigma} & V_\Lambda{}^\Sigma
\end{array} \) \( \begin{array}{c} Y^{ij \Sigma}\\ Z^{ij}{}_\Sigma
\end{array} \)\ ,
\end{eqnarray}
with
\begin{eqnarray}\lb{defizc}
Z^{ij}{}_{\Lambda} = \bF_{\Lambda \Sigma} Y^{ij \Sigma} - \half
\bF_{\Lambda \Sigma \Gamma} \bOm^{i \Sigma} \Om^{j \Gamma} = - 4 i
\vep^{ik} \vep^{jl} \frac{\delta \L_{\mathrm{Y}}}{\delta Y^{kl
\Lambda}}\ ,
\end{eqnarray} which is obviously incompatible with the reality
conditions (\ref{ry}).

How to understand this? As mentioned earlier, similar to the
Bianchi identity of $F_{\mu \nu}{}^\Lambda$, the reality condition
on $Y_{ij}{}^\Lambda$ arises in the reduction of a chiral
multiplet to a vector multiplet. As the Bianchi identity of
$F_{\mu \nu}{}^\Lambda$ transforms under electric/magnetic
duality, so does the embedding of the vector multiplet in the
unreduced chiral multiplet, and therefore the reality condition on
$Y_{ij}{}^\Lambda$ should transform as well.

In fact, the reality conditions on $Y_{ij}{}^\Lambda$ behave very
similar to the Bianchi identities of $F_{\mu \nu}{}^\Lambda$. To
show this, we write them in combination with the field equations
of $Y_{ij}{}^\Lambda$ as
\begin{eqnarray}\lb{ryfy}
\( \begin{array}{c} Y_{ij}{}^{\Lambda}\\ Z_{ij \Lambda}
\end{array} \) - \vep_{ip} \vep_{jq} \( \begin{array}{c}
Y^{pq \Lambda}\\ Z^{pq}{}_{\Lambda} \end{array} \) = 0\ .
\end{eqnarray}
From (\ref{ryfy}) it follows that the transformations (\ref{yenz})
and (\ref{yenzc}) rotate the set of equations formed by the
reality conditions on $Y_{ij}{}^\Lambda$ and the set of field
equations of $Y_{ij}{}^\Lambda$, whereas the combined set remains
equivalent. Observe the similarities with the action of
electric/magnetic duality on the vector field strengths: The role
of the Bianchi identities of $F_{\mu \nu}{}^\Lambda$ / equations
of motion of $A_\mu{}^\Lambda$ in the gauge field sector, is
played by the reality conditions on $Y_{ij}{}^\Lambda$ / equations
of motion of $Y_{ij}{}^\Lambda$ in the $Y_{ij}{}^\Lambda$-sector.

Also like in the gauge field sector, the dual version of
$\L_{\mathrm{Y}}$ follows from $\tZ_{ij \Lambda} \equiv 4 i
\vep_{ik} \vep_{jl} \frac{\delta \tL_{\mathrm{Y}}}{\delta
\tY_{kl}{}^{\Lambda}}$. It takes the same form as (\ref{abely}),
but then in terms of the fields $\tX^\Lambda$, $\tOm_i{}^\Lambda$
and $\tY_{ij}{}^\Lambda$.

\vspace{5mm}

\textbf{Superconformal vector multiplets}

\vspace{5mm}

To obtain the vector multiplet as a representation of the $N=2$
superconformal algebra, one needs to introduce appropriate
transformation rules for $X$, $A_\mu$, $\Om_i$ and $Y_{ij}$ under
dilations, $U(1)_R$, special supersymmetry and special conformal
transformations (recall that the $SU(2)_R$ transformation rules
are already given below (\ref{ry})).

The resulting weights of the fields under dilatations and chiral
$U(1)_R$ can be found in Appendix \ref{supcon}. The special
supersymmetry transformations act on $\Om_i$ only,
\begin{eqnarray}
\de \Om_i = 2 X \eta_i\ ,
\end{eqnarray}
where $\eta_i$ is the special supersymmetry parameter, while the
special conformal transformations turn out to be trivial.

The Lagrangian corresponding to $n$ (rigidly) superconformal
vector multiplets is also given by (\ref{vecsh}). However, to have
the dilatational and $U(1)_R$ symmetry realized, the function
$F(X)$ should be a homogeneous function of second degree,
\begin{eqnarray}
F (\lambda X) = \lambda^2 F(X)\ .
\end{eqnarray}
This implies the following relations between $F$ and its
derivatives,
\begin{eqnarray}
F(X) & = & \half F_{\Lambda} X^{\Lambda}\ ,\nn\\
F_{\Lambda} & = & F_{\Lambda \Sigma} X^{\Sigma}\ ,\nn\\
F_{\Lambda \Sigma \Gamma} X^{\Gamma} & = & 0\ ,\nn\\
F_{\Lambda \Sigma \Gamma} & = & - F_{\Lambda \Sigma \Gamma \Xi}
X^{\Xi}\ .
\end{eqnarray}

Using the fact that $F(X)$ is a homogeneous function of second
degree, it can be shown that the K\"{a}hler space parameterized by
the scalars admits a homothetic Killing vector of weight two,
$(\chi^{\Lambda}, \bar{\chi}^{\Lambda})$,
\begin{eqnarray}\lb{hkv}
\chi_{\Lambda} \equiv \p_{\Lambda} K = N_{\Lambda \Sigma}
\bar{X}^{\Sigma}\ ,\quad D_{\Lambda} \bar{\chi}_{\Sigma} +
D_{\bar{\Sigma}} \chi_{\Lambda} = 2 N_{\Lambda \Sigma}\ ,
\end{eqnarray}
where the derivatives $D_\Lambda$ contain the hermitian
connection,
\begin{eqnarray}
\Gamma_{\Sigma \Gamma}{}^\Lambda = N^{\Lambda \Xi} \p_{\Sigma}
N_{\Gamma \Xi} = -i N^{\Lambda \Xi} F_{\Sigma \Gamma \Xi}\ ,\quad
\Gamma_{\bar{\Sigma} \bar{\Gamma}}{}^{\bar{\Lambda}} = N^{\Lambda
\Xi} \p_{\bar{\Sigma}} N_{\Xi \Gamma} = i N^{\Lambda \Xi}
\bF_{\Sigma \Gamma \Xi}\ .
\end{eqnarray}
The existence of the homothetic Killing vector implies that the
sigma model is scale invariant, which is part of the
superconformal invariance of the action as a whole.

Using the complex structure, $J^\Lambda{}_\Sigma = i
\de^\Lambda{}_\Sigma$, one can also construct the Killing vector
$(k^{\Lambda}, \bar{k}^{\Lambda})$,
\begin{eqnarray}
k^{\Lambda} = - J^\Lambda{}_\Sigma \chi^{\Sigma} = -i X^\Lambda\
,\quad \bar{k}^{\Lambda} = - J^{\bar{\Lambda}}{}_{\bar{\Sigma}}
\chi^{\bar{\Sigma}}= i \bX^\Lambda\ ,\nn\\
D_{\Lambda} \bar{k}_{\Sigma} + D_{\bar{\Sigma}} k_{\Lambda} = 0\
,\quad D_{\Lambda} k_{\Sigma} + D_{\Sigma} k_{\Lambda} = 0\ ,
\end{eqnarray}
which is associated to the chiral $U(1)_R$ transformations of the
scalar fields and the $U(1)_R$ symmetry of the action.

In fact, the $2n$-dimensional target-space of the scalar fields of
$n$ rigidly superconformal vector multiplets is a cone over a
$(2n-1)$-dimensional so-called Sasakian space \cite{GR}, which is
itself a $U(1)$ fibration over a $(2n-2)$-dimensional
\emph{special K\"{a}hler} manifold. This special K\"{a}hler
manifold is important in the context of Poincar\'{e} supergravity.
The target-space of the vector multiplet scalars of the rigidly
superconformal model is referred to as the \emph{special
K\"{a}hler cone}.

The constraints following from demanding superconformal
invariance, have no influence on the electric/magnetic duality
relations contained in (\ref{vecsh}). Only note that the
symplectic vector of the complex scalars now is of the form
$(X^\Lambda, F_\Lambda) = (X^\Lambda, F_{\Lambda \Sigma}
X^\Sigma)$.

\vspace{5mm}

This concludes our treatment of the vector multiplet as a
representation of $N=2$ supersymmetry. We have analyzed both its
rigidly supersymmetric and its rigidly superconformal version.
Later on we consider its role in Poincar\'{e} supergravity.
However, we first turn to another representation of $N=2$
supersymmetry, the hypermultiplet.

\section{Hypermultiplets}\label{hypm}

A hypermultiplet comprises four real scalar fields and two
Majorana spinors. On-shell this adds up to $4+4$ degrees of
freedom. Off-shell two Majorana spinors have $8$ fermionic degrees
of freedom, so at least four additional bosonic degrees of freedom
would be needed to have an off-shell multiplet. However, as it
turns out, there exists no unconstrained off-shell formulation of
a hypermultiplet in terms of a finite number of degrees of
freedom. Therefore we restrict ourselves to its on-shell version,
following \cite{dWKV}.

\vspace{5mm}

\textbf{Rigid hypermultiplets}

\vspace{5mm}

The supersymmetry transformation rules of the rigid hypermultiplet
are given by
\begin{eqnarray}\lb{trh}
\delta \phi^A & = & 2 ( \ga_{i \bal}^A \bep^i \ze^{\bal} + \bga_{
\al}^{A i} \bep_i \ze^{\al})\ ,\nn\\
\delta \ze^{\al} & = & V_{A i}^{\al} \spa \phi^A \ep^i - \de
\phi^A \Gamma_A{}^\al{}_\be \ze^{\be}\ ,\nn\\
\delta \ze^{\bal} & = & \bV_A^{i \bal} \spa \phi^A \ep_i - \de
\phi^A \Gamma_A{}^{\bal}{}_{\bbe} \ze^{\bbe}\ .
\end{eqnarray}
Here $\phi^A$ are the scalar fields, so $A$ runs from 1 to $4m$ in
case of a model with $m$ hypermultiplets. $\ze^{\al}$ $(\al = 1
,..., 2m)$ are the negative-chiral parts of the Majorana spinors.
Complex conjugation gives their positive chiral counterparts. The
latter are denoted by $\ze^{\bal}$. $\ga^A$, $V_A$ and
$\Gamma_A{}^\al{}_\be$ are $\phi$-dependent quantities. Their role
becomes clear shortly.

The rigidly supersymmetric Lagrangian formed by the hypermultiplet
degrees of freedom reads as
\begin{eqnarray}\lb{lrh}
\L & = & - \half g_{AB} \p_{\mu} \phi^A \p^{\mu} \phi^B\nn\\
& & - G_{\bal \be}(\bze^{\bal} \SDm \ze^{\be} + \bze^{\be} \SDm
\ze^{\bal}) - \vier W_{\bal \be \bga \de} \bze^{\bal} \ga_{\mu}
\ze^{\be} \bze^{\bga} \ga^{\mu} \ze^{\de}\ ,
\end{eqnarray}
where we used
\begin{eqnarray}
\D_{\mu} \ze^{\al} = \p_{\mu} \ze^{\al} + \p_{\mu} \phi^{A}
\Gamma_A{}^{\al}{}_\be \ze^{\be}\ .
\end{eqnarray}

The Lagrangian (\ref{lrh}) and transformation rules (\ref{trh})
come with two sets of target-space equivalence transformations.
These are the target-space diffeomorphisms $\phi \rightarrow
\phi'(\phi)$ on the one hand and the reparameterizations of the
fermion `frame' $\ze^{\al} \rightarrow S^{\al}{}_{\be} (\phi)
\ze^{\be}$ on the other hand (the latter are accompanied by
appropriate redefinitions of other quantities carrying indices
$\al$ or $\bal$, i.e for instance $G_{\bal \be} \rightarrow
[\bar{S}^{-1}]^{\bga}{}_{\bal} [S^{-1}]^{\de}{}_{\be} G_{\bga
\de}$). This explains the appearance of the objects
$\Gamma_A{}^\al{}_\be$: They are the connections associated with
the reparameterizations of the fermion frame. Furthermore, observe
that the Lagrangian is invariant under the $U(1)_R$ symmetry
group, which acts by chiral transformations on the fermion fields,
while the $SU(2)_R$ symmetry can only be realized when the
target-space has an $SU(2)$ isometry.

The target-space metric of the non-linear sigma model
parameterized by the scalars is a hyper-K\"{a}hler space
\cite{BW,AF,Ba}, which, by definition, allows the existence of
three anticommuting complex structures that are covariantly
constant with respect to the Levi-Civita connection \cite{S,I,YK}.

The tensor $W$ is defined by
\begin{eqnarray}
W_{\bal \be \bga \de} = R_{AB}{}^{\bep}{}_{\bga} \ga_{i \bal}^A
\bga_{\be}^{B i} G_{\bep \de} = \half R_{ABCD} \ga_{i \bal}^A
\bga_{\be}^{B i} \ga_{j \bga}^C \bga_{\de}^{D j}\ ,
\end{eqnarray}
where $R_{AB}{}^{\al}{}_{\be}$ and $R_{ABCD}$ are the curvatures
corresponding to $\Gamma_A{}^{\al}{}_{\be}$ and the Levi-Civita
connection $\Gamma_{AC}{}^B$. The curvature
$R_{AB}{}^{\al}{}_{\be}$ takes its values in $Sp(m) \sim u Sp (2m,
\mathbb{C})$.

The target-space metric $g_{AB}$, the tensors $\ga^A$, $V_A$ and
the fermionic hermitian metric $G_{\bal \be}$ are all covariantly
constant with respect to the Christoffel connection and the
connections $\Gamma_A{}^{\al}{}_{\be}$ and $\Gamma_{AC}{}^B$.
There are the following relations amongst them
\begin{eqnarray}
\ga_{i \bal}^A \bV^{j \bal}_B + \bga_{\al}^{A j} V_{B i}^{\al} =
\de_I^J \de^A_B\ ,\nn\\
g_{AB} \ga_{i \bal}^B = G_{\bal \be} V^{\be}_{A i}\ ,\quad
\bV_A^{i \bal} \ga_{j \bbe}^A = \de_i^j \de_{\bbe}^{\bal}\ .
\end{eqnarray}

The complex structures of the hyper-K\"{a}hler target-space are
spanned by the antisymmetric covariantly constant target-space
tensors
\begin{eqnarray}
J_{AB}^{ij} = \ga_{A k \bal} \vep^{k(i} \bV_B^{j)\bal}\ ,
\end{eqnarray}
which are symmetric in $i, j$ and satisfy
\begin{eqnarray}
(J_{ij})_{AB} \equiv (J_{AB}^{ij})^* = \vep_{ik} \vep_{jl}
J_{AB}^{kl}\ ,\quad J_A^{ij C} J_{CB}^{kl} = \half \vep^{i(k}
\vep^{l)j} g_{AB} + \vep^{(i(k} J_{AB}^{l)j)}\ .
\end{eqnarray}
In addition we note the following useful identities,
\begin{eqnarray}
\ga_{A i \bal} \bV_B^{j \bal} = \vep_{ik} J_{AB}^{kj} + \half
g_{AB} \de_j^i\ ,\quad J_{AB}^{ij} \ga_{\bal k}^B = - \de_k^{(i}
\vep^{j)l} \ga_{A l \bal}\ .
\end{eqnarray}

Other important objects are the covariantly constant antisymmetric
tensors
\begin{eqnarray}
\Omega_{\bal \bbe} = \half \vep^{ij} g_{AB} \ga^A_{i \bal}
\ga^B_{j \bbe}\ ,\quad \bOm^{\bal \bbe} = \half \vep_{ij} g^{AB}
\bV^{i \bal}_A \bV_B^{j \bbe}\ .
\end{eqnarray}
Using these objects we can derive a reality condition on $V$ and
$\ga$,
\begin{eqnarray}
\vep_{ij} \Om_{\bal \bbe} \bV^{j \bbe}_A = g_{AB} \ga_{i \bal}^B =
G_{\bal \be} V_{A i}^{\be}\ .
\end{eqnarray}
This leads to
\begin{eqnarray}
g^{AB} V_{Ai}^\al V_{B j}^\be = \vep_{ij} \Om^{\al \be}\ ,\quad
g_{AB} \ga_{i \bal}^A \ga_{j \bbe}^B = \vep_{ij} \Om_{\bal \bbe}\
,
\end{eqnarray}
and the relation,
\begin{eqnarray}
\vep_{ij} \Omega_{\bal \bbe} \bV_A^{i \bal} \bV_B^{j \bbe} =
g_{AB}\ ,
\end{eqnarray}
which makes that $V_A$ can be interpreted as the quaternionic
vielbein of the target-space, with $\ga^A$ being the inverse
vielbein.

\vspace{5mm}

\textbf{Superconformal hypermultiplets}

\vspace{5mm}

Next we consider the superconformal version of the hypermultiplet.
The dilatational and $U(1)$ weights of $\phi^A$ and $\ze^{\al}$
can be found in Appendix \ref{supcon}. $SU(2)_R$ acts as a set of
isometries of the target-space (as we will see below).
Furthermore, the special conformal transformations are again
trivial and special supersymmetry works as
\begin{eqnarray}\lb{trsh}
\de \ze^{\al} = \chi^B (\phi) V^{\al}_{B i} (\phi) \eta^i\ .
\end{eqnarray}
$\chi^B$ is a (real) homothetic Killing vector of the space
parameterized by the scalar fields, as we will see shortly.

The corresponding Lagrangian is of the same form as in the rigidly
supersymmetric case (\ref{lrh}), however, the superconformal
couplings imply that the manifold parameterized by the
hypermultiplet scalars is a special type of hyper-K\"{a}hler
manifold, a \emph{hyper-K\"{a}hler cone}.

The (real) metric of a hyper-K\"{a}hler cone is the second
derivative of a function $\chi$,
\begin{eqnarray}
D_A \p_B \chi = g_{AB}\ .
\end{eqnarray}
This function is sometimes called the hyper-K\"{a}hler potential.
As mentioned above, the vector $\chi^A \equiv g^{AB} \p_B \chi$ is
a homothetic Killing vector of weight two,
\begin{eqnarray}
D_A \chi_B + D_B \chi_A = 2 g_{AB}\ ,
\end{eqnarray}
implying that the associated sigma model is indeed invariant under
dilatations. Furthermore, there exist three Killing vectors,
satisfying
\begin{eqnarray}
k^A_{ij} = J_{ij}^{AB} \chi_B\ ,
\end{eqnarray}
which realize the $SU(2)_R$ symmetry of the model.

Altogether it can be shown that the $4m$-dimensional
hyper-K\"{a}hler manifold parameterized by the scalars of $m$
superconformal hypermultiplets is a cone over a
$(4m-1)$-dimensional $3$-Sasakian manifold,  which is itself a
$Sp(1)$ fibration over a $(4m-4)$-dimensional so-called
\emph{quaternionic-K\"{a}hler} (QK) space \cite{dWKV}. As we will
see in the next section, this latter space is the one
parameterized by the scalars of the corresponding Poincar\'{e}
supergravity theory.

\vspace{5mm}

Observe the similarities between the target-spaces of the vector
multiplet and the hypermultiplet scalars. In both cases we find in
the superconformal framework a cone over a fibration ($U(1)_R$ for
the vector multiplet scalars and $Sp(1) \sim SU(2)_R$ for the
hypermultiplet scalars) of the manifold that becomes relevant in
the Poincar\'{e} supergravity context.

\section{Poincar\'{e} supergravity}\label{poin}

In the last two sections we considered the rigidly supersymmetric
and the superconformal version of the vector and the
hypermultiplet. The construction of the superconformal multiplets
was only a first step towards the formulation of $N=2$
Poincar\'{e} supergravity. How to finish this procedure is the
subject of this section.

First we gauge the superconformal symmetries present in the vector
and hypermultiplet sectors by introducing couplings to the
associated gauge fields of the Weyl multiplet. The vector and
hypermultiplet transformation rules this gives rise to can be
found in Appendix \ref{supcon}. The corresponding action is given
in \cite{deWit:1984pk} (the vector multiplet sector) and
\cite{dWKV} (the hypermultiplet sector). Here we simplify matters
and take only the bosonic sector into account. This way we obtain
\cite{dWRVs}
\begin{eqnarray}\lb{scb}
\L & = & N_{\Lambda \Sigma} \mathbb{D}^{\mu} X^{\Lambda}
\mathbb{D}_{\mu} \bX^{\Sigma} + \half g_{AB} \mathbb{D}^{\mu}
\phi^A
\mathbb{D}_{\mu} \phi^B\nn\\
& & - \frac{1}{6} K R - \frac{1}{6} \chi R + D(K - \half \chi)\nn\\
& & + [\vier i \bF_{\Lambda \Sigma} \F^+_{\mu \nu}{}^{\Lambda}
\F^{+
\mu \nu \Sigma}\nn\\
& & + \frac{1}{8} i \bF_{\Lambda} \F^+_{\mu \nu}{}^{\Lambda}
T^{\mu \nu}_{ij} \vep^{ij} + \frac{1}{32}i \bF T_{ij \mu \nu}
T^{\mu \nu}_{kl} \vep^{ij} \vep^{kl} + \mathrm{h.c.}]\ ,
\end{eqnarray}
where
\begin{eqnarray}
\F_{\mu \nu}{}^\Lambda & = & 2 \p_{[\mu} A_{\nu]}{}^\Lambda -
(\vier \vep_{ij} \bX^\Lambda T^{ij}_{\mu \nu} + \mathrm{h.c.})\ .
\end{eqnarray}
In (\ref{scb}) we omitted a term quadratic in $Y_{ij}^{\Lambda}$
as it is irrelevant for the present discussion. Furthermore, we
changed the overall sign for convenience. The covariant
derivatives in (\ref{scb}) are given by
\begin{eqnarray}
\mathbb{D}_{\mu} X^{\Lambda}  & = & \p_{\mu} X^{\Lambda} - b_{\mu}
\chi^{\Lambda}
- W_{\mu} k^{\Lambda}\ ,\nn\\
\mathbb{D}_{\mu} \phi^A & = & \p_{\mu} \phi^A - b_{\mu} \chi^A +
\half \V_\mu{}^i{}_k \vep^{jk} k_{ij}^A\ .
\end{eqnarray}
We recall that $b_{\mu}$, $W_{\mu}$ and $\V_{\mu}$ are the gauge
fields corresponding to dilatations, $U(1)_R$ and $SU(2)_R$
transformations respectively, whereas $\chi^{\Lambda}$ and
$\chi^A$, $k^{\Lambda}$ and $k_{ij}^A$ are the homothetic Killing
and Killing vectors associated with these symmetry
transformations.

The locally superconformal theory (\ref{scb}) is gauge equivalent
to $N=2$ Poincar\'{e} supergravity. To make this more explicit,
the gauge fields of $U(1)_R$ and $SU(2)_R$ (whose generators are
not part of the Poincar\'{e} supergroup), are eliminated
\footnote{Since the Lagrangian is invariant under special
conformal transformations and the dilatational gauge field
$b_{\mu}$ is the only field transforming non-trivially under this
symmetry, the Lagrangian must be independent of $b_{\mu}$.}, as
are the auxiliary fields $T_{\mu \nu}^{ij}$ and $D$. One then gets
\begin{eqnarray}\lb{lps}
\L & = & K \M_{\Lambda \bar{\Sigma}} \p_{\mu} X^{\Lambda} \p^{\mu}
\bX^{\Sigma} + \half
\chi G_{AB} \p_{\mu} \phi^A \p^{\mu} \phi^B\nn\\
& & - K (\frac{1}{6} R - \vier (\p_{\mu} \ln K)^2) -
\chi (\frac{1}{6} R - \vier (\p_{\mu} \ln \chi)^2)\nn\\
& & + (\vier i \N_{\Lambda \Sigma} F^+_{\mu \nu}{}^{\Lambda} F^{+
\mu \nu \Sigma} + \mathrm{h.c.})\ ,
\end{eqnarray}
while the potentials of the hyper-K\"{a}hler cone and the special
K\"{a}hler cone become equal,
\begin{eqnarray}\lb{chx}
\chi = 2K\ .
\end{eqnarray}

$\M_{\Lambda \bar{\Sigma}}$, $G_{AB}$ and $\N_{\Lambda \Sigma}$
are given by
\begin{eqnarray}\lb{matr}
\M_{\Lambda \bar{\Sigma}} & = & \frac{1}{K} (N_{ \Lambda \Sigma} -
\frac{1}{2K} \chi_{\Lambda} \bar{\chi}_{\Sigma} - \frac{1}{2K}
k_{\Lambda} \bar{k}_{\Sigma}) \ ,\nn\\
G_{AB} & = & \frac{1}{\chi} (g_{AB} - \frac{1}{2 \chi} \chi_A
\chi_B -  \frac{1}{\chi} k_{A ij} k^{ij}_B)\ ,\nn\\
\N_{\Lambda \Sigma} & = & \bF_{\Lambda \Sigma} + i
\frac{N_{\Lambda \Gamma} X^{\Gamma} N_{\Sigma \Xi}
X^{\Xi}}{N_{\Delta \Omega} X^{\Delta} X^{\Omega}}\ .
\end{eqnarray}
The factor $\chi$ in (\ref{lps}) can be absorbed in the vierbein
by the rescaling $e_{\mu}{}^a \longrightarrow \sqrt{2 / \chi} \,
e_{\mu}{}^a$, such that a scale invariant metric is obtained. The
Lagrangian then takes the Poincar\'{e} supergravity form
\begin{eqnarray}\label{poinact}
\L & = & - \half R + \M_{\Lambda \bar{\Sigma}} \p_{\mu}
X^{\Lambda} \p^{\mu} \bX^{\Sigma} + G_{AB} \p_{\mu} \phi^A
\p^{\mu} \phi^B\nn\\
& & + \vier i \N_{\Lambda \Sigma} F^+_{\mu \nu}{}^{\Lambda} F^{+
\mu \nu \Sigma} + \mathrm{h.c.}\ .
\end{eqnarray}
To have the correct signs of the kinetic terms, the special
K\"{a}hler metric $\M_{\Lambda \bar{\Sigma}}$ and the QK metric
$G_{AB}$ should be negative definite.

The homothetic Killing and Killing vectors of the cones correspond
to null-vectors of $\M_{\Lambda \bar{\Sigma}}$ and $G_{AB}$,
\begin{eqnarray}
\M_{\Lambda \bar{\Sigma}} \chi^{\Lambda} = \M_{\Lambda
\bar{\Sigma}}
k^{\Lambda} = 0\ ,\nn\\
G_{AB} \chi^B = G_{AB} k_{ij}^B = 0\ .
\end{eqnarray}

Note that the vierbein rescaling implies that we have taken
positive cone potentials. It follows that the cone metrics
$g_{AB}$ and $N_{\Lambda \Sigma}$ are mostly negative, but
positive in the (homothetic) Killing directions. Furthermore, in
\cite{CKvPDFdW} it is shown that in case $\M_{\Lambda
\bar{\Sigma}}$ is negative definite the vector fields come with
positive kinetic energy.

Starting from $n$ superconformal vector multiplets $\M_{\Lambda
\bar{\Sigma}}$ thus describes a $(2n-2)$-dimensional space. This
is a special K\"{a}hler manifold (by definition). It can be
parameterized in terms of $n-1$ complex coordinates $z^A (A = 1,
..., n-1)$ by letting $X^{\Lambda}$ be proportional to some
holomorphic sections $Z^{\Lambda} (z)$ of the projective space $P
C^n$ \cite{CDF,DFF}. Similarly, from $m$ superconformal
hypermultiplets the metric $G_{AB}$ arises, which corresponds to a
$(4m -4)$-dimensional space being necessarily of the
quaternionic-K\"{a}hler type. A procedure to describe this QK
space in terms of $4m-4$ coordinates was explained in
\cite{dWRVf01}. It involves the fixing of the dilatational and
$SU(2)_R$ symmetries.

Altogether we find that the resulting Poincar\'{e} supergravity
model, descending from $n$ vector and $m$ hypermultiplets coupled
to the Weyl multiplet, contains $2n$ (from the vector gauge
fields) $+(2n -2)$ (from the complex vector multiplet scalars)
$+(4m-4)$ (from the real hypermultiplet scalars) $+2$ (from the
metric) $= 4(n+m-1)$ bosonic degrees of freedom. A more complete
analysis - including the fermions - shows that they constitute the
bosonic sector of $n-1$ vector, $m-1$ hyper- and the supergravity
multiplet of $N=2$ Poincar\'{e} supergravity (the latter contains
one of the vector fields, called the graviphoton).

The fields disappearing in the process of going from a rigidly
superconformal model to Poincar\'{e} supergravity are compensators
for the symmetries that are in the superconformal group, but not
in the Poincar\'{e} supergroup.

\vspace{3mm}

To finish with, let us discuss the electric/magnetic duality
properties of the Lagrangian (\ref{lps}). Being of the form
(\ref{acsie}) it obviously exhibits such a type of duality. In
terms of the objects of (\ref{acsie}) we have
\begin{eqnarray}\lb{taup}
\tau_{\Lambda \Sigma} = \bmN_{\Lambda \Sigma}\ ,
\end{eqnarray}
whereas $O_{\mu \nu}^{\Lambda}$ vanishes in the absence of
fermions.

Compared to the rigidly superconformal case the complex coupling
matrix has acquired a non-holomorphic part, which is due to the
coupling to, and elimination of, the auxiliary field $T_{\mu
\nu}^{ij}$. This turns out to have some consequences.

First we recall from chapter \ref{ch1} that $\tau_{\Lambda
\Sigma}$ transforms as
\begin{eqnarray}
\tau_{\Lambda \Sigma} & \longrightarrow & \tilde{\tau}_{\Lambda
\Sigma} \equiv ((W+V \tau )(U+ Z \tau )^{-1})_{\Lambda \Sigma}\ ,
\end{eqnarray}
which implies that $(U+ Z \tau )$ should be invertible to have a
well-defined electric/magnetic duality transformation. In the
superconformal case $\tau_{\Lambda \Sigma} = F_{\Lambda \Sigma}$,
such that
\begin{eqnarray}
U^\Lambda{}_\Sigma + Z^{\Lambda \Gamma} \tau_{\Gamma \Sigma} =
U^\Lambda{}_\Sigma + Z^{\Lambda \Gamma} F_{\Gamma \Sigma} =
(\frac{\de \tX}{\de X})^\Lambda{}_\Sigma\ .
\end{eqnarray}
As invertibility of $(\frac{\de \tX}{\de X})^\Lambda{}_\Sigma$
suffices (and is needed) to have a formulation of the dual theory
in terms of a dual function $\tF (\tX)$, we thus find in the
superconformal case that all allowed electric/magnetic duality
rotations yield models that can be formulated in terms of a dual
function.

We then consider the Poincar\'{e} supergravity case, so with the
matrix $\tau_{\Lambda \Sigma}$ given by (\ref{taup}). Also in this
case electric/magnetic duality transformations give models
determined by a dual function when the matrix $U^\Lambda{}_\Sigma
+ Z^{\Lambda \Gamma} F_{\Gamma \Sigma}$ is invertible. However,
the condition to have a well-defined electric/magnetic duality
rotation in supergravity is the requirement of invertibility of
the \emph{different} matrix $U^\Lambda{}_\Sigma + Z^{\Lambda
\Gamma} \bmN_{\Gamma \Sigma}$. Furthermore,
\begin{eqnarray}
[(U + Z F)^{-1}]^\Lambda{}_\Sigma \ \mathrm{exists} \rightarrow
[(U + Z \bmN)^{-1}]^\Lambda{}_\Sigma \ \mathrm{exists}\ .
\end{eqnarray}
whereas,
\begin{eqnarray}
[(U + Z \bmN)^{-1}]^\Lambda{}_\Sigma \ \mathrm{exists}
\nrightarrow [(U + Z F)^{-1}]^\Lambda{}_\Sigma \ \mathrm{exists}\
.
\end{eqnarray}
From this it follows that in supergravity, starting from a theory
fixed by a function $F(X)$, electric/magnetic duality rotations
can be performed that give duality equivalent theories \emph{not}
determined by a dual function.

The above calls for a formulation of the general $N=2$
Poincar\'{e} supergravity theory that does not presuppose the
existence of a function $F(X)$. Such a framework indeed exists. It
starts from the vector $(X^{\Lambda}, F_{\Lambda})$ instead of the
function $F(X)$ \cite{Stromi}. We refrain from explaining this
formulation in detail as it does not fit in the superconformal
framework we adopted in this thesis. Moreover, it does not
incorporate new physically inequivalent models as all models for
which no function exist are electric/magnetically dual to systems
that do have a formulation in terms of such a function.

\section{The c-map}\lb{twvijf}

In section \ref{sctd} we treated scalar-tensor theories and the
accompanying dualities, which, we argued, could be related to
theories of vector gauge fields and their electric/magnetic
dualities through a dimensional reduction. In (ungauged)
supergravity we have a vector and a hypermultiplet sector. When
there are isometries in the manifold parameterized by the
hypermultiplet scalars, the hypermultiplets can be dualized to
tensor multiplets. As alluded to earlier, this tensor multiplet
sector is indeed related to the vector multiplet sector via a
dimensional reduction (although this is not true for every tensor
multiplet theory). Furthermore, the electric/magnetic dualities on
the vector side turn out to be related to similar dualities in the
scalar-tensor sector. This map from the vector multiplet to the
tensor multiplet sector of $N=2$ supergravity is called the
($N=2$) c-map. From a string theory perspective it has its origin
in the T-duality between type IIA and type IIB.

Besides the vector and tensor multiplets the c-map also includes
the gravitational sector of the theory. In fact, in its most basic
form it appears in a ($N=1$) model of a scalar and a tensor
coupled to gravity (so without vector fields). To introduce the
concept conveniently and to set the notation, we first consider
this simple model. After that we perform the $N=2$ supergravity
c-map. In the latter, to prepare for the next chapter, we take a
Euclidean setting.

\subsection{A prototype model}\label{simol}

We consider the following Lagrangian
\begin{equation}\label{N=1action2}
{\cal L}  = - R(e) - \frac{1}{2} \p_\mu \phi \p^\mu \phi +
\frac{1}{2}e^{2\phi} H_\mu H^\mu \ .
\end{equation}
It appears as a sub-sector of $N=1$ low-energy effective actions
in which gravity is coupled to $N=1$ tensor multiplets. In our
case we have one tensor multiplet only, which, as seen from string
theory, consists of the dilaton and the NS-NS tensor $B_{\mu
\nu}$.

To perform the c-map, we dimensionally reduce the action
\eqref{N=1action2} and assume that all the fields are independent
of one coordinate. This can most conveniently be done by first
choosing an upper triangular form of the vierbein, in coordinates
$(x^m,x^3\equiv \tau), m=0,1,2$,
\begin{equation}\label{vierbein}
e_\mu{}^a=\begin{pmatrix}e^{-{\tilde \phi}/2}{\hat e}_m{}^i &
e^{{\tilde \phi}/2}{\tilde B}_m \cr 0 & e^{{\tilde
\phi}/2}\end{pmatrix}\ .
\end{equation}
The metric then takes the form
\begin{equation}\label{grav-metric}
d s^2 =  e^{{\tilde \phi}} ( d \tau + {\tilde B}_m d x^m)^{2} +
e^{-{\tilde \phi}}\hat{g}_{mn} d x^{m} d x^{n}\ ,
\end{equation}
and we demand ${\tilde \phi},{\tilde B}_m$ and ${\hat g}_{mn}$ to
be independent of $\tau$. For the moment, $\tau$ is one of the
spatial coordinates, but in the next subsection we will apply our
results to the case when $\tau$ is the Euclidean time.

We get $e=e^{-{\tilde \phi}}{\hat e}$, and the scalar curvature
decomposes as
\begin{equation}
- e R(e)= - {\hat e} R (\hat e) - \frac{1}{2} \hat{e} \p_m {\tilde
\phi} \p^m {\tilde \phi} + \frac{1}{2} \hat{e} e^{2{\tilde \phi}}
\tilde{H}_m \tilde{H}^m\ .
\end{equation}
Similarly, we require the dilaton and the tensor to be independent
of $\tau$. The three-dimensional Lagrangian then is
\begin{equation}\label{lns}
{\cal L}_3 = - R(\hat e) - \frac{1}{2}
\p_m {\tilde \phi} \p^m {\tilde \phi} + \frac{1}{2}e^{2{\tilde
\phi}} {\tilde H}_m {\tilde H}^m - \frac{1}{2} \p_m \phi \p^m \phi
+ \frac{1}{2}e^{2{\phi}} H_m H^m\ ,
\end{equation}
where $H^m = - \half i \vep^{mnl} H_{nl} = - i \vep^{mnl} \p_n
B_l$ and $B_l = B_{l \tau}$. In addition, there is an extra term
in the Lagrangian,
\begin{equation}\label{L-aux}
{\cal L}_3^{aux}= \frac{1}{12} e^{2 (\phi + \tilde{\phi})}
(H_{mnl} - 3 B_{[m} H_{nl]}) (H^{mnl} - 3 B^{[m} H^{nl]})\ ,
\end{equation}
which plays no role in the three-dimensional theory. Being of rank
three in three dimensions $H_{mnl}$ is an auxiliary field.
(\ref{L-aux}) can therefore trivially be eliminated by its field
equation.

Note that the Lagrangian ${\cal L}_3$ has the symmetry
\begin{equation}\label{c:phi}
\phi \longleftrightarrow {\tilde \phi}\ ,\qquad B_m
\longleftrightarrow {\tilde B}_m\ .
\end{equation}
In fact, careful analysis shows that also ${\cal L}_3^{aux}$ is
invariant, provided we transform
\begin{equation}\label{c:B}
B_{mn} \rightarrow  \tilde{B}_{mn} \equiv B_{mn} - {\tilde B}_{[m}
B_{n]}\ .
\end{equation}

The resulting theory can now be reinterpreted as a dimensional
reduction of a four-dimensional theory of gravity coupled to a
scalar $\tilde \phi $ and  a tensor ${\tilde B}_{\mu \nu}$ and
with the vierbein given by
\begin{equation}\label{vierbein2}
{\tilde e_\mu{}^a}=\begin{pmatrix}e^{-{\phi}/2}{\hat e}_m{}^i &
e^{{\phi}/2}{B}_m \cr 0 & e^{{\phi}/2}\end{pmatrix}\ .
\end{equation}

The map from (\eqref{N=1action2}) to the latter model is the c-map
in its simplest form. The symmetry transformations involved,
(\ref{c:phi}) and (\ref{c:B}), are related to the Buscher rules
for T-duality \cite{Bus}. We here derived these rules from an
effective action approach in Einstein frame, similar to
\cite{T-dual}.

In case of a Euclidean setting, the dimensional reduction is still
based on the decomposition of the vierbein (\ref{vierbein}) with
$\tau$ the Euclidean time. After dimensional reduction over
$\tau$, the Einstein-Hilbert term gives
\begin{equation}\label{Eucl-red}
e R(e)= {\hat e} R(\hat e) +\frac{1}{2} \p_\mu \tilde{\phi} \p^\mu
\tilde{\phi} + \frac{1}{2}e^{2{\tilde \phi}} {\tilde H}_\mu
{\tilde H}^\mu,
\end{equation}
such that the symmetry \eqref{c:phi} still holds.

\subsection{The c-map in N=2 supergravity}\label{nistcm}

Having introduced the main idea in the last subsection we are now
ready to consider the c-map in $N=2$ supergravity. In view of the
next chapter, we first treat the case of one double-tensor
multiplet coupled to $N=2$ supergravity, after which we perform
the c-map on the $N=2$ supergravity model with an arbitrary number
of multiplets.

\vspace{5mm}

\textbf{The double-tensor multiplet}

\vspace{5mm}

The (Minkowskian) Lagrangian of one double-tensor multiplet
coupled to $N=2$ supergravity can be written as \cite{TV1,TV2}
\begin{equation} \label{DTM-action2}
  {\cal L} = -R  -\vier F_{\mu \nu} F^{\mu \nu} -
\half \p_\mu \phi \p^\mu \phi - \half\,
  e^{-\phi} \p_\mu \chi \p^\mu \chi + \half M_{ab} H_\mu^a H^{\mu b} \ .
\end{equation}
The $N=2$ pure supergravity sector contains the metric and the
graviphoton field strength $F_{\mu \nu}$, whereas the matter
sector consists of two scalars and a doublet of tensors, $H_\mu^a=
- \half i \vep_{\mu \nu \rho \sigma} \p^\nu B^{\rho \sigma a}$
(with $a = 1,2$). The self-interactions in the double-tensor
multiplet are encoded in the matrix
 \begin{equation}\label{M-matrix2}
  M(\phi,\chi) = e^{\phi} \begin{pmatrix} 1 & - \chi \\[2pt] - \chi &
e^{\phi}  + \chi^2 \end{pmatrix}\ .
 \end{equation}
(\ref{DTM-action2}) can be obtained (up to a factor $2$) from
(\ref{poinact}) as a special example of the case of one
hypermultiplet (with the appropriate isometries to dualize two
scalars to tensors) and function $F(X) = \vier i (X^0)^2$ (which
leads to $N_{00} = 1$, $\M_{00} = 0$ and $\N_{00} = \frac{i}{2}$).
Note that (\ref{DTM-action2}) contains the model
(\ref{N=1action2}) as a subsector; we reobtain it when we put
$F_{\mu \nu}$, $\chi$ and $H_\mu^1$ to zero.

We then perform a standard Wick rotation (see Appendix
\ref{notcon}), use Euclidean metrics and dimensionally reduce over
$\tau = it$. Doing so we decompose the vierbein as in
(\ref{vierbein}); this yields a three-dimensional metric, a vector
$\tilde B$ and a scalar ${\tilde \phi}$. The vector gauge
potential decomposes in the standard way
\begin{equation}
A_\mu = (-{\tilde \chi}, {\tilde A}_m - {\tilde \chi}{\tilde
B}_m)\ .
\end{equation}
The result after dimensional reduction is \footnote{We are
suppressing here terms like \eqref{L-aux}, which are irrelevant
for our purpose.}
 \begin{eqnarray} \label{3dDTM-action2}
  {\cal L}_3^e &=& R({\hat e}) + \half \p_m \phi \p^m \phi + \half\,
  e^{-\phi} \p_m \chi \p^m \chi + \half M_{ab}(\phi,\chi) H^a_m H^{m b} \nonumber\\
&& + \half \p_m \tilde{\phi} \p^m \tilde{\phi} + \half \,
e^{-{\tilde\phi}} \p_m \tilde{\chi} \p^m \tilde{\chi} + \half
M_{ab}({\tilde \phi}, {\tilde \chi}) \tilde{H}_m^a \tilde H^{m b}\
.
 \end{eqnarray}
Here we have combined the two vectors in a doublet ${\tilde
B^a_m}=(\tilde{A}_m,\tilde B_m)$ that defines the (dual) field
strengths ${\tilde H}^a_m$ in three dimensions. The matrix
multiplying their kinetic energy is exactly the same as in
\eqref{M-matrix2}, but now with the tilde-fields. Therefore, the
Lagrangian has the symmetry
\begin{equation}\label{tedu2}
\phi  \longleftrightarrow  \tilde{\phi}\ ,\qquad \chi
\longleftrightarrow  \tilde{\chi}\ ,\qquad B_m^a
\longleftrightarrow {\tilde B}_m^a\ ,
\end{equation}
where $B_m^a = B^a_{m \tau}$. Similar to the last subsection, the
symmetry (\ref{tedu2}) makes that (\ref{3dDTM-action2}) can be
reinterpreted as the result of a dimensional reduction of a theory
of the same form as (\ref{DTM-action2}), but with a double-tensor
multiplet consisting of $\tilde{\phi}$, $\tilde{\chi}$ and
$\tilde{H}_\mu^a$, a vierbein given by
\begin{equation}\label{vierbein3}
{\tilde e_\mu{}^a}=\begin{pmatrix}e^{-{\phi}/2}{\hat e}_m{}^i &
e^{{\phi}/2}{B}^2_m \cr 0 & e^{{\phi}/2}\end{pmatrix}\ ,
\end{equation}
and a graviphoton with components
\begin{equation}
A_\mu = (- \chi, B^1_m - \chi B^2_m)\ .
\end{equation}

\vspace{5mm}

\textbf{The general model}

\vspace{5mm}

We then consider general $N=2$ supergravity systems of the form
(\ref{poinact}). Nevertheless, in the models we start from we
suppress the hypermultiplets (in case these hypermultiplets allow
for a tensor multiplet formulation, they could be taken into
account, but for our purposes it suffices to neglect them). So our
starting point is
\begin{eqnarray}\label{cvnt}
\L & = & - R + 2 \M_{\Lambda \bar{\Sigma}} \p_{\mu} X^{\Lambda}
\p^{\mu} \bX^{\Sigma} + \frac{1}{4} i e^{-1} \vep^{\mu \nu \rho
\sigma} F_{\mu \nu}{}^{\Lambda} G_{\rho \sigma \Lambda}\ ,
\end{eqnarray}
where $\Lambda$ runs from $0$ to $n$, i.e. there are $n$ vector
multiplets involved. We put in a factor of $2$ for convenience.
$G_{\mu \nu \Lambda}$ is given by
\begin{eqnarray}
G_{\mu \nu \Lambda} & = & - \half i e \vep_ {\mu \nu \rho \sigma}
\frac{\de \L}{\de F_{\rho \sigma}{}^{\Lambda}} = \half i e
\vep_{\mu \nu \rho \sigma} \Im \N_{\Lambda \Sigma} F^{\rho \sigma
\Sigma} + \Re \N_{\Lambda \Sigma} F_{\mu \nu}{}^{\Sigma}\ .\nn\\
\end{eqnarray}

After a Wick rotation we perform a dimensional reduction over
Euclidean time. The vielbein we parameterize as before,
\begin{equation}\label{vierbein4}
e_\mu{}^a=\begin{pmatrix}e^{-\phi/2}{\hat e}_m{}^i & e^{\phi/2}
B_m \cr 0 & e^{\phi/2}\end{pmatrix}\ ,
\end{equation}
while the vector gauge fields decompose as
\begin{eqnarray}
A_\mu{}^\Lambda & = & (- \chi^{\Lambda}, B_m{}^\Lambda -
\chi^{\Lambda} B_m)\ .
\end{eqnarray}

This way we obtain
\begin{eqnarray}\lb{ldrd}
\L^e_3 & = & \hat{R} + \half \p_m \phi \p^m \phi + \half e^{2
\phi} H_m H^m - 2 \M_{\Lambda \bar{\Sigma}}
\p_{m} X^{\Lambda} \p^{m} \bX^{\Sigma}\nn\\
& & - \half  i \hat{e}^{-1} \vep^{mnl} G_{m \tau \Lambda}
F_{nl}{}^{\Lambda} - \half i \hat{e}^{-1} \vep^{mnl} F_{m
\tau}{}^{\Lambda} G_{nl \Lambda}\ ,
\end{eqnarray}
where $(i F_{m \tau}{}^{\Lambda}, i G_{m \tau \Lambda})$
($=(F_{mt}{}^\Lambda, G_{mt \Lambda})$) and $(F_{mn}{}^{\Lambda},
G_{mn \Lambda})$ are the components of $(F_{\mu \nu}{}^\Lambda,
G_{\mu \nu \Lambda})$ with and without a time-index respectively.
In terms of the three-dimensional fields in which we decomposed
the vielbein and the vector fields, they read as
\begin{eqnarray}\lb{fgde}
\( \begin{array}{c} i F_{m \tau}{}^{\Lambda}\\ i G_{m \tau
\Lambda}
\end{array} \) & = & \( \begin{array}{c} - i \p_m \chi^{\Lambda}\\ - \half
\hat{e} \vep_{mnl} e^{\phi} Im \N_{\Lambda \Sigma} (H^{\Sigma} -
\chi^{\Sigma} H)^{nl} - i Re \N_{\Lambda \Sigma} \p_m
\chi^{\Sigma} \end{array} \)\ ,\nn\\
\( \begin{array}{c} F_{mn}{}^{\Lambda}\\ G_{mn \Lambda}
\end{array} \) & = & \( \begin{array}{c} (H^{\Lambda} - \chi^{\Lambda} H)_{mn}\\
i \hat{e} \vep_{mnl} e^{- \phi} Im \N_{\Lambda \Sigma} \p^l
\chi^{\Sigma}  + Re \N_{\Lambda \Sigma} (H^{\Sigma} -
\chi^{\Sigma} H)_{mn}
\end{array} \)\nn\\
& & + 2 \( \begin{array}{c} F_{[m \tau}{}^{\Lambda} B_{n]}\\
G_{[m \tau \Lambda} B_{n]} \end{array} \)\nn\\
& \equiv & \( \begin{array}{c} \hH_{mn}{}^\Lambda\\ G^{\chi}_{mn
\Lambda}
\end{array} \)
+ 2 \( \begin{array}{c} F_{[m \tau}{}^{\Lambda} B_{n]}\\
G_{[m \tau \Lambda} B_{n]} \end{array} \)\ .
\end{eqnarray}
Here we used $H_{mn}{}^\Lambda = 2 \p_{[m} B_{n]}{}^\Lambda$ and
$H_{mn} = 2 \p_{[m} B_{n]}$. Why we utilize the superscript $\chi$
in $G_{mn \Lambda}^{\chi}$ will become clear shortly. The second
vector on the right hand side of $(F_{mn}{}^\Lambda, G_{mn
\Lambda})$ in fact drops out when plugging in (\ref{fgde}) in
(\ref{ldrd}). So the vector fields $B_m{}^\Lambda$ and $B_m$ only
appear through their field strengths.

Compared with the model of subsection \ref{simol} and the
double-tensor multiplet we considered earlier, no symmetries of
the type (\ref{c:phi}) and (\ref{tedu2}) do appear. Nevertheless,
similar to the case of (\ref{lns}) and (\ref{3dDTM-action2}),
(\ref{ldrd}) can be uplifted to a new four-dimensional model. In
case of (\ref{lns}) and (\ref{3dDTM-action2}), this yielded a
four-dimensional model of the same form as the original one. Now
the models on both sides are different: While we started with the
vector multiplet sector we end up with the hypermultiplet sector
of $N=2$ supergravity.

More precisely, we obtain a four-dimensional (Euclidean) theory of
$n$ tensor multiplets and $1$ double-tensor multiplet,
\begin{eqnarray}\lb{ldrdv}
\L^e & = & R + \half \p_{\mu} \phi \p^{\mu} \phi + \half e^{2
\phi} H_\mu H^\mu - 2 \M_{\Lambda
\bar{\Sigma}} \p_{\mu} X^{\Lambda} \p^{\mu} \bX^{\Sigma}\nn\\
& & - e^{-1} \hH^{\mu \Lambda} G^{\hH}_{\mu \Lambda}  - e^{-1}
G^{\chi \mu}_\Lambda F^{\chi \Lambda}_{\mu} \ ,
\end{eqnarray}
where $H_{\mu \nu \rho} = 3 \p_{[\mu} B_{\nu \rho]}$, $H^\mu =
\frac{1}{6} \vep^{\mu \nu \rho \sigma} H_{\nu \rho \sigma}$ and
$H_{\mu \nu \rho}{}^\Lambda = 3 \p_{[\mu} B_{\nu
\rho]}{}^\Lambda$. Also $\hH_{\mu \nu \rho}{}^\Lambda = H_{\mu \nu
\rho}{}^\Lambda - \chi^{\Lambda} H_{\mu \nu \rho}$ and $\hH^{\mu
\Lambda} = \frac{1}{6} \vep^{\mu \nu \rho \sigma} \hH_{\nu \rho
\sigma}{}^\Lambda$. Furthermore, $F^{\chi \Lambda}_\mu = -i \p_\mu
\chi^\Lambda$, while $G^{\hH}_{\mu \Lambda}$ and $G^{\chi}_{\nu
\rho \sigma \Lambda}$ are the four-dimensional versions of $i G_{m
\tau \Lambda}$ and $G^{\chi}_{mn \Lambda}$, so given by
(\ref{fgde}) with $H_{mn}$ and $\hH_{mn}{}^{\Lambda}$ replaced by
$H_{\mu \nu \rho}$ and $\hH_{\mu \nu \rho}{}^\Lambda$. Finally
$G_\Lambda^{\chi \mu} = \frac{1}{6} \vep^{\mu \nu \rho \sigma}
G^\chi_{\nu \rho \sigma \Lambda}$.

For future convenience we also give the Lagrangian in terms of the
tensor multiplet fields
\begin{eqnarray}\lb{ldrdvv}
\L^e & = & R + \half \p_{\mu} \phi \p^{\mu} \phi + \frac{1}{2}
e^{2 \phi} H_\mu H^\mu - 2 \M_{\Lambda
\bar{\Sigma}} \p_{\mu} X^{\Lambda} \p^{\mu} \bX^{\Sigma}\nn\\
& & + e^{- \phi} \Im \N_{\Lambda \Sigma} \p_{\mu} \chi^{\Lambda}
\p^{\mu} \chi^{\Sigma} + e^{\phi} \Im \N_{\Lambda \Sigma}
\hH^{\Lambda}_\mu \hH^{\mu \Sigma}\nn\\
& & + 2 i e^{-1} \Re \N_{\Lambda \Sigma} \p_{\mu} \chi^{\Lambda}
\hH^{\mu \Sigma}\ ,
\end{eqnarray}
where we implemented $F_\mu^{\chi \Lambda} = - i \p_\mu
\chi^\Lambda$ and eliminated $G_{\mu \Lambda}^\chi$ and $G_{\mu
\Lambda}^{\hH}$, using (the four-dimensional version of)
(\ref{fgde}).

Observe that (\ref{ldrdvv}) is invariant under complex rescalings
of the $X^\Lambda$, which can be traced back to the dilatational
and $U(1)_R$ gauge symmetries of the superconformal model
(\ref{scb}).

Dimensionally reducing (\ref{ldrdvv}) would yield (\ref{ldrd})
when we identify
\begin{eqnarray}
B_{m \tau} & = & B_m\ ,\nn\\
B_{m \tau}{}^{\Lambda} & = & B_m{}^{\Lambda}\ ,
\end{eqnarray}
leave out the kinetic terms for the fields descending from the
metric components with one and two $\tau$-indices
\footnote{Reinstalling these terms does not spoil the picture, but
it would make the analysis more involved.} and neglect terms of
the type (\ref{L-aux}).

Note that the tensor multiplet vectors $(F^{\chi \Lambda}_{\mu},
G^{\hH}_{\mu \Lambda})$ and $(\hH^{\Lambda}_{\mu}, G^{\chi}_{\mu
\Lambda})$ are directly related to the symplectic vector $(F_{\mu
\nu}{}^\Lambda, G_{\mu \nu \Lambda})$ of the vector multiplet
side. Actually, these tensor multiplet objects are symplectic
vectors from a purely scalar-tensor perspective as well: They
transform as such under symplectic scalar-tensor duality
transformations of the type described in the last chapter. The
latter follows from the fact that $G^{\hH}_{\mu \Lambda}$ and
$G^{\chi}_{\nu \rho \sigma \Lambda}$ are the functional
derivatives of (\ref{ldrdv}) with respect to $\hH_{\mu \nu
\rho}{}^{\Lambda}$ and $F^{\chi \Lambda}_{\mu}$ respectively,
\begin{eqnarray}\label{heng}
G^{\hH}_{\mu \Lambda} & = & - \half e \vep_{\mu \nu \rho \sigma}
\frac{\de \L^e}{\de \hH_{\alpha \beta \gamma}{}^\Lambda}\ ,\nn\\
G^{\chi}_{\mu \nu \rho \Lambda} & = & \half e \vep_{\mu \nu \rho
\sigma} \frac{\de \L^e}{\de F^{\chi \Lambda}_{\sigma}}\ ,
\end{eqnarray}
which explains the use of the superscripts $\chi$ and $\hH$ in
$G^{\chi}_{\mu \nu \rho \Lambda}$ and $G^{\hH}_{\mu \Lambda}$. The
set of Bianchi identities and equations of motion of (\ref{ldrdv})
then takes the form
\begin{eqnarray}\lb{eqbisg}
\p_{\mu} \( \begin{array}{c} \hH^{\mu \Lambda}\\
G^{\chi \mu} \end{array} \) & = & - i \( \begin{array}{c} F^{\chi \Lambda}_{\mu}\\
G^{\hH}_{\mu \Lambda} \end{array} \) H^\mu\ ,\nn\\
\vep^{\mu
\nu \rho \sigma} \p_{\rho} \( \begin{array}{c} F^{\chi}_{\sigma}\\
G^{\hH}_{\sigma} \end{array} \) & = & 0\ ,
\end{eqnarray}
which is the $N=2$ supergravity generalization of (\ref{eqbi}).
(\ref{eqbisg}) remains equivalent under the transformations
\begin{eqnarray}\label{emdustsg}
\( \begin{array}{c} \hH_{\mu}{}^\Lambda\\ G^{\chi}_{\mu \Lambda}
\end{array} \)
\longrightarrow \( \begin{array}{c} \tilde{\hH}_{\mu}{}^\Lambda\\
\tG^{\chi}_{\mu \Lambda}
\end{array} \) = \( \begin{array}{cc} U^\Lambda{}_\Sigma & Z^{\Lambda \Sigma}\\
W_{\Lambda \Sigma} & V_\Lambda{}^\Sigma \end{array} \) \(
\begin{array}{c} \hH_{\mu}{}^\Sigma\\ G^{\chi}_{\mu \Sigma}
\end{array} \)\ ,\nn\\
\( \begin{array}{c} F^{\chi \Lambda}_{\mu}\\ G^{\hH}_{\mu \Lambda}
\end{array} \)
\longrightarrow \( \begin{array}{c} \tF^{\chi \Lambda}_{\mu}\\
\tG^{\hH}_{\mu \Lambda}
\end{array} \) = \( \begin{array}{cc} U^\Lambda{}_\Sigma & Z^{\Lambda \Sigma}\\
W_{\Lambda \Sigma} & V_\Lambda{}^\Sigma \end{array} \) \(
\begin{array}{c} F^{\chi \Sigma}_{\mu}\\ G^{H}_{\mu \Sigma}
\end{array} \)\ ,
\end{eqnarray}
just as (\ref{eqbi}) does under (\ref{emdust}). Note that the
non-trivial right-hand side of the first set of equations
(\ref{eqbisg}) does not spoil the duality as it transforms
consistently.

The dual Lagrangian can be obtained from
\begin{eqnarray}
\tG^{\hH}_{\mu \Lambda} = - \half e \vep_{\mu \nu \rho \sigma}
\frac{\de \tL^e}{\de \tilde{\hH}_{\nu \rho \sigma}{}^\Lambda}\
,\quad \tG^{\chi}_{\mu \nu \rho \Lambda} =  \half e \vep_{\mu \nu
\rho \sigma} \frac{\de \tL^e}{\de \tF^{\chi \Lambda}_{\sigma}}\ ,
\end{eqnarray}
which has a consistent solution only when the matrix in
(\ref{emdustsg}) is an element of $Sp(2n, \mathbb{R})$.
Transforming $(X^{\Lambda}, F_{\Lambda})$ as a symplectic vector
as well then gives a dual theory of the same form as
(\ref{ldrdv}), completely similar to the vector multiplet sector.

\newpage

\thispagestyle{empty}

\chapter{Supergravity description of spacetime
instantons}\lb{ch3}

Black holes in superstring theory have both a macroscopic and
microscopic description. On the macroscopic side, they can be
described as solitonic solutions of the effective supergravity
Lagrangian. Microscopically they can typically be constructed by
wrapping $p$-branes over $p$-dimensional cycles in the manifold
that the string theory is compactified on. The microscopic
interpretation is best understood for BPS black holes.

Apart from this solitonic sector, string theory also contains
instantons. Microscopically they arise as wrapped Euclidean
$p$-branes over $p+1$-dimensional cycles of the internal manifold.
The aim of this chapter (which is based on \cite{DDVTV} and
\cite{dVV}) is to present a macroscopic picture of these
instantons as solutions of the Euclidean equations of motion in
the effective supergravity Lagrangian. We focus hereby on
spacetime instantons, whose (non-perturbative) effects are
inversely proportional to the string coupling constant $g_s$.

The models that we will study are type II string theories
compactified on a Calabi-Yau (CY) threefold. The resulting
four-dimensional effective action realizes $N=2$ supergravity,
whose structure we discussed in chapter \ref{ch2}. Both vector and
tensor and hypermultiplets do appear. As explained before, the
tensor multiplets can be dualized to hypermultiplets. The numbers
of multiplets in the four-dimensional action depend on the
topological properties of the CY. The latter are encoded in its
Hodge numbers $h_{1,1}$ and $h_{1,2}$, which give the number of
$(1,1)$ and $(1,2)$ cycles \footnote{In the context of complex
geometry, $(p,q)$ cycles are dual to harmonic tensor fields of
rank $p,q$, where $p$ and $q$ denote the number of holomorphic and
antiholomorphic indices.}. Type IIA(B) string theory
compactifications on a CY with Hodge numbers $h_{1,1}$ and
$h_{1,2}$ yields $h_{1,1}$ $(h_{1,2})$ vector multiplets and
$h_{1,2} + 1$ $(h_{1,1} +1)$ hypermultiplets. The fields in these
multiplets include $h_{1,1}$ \emph{K\"{a}hler} moduli, associated
with deformations of the size of the CY and $h_{1,2}$
\emph{complex structure} moduli, associated with its deformations
in shape.

The geometry of the hypermultiplet moduli space - containing the
dilaton - is known to receive quantum corrections, both from
string loops \cite{RSV} and from instantons \cite{BBS}. The
instanton corrections are exponentially suppressed and are
difficult to compute directly in string theory. Our results yield
some progress in this direction, since within the supergravity
description one finds explicit formulae for the instanton action
\footnote{Instanton actions can also be studied from worldvolume
theories of D-branes. For a discussion on this in the context of
our work, we refer to \cite{MMMS}. It would be interesting to find
the precise relation to our analysis.}. Related work can also be
found in \cite{GS2,BGLMM}, but our results are somewhat different
and contain several new extensions.

Interestingly, there is a relation between black hole solutions in
type IIA/B and instanton solutions in type IIB/A. Microscopically,
this can be understood from T-duality between IIA and IIB.
Macroscopically, this follows from the c-map, discussed at the end
of the last chapter. This makes that (BPS) solutions of the vector
multiplet Lagrangian are mapped to (BPS) solutions of the tensor-
or hypermultiplet Lagrangian. We will use this mapping in
Euclidean spacetimes. Roughly speaking, there are two classes of
solutions on the vector multiplet sector: (Euclidean) Black holes
and Taub-NUT like solutions. These map to D-brane instantons and
NS-fivebrane instantons respectively. The distinguishing feature
is that the corresponding instanton actions are inversely
proportional to $g_s$ or $g_s^2$ respectively. For both type of
instantons, we give the explicit solution and the precise value of
the instanton action.

The D-brane instantons are found to be the solutions to the
equations obtained from c-mapping the BPS equations of
\cite{LWKM}. Their analysis contains also $R^2$ interactions, but
they can be easily switched off. The BPS equations then obtained
are similar, but not identical to the equations derived in
\cite{BLS}. The NS-fivebrane instantons are derived in a different
way, not by using the c-map. This is because the BPS solutions in
Euclidean supergravity coupled to vector multiplets are not fully
classified. We therefore construct the NS-fivebrane instantons by
extending the Bogomol'nyi-bound-formulation of \cite{TV1}.

As explained in the introduction, ultimately, we hope to get a
better understanding of non-perturbative string theory. In
particular, it is expected that instanton effects resolve
conifold-like singularities in the hypermultiplet moduli space of
Calabi-Yau compactifications, see e.g. \cite{OV}. These
singularities are closely related - by the c-map - to the conifold
singularities in the vector multiplet moduli space due to the
appearance of massless black holes \cite{Strom}. Moreover, in
combination with the more recent relation between black holes and
topological strings \cite{OSV}, it would be interesting to study
if topological string theory captures some of the non-perturbative
structure of the hypermultiplet moduli space. For some hints in
this direction, see \cite{RVV}. Finally, we recall that instantons
may play an important role in the stabilization of moduli. For an
example related to our discussion, we refer to \cite{DSTV}.

This chapter is organized as follows: In section \ref{five} we
treat NS-fivebrane instantons in the context of $N=1$
supergravity. We use this simple setup to explain at a basic level
various concepts we use in later sections. Section \ref{univ} is
devoted to a review of instanton solutions in the universal
hypermultiplet of $N=2$ supergravity and their relation to
gravitational solutions of pure $N=2$ supergravity. Then in
section \ref{gen} we consider instanton solutions to the theory
obtained from arbitrary CY compactifications of type II
superstrings.

\section{NS-fivebrane instantons}\label{five}

In this section, we give the $N=1$ supergravity description of the
NS-fivebrane instanton. The main characteristic of this instanton
is that the instanton action is inversely proportional to the
square of the string coupling constant. In string theory, such
instantons appear when Euclidean NS-fivebranes wrap six-cycles in
the internal space, and therefore are completely localized in both
space and (Euclidean) time.

It is well known that Euclidean NS-fivebranes in string theory are
T-dual to Taub-NUT or more generally, ALF geometries \cite{OV}
(see also \cite{Tong}). We here re-derive these results from the
perspective of four-dimensional (super-) gravity in a way that
clarifies the methods used in this chapter.

\subsection{A Bogomol'nyi bound}

We start with the simple system of gravity coupled to a scalar and
tensor in four spacetime dimensions given by (\ref{N=1action2}),
which we repeat for convenience,
\begin{equation}\label{N=1action}
{\cal L}  = - R(e) - \frac{1}{2} \p_\mu \phi \p^\mu \phi +
\frac{1}{2}e^{2\phi} H_\mu H^\mu \ ,
\end{equation}
with
\begin{equation}\label{tensor}
H^\mu = - \frac{1}{2} i \vep^{\mu \nu \rho \sigma} \p_\nu B_{\rho
\sigma}\ .
\end{equation}

The instanton solution can be found by deriving a Bogomol'nyi
bound on the Euclidean Lagrangian \cite{Rey},
\begin{equation}\label{N=1BPSaction}
{\cal L}^e= \half (e^\phi H_\mu \mp e^{\phi} \p_\mu e^{-\phi})
(e^\phi H^\mu \mp e^{\phi} \p^\mu e^{-\phi})  \mp \frac{1}{6} i
e^{-1} \p_\mu (e^{\phi} H^\mu)\ .
\end{equation}
Here, we have left out the Einstein-Hilbert term. It is well known
that this term is not positive definite, preventing us to derive a
Bogomol'nyi bound including gravity. In most cases, our instanton
solutions are purely in the matter sector, and spacetime will be
taken flat. The Bogomol'nyi equation then is
\begin{equation}\label{N=1BPS}
H_\mu = \pm \p_\mu e^{-\phi}\  .
\end{equation}
This implies that $e^{-\phi}$ should be a harmonic function. The
$\pm$ solutions refer to instantons or anti-instantons. Notice
that the surface term in \eqref{N=1BPSaction} is topological in
the sense that it is independent on the spacetime metric. It is
easy to check that the BPS configurations \eqref{N=1BPS} have
vanishing energy-momentum tensor, so that the Einstein equations
are satisfied for any Ricci-flat metric.

One can now easily evaluate the instanton action on this solution.
The only contribution comes from the surface term in
\eqref{N=1BPSaction}. Defining the instanton charge as
\begin{equation}
\int_{S^3}\, d^3 x \,  (\frac{1}{6} \vep^{mnl} H_{mnl}) =Q\ ,
\end{equation}
we find \footnote{In the tensor multiplet formulation, the
instanton action has no imaginary theta-angle-like terms. They are
produced after dualizing the tensor into an axionic scalar, by
properly taking into account the constant mode of the axion. In
the context of NS-fivebrane instantons, this was explained e.g. in
\cite{DTV}.}
\begin{equation}\label{NS5-action}
S_{\mathrm{inst}}=\frac{|Q|}{g_s^2}\ .
\end{equation}
Here we have assumed that there is only a contribution from
infinity, and not from a possible other boundary around the
location of the instanton. It is easy to see this when spacetime
is taken to be flat. In that case the single-centered solution for
the dilaton is
\begin{equation}
e^{-\phi}=e^{-\phi_\infty}+\frac{|Q|}{4\pi^2 r^2}\ ,
\end{equation}
which is the standard harmonic function in flat space with the
origin removed. We have furthermore related the string coupling
constant to the asymptotic value of the dilaton by
\begin{equation}\label{g-string}
g_s\equiv e^{-\phi_\infty/2}\ .
\end{equation}
In our notation, this is the standard convention.

\subsection{Taub-NUT geometries and NS-fivebrane instantons}

As described in subsection \ref{simol}, the model
(\ref{N=1action}) realizes a prototype version of the c-map. We
can make use of this map, i.e. the symmetry transformations
(\ref{c:phi}), to generate scalar-tensor solutions from a
(Euclidean) time independent solution of pure Einstein gravity. In
other words, we do a T-duality over (Euclidean) time (this of
course only makes sense as a solution-generating-technique).
However, such a solution is not an instanton, since it is not
localized in $\tau$. We therefore have to uplift the solution to a
$\tau$-dependent solution in four dimensions. This is easy if the
original solution is in terms of harmonic functions. In that case
there is a natural uplifting scheme, which involves going from
three- to four-dimensional harmonic functions.

We discuss now examples in the class of gravitational instantons
\cite{G-H}. These are vacuum solutions of the Euclidean Einstein
equation, based on a three-dimensional harmonic function $V({\vec
x})$,
\begin{equation}\label{grav-inst}
d s^2=V^{-1}( d \tau + A_m d x^m)^2 + V d {\vec x}\cdot d {\vec
x}\ ,
\end{equation}
with $A_m$ satisfying $\half \vep_{mnl} \p^n A^l = \pm \p_m V$.
The $\pm$ solutions yields selfdual or antiselfdual Riemann
curvatures. In the notation of \eqref{grav-metric}, we have that
$A_m = {\tilde B}_m$, ${\hat e}_m{}^i=\delta_m{}^i$ and
$e^{-{\tilde \phi}}=V$.

Multi-centered gravitational instantons correspond to harmonic
functions of the form
\begin{equation}
V=V_0 + \sum_i \frac{m_i}{|\vec{x}-\vec{x}_i|}\ ,
\end{equation}
for some parameters $V_0$ and $m_i$. For non-zero $V_0$, one can
further rescale $\tau=V_0 {\tilde \tau}$ and $m_i=4V_0{\tilde
m}_i$ such that one can effectively set $V_0=1$. The
single-centered case corresponds to Taub-NUT geometries, or
orbifolds thereof. For $V_0=0$ one obtains smooth resolutions of
ALE spaces (like e.g. the Eguchi-Hanson metric for the
two-centered solution). For more details, we refer to
\cite{Andreas:1998hh}.

Before the c-map, the dilaton and the tensor are taken to be zero.
The three-dimensional $\tau$-independent solution after the
symmetry transformations (\ref{c:phi}) is
\begin{equation}
e^{\phi}=V^{-1}\ ,\qquad H_m = \half \vep_{mnl} \p^n B^l = \pm
\p_m V\ , \qquad H_{mnl} =0\ ,
\end{equation}
and the metric is flat, $g_{mn}=\delta_{mn}$.

We now construct a four-dimensional $\tau$-dependent solution by
taking $V$ a harmonic function in four dimensions. We take the
four-dimensional metric to be flat and $H_{\mu \nu \rho}$ is
determined by $H_\mu = \pm \p_\mu V$.

That this is still a solution for \eqref{N=1action} can directly
be seen from the fact that the Bogomol'nyi equations
\eqref{N=1BPS} are satisfied. The instanton action is again given
by \eqref{NS5-action}. Notice further that the difference between
instantons and anti-instantons for the fivebrane corresponds to
selfdual and antiselfdual gravitational instantons.

Due to our procedure, we are making certain aspects of T-duality
not explicit. We have for instance suppressed any dependence on
the radius of the compactified circle parameterized by $\tau$.
These aspects become important in order to dynamically realize the
uplifting solution in terms of a decompactification limit after
T-duality. It turns out that a proper T-duality of the Taub-NUT
geometry, including world-sheet instanton corrections, produces a
completely localized NS-fivebrane instanton based on the
four-dimensional harmonic function given above. For more details,
we refer to \cite{Tong}.

\section{Membrane and fivebrane instantons}\label{univ}

In the previous section, we have discussed aspects of NS-fivebrane
instantons. Here, we will elaborate further on this, and also
introduce membrane instantons. These appear in M-theory or type
IIA string theory compactifications, and we will be interested in
four-dimensional effective theories with eight supercharges such
as IIA strings compactified on Calabi-Yau manifolds. The main
distinction with the previous section is the presence of RR
fields, and these will play an important role in this section.

General CY compactifications of type IIA strings yield $N=2$
supergravity theories with $h_{1,2}+1$ hypermultiplets (or tensor
multiplets), but in this section we will restrict ourself to the
case of the universal hypermultiplet only, leaving the general
case for the next section. This situation occurs when the CY space
is rigid, i.e. when $h_{1,2}=0$ \footnote{The universal
hypermultiplet cannot be obtained from a geometric
compactification of type IIB strings as there exists no CY with
$h_{1,1} = 0$.}. Then there are only two three-cycles in the CY,
around which the Euclidean membranes can wrap. These are the
membrane instantons, and in this section we give their
supergravity description. The $h_{1,1}$ vector multiplet fields
can be truncated in our setup; it suffices to have pure
supergravity coupled to the universal hypermultiplet.

\subsection{Instantons in the double-tensor multiplet}

We will describe the universal hypermultiplet in the double-tensor
formulation, given by (\ref{DTM-action2}). In this formulation it
contains two tensors and two scalars, which can be thought of as
two $N=1$ tensor multiplets coming from the NS-NS and RR sectors.
Instantons in the double-tensor multiplet were already discussed
in \cite{TV1,DDVTV} (see also \cite{GS1}), and in the context of
the c-map in \cite{BGLMM}. In this section, we reproduce these
results and show the correspondence between instantons and
stationary gravitational solutions using the c-map.

The (Minkowskian) Lagrangian for the double-tensor multiplet, we
recall, reads as
 \begin{equation} \label{DTM-action}
  {\cal L} = -R  -\vier F_{\mu \nu} F^{\mu \nu} -
\half \p_\mu \phi \p^\mu \phi - \half\,
  e^{-\phi} \p_\mu \chi \p^\mu \chi + \half M_{ab} H_\mu^a H^{\mu b} \ ,
 \end{equation}
with
 \begin{equation}\label{M-matrix}
  M(\phi,\chi) = e^{\phi} \begin{pmatrix} 1 & - \chi \\[2pt] - \chi &
e^{\phi}  + \chi^2 \end{pmatrix}\ .
 \end{equation}

From a string theory point of view, the metric, $\phi$ and
$H_\mu^2$ come from the NS-NS sector, while the graviphoton,
$\chi$ and $H_\mu^1$ descend from the RR sector in type IIA
strings. When we truncate to the NS-NS sector, we get the
Lagrangian \eqref{N=1action}, so the results obtained there are
still valid here.

As we are interested in instanton solutions, we consider the
Euclidean version of (\ref{DTM-action}), which can be obtained by
doing a standard Wick rotation (see Appendix \ref{notcon}) and
using Euclidean metrics. The form of the Lagrangian is still given
by (\ref{DTM-action}), but now the matter Lagrangian is positive
definite. In \cite{TV1} and \cite{DDVTV}, Bogomol'nyi equations
were derived and solved for the double-tensor multiplet coupled to
pure $N=2$ supergravity with vanishing graviphoton field strength
and Ricci tensor. The solutions of these equations preserving half
of the supersymmetry can be recasted into the following compact
form,
\begin{eqnarray}\label{bpsinst}
g_{\mu \nu} & = & \delta_{\mu \nu}\ ,\nonumber\\
e^{- \phi} & = & \frac{1}{4} (h^{2} - p^{2})\ ,\nonumber\\
H_\mu^{2} & = & \frac{1}{2}(h \p_\mu p
- p \p_\mu h)\ ,\nonumber\\
\chi & = & - e^{\phi}
p + \chi_{c}\ ,\nonumber\\
H_\mu^{1} - \chi_{c} H_\mu^{2} & = & - \p_\mu h\ ,
\end{eqnarray}
with $h$ and $p$ four-dimensional harmonic functions (satisfying
$|h|\geq |p|$) and $\chi_c$ an arbitrary constant. The cases where
$h$ is negative or positive correspond to instantons or
anti-instantons respectively. We have written here a flat metric
$g_{\mu\nu}$, but it is easy to generalize this to any Ricci flat
metric, as long as it admits harmonic functions. Non-trivial $h$
and $p$ can be obtained when one or more points are taken out of
four-dimensional flat space. The solution is then of the form
\begin{equation}\label{h&p}
h=h_\infty + \frac{Q_h}{4 \pi^2 |\vec{x}- \vec{x}_0|^{2}}\ ,\qquad
p = p_\infty + \frac{Q_p}{4 \pi^2 |\vec{x}- \vec{x}_0|^{2}}\ ,
\end{equation}
or multi-centered versions thereof. It can be easily seen that a
pole in $p$ corresponds to a source with (electric) charge in the
field equation of $\chi$. Similarly a pole in $h$ corresponds to a
source with (magnetic) charge in the Bianchi identity of $H_\mu^1
- \chi_c H_\mu^2$. For single-centered solutions, there are five
independent parameters, two for each harmonic function, together
with $\chi_c$.

\vspace{5mm}

{\bf NS-fivebrane instantons with RR background fields}

\vspace{5mm}

The general solution in \eqref{bpsinst} falls into two classes,
depending on the asymptotic behavior of the dilaton at the origin.
The first class fits into the category of NS-fivebrane instantons.
The solution is characterized by
\begin{eqnarray}\label{pish}
p = \pm (h - \alpha)\ ,
\end{eqnarray}
with $\alpha$ an arbitrary constant. In terms of \eqref{h&p}, this
condition is equivalent to
\begin{equation}\label{QhisQp}
Q_h=\pm \,Q_p\ ,
\end{equation}
such that the solution only has four independent parameters. This
implies that the dilaton behaves at the origin like
\begin{equation}
e^{-\phi} \rightarrow {\cal O}\left(\frac{1}{r^2}\right)\ .
\end{equation}
The condition \eqref{pish} implies that
\begin{equation}\label{BPS-RR-5brane}
H_\mu^{2} = \pm \p_\mu e^{-\phi}\ ,\qquad H_\mu^{1} = \pm \p_\mu
(e^{-\phi}\chi)\ ,
\end{equation}
with $e^{-\phi}$ the harmonic function
\begin{equation}\label{dil-RR-5brane}
e^{-\phi} = \frac{1}{2} \alpha h - \frac{1}{4} \alpha^2\ ,
\end{equation}
and $\chi$ is fixed in terms of $h$ via \eqref{bpsinst}. In this
form, we get back the results of \cite{TV1} and \cite{DDVTV}.
However, as we will see shortly, the equations
(\ref{BPS-RR-5brane}) are not completely equivalent to
(\ref{bpsinst}) and (\ref{pish}).

The prototype example for $e^{-\phi}$ is of the form
\begin{equation}
e^{-\phi} = g_s^2 + \sum_i \frac{|Q_i|}{4 \pi^2 |\vec{x}-
\vec{x}_i|^{2}}\ .
\end{equation}
It is easy to check that, whereas the dilaton diverges, the RR
field $\chi$ remains finite at the excised points. When taking
(\ref{BPS-RR-5brane}) as the starting point, the values of $\chi$
at the points $\vec{x}_i$ are allowed to be different. However, as
pointed out in \cite{DTV}, for the solutions to preserve half of
the supersymmetry, $\chi$ should take equal values ($\chi_0$) at
these points. As the difference with (\ref{BPS-RR-5brane}) we just
alluded to, the equations (\ref{bpsinst}) and (\ref{pish})
\emph{do} include this restriction from supersymmetry.

The solution is characterized by the parameters
$\alpha,\chi_c,g_s$ and the charges
\begin{equation}
Q \equiv \int_{S^3_\infty} d^3 x \, (\frac{1}{6} \vep^{mnl}
H^2_{mnl}) = \mp \sum_i |Q_i|\ .
\end{equation}
The parameters $\alpha$ and $\chi_c$ can be traded for the
boundary values of the RR field, $\chi_\infty$ and $\chi_0$. The
action of the multi-centered instanton was calculated in
\cite{TV1,DDVTV}, and the result is
\begin{eqnarray}\label{dtns}
S_{\mathrm{inst}} = |Q| \Big( \frac{1}{g_{s}^{2}} + \frac{1}{2}
(\Delta \chi)^{2} \Big)\ ,
\end{eqnarray}
with $\Delta \chi \equiv \chi_\infty-\chi_0$.

The solution above describes a generalization of the NS-fivebrane
instanton discussed in section \ref{five}. Notice that the first
term in the instanton action is inversely proportional to the
square of the string coupling constant, as is common for
NS-fivebrane instantons. The second term is the contribution from
the RR background field. Only for constant $\chi$ does one obtain
a local minimum of the action \footnote{Solutions with constant
$\chi$ can be obtained from \eqref{bpsinst} and \eqref{pish} by
taking the limit $\alpha \rightarrow 0$ while both $h_\infty$ and
$Q_h\rightarrow \infty$ in such a way that $\alpha h$ is kept
fixed. Such solutions follow more directly from the Bogomol'nyi
equations considered in \cite{DDVTV}.}.

\newpage

{\bf Membrane instantons}

\vspace{5mm}

The remaining solutions, other than \eqref{QhisQp}, are given by
\eqref{bpsinst} with $Q_h \neq Q_p$. One can see that the
asymptotic behavior of the dilaton around the origin is now
\begin{equation}
e^{-\phi}\rightarrow {\cal O}\left(\frac{1}{r^4}\right)\ .
\end{equation}
Compared to the fivebrane instanton case, this behavior is more
singular. However, the instanton action is still finite. As was
shown in \cite{TV1}, the action reduces to a surface term, and the
only contribution comes from infinity. One way of writing the
instanton action is \cite{DDVTV}
\begin{equation}\label{meact}
S_{\mathrm{inst}}={\sqrt
{\frac{4}{g_s^2}+(\Delta\chi)^2}}\left(|Q_h| \pm \frac{1}{2}\Delta
\chi \, Q\right)\ ,
\end{equation}
with the same convention as for fivebranes, i.e.
\begin{equation}\label{delch}
\Delta \chi \equiv \chi_\infty-\chi_0 = -\frac{p_\infty}{g_s^2}\ ,
\end{equation}
and $Q$ still defined by
\begin{equation}
Q \equiv \int_{S^3_\infty}\, d^3 x \, (\frac{1}{6} \vep^{mnl}
H^2_{mnl}) = -\frac{1}{2}\left(h_\infty Q_p - p_\infty Q_h\right)\
.
\end{equation}
The plus and minus sign in (\ref{meact}) refer to instanton and
anti-instanton respectively. Using the relations given above and
$g_s^2 = \frac{1}{4} (h_{\infty}^2 - p_{\infty}^2)$ one can show
that (\ref{meact}) is always positive, as it should be.

Notice that the instanton action contains both the fivebrane
charge $Q$ and $Q_h$, which we identify with a membrane charge.
For pure membrane instantons, which have vanishing NS-NS field,
the second term in (\ref{meact}) vanishes. When we put $H^2_\mu$
(and its BPS equation) to zero from the start, we can dualize
$\chi$ to a tensor and obtain a ``tensor-tensor" theory. To
perform this dualization we have to replace $- i \p_\mu \chi$ by
the vector $F_\mu^\chi$ in the Euclideanized ($H^2_\mu$-less)
version of (\ref{DTM-action}) and add a Lagrange multiplier term
\begin{eqnarray}
{\cal L}^e (\chi) \longrightarrow {\cal L}^e (F) + e^{-1}
\vep^{\mu \nu \rho \sigma} B_{\mu \nu \, \chi} \p_\rho
F_\sigma^\chi\ .
\end{eqnarray}
Integrating out $B_{\mu \nu \, \chi}$ enforces $\p_{[\mu}
F^\chi_{\nu]} = 0$ and locally $F^\chi_\mu = - i \p_\mu \chi$
again. Subtracting the total derivative $e^{-1} \vep^{\mu \nu \rho
\sigma} \p_\mu (B_{\nu \rho \, \chi} F^\chi_\sigma)$ and
integrating out $F^\chi_\mu$ yields the tensor-tensor theory.
Using this action to evaluate the pure membrane instantons on
gives
\begin{eqnarray}\label{meactt}
S'_{\mathrm{inst}} & = & S_{\mathrm{inst}} + \Delta \chi Q_p\nonumber\\
& = & \frac{2}{g_s} \sqrt{Q_h^2 - Q_p^2}\ .
\end{eqnarray}
The appearance of the second term in the first line is a result of
the subtraction of the boundary term in the dualization procedure.
In going from the first to the second line we used the fact that
$Q=0$, which allowed us to express $\Delta \chi$ in terms of the
charges $Q_h$ and $Q_p$ and $g_s$,
\begin{equation}
\Delta \chi = -\frac{2}{g_s}\frac{Q_p}{\sqrt {Q_h^2-Q_p^2}}\ .
\end{equation}
The $\frac{1}{g_s}$ dependence in the instanton action is typical
for D-brane instantons that arise after wrapping Euclidean
D-branes over supersymmetric cycles in the Calabi-Yau\,
\cite{BBS}.

The microscopic interpretation of the general solution is not so
clear. In the next subsection, we will see how these solutions are
generated from the c-map. In this way, one can give a natural
interpretation in terms of black holes and gravitational
instantons.

The form of the instanton action for both fivebrane (\ref{dtns})
and membrane instantons  (\ref{meact}) and (\ref{meactt}) was
recently re-derived by solving the constraints from supersymmetry
of the effective action \cite{ASV}. This provides an alternative
derivation of the formulas in this section and confirms that the
supergravity method for computing the instanton action is correct.

\subsection{Instantons and gravitational solutions}

In this subsection, we show that our membrane and fivebrane
instanton solutions naturally follow from the c-map.

As considered in subsection \ref{nistcm}, a dimensional reduction
of (\ref{DTM-action}) gives rise to the symmetry transformations
(\ref{tedu2}). Using these symmetry transformations, the general
BPS instanton, given by \eqref{bpsinst}, can be translated back to
stationary BPS solutions of pure $N=2$ supergravity. This involves
replacing four-dimensional by three-dimensional harmonic
functions.

Starting from \eqref{bpsinst}, one thus obtains solutions to the
equations of motion in Euclidean space. Stationary solutions of
the Einstein-Maxwell Lagrangian can however easily be continued
from Minkowski to Euclidean space, and vice versa. If we make the
following decomposition for the metric and graviphoton vector
field in Minkowski space
\begin{eqnarray}\label{minmet}
g_{\mu \nu} d x^{\mu} d x^{\nu} & = & - e^{\tilde{\phi}}(d t +
\omega_m d x^m)^{2} +
e^{-\tilde{\phi}} \hat{g}_{mn} d x^{m} d x^{n}\ ,\nonumber\\
A_\mu& = & (- \tilde{\chi}', \tilde{A}_m - \tilde{\chi}'
\omega_m)\ ,
\end{eqnarray}
then we can analytically continue to Euclidean space by
identifying
\begin{equation}\label{thrwick}
\omega_m  =  -i {\tilde B}_m\ ,\qquad \tilde{\chi}'  =  i
\tilde{\chi}\ .
\end{equation}
Using the inverse of this we generate the following BPS equations
for pure $N=2$ supergravity,
\begin{eqnarray}\label{bpsgra}
\hat{g}_{mn} & = & \delta_{mn}\ ,\nonumber\\
e^{- \tilde{\phi}} & = & \frac{1}{4} (h^{2} +
q^{2})\ ,\nonumber\\
\half \vep_{mnl} \p^n \omega^l & = & - \frac{1}{2} (h \p_m q
- q \p_m h)\ ,\nonumber\\
\tilde{\chi}' & = & - e^{\tilde{\phi}}
q + \chi'_c\ ,\nonumber\\
\tilde{H}^{1}_m - \vep_{mnl} \chi'_{c} \p^n \omega^l & = & - \p_m
h\ ,
\end{eqnarray}
with $h$, $q$ three-dimensional flat space harmonic functions,
$\tilde{H}^1_m = \vep_{mnl} \p^n \tilde{A}^l$ and $\chi'_{c}$ an
arbitrary constant.

BPS equations of pure $N=2$ supergravity were studied in
\cite{LWKM}, \cite{BLS}, \cite{GHU} and \cite{Tod}. (\ref{bpsgra})
can be shown to reproduce the results of \cite{LWKM}.

The line element of (\ref{bpsgra}) falls into the general class of
Israel-Wilson-Perj\'es (IWP) metrics \cite{IW,Perj},
\begin{equation}
d s^2=-|U|^{-2} (d t+ \omega_m d x^m)^2+|U|^2 d {\vec x}\cdot d
{\vec x}\ ,
\end{equation}
where $U$ is any complex solution to the three-dimensional Laplace
equation. Comparing to \eqref{bpsgra} and \eqref{minmet}, we have
that
\begin{equation}
U=\frac{1}{2}(h+iq)\ .
\end{equation}

Let $F_{\mu \nu}$ be the field strength of the four-dimensional
gauge field and $G_{\mu \nu}$ is its dual,
\begin{equation}\label{defig}
G_{\mu \nu} \equiv - \frac{1}{2} i e \vep_{\mu \nu \rho \sigma}
\frac{\de \L}{\de F_{\rho \sigma}}.
\end{equation}
Then the components of $F_{\mu \nu}$ with a time-index are
\begin{equation}\label{drdcomp}
F_{mt}  =  - \p_m \tilde{\chi}'\ ,\qquad G_{mt}  =  - \frac{1}{4}
e^{\tilde \phi} \vep_{mnl}(2 \, \omega^n F^{lt} + F^{nl})\ .
\end{equation}
To derive the second equation in (\ref{drdcomp}) one needs to
decompose the component of (\ref{defig}) with a time-index,
$G_{mt} = - \frac{1}{4} i \varepsilon_{mtnl} g^{n \mu} g^{l \nu}
F_{\mu \nu}$, using the metric parameterization (\ref{minmet}).
The last two equations in \eqref{bpsgra} can now elegantly be
rewritten as
\begin{equation}
F_{mt}  =  \p_m (e^{\tilde{\phi}} q)\ ,\qquad G_{mt}  = -
\frac{1}{2} \p_m (e^{\tilde{\phi}} h)\ .
\end{equation}
In fact, in \cite{LWKM} solutions were given in terms of these
objects. This will become important in the next section.

The class of IWP metrics contains many interesting examples, some
of which we discuss now.

\vspace{5mm}

{\bf Pure membrane instantons and black holes}

\vspace{5mm}

We consider here solutions to (\ref{bpsinst}) with vanishing NS-NS
tensor,
\begin{equation}
H_\mu^2 = 0\ .
\end{equation}
These were the solutions that lead to the pure membrane
instantons. The vanishing of $H^2_\mu$ implies that the two
harmonic functions $h$ and $p$ are proportional to each other,
\begin{equation}
p = c\, h\ ,
\end{equation}
for some real constant $c$. We take $h$ of the form
\begin{equation}
h = h_{\infty} + \sum_i \frac{Q_{h,i}}{4 \pi^2 |\vec{x} -
\vec{x}_i|^2}\ , \qquad Q_{p,i}=c\,Q_{h,i}\ .
\end{equation}
This membrane instanton is in the image of the c-map. The dual
(Minkowskian) gravitational solution is static,
\begin{equation}
\p_{[m} \omega_{n]} = 0\ ,
\end{equation}
and has $q = c'\, h$. The IWP metric now becomes of the
Majumdar-Papapetrou type. These are multi-centered versions of the
extreme Reissner-Nordstr\"{o}m black hole. Our solutions describe
the outer horizon part of spacetime in isotropic coordinates,
\begin{eqnarray}\label{iso}
d s^{2} & = & - \left(\gamma + \sum_i \frac{M_i}{4 \pi |\vec{x} -
\vec{x}_i| }\right)^{-2} d t^{2}\nn\\
& &  + \left(\gamma + \sum_i \frac{M_i}{4 \pi |\vec{x} -
\vec{x}_i|}\right)^{2} (d r^{2} + r^{2} d \Omega^{2})\ ,
\end{eqnarray}
with
\begin{equation}
M_i  =  \frac{1}{2} \sqrt{(Q_{h,i})^{2} + (Q_{q,i})^{2}}\ , \qquad
\gamma  =  \frac{1}{2}  \sqrt{1 + c'^2}\, h_{\infty}\ ,
\end{equation}
and $Q_{q,i} = c'\, Q_{h,i}$ for each charge labelled by $i$. Note
that in the parameterization (\ref{iso}) the event horizons are
located at $\vec{x} = \vec{x}_i$. The metric can be made
asymptotically Minkowski by a rescaling of the coordinates
\begin{equation}
t  =  \gamma t'\ ,\qquad r  =  \frac{r'}{\gamma}\ .
\end{equation}

\vspace{5mm}

{\bf NS-fivebrane instantons and Taub-NUT with selfdual
graviphoton}

\vspace{5mm}

Here we consider the NS-fivebrane instantons with RR background
fields. This solution was specified by equations \eqref{pish},
\eqref{BPS-RR-5brane} and \eqref{dil-RR-5brane}. Using the inverse
c-map, we can relate it to a BPS solution of pure N=2
supergravity, based on the three-dimensional harmonic function
\begin{equation}
e^{-\tilde{\phi}} = V \equiv v +  \sum_i \frac{Q_i}{4 \pi^2
|\vec{x}- \vec{x}_i|^{2}}\ .
\end{equation}
The metric solution of the Taub-NUT geometry (\ref{grav-inst})
then reappears,
\begin{equation}\label{metr}
d s^{2} = V^{-1} (d \tau + \tilde{B}_m d x^m)^{2} + V\,
d\vec{x}\cdot d  {\vec x}\ ,
\end{equation}
with
\begin{equation}
2 \p_{[m} \tilde{B}_{n]} = \pm \vep_{mnl} \p^l V\ .
\end{equation}
Analogously to the NS-fivebrane instanton supporting a non-trivial
$\chi$, the Taub-NUT metric (\ref{metr}) supports a non-trivial
graviphoton,
\begin{eqnarray}\label{grvp}
F_{m \tau} = \pm \frac{1}{2} \alpha V^{-2} \p_m V\ , \qquad -
\half e \vep_{m \tau \mu \nu} F^{\mu \nu} =
\frac{1}{2} \alpha V^{-2} \p_m V\ ,\nonumber\\
F_{mn} = \mp \alpha \p_{[m} \big( V^{-1} \tilde{B}_{n]} \big)\ ,
\qquad - \half e \vep_{mn \mu \nu} F^{\mu \nu} = - \alpha \p_{[m}
\big( V^{-1} \tilde{B}_{n]} \big)\ .
\end{eqnarray}
The solution \eqref{grvp} is (anti)selfdual, $F_{\mu \nu} = \mp
\half e \vep_{\mu \nu \rho \sigma} F^{\rho \sigma}$. In fact, it
is precisely the one found in \cite{EH} (see equation (4.15) in
that reference).

The fact that the graviphoton is (anti)selfdual implies that it
has vanishing energy-momentum, which is consistent with the fact
that the Taub-NUT solution is Ricci-flat. Taub-NUT solutions with
(anti)selfdual graviphoton and their T-duality relation with
NS-fivebranes played an important role in a study of the partition
sum of the NS-fivebrane \cite{DVV}.

\section{Instantons in matter coupled N=2
supergravity}\label{gen}
In the last section we considered instantons in the double-tensor
multiplet coupled to $N=2$ supergravity. Now we are interested in
instanton solutions of the general four-dimensional low energy
effective action which type II superstrings compactified on a
Calabi-Yau gives rise to. In the absence of fluxes, this yields
(ungauged) $N=2$ supergravity coupled to vector and tensor
multiplets (or their dual hypermultiplets). We recall that in type
IIA(B) string theory the number of vector multiplets is $h_{1,1}$
($h_{1,2}$) and the number of tensor multiplets is $h_{2,1} +1$
($h_{1,1} + 1$) (where $h_{1,1}$ and $h_{1,2}$ are Hodge numbers
of the Calabi-Yau).

In section \ref{twvijf} we generated the tensor multiplet model
from the c-map on the gravitational and vector multiplet sector.
This way we obtained from $n$ vector multiplets coupled to $N=2$
supergravity a model of one double-tensor and $n$ tensor
multiplets. These tensor multiplets can be dualized further to
hypermultiplets, but similarly to the previous section, we will
not carry out this dualization. This turns out to be the most
convenient way to describe instanton solutions, i.e. they are
naturally described in the tensor multiplet formulation.

In this section we use the c-map once more, to map the BPS
equations for the vector multiplets as found in \cite{LWKM} (with
the $R^2$-interactions which are present in there switched off) to
instantonic BPS equations for the tensor multiplet theory
\footnote{For some earlier work on vector multiplet BPS equations
see \cite{BLS,De} and references therein.}.

The picture that emerges is that all BPS black hole solutions have
their corresponding instantonic description after the (Euclidean)
c-map. For a generic tensor multiplet theory these solutions all
carry some RR-charge, and the instanton action is inversely
proportional to the string coupling. There should also be
NS-fivebrane instantons whose action is proportional to
$\frac{1}{g_s^2}$. However it is not clear for a generic tensor
multiplet theory how to get these from the Euclidean c-map.
Therefore we derive them in a way independent of the c-map.

\subsection{The tensor multiplet theory}\label{ttmt}

We first discuss the tensor multiplet Lagrangian obtained after
the c-map. Details of the derivation we gave in section
\ref{twvijf}. The result is $N=2$ supergravity coupled to a
double-tensor multiplet and $n$ tensor multiplets, with
$n=h_{1,1}$ or $h_{1,2}$ when starting from the type IIA or IIB
vector multiplet sector respectively. We recall that the bosonic
Lagrangian, in Euclidean space, reads
\begin{eqnarray}\label{eucten}
\mathcal{L}^e & = & R + \frac{1}{2} \p_\mu \phi \p^\mu \phi +
\frac{1}{2} e^{2 \phi} H_\mu H^\mu - 2
\mathcal{M}_{\Lambda \Sigma} \p_\mu X^\Lambda \p^\mu \bar{X}^\Sigma\nn\\
& &  + e^{- \phi}\, Im \mathcal{N}_{\Lambda \Sigma} \p_\mu
\chi^{\Lambda} \p^\mu \chi^{\Sigma} + e^{\phi}\, Im
\mathcal{N}_{\Lambda \Sigma} (H_\mu{}^\Lambda -
\chi^{\Lambda} H_\mu) (H^{\mu \Sigma} - \chi^{\Sigma} H^\mu)\nonumber\\
& &  + 2 i e^{-1} \, Re \mathcal{N}_{\Lambda \Sigma} \p_\mu
\chi^{\Lambda}(H^{\mu \Sigma} - \chi^{\Sigma} H^\mu)\ ,
\end{eqnarray}
as given before in (\ref{ldrdvv}). We have left out the vector
multiplet sector including the graviphoton as it is not relevant
for our purposes. This sector can be easily reinstalled. The NS-NS
part of the bosonic sector of the (double-)tensor multiplets
consists of the dilaton, $\phi$, the tensor $B_{\mu \nu}$ ($H_{\mu
\nu \rho} \equiv 3 \p_{[\mu} B_{\nu \rho]}$, $H^{\mu} =
\frac{1}{6} \vep^{\mu \nu \rho \sigma} H_{\nu \rho \sigma}$) and
the complex scalars $X^\Lambda$ ($\Lambda = 0, 1,..., n$). The RR
part of the bosonic sector of the (double)-tensor multiplets is
formed by the (real) scalars $\chi^\Lambda$ and the tensors
$B_{\mu \nu}{}^\Lambda$ ($H_{\mu \nu \rho}{}^\Lambda \equiv
3\p_{[\mu} B_{\nu \rho]}{}^\Lambda$, $H^{\mu \Lambda} =
\frac{1}{6} \vep^{\mu \nu \rho \sigma} H_{\nu \rho
\sigma}{}^\Lambda$).

The metric $M_{\Lambda \bar{\Sigma}}$ of the manifold
parameterized by the complex scalars $X^\Lambda$ (given by
(\ref{matr})) can be written as
\begin{eqnarray}\label{defm}
\mathcal{M}_{\Lambda \bar{\Sigma}} & \equiv & N_{\Lambda \Sigma}
- \frac{N_{\Lambda \Gamma}N_{\Sigma \Xi}\bar{X}^\Gamma X^\Xi}
{N_{\Omega \Delta} X^{\Omega} \bar{X}^\Delta}\ .\nonumber\\
\end{eqnarray}
The matrix $Im \mathcal{N}_{\Lambda \Sigma}$ appearing in the
quadratic terms of (\ref{eucten}) we recall is determined by
\begin{eqnarray}\label{defn}
\N_{\Lambda \Sigma} & \equiv & \bF_{\Lambda \Sigma} + i
\frac{N_{\Lambda \Gamma} X^\Gamma N_{\Sigma \Xi} X^\Xi}{N_{\Omega
\Delta} X^\Omega X^\Delta}\ .
\end{eqnarray}

Notice that the last term in (\ref{eucten}) is imaginary, similar
to a theta-angle-like term. It will therefore be difficult to find
a Bogomol'nyi bound on the action. We will return to the issue of
a BPS bound in the last subsection. In fact, as we will see, we
need to drop the reality conditions on the fields, as not all
solutions we discuss below respect these reality conditions. For
the moment, we will simply complexify all the fields \footnote{For
instance, this means that we treat $X^\Lambda$ and ${\bar
X}^\Lambda$ as independent complex fields. The action then only
depends on $X^\Lambda$ and ${\bar X}^\Lambda$ in a holomorphic
way.}, and discuss below which instanton solutions respect which
reality conditions.

In the next subsection we derive BPS equations for (\ref{eucten})
by c-mapping BPS equations for stationary solutions of its vector
multiplet counterpart. These latter equations are naturally
formulated in terms of symplectic vectors. Therefore it is useful
to write (\ref{eucten}) in terms of symplectic vectors as well. To
do this we can make use of the results obtained in section
\ref{twvijf}. We get
\begin{eqnarray}\label{clut}
\mathcal{L}^{e} & = & R + \frac{1}{2} e^{2 \phi} H_\mu H^\mu \nonumber\\
& &  + \half e^{2 \phi} (F_\Lambda \p_\mu \bar{Y}^\Lambda -
Y^\Lambda \p_\mu
\bar{F}_\Lambda + c.c.) (F_\Lambda \p^\mu \bar{Y}^\Lambda - Y^\Lambda \p^\mu \bar{F}_\Lambda + c.c.)\nonumber\\
& & - c^{\mu \Lambda} d_{\mu \Lambda} + c^\mu{}_\Lambda d_\mu{}^\Lambda \nn\\
& & - e^{-1} \hH^{\mu \Lambda} G^{\hH}_{\mu \Lambda}  - e^{-1}
G^{\chi \mu}_\Lambda F^{\chi \Lambda}_{\mu}\ .
\end{eqnarray}
Here and below, by $c.c.$ we mean taking the complex conjugate
before dropping the reality conditions, and then treating
$X^\Lambda$ and ${\bar X}^\Lambda$ as independent complex fields.
To obtain (\ref{clut}) we demanded
\begin{eqnarray}\label{conditions}
N_{\Lambda \Sigma} \bar{X}^\Lambda X^\Sigma & = & 1\ ,
\end{eqnarray}
which fixes the norm of the scalar fields $X^\Lambda$ (recall from
chapter \ref{ch2} that (\ref{eucten}) is invariant under complex
rescalings of $X^\Lambda$). Furthermore, we introduced the
$U(1)_R$ invariant variables
\begin{equation}\label{defy}
Y^\Lambda \equiv  e^{- \frac{1}{2} \phi} \bar{h} X^\Lambda\
,\qquad {\bar Y}^\Lambda \equiv  e^{- \frac{1}{2} \phi} h {\bar
X}^\Lambda\ .
\end{equation}
Here $h$ is an arbitrary (space-dependent) phase factor, which
drops out when plugging in (\ref{defy}) in the action. As a
consequence of (\ref{conditions}) and (\ref{defy}), $e^{-\phi}$
should be understood as a function of $Y^\Lambda$ and
$\bar{Y}^\Lambda$,
\begin{equation}
e^{-\phi} = i(Y^\Lambda \bF_\Lambda (\bar{Y}) - \bar{Y}^\Lambda
F_\Lambda (Y))\ .
\end{equation}

$(c_\mu{}^\Lambda, c_{\mu \Lambda})$ and $(d_\mu{}^\Lambda, d_{\mu
\Lambda})$ are symplectic vectors belonging to the NS-NS sector.
They are defined as
\begin{eqnarray}\label{def}
\left( \begin{array}c c_\mu{}^\Lambda\\ c_{\mu \Lambda}
\end{array} \right) \equiv \left(
\begin{array}c + i \p_\mu (Y^\Lambda - \bar{Y}^\Lambda)\\ + i \p_\mu (F_\Lambda -\bar{F}_\Lambda)\end{array}
\right)\ ,\nn\\
\left( \begin{array}c d_\mu{}^\Lambda\\ d_{\mu \Lambda}
\end{array}
\right) \equiv \left( \begin{array}c \p_\mu (e^{\phi} (Y^\Lambda + \bar{Y}^\Lambda))\\
\p_\mu (e^{\phi}(F_\Lambda + \bar{F}_\Lambda))\end{array} \right)\
.
\end{eqnarray}
Furthermore, we repeat that the symplectic vectors from the RR
part of the theory, $(\hH_\mu{}^\Lambda, G^\chi_{\mu \Lambda})$
and $(F_\mu^{\chi \Lambda}, G^{\hH}_{\mu \Lambda})$, are given by
\begin{eqnarray}\label{deff}
\left( \begin{array}{c} \hH_\mu{}^\Lambda\\ G^\chi_{\mu \Lambda}
\end{array} \right)
& \equiv & \left( \begin{array}{c} H_\mu{}^\Lambda - \chi^\Lambda H_\mu\\
- i e e^{- \phi} Im \mathcal{N}_{ \Lambda \Sigma} \p_\mu
\chi^\Sigma + Re
\mathcal{N}_{\Lambda \Sigma} (H_\mu{}^\Sigma - \chi^\Sigma H_\mu)\end{array} \right)\ ,\nonumber\\
\left( \begin{array}{c} F_\mu^{\chi \Lambda}\\ G^{\hH}_{\mu
\Lambda}
\end{array} \right) & \equiv & \left( \begin{array}{c} - i \p_\mu
\chi^\Lambda\\ - e e^{\phi} Im \mathcal{N}_{\Lambda \Sigma}
(H_\mu{}^\Sigma - \chi^\Sigma H_\mu) - i Re \mathcal{N}_{\Lambda
\Sigma} \p_\mu \chi^\Sigma
\end{array} \right)\ .
\end{eqnarray}
$G^\chi_{\mu \Lambda}$ and $G^{\hH}_{\mu \Lambda}$ are the
functional derivatives of the Lagrangian with respect to
$F_\mu^{\chi \Lambda}$ and $\hH_\mu{}^\Lambda$ as given by
(\ref{heng}).

\subsection{BPS equations from the c-map}\label{equations}
In this section we use the c-map to obtain BPS instanton equations
of a $n+1$ tensor multiplet theory from the BPS equations of a $n$
vector multiplet theory. As said before, the latter equations are
known and we use the results of \cite{LWKM}. In here equations
were constructed for stationary solutions preserving half of the
supersymmetry, with parameters satisfying
\begin{equation}
h \epsilon_i = \varepsilon_{ij} \gamma_{0} \epsilon^{j}\ .
\end{equation}
We remind that $h$ is the phase factor appearing in (\ref{defy}).

The metric components, given by
\begin{eqnarray}
g_{\mu \nu} d x^{\mu} d x^{\nu} & = & - e^{\phi}(d t + \omega_m d
x^m)^{2} + e^{-\phi} \hat{g}_{mn} d x^{m} d x^{n}\ ,
\end{eqnarray}
were found to be related to the complex scalars $Y^\Lambda$ in the
following way
\begin{eqnarray}\label{bpslm4}
e^{-\phi} & = & i(Y^\Lambda \bar{F}_\Lambda - \bar{Y}^\Lambda F_\Lambda)\ ,\nonumber\\
\hat{g}_{mn} & = & \delta_{mn}\ ,\nonumber\\
\vep_{mnl} \p^n \omega^l & = &  \bar{F}_\Lambda \p_m Y^\Lambda +
\bar{Y}^\Lambda \p_m F_\Lambda + c.c.\ .
\end{eqnarray}
Furthermore, $-i(Y^\Lambda - \bar{Y}^\Lambda)$ and $-i(F_\Lambda -
\bar{F}_\Lambda)$ are three-dimensional harmonic functions. This
fixes the NS-NS sector completely. Recall that the equation for
$e^{-\phi}$ is identically true, as follows from the definition of
$Y^\Lambda$  and the condition (\ref{conditions}).

For the RR fields the BPS equations of \cite{LWKM} are
\begin{eqnarray}\label{RReq}
\left( \begin{array}c F_{mt}{}^\Lambda\\ G_{mt \Lambda}
\end{array} \right) = \left(
\begin{array}c d_m{}^\Lambda\\ d_{m \Lambda}\end{array} \right)\ .
\end{eqnarray}
These equations can be equivalently formulated as
\begin{eqnarray}\label{RReqa}
\left( \begin{array}c F_{mn}{}^\Lambda\\ G_{mn \Lambda}
\end{array}
\right) = \vep_{mnl} \left( \begin{array}c c^{l \Lambda} \\
c^l{}_\Lambda \end{array} \right)  + 2 \p_{[m} \left( \begin{array}c e^{\phi} (Y^\Lambda + \bar{Y}^\Lambda) \omega_{n]}|_{BPS}\\
e^{\phi} (F_\Lambda + \bar{F}_\Lambda)
\omega_{n]}|_{BPS}\end{array} \right)\ ,
\end{eqnarray}
where $\omega_n|_{BPS}$ is the BPS solution of $\omega_n$.
(\ref{RReqa}) is the form in which the equations for the RR fields
in \cite{BLS} are written, however they do not have the second
term on the right-hand side. Both sets of equations (\ref{RReq})
and (\ref{RReqa}) fix the RR fields completely in terms of the
complex scalars $Y^\Lambda$.

By construction the equations above only have stationary
solutions. When $\omega_n = 0$ one gets static extremal black
holes. This works similar as in the pure supergravity case
discussed in the last section. However there is a difference
between the generic case and pure supergravity, which will become
important later on in the context of NS-fivebrane instantons. We
saw in the last section that the pure $N=2$ supergravity BPS
equations, after an analytic continuation to Euclidean space, gave
rise to Taub-NUT solutions as well. In contrast to this, for
generic functions $F(X)$ it is far from clear if, and if yes how,
this kind of solutions is contained in the general solution.

Using the c-map treated in subsection \ref{nistcm}, the equations
above can be mapped quite easily to instanton equations of the
Euclidean tensor sector. It requires an analytic continuation,
involving $\omega_m = - i B_m$, and a replacement of
three-dimensional by four-dimensional harmonic functions.

We thus find as instanton equations for the NS-NS fields
\begin{eqnarray}\label{bpsensns}
e^{-\phi} & = & i(Y^\Lambda \bar{F}_\Lambda - \bar{Y}^\Lambda F_\Lambda)\ ,\nonumber\\
g_{\mu \nu} & = & \delta_{\mu \nu}\ ,\nonumber\\
H_\mu & = & i(\bar{F}_\Lambda \p_\mu Y^\Lambda - \bar{Y}^\Lambda
\p_\mu F_\Lambda + c.c.)\ ,
\end{eqnarray}
while $-i(Y^\Lambda - \bar{Y}^\Lambda)$ and $-i(F_\Lambda -
\bar{F}_\Lambda)$ are now four-dimensional harmonic functions. The
instanton equations for the RR fields are
\begin{eqnarray}\label{RReqb}
\left( \begin{array}c F_\mu^{\chi \Lambda}\\ G_{\mu \Lambda}^{\hH}
\end{array} \right) = \left(
\begin{array}c d_\mu{}^\Lambda\\ d_{\mu \Lambda}\end{array} \right)\ ,
\end{eqnarray}
or
\begin{eqnarray}\label{RReqaa}
\left( \begin{array}c \hH_\mu{}^\Lambda\\ G^\chi_{\mu \Lambda}
\end{array} \right) = \left(
\begin{array}c c_\mu{}^\Lambda\\ c_{\mu \Lambda}\end{array}
\right) - i \left( \begin{array}c e^{\phi} (Y^\Lambda + \bar{Y}^\Lambda)\\
e^{\phi} (F_\Lambda + \bar{F}_\Lambda)\end{array} \right) H_\mu
|_{\mathrm{inst}} \ .
\end{eqnarray}
Just as on the vector multiplet side both (\ref{RReqb}) and
(\ref{RReqaa}) fix the RR fields completely in terms of the
complex scalars $Y^\Lambda$. For the fields appearing in
(\ref{eucten}) the equations take the form
\begin{eqnarray}\label{BPS-chiI}
\chi^\Lambda & = & i e^{\phi} (Y^\Lambda + \bar{Y}^\Lambda) + \chi^\Lambda_c\ ,\nonumber\\
H_\mu{}^\Lambda  - \chi^\Lambda_c H_\mu& = &  i  \p_\mu (Y^\Lambda
- \bar{Y}^\Lambda)\ ,
\end{eqnarray}
where $\chi^\Lambda_c$ are arbitrary constants.

Recall from subsection (\ref{ttmt}) that all fields are complex.
However, when we take $-i(Y^\Lambda-{\bar Y}^\Lambda)$ and
$-i(F_\Lambda-{\bar F}_\Lambda)$ to be real, then the solutions
for the dilaton and $H^\Lambda$ are real whereas $\chi^\Lambda$
and $H$ become imaginary.

\vspace{3mm}

Let us make contact with the results of section \ref{univ}. When
we take the function $F(Y)$ to be $F(Y) = \frac{1}{4} i (Y^0)^2$,
we see that (\ref{eucten}) reduces to the Euclidean version of
(\ref{DTM-action}). We then get
\begin{equation}\label{dteen}
Y^0 + \bar{Y}^0  =  - 2 i (F_0 - \bar{F}_0)\ ,\qquad F_0 +
\bar{F}_0 =  \half i (Y^0 -\bar{Y}^0)\ .
\end{equation}
Now we make the following identification of the harmonic functions
$-i(Y^0 - \bar{Y}^0)$ and $-i(F_0 - \bar{F}_0)$ and the harmonic
functions $h$ and $p$ which appeared in the BPS equations of
section \ref{univ},
\begin{equation}\label{dttwee}
-i(Y^0 - \bar{Y}^0)  =  h\ ,\qquad -i(F_0 - \bar{F}_0)  =   \half
i p\ .
\end{equation}
Equations (\ref{bpsinst}) then follow directly. We can in this
case obtain a real solution for $\chi$ and $H$ if we impose that
$-i(F_0-{\bar F}_0)$ is imaginary, such that the harmonic function
$p$ is real.

\subsection{D-brane instantons}

We now discuss the different types of solutions to the equations
we obtained above. Clearly the general solution is a function of
$2n+2$ harmonic functions. In the following we take them
single-centered
\begin{eqnarray}\label{hp}
- i(Y^\Lambda - \bar{Y}^\Lambda) & = & - i(Y^\Lambda -
\bar{Y}^\Lambda)_\infty +
\frac{\hat{Q}^\Lambda}{4 \pi^2 |\vec{x}- \vec{x}_0|^{2}}\ ,\nonumber\\
- i(F_\Lambda - \bar{F}_\Lambda) & = & - i(F_\Lambda -
\bar{F}_\Lambda)_{\infty} + \frac{Q_\Lambda}{4 \pi^2 |\vec{x}-
\vec{x}_0|^{2}}\ .
\end{eqnarray}
However our results are easily generalized to multi-centered
versions of (\ref{hp}).

In section  \ref{univ} we saw that the two different types of
solutions to the BPS equations for the double-tensor multiplet,
membrane and the NS-fivebrane instantons, have different behavior
of $e^{-\phi}$. For membrane instantons (having non-zero
RR-charge) the dilaton behaves towards the excised point(s) as
$e^{-\phi}\rightarrow {\cal O}\left(\frac{1}{|\vec{x} -
\vec{x}_0|^4}\right)$. For NS-fivebrane instantons (having
non-zero NS-NS-, but vanishing RR-charge) $e^{-\phi}$ is a
harmonic function, which implies that towards the excised point(s)
the behavior of the dilaton is $e^{-\phi}\rightarrow {\cal
O}\left(\frac{1}{|\vec{x} - \vec{x}_0|^2}\right)$. The different
behavior of the dilaton in both types of solutions is reflected in
a different dependence of the instanton action on the string
coupling.

Let us now consider solutions to the general equations of last
subsection (i.e. for general functions $F(Y)$). The above seems to
indicate that for a study of the characteristics of the instanton
solutions it is good to start by analyzing the behavior of the
dilaton towards the excised point(s). Doing this analysis we find
that to leading order in $\frac{1}{|\vec{x} - \vec{x}_0|}$
\begin{eqnarray}\label{dilor}
e^{-\phi}|_{\vec{x} \rightarrow \vec{x}_0}  & = & \frac{|Z_0|^2}
{16 \pi^4 |\vec{x} - \vec{x}_0|^4}\ ,
\end{eqnarray}
which is as singular as the membrane instanton of section
\ref{univ}. Here $Z_0$ is defined as
\begin{equation}
Z_0 \equiv (\hat{Q}^\Lambda F_\Lambda (X) - Q_\Lambda
X^\Lambda)|_{\vec{x} \rightarrow \vec{x}_0}\ .
\end{equation}
As seen from the c-map, the function $Z = Q^\Lambda F_\Lambda (X)
- Q_\Lambda X^\Lambda$ is the dual of the central charge function
of the vector multiplet theory \footnote{$Z_{\infty} = (Q^\Lambda
F_\Lambda (X) - Q_\Lambda X^\Lambda)|_{\infty}$ is the dynamically
generated central charge mentioned in the footnote at the
beginning of section \ref{cswe}.}.

\vspace{5mm}

$\mathbf{Z_0 \neq 0}$

\vspace{5mm}

We first consider the case $Z_0 \neq 0$. Generic single-centered
solutions consist of $5n +5$ parameters, $2$ for each harmonic
function and the $n+1$ constants $\chi^\Lambda_c$. The RR scalars
$\chi^\Lambda$ take the values $\chi^\Lambda_c$ at $\vec{x} =
\vec{x}_0$. The constants $\hat{Q}^\Lambda$ and $Q_\Lambda$
appearing in (\ref{hp}) can be identified with magnetic and
electric charges of sources appearing in Bianchi identities and
field equations respectively. $\hat{Q}^\Lambda$ is equal to the
charge of the source in the Bianchi identity of $(H^\Lambda -
\chi_c^\Lambda H)_\mu$,
\begin{equation}\label{ambi}
\hat{Q}^\Lambda = \int_{S^3_{\infty}} \, d^3 x \, (\frac{1}{6}
\vep^{mnl} (H_{mnl}{}^\Lambda - \chi_c^\Lambda H_{mnl}))\ .
\end{equation}
$Q_\Lambda$ is up to a factor $-2i$ the charge of the source in
the field equation of $\chi^\Lambda$,
\begin{equation}
- 2i Q_\Lambda = \int_{\mathbb{R}^4} d^4 x \, e \, (\frac{\delta
{\cal L}}{\delta \chi^\Lambda} -
\partial_{\mu} \frac{\delta {\cal L}}{\delta \partial_{\mu} \chi^\Lambda})\ .
\end{equation}
Notice that this is consistent with the fact that the solutions
for $\chi^\Lambda$ are imaginary. As there are non-vanishing RR
charges we can identify these solutions as D-brane instantons,
generalizing the membrane instantons found in section \ref{univ}.
Also the Bianchi identity of $H_\mu$ is sourced. The corresponding
charge can be expressed in terms of the parameters appearing in
the $2n+2$ harmonic functions,
\begin{equation}
Q \equiv \int_{S^3_{\infty}} \, d^3 x \, (\frac{1}{6} \vep^{mnl}
H_{mnl}) = (Y^\Lambda - \bar{Y}^\Lambda)_{\infty} Q_\Lambda -
(F_\Lambda - \bar{F}_\Lambda)_{\infty} \hat{Q}^\Lambda\ .
\end{equation}

Evaluating (\ref{clut}) on these instantons gives
\begin{eqnarray}\label{tenact}
S_{\mathrm{inst}} & = & \int_{\mathbb{R}^4} d^4 x (- c^{\mu
\Lambda} d_{\mu \Lambda} + i (e^{\phi} (Y^\Lambda +
\bar{Y}^\Lambda) d_{\mu
\Lambda}\nn\\
& & + e^{\phi} (F_\Lambda (Y)
+ \bar{F}_\Lambda (\bar{Y})) d_\mu{}^\Lambda) H^\mu )|_{BPS} \nonumber\\
& = & - \frac{2}{g_s^2} (F_\Lambda (Y) + \bar{F}_\Lambda
(\bar{Y}))_{\infty}
\hat{Q}^\Lambda\nn\\
& &  + \frac{i}{g_s^4} (F_\Lambda (Y) + \bar{F}_\Lambda
(\bar{Y}))_{\infty} (Y^\Lambda + \bar{Y}^\Lambda)_{\infty} Q\ .
\end{eqnarray}

Applying (\ref{tenact}) to the double-tensor multiplet theory of
section \ref{univ}, we have to take again $F(Y) = \frac{1}{4} i
(Y^0)^2$. Then using (\ref{dteen}), (\ref{dttwee}), (\ref{delch})
and the double-tensor multiplet relation $g_s^2 = \frac{1}{4}
(h_{\infty}^2 - p_{\infty}^2)$ we re-obtain (\ref{meact}).

Defining $\Delta\varphi_\Lambda \equiv \frac{i}{g^2_s}
(F_\Lambda(Y) +  \bar{F}_\Lambda (\bar{Y}))_\infty$ and $\Delta
\sigma \equiv \half \frac{1}{g_s^4} (Y^\Lambda +
\bY^\Lambda)_\infty (F_\Lambda (Y) + \bF_\Lambda (\bY))_\infty$,
we can rewrite (\ref{tenactt}) as
\begin{equation}
S_{\mathrm{inst}} =  2i \Delta \varphi_\Lambda \hat{Q}^\Lambda +
2i \Delta \sigma Q\ .
\end{equation}
In fact, one can show that
$\Delta\varphi_\Lambda=\varphi_{\Lambda\infty}-\varphi_{\Lambda0}$,
and $\Delta \sigma = \sigma_\infty - \sigma_0$, where
$\varphi_\Lambda$ is the dual (RR) scalar of $\hat{H}_\mu^\Lambda$
and $\sigma$ is the dual (NS-NS) scalar of $H_\mu$ and
$\varphi_{\Lambda\infty}$, $\varphi_{\Lambda0}$, $\sigma_\infty$,
and $\sigma_0$ are the asymptotic values of $\varphi_\Lambda$ and
$\sigma$ evaluated on the BPS solution
\begin{eqnarray}
\varphi_\Lambda & = & ie^\phi(F_\Lambda +{\bar F}_\Lambda)
+\varphi_{\Lambda c}\ ,\nn\\
\sigma & = & \half e^{2 \phi} (Y^\Lambda + \bY^\Lambda) (F_\Lambda
+ \bF_\Lambda) + \sigma_c\ .
\end{eqnarray}
Here $\varphi_{\Lambda c}$ and $\sigma_c$ are integration
constants, which coincide with $\varphi_{\Lambda 0}$ and
$\sigma_0$, the values of $\varphi_\Lambda$ and $\sigma$ at the
point $\vec{x}=\vec{x}_0$.

The BPS equation for $\varphi_\Lambda$ is in fact implicitly
stated already in the bottom equation in (\ref{RReqb}). Observe
furthermore that the BPS solutions for $\chi^\Lambda$, as in
(\ref{BPS-chiI}), and $\varphi_\Lambda$ are consistent with
symplectic transformations, so we can write
\begin{eqnarray}\label{chi-phi}
\left( \begin{array}c \chi^\Lambda\\ \varphi_\Lambda\end{array}
\right) =
i e^\phi \left( \begin{array}c Y^\Lambda + \bar{Y}^\Lambda\\
F_\Lambda + \bar{F}_\Lambda\end{array} \right) + \left(
\begin{array}c \chi^\Lambda_c\\ \varphi_{\Lambda c}\end{array} \right) \ .
\end{eqnarray}

For $Q=0$ the second term in (\ref{tenact}) vanishes and we find
\begin{eqnarray}\label{tenactt}
S_{\mathrm{inst}}  = 2i \Delta \varphi_\Lambda \hat{Q}^\Lambda = -
\frac{2}{g_s} (\bar{h}F_\Lambda (X) + h \bar{F}_\Lambda
(\bar{X}))_{\infty} \hat{Q}^\Lambda\ .
\end{eqnarray}
This is the action for pure D-brane instantons of which the pure
membrane instanton of section \ref{univ} is a specific example. We
have reintroduced the variables $X^\Lambda$ to make explicit the
typical $\frac{1}{g_s}$ dependence of D-brane instanton actions.
From the c-map point of view pure D-brane instantons are the duals
of static BPS black holes living in the vector multiplet sector.
Microscopically D-brane instantons come from wrapping even/odd
branes over odd/even cycles in the Calabi-Yau in type IIA/B string
theory.

Like in section \ref{univ}, when we put $H_\mu$ (and its BPS
equation) to zero from the start, we can dualize all RR-scalars to
tensors. This way we obtain a formulation of the theory consisting
of $2n+2$ tensors, the ``tensor-tensor" theory. The dualization
procedure works similar as the one described in section
(\ref{univ}). First we write $F_\mu^{\chi \Lambda}$ instead of $-
i \p_\mu \chi^\Lambda$ in (\ref{eucten}) (without $H_\mu$) and add
a Lagrange multiplier term
\begin{eqnarray}
{\cal L}^e (\chi) \longrightarrow {\cal L}^e (F) + e^{-1}
\vep^{\mu \nu \rho \sigma} B_{\mu \nu \Lambda} \p_\rho
F_\sigma^{\chi \Lambda}\ .
\end{eqnarray}
Integrating out $B_{\mu \nu \Lambda}$ enforces $\p_{[\mu} F^{\chi
\Lambda}_{\nu]} = 0$, giving back (locally) $F^{\chi \Lambda}_\mu
= - i \p_\mu \chi^\Lambda$. Subtracting the total derivative
$\vep^{\mu  \nu \rho \sigma} \p_\mu (B_{\nu \rho \Lambda}
F_\sigma^{\chi \Lambda})$ and integrating out $F^{\chi
\Lambda}_\mu$ yields the tensor-tensor theory. When we evaluate
this action on the pure D-brane instantons we get
\begin{eqnarray}\label{acttt}
S'_{\mathrm{inst}} & = & 2 i \Delta \varphi_\Lambda Q^\Lambda
- 2 i\Delta \chi^\Lambda Q_\Lambda\nonumber\\
& = & - \frac{2}{g_s} (\bar{h} F_\Lambda(X) + h \bar{F}_\Lambda
(\bar{X}))_{\infty} \hat{Q}^\Lambda + \frac{2}{g_s} (\bar{h}
X^\Lambda + h
\bar{X}^\Lambda)_{\infty} Q_\Lambda\nonumber\\
& = & \frac{4}{g_s} |Z|_{\infty}\ .
\end{eqnarray}
The second term in the first line is due to the subtraction of the
boundary term in the dualization procedure. To arrive at the last
line we have used that $Q = 0$, which is a consequence of the fact
that we have put $H_\mu$ to zero. The expression in the last line
is (up to a factor of $4$) the value of the real part of the pure
D-brane instanton action as suggested in (a five-dimensional
context) in \cite{GS2}.

\vspace{5mm}

$\mathbf{Z_0 = 0}$

\vspace{5mm}

Since the behavior of the dilaton is different, the case $Z_0=0$
needs to be analyzed separately. For the double-tensor multiplet
of section \ref{univ} it yields NS-fivebrane instantons, which
have a harmonic $e^{-\phi}$. This can most easily be understood
from the fact that in the double-tensor multiplet case we have
$|Z_0| = \frac{1}{2} \sqrt{Q_h^2 - Q_p^2}$ (for single-centered
instantons, as can be derived using (\ref{defy}), (\ref{dteen})
and (\ref{dttwee})). Requiring $|Z_0|$ to vanish then gives the
NS-fivebrane relation (\ref{QhisQp}).

However, for generic functions $F(X)$ NS-fivebrane instantons do
not arise from taking $Z_0=0$. In fact, in these cases the $Z_0=0$
solution only differs qualitatively from the $Z_0 \neq 0$ solution
close to the excised points, which is directly related to the fact
that only the asymptotic behavior of the dilaton is different. Now
recall that $Z$ is the dual of the central charge function of the
vector multiplet theory. So $Z_0=0$ solutions are the duals of
vector multiplet solutions with vanishing central charge function
at $\vec{x} = \vec{x}_0$. In case $Q=0$ these are zero-horizon
black holes. Just as higher derivative corrections lift
zero-horizon black holes at the two-derivative-level to finite
horizon black holes \cite{DKM}, we expect that for
$Z_0=0$-instantons higher derivative corrections have a
qualitative effect on the behavior of $e^{-\phi}$ in the limit
$\vec{x} \rightarrow \vec{x}_0$. If this is the case the
(two-derivative) differences between these solutions and the $Z
\neq 0$ instanton have no real physical significance.

\subsection{NS-fivebrane instantons}

In the previous section, we have discussed D-brane instantons.
These were obtained from the c-map of the BPS solutions of [45],
analytically continued to Euclidean space. We also saw that for
generic functions $F(X)$ NS-fivebrane instantons did not appear as
a limiting case in a similar way as in the double-tensor multiplet
theory of section \ref{univ}. In fact, it is not clear if they are
contained at all in the general solution to the equations in
subsection \ref{equations}, just as was the case for their
supposedly dual Taub-NUT solutions on the vector multiplet side.

However, we expect there to be (BPS) NS-fivebrane instantons in
the general theory as well. That we have missed them so far could
be understood from the fact that not all solutions in the
Euclidean theory can be obtained from Wick rotating real solutions
in the Lorentzian theory.  Therefore, we will follow a different
strategy and work directly in the Euclidean tensor multiplet
Lagrangian, using a similar method as in \cite{TV1}. This way we
indeed find a class of NS-fivebrane instanton solutions.

We first write (\ref{eucten}) as
\begin{eqnarray}\label{bbound}
{\cal L}^e & = & R - 2 \mathcal{M}_{\Lambda \Sigma} \p_\mu
X^\Lambda
\p^\mu \bar{X}^\Sigma\nonumber\\
& & + (N \H_\mu + OE_\mu) A (N \H^\mu + OE^\mu) - 2 e^{-1} \H^t_\mu N^t A O E^\mu \nonumber\\
& & + 2 i e^{-1} Re \N_{\Lambda \Sigma} \p_\mu \chi^\Lambda
(H^{\mu \Lambda} - \chi^\Lambda H^\mu)\ .
\end{eqnarray}
Here we have defined the vectors
\begin{eqnarray}
\mathcal{H}_\mu = \left( \begin{array}{c} H_\mu\\ H_\mu{}^\Lambda
\end{array} \right)\
,\quad E_\mu = \left( \begin{array}{c} \p_\mu \phi\\
e^{-\frac{\phi}{2}} \p_\mu \chi^\Lambda \end{array} \right)\ ,
\end{eqnarray}
and the matrices
\begin{eqnarray}
N = e^{\frac{\phi}{2}} \left( \begin{array}{cc} e^{\frac{\phi}{2}} & 0\\
- \chi^\Lambda & \delta^\Lambda{}_\Sigma \end{array} \right)\
,\quad A = \left(
\begin{array}{cc}
 \frac{1}{2} & 0\\
0 & Im \mathcal{N}_{\Lambda \Sigma}\end{array} \right)\ ,
\end{eqnarray}
$O$ is a matrix as well, satisfying $O^t A O = A$. When all fields
are taken real, clearly the real part of (\ref{bbound}) is bounded
from below by
\begin{eqnarray}\label{bbo}
Re {\cal L}^e & \geq & R
- 2 \mathcal{M}_{\Lambda \Sigma} \p_\mu X^\Lambda \p^\mu \bar{X}^\Sigma\nonumber\\
& & - 2 e^{-1} \H^t_\mu N^t A O E^\mu\ .
\end{eqnarray}
Next we take the matrix $O$ to be
\begin{eqnarray}
O_{1,2} = \pm \left( \begin{array}{cc} 1 & 0\\ 0 & \epsilon
\end{array} \right)\ ,
\end{eqnarray}
with $\epsilon = \delta^\Lambda{}_\Sigma$ in the $O_1$-case and
$\epsilon = - \delta^\Lambda{}_\Sigma$ in the $O_2$-case. The plus
and minus signs refer to the instanton and the anti-instanton
respectively.

We now consider configurations for which the square in
(\ref{bbound}) is zero (i.e that saturate the bound (\ref{bbo}) in
case all fields are taken real). It is easy to show that for
constant $X^\Lambda$ these configurations satisfy the field
equations of $\phi$, $\chi^\Lambda$ and the tensors (to the field
equations of $X^\Lambda$ we come back at a later stage).
Furthermore, these configurations can be shown to have vanishing
energy-momentum. Therefore the gravitational background should be
flat and (\ref{bbound}) reduces to a total derivative. In the
following we need the explicit form of this total derivative in
the $O_2$-case
\begin{eqnarray}
{\cal L}^{i.}_{2} & = & - e^{-1} \p_\mu \Big( e^{\phi} H^\mu \Big)
+ 2i e^{-1} \p_\mu \Big( \bar{\mathcal{N}}_{\Lambda \Sigma}
\chi^\Lambda (H^{\mu \Sigma} - \frac{1}{2} \chi^\Sigma H^\mu) \Big)\ ,\nonumber\\
{\cal L}^{a.i.}_{2} & = & + e^{-1} \p_\mu \Big( e^{\phi} H^\mu
\Big) + 2i e^{-1} \p_\mu \Big( \mathcal{N}_{\Lambda \Sigma}
\chi^\Lambda (H^{\mu \Sigma} - \frac{1}{2} \chi^\Sigma H^\mu)
\Big)\ ,
\end{eqnarray}
where the upper equation corresponds to the instanton and the
lower one to the anti-instanton. In the $O_1$-case we get similar
expressions.

Again we can make contact with the double-tensor multiplet theory
by taking the function $F$ to be $F(X) = + \frac{1}{4} i (X^0)^2$.
The analysis above then reduces to the analysis of \cite{TV1},
with the matrices $O_{1,2}$ corresponding to their matrices
$O_{1,2}$. The instantons related to these matrices are the
NS-fivebrane instantons discussed in section \ref{univ}.

\vspace{5mm}

Let us now consider the conditions which follow from requiring the
square in (\ref{bbound}) to vanish. Firstly, the $O_1$ matrix
gives
\begin{eqnarray}\label{o1}
\mathcal{H}_\mu = \pm \left( \begin{array}{c} \p_\mu e^{-\phi}\\
\chi^\Lambda \p_\mu e^{-\phi} - e^{-\phi} \p_\mu \chi^\Lambda
\end{array} \right)\ .
\end{eqnarray}
These equations are very similar to the $O_1$ equations of
\cite{TV1}. Note in particular the relation $H_\mu = \pm \p_\mu
e^{-\phi}$, which is contained in both. Similarly to \cite{TV1} we
find that the finite-action-solution to (\ref{o1}) has a harmonic
$e^{-\phi}$ and constant $\chi^\Lambda = \chi^\Lambda_0 =
\frac{Q^\Lambda}{Q}$. Here
\begin{equation}
Q^\Lambda = \int_{S^3_{\infty}} \, d^3x \, (\frac{1}{6} \vep^{mnl}
H^\Lambda_{mnl})\ ,\qquad Q = \int_{S^3_{\infty}} \, d^3x \,
(\frac{1}{6} \vep^{mnl} H_{mnl})\ ,
\end{equation}
consistent with the notation we used in our treatment of D-brane
instantons.

\vspace{3mm}

The conditions following from taking the matrix $O_2$ in
(\ref{bbound}) are
\begin{eqnarray}\label{o2}
\mathcal{H}_\mu = \pm \p_\mu \left( \begin{array}{c} e^{-\phi}\\
e^{-\phi} \chi^\Lambda \end{array} \right)\ .
\end{eqnarray}
These equations are very similar to the $O_2$ equations of
\cite{TV1}, with once more $H_\mu = \pm \p_\mu e^{-\phi}$
contained in both sets. The latter equation implies that
$e^{-\phi}$ is again harmonic. The remaining equations in
(\ref{o2}) tell us that the same is true for $e^{-\phi}
\chi^\Lambda$. For single-centered solutions this allows us to
write $\chi^\Lambda$ as
\begin{equation}\label{eqchi}
\chi^\Lambda = \chi^\Lambda_1 e^{\phi} + \chi^\Lambda_0\ ,
\end{equation}
where the $\chi^\Lambda_1$ are arbitrary constants. Note that
$\chi^\Lambda_0$ is the value $\chi^\Lambda$ takes at the excised
point(s). Putting (\ref{eqchi}) back into (\ref{o2}) we find again
$\chi^\Lambda_0 = \frac{Q^\Lambda}{Q}$. Observe that the
finite-action $O_1$-solution is contained in this $O_2$-solution;
we re-obtain it when we put $\chi^\Lambda_1$ to zero. The action
becomes for the single-centered instanton
\begin{equation}\label{insaco2}
S_{\mathrm{inst}} = \frac{|Q|}{g_s^2} - i
\bar{\mathcal{N}}_{\Lambda \Sigma}^c \Delta \chi^\Lambda \Delta
\chi^\Sigma Q\ ,
\end{equation}
where we have (again) defined $\Delta \chi^\Lambda \equiv
\chi^\Lambda_{\infty} - \chi^\Lambda_0$.
$\mathcal{\bar{N}}_{\Lambda \Sigma}^c$ is the (constant) solution
of $ \mathcal{\bar{N}}_{\Lambda \Sigma}$. For the anti-instanton
$\bar{\mathcal{N}}^c$ should be replaced by $\mathcal{N}^c$.

Similarly to the case of D-brane instantons, we can rewrite
(\ref{insaco2}) as
\begin{eqnarray}
S_{\mathrm{inst}} = 2i \Delta \sigma Q\ .
\end{eqnarray}
$\Delta \sigma$ is now defined as $\Delta \sigma \equiv \half
\frac{i}{g_s^2} - \half \mathcal{\bar{N}}_{\Lambda \Sigma}^c
\Delta \chi^\Lambda \Delta \chi^\Sigma$ and satisfies $\Delta
\sigma = \sigma_\infty - \sigma_0$, where $\sigma$ is the dual
scalar of $H$ and $\sigma_\infty$ and $\sigma_0$ are the
asymptotic values of its solution
\begin{eqnarray}
\sigma = \half i e^{\phi} - \half \mathcal{\bar{N}}_{\Lambda
\Sigma} (\chi^\Lambda - \chi_0^\Lambda) (\chi^\Sigma -
\chi_0^\Sigma) + \sigma_c\ .
\end{eqnarray}
$\sigma_c$ is an integration constant, which coincides with
$\sigma_0$, the value of the solution of $\sigma$ at the excised
point.

The equations of motion of $X^\Lambda$ are not automatically
satisfied. Requiring this gives the extra condition that the last
term in (\ref{insaco2}) should be extremized with respect to (the
constants) $X^\Lambda$. Consequently the $\chi^\Lambda_1$ and the
$X^\Lambda$ become related, unless $\mathcal{N}_{\Lambda \Sigma}$
is a constant matrix. The latter is for example the case in the
double-tensor multiplet theory, for which we have
$\mathcal{N}_{00} = \frac{i}{2}$ (in that case (\ref{insaco2}) can
be seen to reduce to (\ref{dtns})). The precise relations between
$\chi^\Lambda_1$ and the $X^\Lambda$ depend on the function
$F(X)$. This implies that there is no general prescription for
obtaining real solutions.

{}From $\chi^\Lambda_0 = \frac{Q^\Lambda}{Q}$ it directly follows
that the charges $\hat{Q}^\Lambda \equiv Q^\Lambda -
\chi^\Lambda_0 Q$ are zero. Furthermore, one can show that there
are no sources in the field equations of $\chi^\Lambda$. So there
are no RR charges at all in the solutions. This means that they
can be identified as (generalized) NS-fivebrane instantons. On the
basis of what we know about NS-fivebrane instantons in the
double-tensor multiplet \cite{DTV} we expect these solutions (or
at least all single-centered ones) to preserve half of the
supersymmetry.

Let us finish our treatment of (generalized) NS-fivebrane
instantons by considering its image under the (inverse) c-map. We
find that this is a Taub-NUT geometry with $n$ (anti)selfdual
vector fields, all of the form (\ref{grvp}). It would be
interesting to find out if there are more general Euclidean BPS
solutions of this type. We leave this for further study.

\chapter{N=2 supersymmetric gaugings with dyonic charges
}\label{ch4}

In chapter \ref{ch2} we introduced $N=2$ supersymmetric theories
based on a variety of supermultiplets. The vector multiplets
contain gauge fields which can be associated with a certain gauge
group. So far we only considered gauge groups that are abelian. In
this chapter (which is based on \cite{dVdW}) we present the
extension to non-abelian gauge groups. Because of supersymmetry,
all the fields of the vector supermultiplet will transform under
this non-abelian group. In particular the target-space
parameterized by the vector multiplet scalars must possess
associated isometries. In addition, we study models in which a
subgroup of the isometries of the hypermultiplet target-space is
associated with the local gauge group (which can be both abelian
and non-abelian). This introduces a coupling between vector and
hypermultiplets. In principle, these $N=2$ supersymmetric gauge
theories are well known for the case where all the charges are
electric. A distinct feature of the approach followed in this
chapter is that we consider theories that may have both electric
\emph{and} magnetic (dyonic) charges.

As explained in chapter \ref{ch2}, the rigid invariance group of
the abelian supersymmetric gauge theories is not necessarily an
invariance of the action (in the sense that the action does not
need to transform as an invariant function) but of the set of
Bianchi identities and equations of motion. The group that can be
realized on this set is a subgroup of the electric/magnetic
duality group, $Sp(2n, \mathbb{R})$, where $n$ denotes the number
of independent vector multiplets. Because this group rotates
electric into magnetic fields and vice versa, there is, for a
given Lagrangian, the option of having both electric and magnetic
charges. Because vector multiplets only provide the gauge fields
that couple to electric charges, the standard approach is
therefore to apply an electric/magnetic duality transformation to
the ungauged Lagrangian, so that all the charges that one intends
to introduce will be electric. In other words, one first converts
the Lagrangian to a suitable electric/magnetic duality frame after
which one switches on purely electric charges corresponding to a
certain gauge group.

This approach is somewhat inconvenient. To set up a more general
framework requires the introduction of electric and magnetic gauge
fields on a par. Such a framework has been proposed in \cite{dWST}
and it allows to introduce a gauging irrespective of the choice of
the duality frame. It incorporates both electric and magnetic
charges and their corresponding gauge fields. The former are
encoded in terms of a so-called embedding tensor, which determines
the embedding of the gauge group into the full rigid invariance
group. This embedding tensor is treated as a spurionic object, so
that the electric/magnetic duality structure of the ungauged
theory is preserved after charges are turned on. Besides
introducing a set of dual magnetic gauge fields, the framework
requires the introduction of a number of tensor fields,
transforming in the adjoint representation of the rigid invariance
group. These extra fields carry additional off-shell degrees of
freedom. The number of physical degrees of freedom remains the
same, owing to extra gauge transformations that are associated
with the tensor fields.

Besides avoiding the need for performing duality transformations
of the Lagrangian prior to switching on the gauging, the more
general framework is important for a variety of other reasons. For
instance, the scalar potential (and other, masslike, terms) that
accompany the gaugings can be formulated in a way that is
independent of the electric/magnetic duality frame. By introducing
both electric and magnetic charges the potential will thus fully
exhibit the duality invariances.  This is of interest, when
studying flux compactifications in string theory, because the
underlying fluxes are usually subject to integer-valued rotations
associated to the non-trivial cycles of the underlying internal
manifold. Furthermore, the fact that tensor gauge fields are
involved in the procedure relates to earlier examples of more
general gaugings (see for instance \cite{LM}).

In this chapter  we show explicitly how to apply the formalism of
\cite{dWST} to $N=2$ gauge theories based on vector multiplets and
hypermultiplets. In other words, by introducing dyonic charges, we
gauge the isometries in a symplectically covariant way in both the
special K\"{a}hler and the hyper-K\"{a}hler sector of the
target-space parameterized by the scalar fields associated with
the vector multiplets and hypermultiplets.





The supersymmetric Lagrangians we will derive in sections
\ref{sec:rest-supersymm-non} and \ref{sec:hypermultiplets}
introduce gaugings in both the vector and hypermultiplet sectors.
Although the vector multiplets are off-shell multiplets, the
presence of the magnetic charges introduces a breakdown of
off-shell supersymmetry. The hypermultiplets are also sensitive to
this, but they are not based on an off-shell representation of the
supersymmetry algebra prior to introducing the charges. It is an
interesting question whether the results of this chapter can be
reformulated in an off-shell form and we will reflect on this in
section \ref{sec:summary-discussion}.

In section \ref{sec:summary-discussion} we also discuss some
applications of our results, concerning for instance
Fayet-Iliopoulos terms. Other possible applications are mainly in
supergravity, where our work may be useful in constructing low
energy effective actions corresponding to string theory flux
compactifications.


This chapter is organized as follows. In section \ref{vmnac} we
repeat the relevant features of $N=2$ vector multiplets and their
behavior under electric/magnetic duality. Furthermore, we explain
how dyonic non-abelian charges are introduced in models of this
type. In section \ref{sec:embedding-tensor} we discuss how the
gauge group is embedded into the rigid invariance group by means
of the embedding tensor. Section \ref{sec:rest-supersymm-non}
deals with the restoration of supersymmetry in vector multiplet
models after gauging, while section \ref{sec:hypermultiplets}
gives the extension with hypermultiplets. In section
\ref{sec:summary-discussion} we summarize the results obtained,
and as we mentioned, indicate some of their applications and
discuss some features related to the off-shell structure of these
theories.

\section{Vector multiplets and non-abelian charges}\label{vmnac}

As explained in section \ref{vmul}, an off-shell $N=2$ vector
multiplet consists of a vector gauge field, $A_\mu$, a complex
scalar $X$, two Majorana fermions, $\Om_i$, and an auxiliary
bosonic field $Y_{ij}$, satisfying the reality condition
$(Y_{ij})^* = \vep^{ik} \vep^{jl} Y_{kl}$. The supersymmetric
Lagrangian of $n$ of such multiplets is encoded in terms of a
function $F(X)$. Its rigid version is given by (\ref{vecsh}).

The Lagrangian of this model is subject to electric/magnetic
duality transformations, which, we remind, act as
\begin{eqnarray}\label{emduuu}
\( \begin{array}{c} F^\pm_{\mu \nu}{}^\Lambda\\ G_{\mu \nu
\Lambda}^\pm
\end{array} \)
\longrightarrow \( \begin{array}{c} \tilde{F}^\pm_{\mu \nu}{}^{\Lambda}\\
\tilde{G}_{\mu \nu \Lambda}^\pm
\end{array} \) = \( \begin{array}{cc} U^\Lambda{}_\Sigma & Z^{\Lambda \Sigma}\\
W_{\Lambda \Sigma} & V_\Lambda{}^\Sigma
\end{array} \) \( \begin{array}{c} F^\pm_{\mu \nu}{}^{\Sigma}\\ G_{\mu \nu
\Sigma}^\pm
\end{array} \)\ ,
\end{eqnarray}
where the matrix involved is from the group $Sp(2n, \mathbb{R})$.
Here
\begin{eqnarray}
G_{\mu \nu \Lambda}^+ = i \vep_{\mu \nu \rho \sigma} \frac{\delta
\L_{\mathrm{vector}}}{\de F^+_{\rho \sigma}{}^{\Lambda}}\ .
\end{eqnarray}
Except for the $Y_{ij}{}^\Lambda$-sector \footnote{As we saw in
section \ref{vmul}, to reobtain an expression of the form
(\ref{vecsh}) including the $Y_{ij}{}^\Lambda$-sector,
$(Y_{ij}{}^\Lambda, Z_{ij \Lambda})$ and $(Y^{ij \Lambda},
Z^{ij}{}_\Lambda)$ (with $Z_{ij \Lambda}$ and $Z^{ij}{}_\Lambda$
given by (\ref{defiz}) and (\ref{defizc})) should transform as
symplectic vectors as well. This comes down to a rotation of the
the reality conditions on $Y_{ij}{}^\Lambda$ and the equations of
motion of these fields.}, the Lagrangian obtained after such a
duality transformation can be written back in the form
(\ref{vecsh}), using the function $\tF$, the scalars $\tX^\Lambda$
and the fermions $\tOm_i{}^\Lambda$, as obtained from a similar
symplectic transformation on $(X^\Lambda, F_\Lambda)$ and
$(\Om_i{}^\Lambda, F_{\Lambda \Sigma} \Om_i{}^\Sigma)$ .

In the following we find it convenient to use the notation
$\alpha^M = (\al^\Lambda, \al_\Lambda)$ for symplectic vectors. So
we get
\begin{eqnarray}\label{symvec}
G^\pm_{\mu \nu}{}^M & = & (F^\pm_{\mu \nu}{}^{\Lambda}, G_{\mu \nu
\Lambda}^\pm)\ ,\nn\\
X^M & = & (X^\Lambda, F_\Lambda)\ ,\nn\\
\Om_i{}^M & = & (\Om_i{}^\Lambda, F_{\Lambda \Sigma}
\Om_i{}^\Sigma)\ .
\end{eqnarray}
Likewise we use vectors with lower indices, $\beta_M =
(\beta_\Lambda, \beta^\Lambda)$, transforming according to the
conjugate representation such that $\al^M \be_M$ is invariant.

We are especially interest in the subgroup of electric/magnetic
duality transformations that leaves the function $F$ (and
therefore the Lagrangian) invariant,
\begin{eqnarray}\label{invf}
\tF (\tX) = F (\tX)\ ,
\end{eqnarray}
as these are the ones that can be gauged. We note that for such an
invariance one gets
\begin{eqnarray}
  \label{eq:dual-symm-F-der}
  F_\Lambda (\tilde X) &=&
  V_\Lambda{}^\Sigma F_\Sigma(X) + W_{\Lambda\Sigma} X^\Sigma \ ,
  \nonumber\\
  F_{\Lambda\Sigma}(\tilde X) &=& (V_\Lambda{}^\Gamma
  F_{\Gamma\Xi} + W_{\Lambda\Xi} )\, [\mathcal{S}^{-1}]^\Xi{}_\Sigma\ ,
  \nonumber\\
  F_{\Lambda\Sigma\Gamma}(\tilde X) &=& F_{\Xi\Delta\Omega}\,
  [\mathcal{S}^{-1}]^\Xi{}_\Lambda\,[\mathcal{S}^{-1}]^\Delta{}_\Sigma\,
  [\mathcal{S}^{-1}]^\Omega{}_\Gamma\ .
\end{eqnarray}
where we recall that $\mathcal{S}^\Lambda{}_\Sigma = \partial
\tilde X^\Lambda/\partial X^\Sigma = U^\Lambda{}_\Sigma +
Z^{\Lambda\Gamma} F_{\Gamma\Sigma}$.

We elucidate these invariances for the subgroup that acts linearly
on the gauge fields $A_\mu{}^\Lambda$. These symmetries are
characterized by the fact that the matrix in (\ref{emduuu}) has a
block-triangular form with $V= [U^{\mathrm{T}}]^{-1}$ and $Z=0$.
Hence this is not a general duality as the Lagrangian is still
based on the same gauge fields, up to the linear transformation
$A_\mu{}^\Lambda\to \tilde A_\mu{}^\Lambda= U^\Lambda{}_\Sigma
A_\mu{}^\Sigma$. All fields in the Lagrangian (\ref{vecsh}) carry
upper indices and are thus subject to the same linear
transformation. The function $F(X)$ changes with an additive term
which is a quadratic polynomial with real coefficients.
\begin{equation}
  \label{eq:F-PQ}
  \tilde F(\tilde X) = F (U^\Lambda{}_\Sigma X^\Sigma) = F(X) + \half
  (U^\mathrm{T}   W)_{\Lambda\Sigma}\, X^\Lambda X^\Sigma \ .
\end{equation}
This term induces a total derivative term in the Lagrangian, equal
to
\begin{equation}
  \label{eq:variationPQ}
  \mathcal{L} \to  \mathcal{L} - \frac{1}{8} i
  \varepsilon^{\mu\nu\rho\sigma}
  (U^\mathrm{T} W)_{\Lambda\Sigma}\,
  F_{\mu\nu}{}^\Lambda F_{\rho\sigma}{}^\Sigma\ .
\end{equation}

\subsection{Gauge transformations}
\label{sec:gauge-transformations}
Non-abelian gauge groups will act non-trivially on the vector
fields and must therefore involve a subgroup of the duality group.
The electric gauge fields $A_\mu{}^\Lambda$ associated with this
gauge group are provided by vector multiplets. Because the duality
group acts on both electric and magnetic charges, in view of the
fact that it mixes field strengths with dual field strengths as
shown by (\ref{emduuu}), we will eventually have to introduce
magnetic gauge fields $A_{\mu \Lambda}$ as well, following the
procedure explained in \cite{dWST}. The $2n$ gauge fields
$A_\mu{}^M$ will then comprise both type of fields, $A_\mu{}^M=
(A_\mu{}^\Lambda, A_{\mu\Lambda})$. The role played by the
magnetic gauge fields will be clarified later. For the moment one
may associate $A_{\mu\,\Lambda}$ with the dual field strengths
${G}_{\mu\nu \,\Lambda}$, by writing ${G}_{\mu\nu\,\Lambda} \equiv
2 \,\partial_{[\mu} A_{\nu]\Lambda}$.

The generators (as far as their embedding in the duality group is
concerned) are defined as follows. The generators of the subgroup
that is gauged, are $2n$-by-$2n$ matrices $T_M$, where we are
assuming the presence of both electric and magnetic gauge fields,
so that the generators decompose according to
$T_M=(T_{\Lambda},T^\Lambda)$. Obviously $T_{\Lambda N}{}^P$ and
$T^\Lambda{}_N{}^P$ can be decomposed into the generators of the
duality group and are thus of the form specified in
(\ref{emduuu}). Denoting the gauge group parameters by
$\Lambda^M(x) = (\Lambda^\Lambda(x), \Lambda_\Lambda(x))$,
$2n$-dimensional ${\rm Sp}(2n;\mathbb{R})$ vectors $\al^M$ and
$\be_M$ transform according to
\begin{equation}
  \label{eq:gauge-tr-Y-Z}
  \delta \al^M = -g \Lambda^N \,T_{NP}{}^M \,\al^P\ ,\qquad
    \delta \be_M = g \Lambda^N \,T_{NM}{}^P \,\be_P\ ,
\end{equation}
where $g$ denotes a universal gauge coupling constant. Covariant
derivatives thus take the form,
\begin{eqnarray}
  \label{eq:cov-derivative}
  D_\mu \al^M &=& \partial_\mu \al^M + g A_\mu{}^N\, T_{NP}{}^M\,\al^P
  \nonumber \\
  &=& \partial_\mu \al^M + g A_\mu{}^\Lambda\, T_{\Lambda P}{}^M\,\al^P +
  g A_{\mu\Lambda}\, T^\Lambda{}_{P}{}^M\,\al^P \ ,
\end{eqnarray}
and similarly for $D_\mu \be_M$. The gauge fields then transform
according to
\begin{equation}
  \label{eq:gauge-tr-A}
  \delta A_\mu{}^M = \partial_\mu \Lambda^M + g\, T_{PQ}{}^M
  A_\mu{}^P\, \Lambda^Q \ .
\end{equation}

\vspace{5mm}

\textbf{Electric charges}

\vspace{5mm}

For clarity we first consider electric gaugings where the gauge
transformations have a block-triangular form and there are only
electric gauge fields. Hence we ignore the fields $A_{\mu\Lambda}$
and assume $T^\Lambda{}_N{}^P=0$ and $T_\Lambda{}^{\Sigma\Gamma}=
0$. All the fields in the Lagrangian carry upper indices, so that
they will transform as in $\delta X^\Lambda = - g \Lambda^\Gamma
T_{\Gamma\Sigma}{}^\Lambda \,X^\Sigma$.  The transformation rule
for $A_\mu{}^\Lambda$ given above is in accord with this
expression, provided we assume that $T_{\Gamma\Sigma}{}^\Lambda$
is antisymmetric in $\Gamma$ and $\Sigma$. This has to be the case
here as consistency requires that the $T_{\Gamma\Sigma}{}^\Lambda$
are structure constants of the non-abelian group.  In the more
general situation discussed in later sections, this is not
necessarily the case. The embedding into
$\mathrm{Sp}(2n,\mathbb{R})$ implies furthermore that
$T_{\Lambda\Sigma}{}^\Gamma = - T_\Lambda{}^\Sigma{}_\Gamma$,
while the non-vanishing left-lower block $T_{\Lambda\Sigma\Gamma}$
is symmetric in $\Sigma$ and $\Gamma$.

Furthermore, we note that (\ref{eq:F-PQ}) implies
\begin{equation}
  \label{eq:delta-F}
  F_\Lambda(X)\,\delta X^\Lambda = -g \Lambda^\Gamma
  T_{\Gamma\Sigma}{}^\Lambda\,
  F_\Lambda(X)\,X^\Sigma = - \half g\,\Lambda^\Lambda
\,T_{\Lambda\Sigma\Gamma} X^\Sigma X^\Gamma \ .
\end{equation}
Upon replacing $\Lambda^\Lambda$ with $X^\Lambda$ we conclude that
the fully symmetric part of $T_{\Lambda\Sigma\Gamma}$ vanishes.
This, and the closure of the gauge group, leads to the following
three equations,
\begin{eqnarray}
  \label{eq:T-triangle}
  &&
  T_{(\Lambda\Sigma\Gamma)}= 0\ , \nonumber\\
  &&
  T_{[\Lambda\Sigma}{}^\Delta\,T_{\Gamma]\Delta}{}^\Xi = 0 \ ,
  \nonumber\\
  &&
  4\,T_{(\Gamma[\Lambda}{}^\Delta \, T_{\Sigma]\Xi)\Delta}
  - T_{\Lambda\Sigma}{}^\Delta T_{\Delta\Gamma\Xi} = 0\ .
\end{eqnarray}
The variation of the Lagrangian under gauge transformations now
takes the form
\begin{equation}
  \label{eq:gauge-variationPQ}
  \mathcal{L} \to  \mathcal{L} + \frac{1}{8} i \,
  \varepsilon^{\mu\nu\rho\sigma}  \,\Lambda^\Lambda\,
  T_{\Lambda\Sigma\Gamma}\,
  \mathcal{F}_{\mu\nu}{}^\Sigma \mathcal{F}_{\rho\sigma}{}^\Gamma\ .
\end{equation}
where the tensors $\mathcal{F}_{\mu\nu}{}^\Lambda$ denote the
non-abelian field strengths,
\begin{equation}
  \label{eq:non-abelian-FS}
  \mathcal{F}_{\mu\nu}{}^\Lambda = \partial_\mu A_\nu{}^\Lambda -
  \partial_\nu A_\mu{}^\Lambda  + g\, T_{\Sigma\Gamma}{}^\Lambda\,
  A_\mu{}^\Sigma A_\nu{}^\Gamma \ .
\end{equation}
This result implies that (\ref{eq:gauge-variationPQ}) no longer
constitutes a total derivative in view of the spacetime dependent
transformation parameters $\Lambda^\Lambda(x)$. Therefore its
cancellation requires to add a new type of term
\cite{deWit:1984pk},
\begin{equation}
  \label{eq:cs-electric}
  \mathcal{L} =  \frac{1}{3} i g\,
  \varepsilon^{\mu\nu\rho\sigma}\,T_{\Lambda\Sigma\Gamma}
  \,A_\mu{}^\Lambda A_\nu{}^\Sigma (\partial_\rho A_\sigma{}^\Gamma
  + \frac{3}{8} g\, T_{\Xi\Delta}{}^\Gamma \,A_\rho{}^\Xi A_\sigma{}^\Delta)
  \ .
\end{equation}
No other terms in the action will depend on
$T_{\Lambda\Sigma\Gamma}$. At this point we should remind the
reader that the gauging breaks supersymmetry, unless one adds the
standard masslike and potential terms to the Lagrangian
(\ref{vecsh}), which involve the $T_{\Lambda \Sigma}{}^\Gamma$. We
present them below for completeness,
\begin{eqnarray}
  \label{eq:electric-masslike-potential}
  \mathcal{L}_g &=& - \half  g \,N_{\Lambda\Sigma}
  T_{\Gamma\Xi}{}^\Sigma  \Big[ \varepsilon^{ij}\, \bar\Omega_i{}^\Lambda
  \Omega_j{}^\Gamma \bar X^\Xi +
  \varepsilon_{ij} \, \bar\Omega^{i\Lambda} \Omega^{j\Gamma} X^\Xi
  \Big] \ ,   \nonumber \\
  \mathcal{L}_{g^2} &=& g^2\, N_{\Lambda\Sigma}\,
  T_{\Gamma\Xi}{}^\Lambda \bar X^\Gamma   X^\Xi \,
  T_{\Delta\Omega}{}^\Sigma \bar X^\Delta  X^\Omega\ .
\end{eqnarray}
In later sections we will exhibit the generalization of these
terms to the case where both electric and magnetic charges are
present.

\vspace{5mm}

\textbf{Electric and magnetic charges}

\vspace{5mm}

We now consider more general gauge groups without restricting
ourselves to electric charges. Therefore we have to include both
electric gauge fields $A_\mu{}^\Lambda$ and magnetic gauge fields
$A_{\mu\,\Lambda}$.  Only a subset of these fields is usually
involved in the gauging, but the additional magnetic gauge fields
could conceivably lead to new propagating degrees of freedom. We
will discuss in due course how this is avoided. In the remainder
of this section we will consider the scalar and spinor fields. The
treatment of the vector fields is more involved and is explained
in section \ref{sec:embedding-tensor}.

The charges $T_{MN}{}^P$ correspond to a more general subgroup of
the duality group. Hence they must take values in the Lie algebra
associated with $\mathrm{Sp}(2n,\mathbb{R})$, which implies,
\begin{equation}
  \label{eq:sp-constraint}
  T_{M[N}{}^Q\,\Omega_{P]Q} =0\ .
\end{equation}
Combining the two equations (\ref{fx}) and (\ref{invf}) leads to
the condition \cite{deWit:1984pk},
\begin{equation}
  \label{eq:symplectic-invariance}
  T_{MN}{}^Q \Omega_{PQ} \,X^NX^P =
  T_{M\Lambda\Sigma} X^\Lambda X^\Sigma -2 T_{M\Lambda}{}^\Sigma X^\Lambda
  F_\Sigma - T_M{}^{\Lambda\Sigma}F_\Lambda F_\Sigma =0\ .
\end{equation}
This result can also be written as
\begin{equation}
  \label{eq:dlta-F}
  F_\Lambda \delta X^\Lambda = -\half \Lambda^M \Big( T_{M\Lambda\Sigma}
  X^\Lambda X^\Sigma
  +T_M{}^{\Lambda\Sigma} F_\Lambda F_\Sigma\Big) \ ,
\end{equation}
which generalizes (\ref{eq:delta-F}). Furthermore, we impose the
so-called representation constraint \cite{dWST}, which implies
that we suppress a representation of the rigid symmetry group in
$T_{MN}{}^P$,
\begin{equation}
  \label{eq:lin}
  T_{(MN}{}^{Q}\,\Omega_{P)Q} =0
\Longrightarrow  \left\{
\begin{array}{l}
T^{(\Lambda\Sigma\Gamma)}=0\ ,\\[.2ex]
2T^{(\Gamma\Lambda)}{}_{\Sigma}=
T_{\Sigma}{}^{\Lambda\Gamma}\ , \\[.2ex]
T_{(\Lambda\Sigma\Gamma)}=0\ ,\\[.2ex]
2T_{(\Gamma\Lambda)}{}^{\Sigma}= T^{\Sigma}{}_{\Lambda\Gamma}\ .
\end{array}
\right.
\end{equation}
This constraint is a generalization of the first equation
(\ref{eq:T-triangle}). Observe that the generators
$T_{\Lambda\Sigma}{}^\Gamma$ are no longer antisymmetric in
$\Lambda$ and $\Sigma$, a feature that we will discuss in more
detail in the following section.

Using (\ref{symvec}) we can rewrite the Lagrangian (\ref{abelm1})
in a compact form,
\begin{equation}
  \label{eq:lagrangian-pieces}
  \mathcal{L}_{\mathrm{matter}} = -i \Omega_{MN}\,
  \partial_\mu X^M \,\partial^\mu \bar   X^N +
  \vier i\Omega_{MN}\Big[\bar
  \Omega^{iM} \slash{\partial} \Omega_i{}^N
  -\bar \Omega_i{}^M \slash{\partial} \Omega^{iN} \Big] \ .
\end{equation}
In the expressions on the right-hand side it is straightforward to
 replace the ordinary derivatives by the covariant ones defined in
 (\ref{eq:cov-derivative}), i.e.,
\begin{eqnarray}
  \label{eq:covariant-symplectic}
  D_\mu X^M&=&\partial_\mu X^M +g\, A_\mu{}^N \,T_{NP}{}^M \,X^P\ ,
  \nonumber \\
  D_\mu \Omega_i{}^M &=&\partial_\mu\Omega_i{}^M +g\, A_\mu{}^N
  \,T_{NP}{}^M \,\Omega_i{}^P \ ,
\end{eqnarray}
and evaluate the gauge couplings. In particular we can then
compare to the results of subsection
\ref{sec:gauge-transformations}, where we considered only electric
gauge fields with charges restricted by
$T_{\Lambda}{}^{\Sigma\Gamma}=0$. To do this systematically we
note the identity,
\begin{equation}
  \label{eq:derivative-invariance}
  T_{MN\Lambda} X^N - F_{\Lambda\Sigma} \,T_{MN}{}^\Sigma X^N =0\ .
\end{equation}
This equation can also be written as $F_{\Lambda\Sigma}\, \delta
X^\Sigma = - \Lambda^M T_{MN\Lambda} X^N$, which is the
infinitesimal form of the first equation
(\ref{eq:dual-symm-F-der}). Alternatively it can be derived from
(\ref{eq:symplectic-invariance}) upon differentiation with respect
to $X^\Lambda$.

It is possible to cast (\ref{eq:derivative-invariance}) in a
symplectic covariant form by introducing a vector $U^M =
(U^\Lambda, F_{\Sigma\Gamma} U^\Gamma)$, so that
\begin{equation}
  \label{eq:cov-der-invariance}
  \Omega_{MQ} T_{NP}{}^Q \,X^P \,U^M=0\ ,
\end{equation}
for any such vector $U^M$. This form is convenient in calculations
presented later.

From (\ref{eq:derivative-invariance}) one easily derives that
$D_\mu X_\Lambda = D_\mu F_\Lambda = F_{\Lambda\Sigma} \, D_\mu
X^\Sigma$, which enables one to derive
\begin{equation}
  \label{eq:scalar-kinetic}
  -i \Omega_{MN}\, D_\mu X^M \,D^\mu \bar
  X^N = -N_{\Lambda\Sigma} \, D_\mu X^\Lambda \, D^\mu \bar
  X^\Sigma\ ,
\end{equation}
This result shows that the generators $T_{M\Lambda\Sigma}$ are
absent, in accord with what was found in subsection
\ref{sec:gauge-transformations}.

Next we consider the gauge field interactions with the fermions.
It is convenient to first derive an additional identity, which
follows from taking a supersymmetry variation of
(\ref{eq:derivative-invariance}),
\begin{equation}
  \label{eq:fermion-der-invariance}
  T_{MN\Lambda} \Omega_i{}^N = F_{\Lambda\Sigma} \,T_{MN}{}^\Sigma
    \Omega_i{}^N +F_{\Lambda\Sigma\Gamma}\,\Omega_i{}^\Sigma
    \,T_{MN}{}^\Gamma X^N   \ .
\end{equation}
This result can be obtained from the infinitesimal form of the
third equation of (\ref{eq:dual-symm-F-der}). Using this equation
one verifies that $D_\mu \Omega_{i\Lambda} = F_{\Lambda\Sigma}
\,D_\mu \Omega_i{}^\Sigma + F_{\Lambda\Sigma\Gamma}\,
\Omega_i{}^\Gamma D_\mu X^\Sigma$, which leads to
\begin{eqnarray}
  \label{eq:fermion-kinetic}
    \vier i\Omega_{MN}\Big[\bar
    \Omega^{iM} \Slash{D} \Omega_i{}^N
    -\bar \Omega_i{}^M \Slash{D} \Omega^{iN} \Big]  &=& -\vier
    N_{\Lambda\Sigma}\Big(\bar\Omega^{i\Lambda}\Slash{D}
    \Omega_i{}^\Sigma
      +\bar \Omega_i{}^\Lambda \Slash{D}\Omega ^{i\Sigma}\Big)
      \nonumber\\
      &&
      -\vier i\Big( F_{\Lambda\Sigma\Gamma}\bar
    \Omega_i{}^\Lambda\Slash{D} X^\Sigma \Omega ^{i\Gamma}
      \nn\\
      &&
      - \bar F_{\Lambda\Sigma\Gamma} \bar\Omega^{i
    \Lambda} \Slash{D} \bar X^\Sigma \Omega _i{}^\Gamma\Big) \ .
\end{eqnarray}
Again the generator $T_{M\Lambda\Sigma}$ is absent in the
expression above. The results of this subsection explain how to
introduce the electric and magnetic charges, but in no way ensure
the gauge invariance or the supersymmetry of the Lagrangian. To
obtain such a result we first need to explain some more general
features of theories with both electric and magnetic gauge fields
in four spacetime dimensions. This is the topic of the following
section.

As a side remark we note that the moment map associated with the
isometries considered above, takes the form,
\begin{equation}
  \label{eq:U(1)-moment-map}
  \nu_M = T_{MN}{}^Q \Omega_{PQ} \bar X^N X^P \ .
\end{equation}
Indeed, making use again of (\ref{eq:derivative-invariance}), one
straightforwardly derives $\partial_\Lambda \nu_M = i
N_{\Lambda\Sigma} \,\delta\bar X^\Sigma$.

\section{The gauge group and the embedding tensor}
\label{sec:embedding-tensor}
Here we follow \cite{dWST} and discuss the embedding of possible
gauge groups into the rigid invariance group
$\mathrm{G}_{\mathrm{rigid}}$ of the theory. In the context of our
work, the latter is often a product group as the vector multiplets
and the hypermultiplets are invariant under independent symmetry
groups. As explained in the previous section the non-abelian gauge
transformations on the vector multiplets are necessarily embedded
into the electric/magnetic duality group.

It is convenient to discuss group embeddings in terms of a
so-called embedding tensor $\Theta_M{}^{\sf a}$ which specifies
the decomposition of the gauge group generators $T_M$ into the
generators associated with the full rigid invariance group
 $\mathrm{G}_{\mathrm{rigid}}$,
\begin{equation}
  \label{eq:T-into-t}
  T_M = \Theta_M{}^{\sf a} \,t_{\sf a} \ .
\end{equation}
Not all the gauge fields have to be involved in the gauging, so
generically the embedding tensor projects out certain combinations
of gauge fields; the rank of the tensor determines the dimension
of the gauge group, up to central extensions associated with
abelian factors. Decomposing the embedding tensor as
$\Theta_M{}^{\sf a} = (\Theta_\Lambda{}^{\sf a},
\Theta^{\Lambda\,{\sf a}})$, covariant derivatives take the form,
\begin{equation}
  \label{eq:cov-der}
   D_\mu\equiv
  \partial_{\mu}-g A_{\mu}{}^{M} T_M = \partial_{\mu}-g
  A_{\mu}{}^{\Lambda}\Theta_{\Lambda}{}^{\sf a} \,t_{\sf a} -g
  A_{\mu\,\Lambda}\Theta^{\Lambda\,\sf{a}} \,t_{\sf a} \;.
\end{equation}
The embedding tensor will be regarded as a spurionic object which
can be assigned to a (not necessarily irreducible) representation
of the rigid invariance group $\mathrm{G}_{\mathrm{rigid}}$.

It is known that a number of
($\mathrm{G}_{\mathrm{rigid}}$-covariant) constraints must be
imposed on the embedding tensor. We already encountered the
representation constraint (\ref{eq:lin}), which is linear in the
embedding tensor. Two other constraints are quadratic in the
embedding tensor and read,
\begin{eqnarray}
  f_{\sf ab}{}^{\sf c}\, \Theta_{M}{}^{\sf a}\,\Theta_{N}{}^{\sf b}
+(t_{\sf a})_{N}{}^{P}\,\Theta_{M}{}^{\sf a}\Theta_{P}{}^{\sf c}
&=&0\ ,
  \label{eq:clos}  \\[1ex]
\Omega^{MN}\,\Theta_{M}{}^{\sf a}\Theta_{N}{}^{\sf b}~=~0
\;\;\Longleftrightarrow\; \; \Theta^{\Lambda\,[\sf
a}\Theta_{\Lambda}{}^{\sf b]} &=&0 \;, \label{eq:quad}
\end{eqnarray}
where the $f_{{\sf a} {\sf b}}{}^\gamma$ are the structure
constants associated with the group $\mathrm{G}$.  The first
constraint is required by the closure of the gauge group
generators. Indeed, from (\ref{eq:clos}) it follows that the gauge
algebra generators close according to
\begin{equation}
  \label{eq:closure}
  {}[T_{M},T_{N}] = -T_{MN}{}^{P}\,T_{P} \;,
\end{equation}
where the structure constants of the gauge group coincide with
$T_{MN}{}^{P}\equiv \Theta_{M}{}^{\sf a} \,(t_{\sf a})_{N}{}^{P}$
up to terms that vanish upon contraction with the embedding tensor
$\Theta_P{}^{\sf a}$.  We recall that the $T_{MN}{}^P$ generate a
subgroup of $\mathrm{Sp}(2n,\mathbb{R})$ in the $(2n)$-dimensional
representation, so that they are subject to the condition
(\ref{eq:sp-constraint}).  In electric/magnetic components the
latter condition corresponds to $T_{M\Lambda}{}^\Sigma=
-T_{M}{}^\Sigma{}_\Lambda$, $T_{M
\Lambda\Sigma}=T_{M\Sigma\Lambda}$ and
$T_{M}{}^{\Lambda\Sigma}=T_{M}{}^{\Sigma\Lambda}$.

Note that (\ref{eq:clos}) implies that the embedding tensor is
gauge invariant, while the second quadratic constraint
(\ref{eq:quad}) implies that the charges are mutually local, so
that an electric/magnetic duality exists that converts all the
charges to electric ones.  These two quadratic constraints are not
completely independent, as can be seen from symmetrizing the
constraint (\ref{eq:clos}) in $(MN)$ and making use of the linear
conditions (\ref{eq:lin}) and (\ref{eq:sp-constraint}). This leads
to
\begin{eqnarray}
  \label{eq:constraint-eq}
\Omega^{MN}\,\Theta_{M}{}^{\sf a}\Theta_{N}{}^{\sf b}\, (t_{\sf
b})_{P}{}^{Q}&=&0 \;.
\end{eqnarray}
This shows that, for non-vanishing $(t_{\sf b})_{P}{}^{Q}$, the
second quadratic constraint (\ref{eq:quad}) is in fact a
consequence of the other constraints. The constraint
(\ref{eq:quad}) is only an independent constraint when ${\sf a}$
and ${\sf b}$ do not refer to generators that act on the vector
multiplets. This issue is relevant here as
$\mathrm{G}_{\mathrm{rigid}}$ may contain independent generators
that act exclusively in the matter (i.e., hypermultiplet) sector.

A further consequence of~(\ref{eq:lin}) is the equation
\begin{equation}
  \label{eq:Z-d}
  T_{(MN)}{}^{P}= Z^{P,{\sf a}}\, d_{{\sf a} \,MN} \;,
\end{equation}
with
\begin{eqnarray}
  \label{eq:def-Z-d}
d_{{\sf a} \, MN} &\equiv& (t_{\sf a})_M{}^P\, \Omega_{NP}\ ,\nonumber\\
Z^{M, {\sf a}}&\equiv&\half\Omega^{MN}\Theta_{N}{}^{\sf a} \quad
\Longrightarrow \quad \left\{
\begin{array}{rcr}
Z^{\Lambda {\sf a}} &=& \half\Theta^{\Lambda {\sf a}} \ ,\\[1ex]
Z_{\Lambda}{}^{\sf a} &=& -\half\Theta_{\Lambda}{}^{\sf a} \ ,
\end{array}
\right.
\end{eqnarray}
so that $d_{{\sf a} \,MN}$ defines a
$\mathrm{G}_{\mathrm{rigid}}$-invariant tensor symmetric in
$(MN)$. The gauge invariant tensor $Z^{M, {\sf a}}$ will serve as
a projector on the tensor fields to be introduced below
\cite{deWit:2005hv}. We note that the constraint (\ref{eq:quad})
can now be written as,
\begin{equation}
  \label{eq:Z-Theta}
  Z^{M, {\sf a}} \,\Theta_M{}^{\sf b} =0\ .
\end{equation}

Let us return to the closure relation (\ref{eq:closure}). Although
the left-hand side is antisymmetric in $M$ and $N$, this does not
imply that $T_{MN}{}^P$ is antisymmetric as well, but only that
its symmetric part vanishes upon contraction with the embedding
tensor. Indeed, this is reflected by (\ref{eq:Z-d}) and
(\ref{eq:Z-Theta}). Consequently, the Jacobi identity holds only
modulo terms that vanish upon contraction with the embedding
tensor, as is shown explicitly by
\begin{equation}
  \label{Jacobi-X}
  {T_{[MN]}{}^P\, T_{[QP]}{}^R + T_{[QM]}{}^P\, T_{[NP]}{}^R  +
 T_{[NQ]}{}^P \, T_{[MP]}{}^R} = - Z^{R, {\sf a}}\,d_{{\sf a}\,P[Q}\,
 T_{MN]}{}^P \ .
\end{equation}
To compensate for this lack of closure and, at the same time, to
avoid unwanted degrees of freedom, we introduce an extra gauge
invariance for the gauge fields, in addition to the usual
non-abelian gauge transformations,
\begin{equation}
  \label{eq:A-var}
  \delta A_\mu{}^M =  D_\mu\Lambda^M-
  g\,Z^{M,{\sf a}}\,\Xi_{\mu\,{\sf a}} \ ,
\end{equation}
where the $\Lambda^M$ are the gauge transformation parameters and
the covariant derivative reads, $D_\mu\Lambda^M
=\partial_\mu\Lambda^M + g\, T_{PQ}{}^M\,A_\mu{}^P\Lambda^Q$. The
transformations proportional to $\Xi_{\m\,{\sf a}}$ enable one to
gauge away those vector fields that are in the sector of the gauge
generators $T_{MN}{}^P$ where the Jacobi identity is not satisfied
(this sector is perpendicular to the embedding tensor by virtue of
(\ref{eq:Z-Theta})). Note that the covariant derivative is
invariant under the transformations parameterized by
$\Xi_{\mu\,{\sf a}}$, because of the contraction of the gauge
fields $A_\mu{}^M$ with the generators $T_M$. The gauge symmetries
parameterized by the functions $\Lambda^M(x)$ and $\Xi_{{\sf
a}\mu}(x)$ form a group, as follows from the commutation
relations,
\begin{eqnarray}
  \label{eq:gauge-commutators}
  {}[\delta(\Lambda_1),\delta(\Lambda_2)] &=& \delta(\Lambda_3) +
  \delta(\Xi_3) \ ,  \nonumber \\
  {}[\delta(\Lambda),\delta(\Xi)] &=& \delta(\tilde\Xi) \ ,
\end{eqnarray}
where
\begin{eqnarray}
  \label{eq:gauge-parameters}
  \Lambda_3{}^M &=& g\,T_{[NP]}{}^M \Lambda_2^N\Lambda_1^P\ , \nonumber\\
  \Xi_{3 \mu{\sf a}} &=& d_{{\sf a} NP}( \Lambda_1^N
  D_\mu\Lambda_2^P - \Lambda_2^N D_\mu \Lambda_1^P) \ , \nonumber\\
  \tilde\Xi_{\mu{\sf a}} &=& g\Lambda^P( T_{P{\sf a}}{}^{\sf b} + 2 d_{{\sf a}
  PN} Z^{N,{\sf b}}) \Xi_{\mu{\sf b}} \ .
\end{eqnarray}

The field strengths follow from the Ricci identity,
$[D_\mu,D_\nu]= - g \mathcal{F}_{\mu\nu}{}^M\,T_M$, and depend
only on the antisymmetric part of $T_{MN}{}^P$,
\begin{equation}
  \label{eq:field-strength}
  {\cal  F}_{\mu\nu}{}^M =\p_\m A_\n{}^M -\p_\n A_\m{}^M + g\,
  T_{[NP]}{}^M \,A_\m{}^N A_\n{}^P \ .
\end{equation}
Because of the lack of closure expressed by (\ref{Jacobi-X}), they
do not satisfy the Palatini identity,
\begin{equation}
  \label{eq:Palatini}
  \delta\mathcal{F}_{\mu\nu}{}^M = 2\, D_{[\mu}\delta A_{\nu]}{}^M -
  2 g\, T_{(PQ)}{}^M \,A_{[\mu}{}^P \,\delta A_{\nu]}{}^Q\ ,
\end{equation}
under arbitrary variations $\delta A_\mu{}^M$. Note that the last
term cancels upon multiplication with the generators $T_M$. The
result (\ref{eq:Palatini}) shows that $\mathcal{F}_{\mu\nu}{}^M$
transforms under gauge transformations as
\begin{equation}
  \label{eq:delta-cal-F}
  \delta\mathcal{F}_{\mu\nu}{}^M= g\, \Lambda^P T_{NP}{}^M
  \,\mathcal{F}_{\mu\nu}{}^N - 2 g\, Z^{M,{\sf a}} (D_{[\mu}
  \Xi_{\nu]{\sf a}} +d_{{\sf a} PQ} \,A_{[\mu}{}^P\,\delta A_{\nu]}{}^Q)
  \ ,
\end{equation}
and is therefore not covariant. The standard strategy is therefore
to define modified field strengths,
\begin{equation}
  \label{eq:modified-fs}
  {\cal H}_{\m\n}{}^M= {\cal F}_{\mu\nu}{}^M
  + g\, Z^{M,{\sf a}} \,B_{\m\n {\sf a}}\;,
\end{equation}
by introducing new tensor fields $B_{\mu\nu {\sf a}}$ with
suitably chosen gauge transformation rules, so that covariant
results can be obtained.

At this point we remind the reader that the invariance
transformations in the rigid case implied that the field strengths
$G_{\mu\nu}{}^M$ transform under a subgroup of
$\mathrm{Sp}(2n,\mathbb{R})$ (c.f. (\ref{emduuu})). Our aim is to
find a similar symplectic vector of field strengths so that these
transformations are generated in the non-abelian case as well.
This is not possible based on the variations of the vector fields
$A_\mu{}^M$, which will never generate the type of fermionic terms
contained in $G_{\mu\nu\Lambda}$. However, the presence of the
tensor fields enables us to achieve our objectives, at least in
part. Just as in the abelian case, we define an
$\mathrm{Sp}(2n,\mathbb{R})$ vector of field strengths
$\mathcal{G}_{\mu\nu}{}^M$ by
\begin{eqnarray}
  \label{eq:cal-G}
  \mathcal{G}^-_{\mu\nu}{}^\Lambda &=& \mathcal{H}^-_{\mu\nu}{}^\Lambda\ ,
  \nonumber\\
  \mathcal{G}^-_{\mu\nu\Lambda} &=& F_{\Lambda\Sigma}
  \,\mathcal{H}^-_{\mu\nu}{}^\Sigma -
  \frac{1}{8} F_{\Lambda\Sigma\Gamma} \, \bar \Omega_i{}^{\Sigma}
  \gamma_{\mu\nu} \Omega_j{}^{\Gamma}\,\varepsilon^{ij} \ .
\end{eqnarray}
Note that the expression for $\mathcal{G}_{\mu\nu\Lambda}$ is the
analogue of (\ref{ggedef}), with $F_{\mu\nu}{}^\Lambda$ replaced
by $\mathcal{H}_{\mu\nu}{}^\Lambda$.

Following \cite{dWST} we introduce the following transformation
rule for $B_{\mu\nu{\sf a}}$ (contracted with $Z^{M,{\sf a}}$,
because only these combinations will appear in the Lagrangian),
\begin{equation}
  \label{eq:B-transf-0}
   Z^{M,{\sf a}}\, \delta B_{\mu\nu {\sf a}} = 2\,Z^{M,{\sf a}}
   (D_{[\mu} \Xi_{\nu]{\sf a}} + d_{{\sf a}\,NP} A_{[\mu}{}^N \delta
   A_{\nu]}{}^P)  - 2\,T_{(NP)}{}^M  \Lambda^P
   \mathcal{G}_{\mu\nu}{}^N    \ ,
\end{equation}
where $D_\mu \Xi_{\nu{\sf a}}= \partial_\mu \Xi_{\nu{\sf a}} - g
A_\mu{}^M T_{M{\sf a}}{}^{\sf b} \Xi_{\nu{\sf b}}$ with $T_{M{\sf
a}}{}^{\sf b}= -\Theta_M{}^\gamma f_{\gamma{\sf a}}{}^{\sf b}$ the
gauge group generator in the adjoint representation of
$\mathrm{G}_{\mathrm{rigid}}$.  With this variation the modified
field strengths (\ref{eq:modified-fs}) are invariant under tensor
gauge transformations. Under the vector gauge transformations we
derive the following result,
\begin{eqnarray}
  \label{eq:delta-G/H}
  \delta \mathcal{G}^-_{\mu\nu}{}^\Lambda&=& - g\,\Lambda^P
  T_{PN}{}^\Lambda \,\mathcal{G}^-_{\mu\nu}{}^N  - g\,\Lambda^P
  T^\Gamma{}_P{}^\Lambda \,
  (\mathcal{G}^-_{\mu\nu} - \mathcal{H}^-_{\mu\nu})_\Gamma \ ,
\nonumber\\
  \delta \mathcal{G}^-_{\mu\nu\Lambda} &=& - g\,\Lambda^P
  T_{PN\Lambda} \, \mathcal{G}^-_{\mu\nu}{}^N  -g \,
  F_{\Lambda\Sigma}\,\Lambda^P T^\Gamma{}_P{}^\Sigma\,
  (\mathcal{G}^-_{\mu\nu} - \mathcal{H}^-_{\mu\nu})_\Gamma\ ,
\nonumber\\
  \delta(\mathcal{G}^-_{\mu\nu} - \mathcal{H}^-_{\mu\nu})_\Lambda
  &=&  g \, \Lambda^P(  T^\Gamma{}_{P\Lambda} -T^\Gamma{}_P{}^\Sigma
  \, F_{\Sigma\Lambda})\,
  (\mathcal{G}^-_{\mu\nu} - \mathcal{H}^-_{\mu\nu})_\Gamma\ .
\end{eqnarray}
Hence $\delta\mathcal{G}_{\mu\nu}{}^M=-g\,\Lambda^PT_{PN}{}^M\,
\mathcal{G}_{\mu\nu}^N$, just as the variation of the abelian
field strengths $G_{\mu\nu}{}^M$ in the absence of charges, up to
terms that are proportional to $\Theta^{\Lambda,{\sf
a}}(\mathcal{G}_{\mu\nu} -\mathcal{H}_{\mu\nu})_\Lambda$.
According to \cite{dWST}, the latter terms represent a set of
field equations. In that case the last equation of
(\ref{eq:delta-G/H}) expresses the well-known fact that, under a
symmetry, field equations transform into field equations.  As a
result the gauge algebra on these tensors closes according to
(\ref{eq:gauge-commutators}), up to the same field equations.

In order that the Lagrangian becomes invariant under the vector
and tensor gauge transformations, we have to make a number of
changes.  First of all, we replace the abelian field strengths
$F_{\mu\nu}{}^\Lambda$ in (\ref{abelf}) by
$\mathcal{H}_{\mu\nu}{}^\Lambda$, so that
\begin{equation}
  \label{eq:def-cal-G}
  \mathcal{G}_{\mu\nu\,\Lambda} = i
  \varepsilon_{\mu\nu\rho\sigma}\,
  \frac{\de \mathcal{L}_{\mathrm{vector}}}
  {\de {\mathcal{H}}_{\rho\sigma}{}^\Lambda}   \;.
\end{equation}
Under general variations of the vector and tensor fields we then
obtain the result,
\begin{equation}
  \label{eq:var-L-vector}
  \delta\mathcal{L}_{\mathrm{vector}} = -i
  \mathcal{G}^{+\mu\nu}{}_\Lambda \Big[ D_\mu\delta A_\nu{}^\Lambda +
  \vier g\Theta^{\Lambda{\sf a}} (\delta B_{\mu\nu{\sf a}} - 2d_{{\sf a}
  PQ} A_\mu{}^P \delta A_\nu{}^Q)\Big ] + \mathrm{h.c.}  \ .
\end{equation}
The reader can check that the Lagrangian (\ref{abelf}) (with
$F_{\mu\nu}{}^\Lambda$ replaced by
$\mathcal{H}_{\mu\nu}{}^\Lambda$) is indeed invariant under the
tensor gauge transformations. Even when we include the
transformations of the scalar and spinor fields, the Lagrangian
is, however, not yet invariant under the vector gauge
transformations. For that it is necessary to introduce the
following universal terms to the Lagrangian \cite{dWST},
\begin{eqnarray}
  \label{eq:Ltop}
  {\cal L}_{\rm top} &=&
  \frac{1}{8}i g\, \varepsilon^{\mu\nu\rho\sigma}\,
  \Theta^{\Lambda{\sf a}}\,B_{\mu\nu{\sf a}} \,
  \Big(2\,\partial_{\rho} A_{\sigma\,\Lambda} + g
  T_{MN\,\Lambda} \,A_\rho{}^M A_\sigma{}^N
  -\vier g\Theta_{\Lambda}{}^{\sf b}B_{\rho\sigma{\sf b}}\Big)
  \nonumber\\[.9ex]
  &&{}
  + \frac{1}{3}i g\, \varepsilon^{\mu\nu\rho\sigma} T_{MN\,\Lambda}\,
  A_{\mu}{}^{M} A_{\nu}{}^{N}
  \Big(\partial_{\rho} A_{\sigma}{}^{\Lambda}
  +\vier gT_{PQ}{}^{\Lambda} A_{\rho}{}^{P}A_{\sigma}{}^{Q}\Big)
  \nonumber\\[.9ex]
  &&{}
  + \frac{1}{6} i g\,\varepsilon^{\mu\nu\rho\sigma}T_{MN}{}^{\Lambda}\,
  A_{\mu}{}^{M} A_{\nu}{}^{N}
  \Big(\partial_{\rho} A_{\sigma}{}_{\Lambda}
  +\vier gT_{PQ\Lambda} A_{\rho}{}^{P}A_{\sigma}{}^{Q}\Big)
  \;.
\end{eqnarray}
The first term represents a topological coupling of the
antisymmetric tensor fields with the magnetic gauge fields, and
the last two terms are a generalization of the Chern-Simons-like
terms (\ref{eq:cs-electric}) that we encountered in subsection
\ref{sec:gauge-transformations}. Under variations of the vector
and tensor fields, this Lagrangian varies into (up to total
derivative terms)
\begin{equation}
  \label{eq:var-L-top}
  \delta\mathcal{L}_{\mathrm{top}} = i
  \mathcal{H}^{+\mu\nu\Lambda} \, D_\mu\delta A_{\nu\Lambda}  +
  \vier i g\, \mathcal{H}^{+\mu\nu}{}_\Lambda
  \,\Theta^{\Lambda{\sf a}} (\delta B_{\mu\nu{\sf a}} - 2d_{{\sf a}
  PQ} A_\mu{}^P \delta A_\nu{}^Q) + \mathrm{h.c.} \ .
\end{equation}
Under the gauge transformations associated with the tensor fields
$B_{\mu \nu \sf{a}}$ this variation becomes equal to $(i g\,
\mathcal{H}^{+\mu\nu M}\,\Theta_M{}^{\sf a}\, D_\mu\Xi_{\nu{\sf
a}} + \mathrm{h.c.})$. This expression equals a total derivative
by virtue of the invariance of the embedding tensor, the Bianchi
identity - which reads $D_{[\mu}{\cal H}_{\nu\rho]}{}^{M} =
\frac{1}{3}g\,Z^{M,{\sf a}}\,{\cal H}_{\mu\nu\rho\,{\sf a}}$ - and
(\ref{eq:Z-Theta}).

In the Bianchi identity mentioned above, $D_{\mu}{\cal
  H}_{\nu\rho}{}^{M} = \partial_{\mu}{\cal H}_{\nu\rho}{}^{M} +
gA_\mu{}^P T_{PN}{}^M \mathcal{H}_{\nu\rho}{}^N$ and ${\cal
  H}_{\mu\nu\rho\,{\sf a}}$ denotes a field strength associated with
the tensor fields.  The expression for the Bianchi identity given
above is suitable for our purpose here, but we note that it is not
covariant in this form, in view of the fact that the fully
covariant derivative of $\mathcal{H}_{\mu\nu}{}^M$ reads,
\begin{equation}
  \label{eq:new-cov-der-H}
  \mathcal{D}_\rho \mathcal{H}_{\mu\nu}{}^M =
  \partial_\rho \mathcal{H}_{\mu\nu}{}^M + gA_\rho{}^P
  T_{PN}{}^M \,\mathcal{G}_{\mu\nu}{}^N  + gA_\rho{}^P
  T_{NP}{}^M \,
  (\mathcal{G}_{\mu\nu} - \mathcal{H}_{\mu\nu})_N \ ,
\end{equation}
and the covariant field strength of the tensor fields equals
\begin{eqnarray}
  \label{eq:tensor-H}
  {\cal H}_{\mu\nu\rho\,{\sf a}} & \equiv & 3\, D_{[\mu}
  B_{\nu\rho]{\sf a}} +6 \,d_{{\sf a}\, MN}\,A_{[\mu}{}^{M}
  \Big(\partial_{\nu} A_{\rho]}{}^N\nn\\
  && + \frac{1}{3} g T_{[RS]}{}^{N}
  A_{\nu}{}^{R}A_{\rho]}{}^{S} + \mathcal{G}_{\nu\rho]}{}^N -
  \mathcal{H}_{\nu\rho]}{}^N \Big) \;,
\end{eqnarray}
where $D_\rho B_{\mu\nu{\sf a}}= \partial_\rho B_{\mu\nu{\sf a}} -
g A_\rho{}^M T_{M{\sf a}}{}^{\sf b} B_{\mu\nu{\sf b}}$.  With
these definitions the covariant form of the Bianchi identity
holds,
\begin{equation}
  \label{eq:bianchi-H}
  \mathcal{D}_{[\mu}{\cal H}_{\nu\rho]}{}^{M} =
  \frac{1}{3}g\,Z^{M,{\sf a}}\,{\cal H}_{\mu\nu\rho\,{\sf a}} \;.
\end{equation}

These modifications ensure the gauge invariance of the total
Lagrangian $\mathcal{L}_{\mathrm{vector}} +
\mathcal{L}_{\mathrm{top}}$, provided we include the gauge
transformations of the scalar and spinor fields \cite{dWST}.
Furthermore, variation of the tensor fields yields the field
equations identified above,
\begin{equation}
  \label{eq:B-field-eq}
  \delta\mathcal{L}_{\mathrm{vector}} +
  \delta\mathcal{L}_{\mathrm{top}} = -  \vier ig\,
  \, \delta B_{\mu\nu {\sf a}}\;
  \Theta^{\Lambda,{\sf a}} \Big[(\mathcal{G}^{+\mu\nu}
  -\mathcal{H}^{+\mu\nu})_\Lambda - (\mathcal{G}^{-\mu\nu}
  -\mathcal{H}^{-\mu\nu})_\Lambda \Big] \ .
\end{equation}
In spite of the modifications above, supersymmetry will be broken
by the gauging. In the next section we show it can be restored.

\section{Restoring supersymmetry for non-abelian vector multiplets}
\label{sec:rest-supersymm-non}
In this section we show how the supersymmetry can be restored in
the presence of a gauging. In this way we will find the
generalizations of the masslike and potential terms of order $g$
and $g^2$, respectively, which was already exhibited in
(\ref{eq:electric-masslike-potential}) for the case of electric
charges. In addition we determine the corresponding changes in the
transformation rules.

The supersymmetry transformations that leave the ungauged action
(\ref{vecsh}) invariant, we recall, are given by
\begin{eqnarray}
  \label{eq:susyr}
  \delta X^{\Lambda} & = & \bar{\epsilon}^i \Omega_i^{\; \Lambda}\ ,
  \,\nonumber\\
  \delta A_{\mu}{}^{\Lambda} & = & \varepsilon^{ij} \bar{\epsilon}_i
  \gamma_{\mu} \Omega_j{}^{\Lambda} + \varepsilon_{ij}
  \bar{\epsilon}^i \gamma_{\mu} \Omega^{j\, \Lambda}\ ,\nonumber\\
  \delta \Omega_i{}^{\Lambda} & = & 2 \slash{\partial}
  X^{\Lambda} \epsilon_i + \half \gamma^{\mu \nu}
  F^-_{\mu\nu}{}^\Lambda \varepsilon_{ij} \epsilon^j +
  Y_{ij}{}^{\Lambda} \epsilon^j\ , \nonumber\\
  \delta Y_{ij}{}^{\Lambda} & = & 2 \bar{\epsilon}_{(i}
  \slash{\partial}\Omega_{j)}{}^{\Lambda} + 2 \varepsilon_{ik}
  \varepsilon_{jl}\, \bar{\epsilon}^{(k} \slash{\partial}\Omega^{l)
  \Lambda}   \ .
\end{eqnarray}
The extension of these transformations in the presence of electric
charges is known \cite{dWvHvP}. Therefore we will now proceed and
consider the case of electric and/or magnetic charges.

Introducing the charges, with a uniform gauge coupling constant
$g$ as before, we have already discussed some universal changes of
the Lagrangian in the previous section. In
$\mathcal{L}_{\mathrm{matter}}$ we have to covariantize the
derivatives as already discussed in section
\ref{sec:gauge-transformations}. It is convenient to use the
representation (\ref{eq:lagrangian-pieces}). With the
covariantizations included we thus have
\begin{equation}
  \label{eq:lagrangian-matter-cov}
  \mathcal{L}_{\mathrm{matter}} = -i \Omega_{MN}\,
  D_\mu X^M \,D^\mu \bar   X^N +
  \vier i\Omega_{MN}\Big[\bar
  \Omega^{iM} \Slash{D} \Omega_i{}^N
  -\bar \Omega_i{}^M \Slash{D} \Omega^{iN} \Big] \ .
\end{equation}
In $\mathcal{L}_{\mathrm{vector}}$ we must replace the abelian
field strengths $F_{\mu\nu}{}^\Lambda$ by the modified field
strengths $\mathcal{H}_{\mu\nu}{}^\Lambda$, defined in
(\ref{eq:modified-fs}). Therefore we replace (\ref{abelf}) by
\begin{eqnarray}
  \label{eq:vector-H-cov}
  \mathcal{L}_{\mathrm{vector}}  &=&
  (\vier iF_{\Lambda\Sigma}\mathcal{H}^{-}_{\mu\nu}{}^\Lambda
  \mathcal{H}^{- \mu\nu \Sigma}
  -\frac{1}{16} i F_{\Lambda\Sigma\Gamma}\bar\Omega
  _i^\Lambda\,\gamma^{\m\n}
  \mathcal{H}^-_{\m\n}{}^{\Sigma}\,\Omega _j^\Gamma\, \varepsilon^{ij}
      \nonumber\\
  &&
  - \frac{1}{256}i N^{\Delta\Omega} \Big(F_{\Delta\Lambda\Sigma}
      \bar\Omega _i{}^\Lambda\gamma_{\m\n}\Omega _j{}^\Sigma
  \varepsilon^{ij} \Big) \Big(F_{\Gamma\Xi\Omega}
  \bar\Omega _k{}^\Gamma\gamma^{\m\n}\Omega _l{}^\Xi
  \varepsilon^{kl}\Big)+ \mathrm{h.c.})  \ . \nn\\
\end{eqnarray}
Furthermore, one includes the Lagrangians $L_{\mathrm{\Om^4}}$
(\ref{abelm2}), $L_{\mathrm{Y}}$ (\ref{abely}) (which remain
unaltered) and (\ref{eq:Ltop}). Up to an extension of
(\ref{eq:electric-masslike-potential}), whose form we will
establish in this section, we do not expect further modifications.

Also the supersymmetry transformation rules acquire a number of
modifications, extending spacetime derivatives and field strengths
to covariant ones.  Furthermore, one has to take account of the
presence of the new magnetic gauge fields and the tensor fields.
However, one also needs a few additional terms in the
transformation rules, whose form will be established in due
course. For the moment we use the following modified
transformation rules, where we also include the variations of the
magnetic gauge fields, which we denote by $\delta_0$,
\begin{eqnarray}
  \label{eq:susy-1}
  \delta_0 X^{\Lambda} & = & \bar{\epsilon}^i \Omega_i{}^{\Lambda}\ ,
  \,\nonumber\\
  \delta_0 A_{\mu}{}^{\Lambda} & = & \varepsilon^{ij} \bar{\epsilon}_i
  \gamma_{\mu} \Omega_j{}^{\Lambda} + \varepsilon_{ij}
  \bar{\epsilon}^i \gamma_{\mu} \Omega^{j\Lambda}\ ,\nonumber\\
  \delta_0 A_{\mu\Lambda} & = & F_{\Lambda\Sigma}\, \varepsilon^{ij}
  \bar{\epsilon}_i \gamma_{\mu} \Omega_j{}^{\Sigma} +
  \bar F_{\Lambda\Sigma}\, \varepsilon_{ij}
  \bar{\epsilon}^i \gamma_{\mu}\Omega^{j\Sigma} \ , \nonumber\\
  \delta_0\Omega_i{}^\Lambda & = & 2 \Slash{D}X^{\Lambda} \epsilon_i
  + \half \gamma^{\mu \nu} \mathcal{H}^-_{\mu\nu}{}^\Lambda
  \varepsilon_{ij} \epsilon^j + Y_{ij}{}^{\Lambda}
  \epsilon^j\ , \nonumber\\
  \delta_0 Y_{ij}{}^{\Lambda} & = & 2 \bar{\epsilon}_{(i}
  \Slash{D}\Omega_{j)}{}^{\Lambda} + 2 \varepsilon_{ik}
  \varepsilon_{jl}\, \bar{\epsilon}^{(k} \Slash{D}\Omega^{l)
  \Lambda}   \ .
\end{eqnarray}
At this point it is convenient to note that the supersymmetry
variations of the scalar, spinor and vector fields can be written
in the form,
\begin{eqnarray}
  \label{eq:susy-symplectic}
  \delta_0 X^M & = & \bar{\epsilon}^i \Omega_i{}^M \ ,
  \,\nonumber\\
  \delta_0 A_{\mu}{}^M& = & \varepsilon^{ij} \bar{\epsilon}_i
  \gamma_{\mu} \Omega_j{}^M + \varepsilon_{ij}
  \bar{\epsilon}^i \gamma_{\mu} \Omega^{j M}\ ,\nonumber\\
  \delta_0 \Omega_i{}^M & = & 2 \Slash{D}X^M \epsilon_i
  + \half \gamma^{\mu \nu} \mathcal{G}^-_{\mu\nu}{}^M
  \varepsilon_{ij} \epsilon^j + \cdots \ ,
\end{eqnarray}
where the fermions $\Omega_i{}^M$ and the field strengths
$\mathcal{G}_{\mu\nu}{}^M$ were defined in (\ref{symvec}) and
(\ref{eq:cal-G}), respectively. The suppressed terms in
$\delta\Omega_i{}^M$ are proportional to $Y_{ij}{}^\Lambda$ and/or
terms quadratic in the spinor fields and are not of immediate
interest here.

Most of the cancellations required for demonstrating the
supersymmetry of the Lagrangian will still take place when
derivatives are replaced by covariant derivatives. A clear
exception arises when dealing with the commutator of two
derivatives, because it will lead to a field strength upon using
the Ricci identity. This situation occurs for the variations of
the fermion kinetic term. Furthermore, when establishing
supersymmetry for the more conventional Lagrangians, one makes use
of the Bianchi identity for the field strengths, which no longer
applies to the new field strengths. Of course, the gauge fields in
the covariant derivatives will also lead to new variations. To
investigate these issues, we first determine the supersymmetry
variation of $\mathcal{L}_{\mathrm{matter}}$ under the
transformations given above (up to total derivatives),
\begin{eqnarray}
  \label{eq:L-matter-variations}
  \delta_0\mathcal{L}_{\mathrm{matter}}&=&
  i g\, \Omega_{MQ} T_{PN}{}^Q \,\left[D^\mu\bar X^M \, X^N -
  \bar X^M \, D_\mu X^N +\half \bar\Omega^{iM}
  \gamma^{\mu}\Omega_i{}^N \right] \,  \delta A_\mu{}^P \nonumber \\
  &&
  -\half i g\, \Omega_{MQ} T_{PN}{}^Q \,\left[\bar X^M\,
  \bar\Omega_i{}^N   \gamma^{\mu\nu} \epsilon^i\,
  \mathcal{H}^-_{\mu\nu}{}^P - \mathrm{h.c.} \right]
  \nonumber\\
  &&
  +i\, \Omega_{MN} \,\left[\bar\Omega^{iM}\gamma_\nu
  \epsilon^j\,\varepsilon_{ij} \,D_\mu    \mathcal{G}^{-\mu\nu N}
   - \mathrm{h.c.} \right]  \ ,
\end{eqnarray}
where we suppressed variations that involve neither the gauge
coupling constant $g$ nor the modified field strengths. These
variations will cancel as before.

It is now easy to verify that the term of order $g^0$ can be
combined with the result from the variation of
$\mathcal{L}_{\mathrm{vector}} + \mathcal{L}_{\mathrm{top}}$ (c.f.
(\ref{eq:var-L-vector}) and (\ref{eq:var-L-top})),
\begin{equation}
  \label{eq:var-L-vector+top}
  \delta_0(\mathcal{L}_{\mathrm{vector}} +\mathcal{L}_{\mathrm{top}})
  =  i \Omega_{MN} \,
  \mathcal{G}^{+\mu\nu M} \; D_\mu\delta A_\nu{}^N  + \mathrm{h.c.} +
  \cdots   \ .
\end{equation}
The combined result thus leads to a total derivative plus terms
proportional to $D_\mu F_{\Lambda\Sigma}$ and terms cubic in the
fermions. These terms cancel for the abelian theory with an
ordinary derivative and the cancellation proceeds identically when
ordinary derivatives are replaced by covariant ones. Note that
nowhere one needs to use the Bianchi identity. This calculation
confirms the correctness of the transformation rule for the
magnetic gauge fields. Hence we can now concentrate on the
remaining terms of (\ref{eq:L-matter-variations}), which are the
only variations left, up to terms induced by the variation of the
tensor fields which we will need in due course.

To cancel the order-$g$ terms in (\ref{eq:L-matter-variations}) we
need to add new terms in the transformation rules of
$\Omega_i{}^\Lambda$ and $Y_{ij}{}^\Lambda$. Furthermore, new
terms to the Lagrangian are required.  For the case of purely
electric charges these terms are known and the easiest strategy is
to simply generalize these terms. This leads to the expressions,
\begin{eqnarray}
  \label{eq:order-g}
  \delta_g \Omega_i{}^\Lambda &=& - 2 g\, T_{MN}{}^\Lambda \,\bar X^M
  X^N  \,\varepsilon_{ij} \, \epsilon^j\ ,
  \nonumber \\
  \delta_g Y_{ij}{}^\Lambda &=& -4g\, T_{MN}{}^\Lambda\left[ \bar
  \Omega_{(i}{}^M \epsilon^k \,\varepsilon_{j)k}\,\bar X^N  -
  \bar\Omega^{kM} \epsilon_{(i}\, \varepsilon_{j)k} ,X^N\right]\ ,
  \nonumber \\
  \mathcal{L}_g &=&  - \half i g \,\Omega_{MQ}
  T_{PN}{}^Q \,\left[ \varepsilon^{ij}\, \bar\Omega_i{}^M
  \Omega_j{}^P \bar X^N -
  \varepsilon_{ij} \, \bar\Omega^{iM} \Omega^{jP} X^N \right] \ .
\end{eqnarray}
In the case of purely electric charges the expression for
$\mathcal{L}_g$ reduces to the first expression of
(\ref{eq:electric-masslike-potential}) upon using
(\ref{eq:derivative-invariance}).

Collecting the new variations proportional to the field strengths
that arise as a result of (\ref{eq:order-g}), we find, using
(\ref{eq:cal-G}), (\ref{eq:fermion-der-invariance}) and
(\ref{eq:lin}),
\begin{equation}
  \label{eq:g-G-variations}
  \delta_g\mathcal{L}_{\mathrm{vector}} + \delta_0\mathcal{L}_g =
  \half i g\,\Omega_{MQ} T_{PN}{}^Q\, \bar X^M\,\bar
  \Omega_i{}^N\gamma^{\mu\nu}\epsilon^i \,\mathcal{G}^-_{\mu\nu}{}^P
  + \mathrm{h.c.} \ .
\end{equation}
This term is almost identical to the second term of
(\ref{eq:L-matter-variations}) except that it is proportional to
$\mathcal{G}_{\mu\nu}{}^M$ rather than to
$\mathcal{H}_{\mu\nu}{}^M$. However, the combination of these two
terms is cancelled by assigning the following variation to the
tensor fields,
\begin{equation}
  \label{eq:B-susy-transformation}
  \delta B_{\mu \nu{\sf a}} = - 2 t_{{\sf a} M}{}^P\Omega_{PN}
  \left(A_{[\mu}{}^M \,\delta A_{\nu]}{}^N - \bar{X}^M \bar{\Omega}_i{}^N
  \gamma_{\mu\nu} \epsilon^i - X^M \bar\Omega^{iN}
  \gamma_{\mu\nu}\epsilon_i\right)\ .
\end{equation}

At this point one can verify that all supersymmetry variations
linear in the gauge coupling constant $g$ vanish. Here one makes
use of the various results derived in section \ref{vmnac}, and in
particular of (\ref{eq:cov-der-invariance}). What remains are the
order-$g^2$ interactions induced by the order-$g$ transformations
of the spinors, which can be written as,
\begin{equation}
  \label{eq:var-Omega^M-g}
  \delta\Omega_i{}^M = -2g\, T_{NP}{}^M \, \bar X^N X^P
  \,\varepsilon_{ij}\, \epsilon^j \ .
\end{equation}
The order-$g^2$ variation follows from $\delta_g \mathcal{L}_g$,
and can be written proportional to the supersymmetry variation
$\delta X^M$ given in (\ref{eq:susy-symplectic}),
\begin{equation}
  \label{eq:order-g2}
  \delta_g \mathcal{L}_g= - 2i g^2 \,\Omega_{MQ} T_{NP}{}^Q \,
  \bar X^P \delta X^{[M} \; T_{RS}{}^{N]} \, \bar X^R X^S +
  \mathrm{h.c.}  \ .
\end{equation}
Using the Lie algebra relation (\ref{eq:closure}), as well as the
relation (\ref{eq:cov-der-invariance}), we can write this in a
form that can be integrated. This reveals that these variations
can be cancelled by the variation of a scalar potential,
corresponding to
\begin{equation}
  \label{eq:scalar-g2-lagrangian}
  \mathcal{L}_{g^2}= ig^2\, \Omega_{MN} \, T_{PQ}{}^M X^P
  \bar X^Q  \;  T_{RS}{}^N \bar X^R X^S \ .
\end{equation}
This expression reduces to (\ref{eq:electric-masslike-potential})
for purely electric gaugings upon using
(\ref{eq:derivative-invariance}). Observe that the charges
$T_{\Lambda\Sigma\Gamma}$ do not contribute to
(\ref{eq:scalar-g2-lagrangian}), as is well known from previous
constructions.

This concludes the derivation of supersymmetric vector multiplet
Lagrangians with electric and magnetic gauge charges. In the
following section we will consider the coupling to matter by
introducing hypermultiplets. This will lead to a second scalar
potential.

\section{Hypermultiplets}
\label{sec:hypermultiplets}
In this section we give a brief description of the possible
gaugings of isometries in the hyper-K\"{a}hler space parameterized
by the hypermultiplet scalars, following the framework of
\cite{dWKV}.

As we saw in section \ref{hypm}, $n_{\mathrm{H}}$ hypermultiplets
are described by $4n_{\mathrm{H}}$ real scalars $\phi^A$,
$2n_{\mathrm{H}}$ positive-chirality spinors $\zeta^{\bar\alpha}$
and $2n_{\mathrm{H}}$ negative-chirality spinors $\zeta^\alpha$.
Their (rigid) supersymmetry transformations and invariant
Lagrangian are given by (\ref{trh}) and (\ref{lrh}).

The equivalence transformations of the fermions and the
target-space diffeomorphisms associated with this Lagrangian do
not constitute invariances of the theory, unless they leave the
metric $g_{AB}$ and the
$\mathrm{Sp}(n_{\mathrm{H}})\times\mathrm{Sp}(1)$ one-form
$V^\alpha_i$ (and thus the related geometric quantities)
invariant. Therefore invariances are related to isometries of the
hyper-K\"ahler space. A subset of them can be elevated to a group
of local (i.e. spacetime-dependent) transformations, which require
a coupling to corresponding vector multiplets. Such gauged
isometries have been studied in the literature
\cite{dWRVs,DFF,ST2,HKLR,BGIO,ABCDFFM} but only for electric
charges.

Infinitesimal isometries are characterized by Killing vectors and
the ones associated to local transformations will be labelled by
the same index $M$ that labels the electric and magnetic gauge
fields of the previous sections. In principle, the gauged
isometries constitute a subgroup of the full group of isometries,
defined by the embedding tensor. Hence the corresponding Killing
vectors are proportional to the embedding matrix, $k^A{}_M=
\Theta_M{}^{\sf a} \,k^A{}_{\sf a}$, and (\ref{eq:Z-Theta})
implies,
\begin{equation}
  \label{eq:Z-killing}
  Z^{M,{\sf a}} \,k^A{}_M =0\ .
\end{equation}
Without gauge interactions, the hypermultiplets do not couple to
the vector multiplets, so that the full group of invariances
factorizes into separate invariance groups of the vector multiplet
Lagrangian and of the hypermultiplet Lagrangian. The index ${\sf
a}$ refers to all these symmetries, and therefore $k^A{}_{\sf a}$
will vanish whenever the index ${\sf a}$ refers to a generator
acting exclusively on the vector multiplets.

The local gauge transformations are thus generated by the Killing
vectors $k^A{}_M(\phi)= (k^A{}_\Lambda(\phi),
k^{A\Lambda}(\phi))$, with parameters $\Lambda^M$. Under
infinitesimal transformations we have
\begin{equation}
  \label{eq:delta-phi-Killing}
\delta\phi^A=g\,\Lambda^M k^A{}_M(\phi)\ ,
\end{equation}
where $g$ is the coupling constant and the $k^A{}_M(\phi)$ satisfy
the Killing equation,
\begin{equation}
  \label{eq:killing-eq}
  D_Ak_{BM} + D_B k_{AM}=0\ .
\end{equation}
Higher derivatives of Killing vector are not independent, as is
shown by
\begin{equation}
  \label{eq:symmetric}
  D_{A}D_{B} k_{C M}  =  R_{BCAE}\,  k^{\,E}{}_M   \ .
\end{equation}
The isometries close under commutation,
\begin{equation}
  \label{eq:killingclosure}
  k^B{}_M\partial_Bk^A{}_N-k^B{}_N\partial_Bk^A{}_M = T_{MN}{}^P\,
  k^A{}_P \ ,
\end{equation}
where, as before, the antisymmetry in $[MN]$ on the right-hand
side is ensured by (\ref{eq:Z-killing}).

The invariances associated with the target-space isometries act on
the fermions by field dependent matrices, which satisfy the
relation
\begin{equation}
  \label{eq:fermion-t-relation}
  (t_M)^{\alpha}{}_{\!\beta}\, V^\beta_{Ai}  = D_A k^B{}_M\,
  V^\alpha_{Bi}\ ,
\end{equation}
leading to
\begin{equation}
  \label{eq:fermion-t}
  (t_M)^{\alpha}{}_{\!\beta} = \half V_{Ai}^{\alpha} \,
  \bar\gamma^{Bi}_{\beta}\; D_B k^A{}_M\ .
\end{equation}
The result (\ref{eq:fermion-t-relation}) was derived by requiring
that the tensor $V_{Ai}^\alpha$ is invariant under the isometries,
up to a rotation on the indices $\alpha$.  The invariance implies
that target-space scalars satisfy algebraic identities such as
\begin{equation}
  \label{eq:t-G-Omega}
  \bar t_M{}^{\bar\gamma}{}_{\!\bar\alpha} \, G_{\bar\gamma\beta}
  +{t_M}^{\gamma}{}_{\!\beta} \,
  G_{\bar\alpha\gamma}= {t_M}^{\bar\gamma}{}_{\![\bar\alpha} \,
  \Omega_{\bar\beta]\bar\gamma} = 0\ ,
\end{equation}
which establishes that the matrices ${t_M}^\alpha{}_\beta$ take
values in $\mathrm{sp}(n_{\mathrm{H}})$. From
(\ref{eq:killingclosure}) and (\ref{eq:symmetric}), one may derive
\begin{equation}
  \label{eq:t-der}
  D_A t_M{}^{\alpha}{}_{\!\beta}  =
  R_{AB}{}^{\!\alpha}{}_{\!\beta} \,k^B{}_M \,  ,
\end{equation}
for any infinitesimal isometry. From the group property of the
isometries it follows that the matrices $t_M$ satisfy the
commutation relations,
\begin{equation}
  \label{eq:t-comm}
  [\,t_M ,\,t_N\,]^\alpha{}_{\!\beta}   = -T_{MN}{}^P\,
  (t_P)^\alpha{}_{\!\beta}  + k^A{}_M\,k^B{}_N\,
  R_{AB}{}^{\!\alpha}{}_{\!\beta} \ ,
\end{equation}
which takes values in $\mathrm{sp}(n_{\mathrm{H}})$.  This result
is consistent with the Jacobi identity.

The previous results imply that the complex structures
$J_{AB}^{ij}$ are invariant under the isometries,
\begin{equation}
  \label{eq:triholo}
  k^C{}_M\, \partial_C J_{AB}^{ij} - 2 \p_{[A}k^C{}_M \,J_{B]C}^{ij} =
  0\ ,
\end{equation}
implying that the isometries are {\it tri-holomorphic}. From
(\ref{eq:triholo}) one shows that $\partial_A(J^{ij}_{BC}\,
k^C{}_M) -\partial_B(J^{ij}_{AC}\, k^C{}_M)=0$, so that, locally,
one can associate three Killing potentials (or moment maps)
$\mu^{ij}{}_M$ to every Killing vector, according to
\begin{equation}
  \label{eq:tri-Killing-potential}
  \partial_A \mu^{ij}{}_M  = J_{AB}^{ij} \,k^{B}{}_M  \ ,
\end{equation}
which determines $\mu^{ij}{}_M$ up to a constant. These constants
correspond to Fayet-Iliopoulos terms. Up to such constants one
derives the equivariance condition,
\begin{equation}
  \label{eq:equivariance}
  J^{ij}_{AB}\,k^A{}_M\,k^B{}_N= T_{MN}{}^{P}\,\mu^{ij}{}_P \ ,
\end{equation}
which implies that the Killing potentials transform covariantly
under the isometries,
\begin{equation}
  \label{k-potential-isometry}
  \delta\mu^{ij}{}_M  = \Lambda^N \,k^A{}_N\,\partial_A
  \mu^{ij}_M  = \Lambda^N \, T_{NM}{}^{P} \,\mu^{ij}{}_P\, .
\end{equation}

Subsequently we consider the consequences of realizing the
isometry (sub)group generated by the $k^A{}_M$ as local gauge
group. The latter acts on the hypermultiplet fields in the
following way,
\begin{equation}
  \label{eq:fermgaugetr}
  \delta\phi = g\,\Lambda^M\,k^{A}{}_M \ ,\qquad
  \delta\zeta^\alpha=g\, \Lambda^M {t_M}^{\alpha}{}_{\!\beta}
  \,\zeta^\beta - \delta\phi^A \Gamma_A{}^{\!\alpha}{}_{\!\beta}
  \,\zeta^\beta \ ,
\end{equation}
where the parameters $\Lambda^M$ are functions of $x^\mu$. The
relevant covariant derivatives are equal to,
\begin{equation}
  \label{eq:cov-hypermultiplet-der}
  {\cal D}_\mu \phi^A = \partial_\mu \phi^A - g A_\mu{}^M \,k^A{}_M
  \ , \qquad
  {\cal D}_\mu\zeta^\alpha =\partial_\mu \zeta^\alpha+
  \partial_\mu\phi^A\,\Gamma_A{}^{\!\alpha}{}_{\!\beta}\,
  \zeta^\beta -gA_\mu{}^M {t_M}^\alpha{}_{\!\beta}\,\zeta^\beta\, . \;
\end{equation}
These covariant derivatives must be substituted into the
transformation rules (\ref{trh}) and the Lagrangian (\ref{lrh}).
The covariance of ${\cal D}_\mu\zeta^\alpha$,
\begin{equation}
  \label{eq:trasnf-cov-cer}
  \delta{\cal D}_\m \zeta^\alpha=g\,
  \Lambda^M t_M{}^{\alpha}{}_{\!\beta}\,{\cal D}_\m \zeta^\beta -
  \delta\phi^A
  \Gamma_A{}^{\!\alpha}{}_{\!\beta} \,{\cal D}_\m\zeta^\beta \ ,
\end{equation}
follows from (\ref{eq:t-der}) and (\ref{eq:t-comm}).

Just as for the vector multiplets, the introduction of the gauge
covariant derivatives to the Lagrangian breaks the supersymmetry
of the Lagrangian. To restore supersymmetry we follow the same
procedure as in section \ref{sec:rest-supersymm-non}. But in this
case the situation is somewhat simpler because the electric and
magnetic gauge fields couple to standard hypermultiplet
isometries. This means that the initial results will coincide with
those obtained for electric gaugings.

Let us first present the variations of the Lagrangian (\ref{lrh})
with the proper gauge covariantizations and determine the
supersymmetry variation linear in the gauge coupling constant $g$
and linear in the fermion fields,
\begin{equation}
  \label{eq:delta-L0}
  \delta\mathcal{L}_0= g \,k_{AM} \Big[\gamma^A_{i\bar\alpha}
  \,\bar\zeta^{\bar\alpha} \gamma^{\mu\nu}\epsilon^i
  \mathcal{F}^-_{\mu\nu}{}^M
  + \varepsilon^{ij} \,\bar\Omega_i{}^M
  \Slash{\mathcal{D}}\phi^A \epsilon_j + \mathrm{h.c.}\Big] \ .
\end{equation}
The first term originates from the fact that the commutator of two
covariant derivatives acquires an extra field strength in the
presence of the gauging, whereas the second term originates from
the variation of the gauge fields in the covariant derivatives of
the scalars. The first term can be cancelled by a supersymmetry
variation of the following new term,
\begin{equation}
  \label{eq:Lagr-g-1}
  \mathcal{L}_g^{(1)} = 2g\, k_{AM} \left[\bar \gamma^{Ai}_{\alpha}
  \varepsilon_{ij}\,{\bar\zeta}^\alpha \Omega^{jM} +
  \gamma^A_{i\bar\alpha} \varepsilon^{ij}\,{\bar\zeta}^{\bar\alpha}
  \Omega_j{}^{M}\right]  \ .
\end{equation}
The variations of this term proportional to the field strength
$\mathcal{G}_{\mu\nu}{}^M$ cancel against the term proportional to
$\mathcal{H}_{\mu\nu}{}^M$ (the field strength
$\mathcal{F}_{\mu\nu}{}^M$ can be replaced by
$\mathcal{H}_{\mu\nu}{}^M$ by virtue of (\ref{eq:Z-killing})) by
introducing a new term to the variation of the tensor fields
$B_{\mu\nu{\sf a}}$ (\ref{eq:B-susy-transformation}),
\begin{equation}
  \label{eq:delta-B-hyper}
  \delta B_{\mu\nu{\sf a}}=  - 4i k^A{}_{{\sf a}} \;
  \left[ \gamma_{Ai\bar\alpha}
  \,\bar\zeta^{\bar\alpha} \gamma_{\mu\nu}\epsilon^i -
  \bar\gamma^i_{A\alpha}
  \,\bar\zeta^{\alpha} \gamma_{\mu\nu}\epsilon_i  \right] \ .
\end{equation}

Another term in the variation of (\ref{eq:Lagr-g-1}) is
proportional to $X^M$ and its complex conjugate. Their
cancellation requires the following extra variations of the
hypermultiplet spinors,
\begin{equation}
  \label{eq:susyferm}
  \delta\zeta^\alpha =
  2gX^M \,k^A{}_M V^\alpha_{Ai}\,\varepsilon^{ij}\epsilon_j\ ,\qquad
  \delta\zeta^{\bar \alpha}=  2g{\bar X}^M\,k^A{}_M  {\bar V}^{{\bar
  \alpha}i}_{A}\,\varepsilon_{ij}\epsilon^j\ ,
\end{equation}
and an extra term in the Lagrangian equal to
\begin{equation}
    \label{eq:Lagr-g-2}
  \mathcal{L}_g^{(2)} =
   2g\left[ {\bar X}^M{t_M}^{\!\gamma}{}_{\!\alpha}\,\bar
    \Omega_{\beta\gamma}\,{\bar \zeta}^\alpha\zeta^\beta+
    X^M{t_M}^{\!\bar\gamma}{}_{\!\bar\alpha}\,
    \Omega_{\bar\beta\bar\gamma}\,{\bar\zeta}^{\bar\alpha}
    \zeta^{\bar\beta}\right]     \ .
\end{equation}
The remaining variations then take the following form.
\begin{eqnarray}
  \label{eq:delta-0+1+2}
  \delta\mathcal{L}_0 +\delta\mathcal{L}_g^{(1)}
  +\delta\mathcal{L}_g^{(2)} &=& - 2 g\,\partial_A \mu^{ij}{}_M\,
  \bar\Omega_i{}^M \Slash{\mathcal{D}}\phi^A \epsilon_j
  -2 g\,\partial_A \mu_{ijM} \,
  \bar\Omega^{iM} \Slash{\mathcal{D}}\phi^A \epsilon^j
  \nonumber\\
  &&{}
  - 2g\,\left[\partial_A\mu_{ij\Lambda}\, Y^{ij\Lambda} +
  \partial_A\mu_{ij}{}^\Lambda\,\bar{F}_{\Lambda\Sigma}\,Y^{ij\Sigma}\right]
  \, \bar\gamma^{Ak}_\alpha\,\bar\epsilon_k\zeta^\alpha
  \nonumber\\
  &&{}
  - 2g\,\left[\partial_A\mu_{ij\Lambda}\, Y^{ij\Lambda} +
  \partial_A\mu_{ij}{}^\Lambda\,{F}_{\Lambda\Sigma}\,Y^{ij\Sigma}\right]
  \, \gamma^A_{k\bar\alpha}\,\bar\epsilon^k\zeta^{\bar\alpha}\ ,
  \nn\\
\end{eqnarray}
where we restricted ourselves to variations linear in the fermion
fields and linear in $g$.

To cancel these variations we must include the following new term
in the Lagrangian,
\begin{eqnarray}
  \label{eq:Lagr-g-3}
  \mathcal{L}_g^{(3)} &=& g\,Y^{ij\Lambda} \left[ \mu_{ij\Lambda}
    +\half (F_{\Lambda\Sigma}+ \bar
    F_{\Lambda\Sigma})\,\mu_{ij}{}^\Sigma\right]
    \nonumber\\
    &&{}
    -\vier g\,\left[ F_{\Lambda\Sigma\Gamma}\,
  \mu^{ij\Lambda} \,\bar\Omega_i{}^\Sigma \Omega_j{}^\Gamma
   + \bar F_{\Lambda\Sigma\Gamma} \,
  \mu_{ij}{}^\Lambda\,\bar\Omega^{i\Sigma} \Omega^{j\Gamma} \right]\ ,
\end{eqnarray}
as well as assign new variations of the fields
$\Omega_i{}^\Lambda$ and $Y_{ij}^\Lambda$ of the vector multiplet,
\begin{eqnarray}
  \label{eq:delta-g-Omega-2}
  \delta_g\Omega_i{}^\Lambda &=& 2\,ig\, \mu_{ij}{}^\Lambda
  \epsilon^j \ , \nonumber\\
  \delta Y_{ij}{}^\Lambda &=& 4\,i g\,k^{A\Lambda}
  \left[\varepsilon_{k(i}\,
  \gamma_{j)\bar\alpha A} \bar\epsilon^k\zeta^{\bar\alpha} +
  \varepsilon_{k(i}\, \bar\epsilon_{j)} \zeta^{\alpha} \,
  \bar\gamma^{k}_{\alpha A}   \right]\ .
\end{eqnarray}
This completes the discussion of all the variations linear in $g$
and in the fermion fields. The result remains valid for the cubic
fermion variations as well. However, new variations arise in
second order in $g$, by the order-$g$ variations in the new
order-$g$ terms in the Lagrangian. These variations cancel against
the variation of a scalar potential, corresponding to
\begin{equation}
  \label{eq:hyper-pot}
  \mathcal{L}_{g^2} = -2g^2k^A{}_M \,k^B{}_N  \,g_{AB}\,X^M{\bar
    X^N} - \half g^2 \,N_{\Lambda\Sigma} \,\mu_{ij}{}^\Lambda \,
    \mu^{ij\Sigma}\ .
\end{equation}
To prove (\ref{eq:hyper-pot}), one has to make use of the
equivariance condition (\ref{eq:equivariance}). Actually, gauge
invariance, which is prerequisite to supersymmetry, already
depends on (\ref{k-potential-isometry}).

\section{Summary and discussion}
\label{sec:summary-discussion}

In this chapter we presented Lagrangians and supersymmetry
transformations for a general supersymmetric system of vector
multiplets and hypermultiplets in the presence of both electric
and magnetic charges. The results were verified to all orders and
are consistent with results known in the literature that are based
on purely electric charges. We have also verified the closure of
the supersymmetry algebra, which holds up to the field equations
associated with the fields $A_{\mu \Lambda}$, $B_{\mu \nu
\sf{a}}$, $Y_{ij}{}^\Lambda$ and the hypermultiplet spinors
$\ze^{\al}$. In the absence of magnetic charges, the supersymmetry
algebra closes on the vector multiplets without the need for
imposing the field equations. We return to this issue later in
subsection \ref{sec:off-shell-structure}.

Before discussing possible implications of these results, let us
first summarize the terms induced by the gauging. We first present
the combined supersymmetry variations. First of all, we have the
original transformations in the absence of the gauging, where
spacetime derivatives are replaced by gauge-covariant derivatives
and where the abelian field strengths $F_{\mu\nu}{}^\Lambda$ are
replaced by the covariant field strengths
$\mathcal{H}_{\mu\nu}{}^\Lambda$. We will not repeat the
corresponding expressions here, but we present the other terms in
the transformation rules that are induced by the gauging. They
read as follows,
\begin{eqnarray}
  \label{eq:var-g-fields}
  \delta_g\Omega_i{}^\Lambda &=& -2g\, T_{NP}{}^\Lambda \, \bar X^N X^P
  \,\varepsilon_{ij}\, \epsilon^j + 2\,ig\, \mu_{ij}{}^\Lambda
  \epsilon^j \ , \nonumber\\
  \delta_g\zeta^\alpha &=& 2\,gX^M \,k^A{}_M
  V^\alpha_{Ai}\,\varepsilon^{ij}\epsilon_j\ ,\nonumber \\
  \delta_g Y_{ij}{}^\Lambda &=& -4g\, T_{MN}{}^\Lambda\left[ \bar
  \Omega_{(i}{}^M \epsilon^k \,\varepsilon_{j)k}\,\bar X^N  -
  \bar\Omega^{kM} \epsilon_{(i}\, \varepsilon_{j)k} \,X^N\right]
  \nonumber \\
  &&{}
  + 4\,i g\,k^{A\Lambda}
  \left[\varepsilon_{k(i}\,
  \gamma_{j)\bar\alpha A} \bar\epsilon^k\zeta^{\bar\alpha} +
  \varepsilon_{k(i}\, \bar\epsilon_{j)} \zeta^{\alpha} \,
  \bar\gamma^{k}_{\alpha A}   \right]\ , \nonumber\\
  \delta B_{\mu\nu{\sf a}}&=& - 2 t_{{\sf a} M}{}^P\Omega_{PN}
  \left(A_{[\mu}{}^M \,\delta A_{\nu]}{}^N - \bar{X}^M \bar{\Omega}_i{}^N
  \gamma_{\mu\nu} \epsilon^i - X^M \bar\Omega^{iN}
  \gamma_{\mu\nu}\epsilon_i\right)\nonumber\\
   &&{}
   - 4i k^A{}_{{\sf a}} \;
  \left[ \gamma_{Ai\bar\alpha}
  \,\bar\zeta^{\bar\alpha} \gamma_{\mu\nu}\epsilon^i -
  \bar\gamma^i_{A\alpha}
  \,\bar\zeta^{\alpha} \gamma_{\mu\nu}\epsilon_i  \right] \ .
\end{eqnarray}

Likewise we will not repeat the original Lagrangians (\ref{vecsh})
and (\ref{lrh}) for the vector multiplets and hypermultiplets,
respectively, which are only modified by replacing spacetime
derivatives by gauge-covariant ones, and field strengths by the
covariant field strengths $\mathcal{H}_{\mu\nu}{}^\Lambda$. The
Lagrangian (\ref{eq:Ltop}) remains unchanged. The additional terms
induced by the gauging that are linear in $g$ take the following
form,
\begin{eqnarray}
  \label{eq:order-g-Lagrangian}
  \mathcal{L}_g &=& - \half i g \,\Omega_{MQ}
  T_{PN}{}^Q \,\left[ \varepsilon^{ij}\, \bar\Omega_i{}^M
  \Omega_j{}^P \bar X^N -
  \varepsilon_{ij} \, \bar\Omega^{iM} \Omega^{jP} X^N \right]
  \nonumber\\
  &&{}
    -\vier g\,\left[ F_{\Lambda\Sigma\Gamma}\,
  \mu^{ij\Lambda} \,\bar\Omega_i{}^\Sigma \Omega_j{}^\Gamma
   + \bar F_{\Lambda\Sigma\Gamma} \,
  \mu_{ij}{}^\Lambda\,\bar\Omega^{i\Sigma} \Omega^{j\Gamma} \right]
    \nonumber\\
   &&{}
   +2g\, k_{AM} \left[\bar \gamma^{Ai}_{\alpha}
  \varepsilon_{ij}\,{\bar\zeta}^\alpha \Omega^{jM} +
  \gamma^A_{i\bar\alpha} \varepsilon^{ij}\,{\bar\zeta}^{\bar\alpha}
  \Omega_j{}^{M}\right]  \nonumber\\
   &&{}
   +2g\left[ {\bar X}^M{t_M}^{\!\gamma}{}_{\!\alpha}\,\bar
    \Omega_{\beta\gamma}\,{\bar \zeta}^\alpha\zeta^\beta+
    X^M{t_M}^{\!\bar\gamma}{}_{\!\bar\alpha}\,
    \Omega_{\bar\beta\bar\gamma}\,{\bar\zeta}^{\bar\alpha}
    \zeta^{\bar\beta}\right]     \nonumber\\
   &&{}
    +g\,Y^{ij\Lambda} \left[ \mu_{ij\Lambda}
    +\half (F_{\Lambda\Sigma}+ \bar
    F_{\Lambda\Sigma})\,\mu_{ij}{}^\Sigma\right] \ .
\end{eqnarray}
The terms of order $g^2$ correspond to a scalar potential
proportional to $g^2$ and are given by
\begin{eqnarray}
  \label{eq:full-g2-lagrangian'}
  \mathcal{L}_{g^2} &=& ig^2\, \Omega_{MN}\,T_{PQ}{}^M X^P
  \bar X^Q  \;  T_{RS}{}^N \bar X^R X^S \nonumber\\
  &&{}
  -2g^2k^A{}_M \,k^B{}_N  \,g_{AB}\,X^M{\bar
    X^N} - \half g^2 \,N_{\Lambda\Sigma} \,\mu_{ij}{}^\Lambda \,
    \mu^{ij\Sigma}\ .
\end{eqnarray}

\subsection{Applications}
\label{sec:applications}
The above results have many applications. A relatively simple one
concerns the Fayet-Iliopoulos terms, which are the integration
constants of the Killing potentials $\mu^{ij}{}_M$.  This enables
us to truncate the above expressions by setting the embedding
tensor to zero, while still retaining the constants
$g\mu^{ij}{}_M$. In that case all effects of the gauging are
suppressed and one is left with a potential accompanied by
fermionic masslike terms,
\begin{eqnarray}
  \label{eq:FI}
  \mathcal{L}_{\mathrm{FI}} &=&   \frac{1}{8}
     N^{\Lambda\Sigma}\,\left (N_{\Lambda\Gamma}Y_{ij}{}^\Gamma +
     \half i (F_{\Lambda\Gamma\Omega}\,
     \bar\Omega_i{}^\Gamma\Omega_j{}^\Omega - \bar F_{\Lambda\Gamma\Omega}
     \,\bar\Omega^{k\Gamma}\Omega^{l\Omega}\varepsilon_{ik}\varepsilon_{jl}
     )\right)  \nonumber\\
     &&
    \times\left(N_{\Sigma\Xi}Y^{ij\Xi}  +
     \half i (F_{\Sigma\Xi\Delta}
     \,\bar\Omega_m{}^\Xi\Omega_n{}^\Delta
     \varepsilon^{im}\varepsilon^{jn} -  \bar F_{\Sigma\Xi\Delta}
     \,\bar\Omega^{i\Xi}\Omega^{j\Delta} )  \right) \nonumber\\
  &&{}
  - \half g^2 \,N_{\Lambda\Sigma} \,\mu_{ij}{}^\Lambda \,
    \mu^{ij\Sigma} +g\,Y^{ij\Lambda} \left[ \mu_{ij\Lambda}
    +\half (F_{\Lambda\Sigma}+ \bar
    F_{\Lambda\Sigma})\,\mu_{ij}{}^\Sigma\right]
      \nonumber\\
    &&{}
    -\vier g\,\left[ F_{\Lambda\Sigma\Gamma}\,
  \mu^{ij\Lambda} \,\bar\Omega_i{}^\Sigma \Omega_j{}^\Gamma
   + \bar F_{\Lambda\Sigma\Gamma} \,
  \mu_{ij}{}^\Lambda\,\bar\Omega^{i\Sigma} \Omega^{j\Gamma}
     \right]\ .
\end{eqnarray}
Eliminating the auxiliary fields $Y_{ij}{}^\Lambda$ gives rise to
the following expression,
\begin{eqnarray}
  \label{eq:FI-2}
    \mathcal{L}_{\mathrm{FI}} &=& -\half iN^{\Lambda\Sigma}\,
    F_{\Sigma\Gamma\Xi}\,
    \,\bar\Omega_i{}^\Gamma\Omega_j{}^\Xi \,\left[\mu^{ij}{}_\Lambda +
    \bF_{\Lambda\Delta}\,\mu^{ij\Delta} \right] \nonumber\\
    &&{}
    +\half i N^{\Lambda\Sigma}\,
    \bar F_{\Sigma\Gamma\Xi}\,
     \Omega^{i\Gamma}\Omega^{j\Xi} \,\left[\mu_{ij\Lambda} +
    F_{\Lambda\Delta}\,\mu_{ij}{}^\Delta \right] \nonumber\\
    &&{}
    -2\,g^2 \left[\mu^{ij}{}_\Lambda +
    F_{\Lambda\Gamma}\,\mu^{ij\Gamma} \right] \,N^{\Lambda\Sigma}
    \left[\mu_{ij\Sigma} +\bar F_{\Sigma\Xi}\,\mu_{ij}{}^\Xi\right]
    \ .
\end{eqnarray}
The above expression transforms as a function under
electric/magnetic duality provided that the $\mu^{ij}{}_M$ are
treated as spurionic quantities. The last term in (\ref{eq:FI-2})
corresponds to minus the potential, which is positive definite
(assuming positive $N_{\Lambda\Sigma}$). The Lagrangian is a
generalization of the Lagrangian presented in
\cite{Antoniadis:1996}, where it was also shown how the potential
can lead to spontaneous partial supersymmetry breaking when
$\mu_{ij}{}^\Lambda \neq 0$. Note that the hypermultiplets play
only an ancillary role here, as they decouple from the vector
multiplets.

Most of the possible applications can be found in the context of
supergravity, where they will be useful for constructing
low-energy effective actions associated with string
compactifications in the presence of fluxes. In principle it is
straightforward to extend our results to the case of local
supersymmetry. The target-space of the vector multiplets should
then be restricted to a special K\"ahler cone (as we saw in
chapter \ref{ch2}, this requires that $F(X)$ be a homogeneous
function of second degree), and the hypermultiplet scalars should
coordinatize a hyper-K\"ahler cone. Furthermore, the various
formulae for the action and the supersymmetry transformation rules
should be evaluated in the presence of a superconformal
background, so that the action and transformation rules will also
involve the superconformal fields. This has not yet been worked
out in detail for $N=2$ supergravity, although it is in principle
straightforward. In view of the fact that gaugings of $N=4$ and
$N=8$ supergravity have already been worked out using the same
formalism as in our work \cite{Schon,dWST-8}, no complications are
expected. Note that Fayet-Iliopoulos terms do not exist in $N=2$
supergravity, because the Killing potentials cannot contain
arbitrary constants as those would break the scale invariance of
the hyper-K\"ahler cone.

The potential is rather independent of all these details, although
it must be rewritten in terms of the proper quantities, as was for
instance demonstrated in \cite{dWRVs}. It was already shown in
\cite{dWST} that the theory simplifies considerably for abelian
gaugings where $T_{MN}{}^P=0$ and where the potential is
exclusively generated by the hypermultiplet charges. Making use of
the steps described in \cite{dWRVs}, it is rather straightforward
to derive the potential (as was already foreseen in \cite{dWST}),
which takes precisely the form conjectured quite some time ago in
\cite{Mi}. The results can also be compared to the work of
\cite{dASVal}.

\subsection{Off-shell structure} \label{sec:off-shell-structure}

In the absence of magnetic charges, the vector multiplets
constitute off-shell representations of the supersymmetry algebra.
On the hypermultiplets the supersymmetry algebra is only realized
up to the fermionic field equations. The tensor fields decouple
from the theory. However, when magnetic charges are present, there
are no longer any off-shell multiplets and the supersymmetry
algebra is only realized when the fields satisfy the field
equations of the hypermultiplet spinors and of the fields $A_{\mu
\Lambda}$, $Y_{ij}{}^\Lambda$ and $B_{\mu \nu \sf{a}}$. In this
subsection we discuss how the off-shell closure can possibly be
regained for the vector multiplets when magnetic charges are
switched on.

We start by introducing $2n$ \emph{independent} vector multiplets,
associated with the electric and magnetic gauge fields,
$A_\mu{}^\Lambda$ and $A_{\mu \Lambda}$, and collectively denoted
by $A_\mu{}^M$. In the absence of charges, these fields are
subject to the standard off-shell transformation rules,
\begin{eqnarray}\label{emvm}
\delta X^M & = & \bar{\epsilon}^i \Om_i{}^M \ ,\nonumber\\
  \delta A_{\mu}{}^M& = & \varepsilon^{ij} \bar{\epsilon}_i
  \gamma_{\mu} \Om_j{}^M + \varepsilon_{ij}
  \bar{\epsilon}^i \gamma_{\mu} \Om^{j M}\ ,\nonumber\\
  \delta \Om_i{}^M & = & 2 \spa X^M \epsilon_i
  + \half \gamma^{\mu \nu} F^-_{\mu\nu}{}^M
  \varepsilon_{ij} \epsilon^j + Y_{ij}{}^M \ep^j \ ,\nn\\
  \delta Y_{ij}{}^M & = & 2 \bep_{(i} \spa \Om_{j)}{}^M + 2
  \vep_{ik} \vep_{jl} \bep^{(k} \spa \Om^{l) M}\ .
\end{eqnarray}
We stress once more that, unlike previously, the $2n$ vector
multiplets are independent. In due course we shall see how to make
contact with the previous description.

We also introduce $p$ off-shell tensor multiplets,
\begin{eqnarray}\label{tens}
\de G_{\ra} & = & - 2 \bep_i \spa \varphi^i{}_{\ra}\ ,\nn\\
\de B_{\mu \nu \ra} & = & \half(i \bep^i \ga_{\mu \nu}
\varphi^j{}_{\ra} \vep_{ij} - i \bep_i \ga_{\mu \nu} \varphi_{j
\ra} \vep^{ij})\
,\nn\\
\de \varphi^i{}_{\ra} & = & \spa (l^{ij}{}_{\ra} + \vep^{ik}
\vep^{jl} l_{kl \ra}) \ep_j + 2 \vep^{ij}
\SH_{\ra} \ep_j - G_{\ra} \ep^i\ ,\nn\\
\de l_{ij \ra} & = & 2 \vep_{ik} \vep_{jl} \bep^{(k}
\varphi^{l)}{}_{\ra}\ ,
\end{eqnarray}
where ${\sf a} = 1,...,p$. $\H_{\mu \ra} = \frac{1}{6} i \vep_{\mu
\nu \rho \sigma} \H^{\nu \rho \sigma}{}_{\ra}$ and $\H_{\mu \nu
\rho \, \ra} = 3 \p_{[\mu} B_{\nu \rho] \ra}$. The $G_a$ are
complex scalars fields, the $l_{ij \sf{a}}$ triplets of complex
scalar fields and the $\varphi^i{}_a$ are the right-handed parts
of sets of two Majorana spinors. $B_{\mu \nu \ra}$ and $l_{ij
\ra}$ are subject to the gauge transformations
\begin{eqnarray}\label{gabl}
\de B_{\mu \nu \ra} = 2 \p_{[\mu} \Xi_{\nu] \ra}\ ,\quad \de l_{ij
\ra} = \ga_{ij \ra}\ ,
\end{eqnarray}
where $\Xi_{\mu \ra}$ is real and $\ga_{ij \ra}$ is imaginary (in
the sense that $(\ga_{ij \ra})^* = - \vep^{ik} \vep^{jl} \ga_{kl
\ra}$). When $l_{ij \ra} - \vep_{ik} \vep_{jl} l^{kl}{}_{\ra} = 0$
(\ref{tens}) reduces to the set of transformation rules for
off-shell tensor multiplets of \cite{dWvHvP2}.

Next charges are turned on. We restrict ourselves to abelian gauge
groups and leave non-abelian gaugings for further study. In the
presence of charges the tensor multiplets appear in the
supersymmetry transformations of the vector multiplets. As before,
the field strengths $F_{\mu\nu}{}^M$ in (\ref{emvm}) are replaced
by the covariant field strengths $\H_{\mu\nu}{}^M = F_{\mu\nu}{}^M
+ g Z^{M, \sf{a}} B_{\mu \nu \sf{a}}$. Furthermore, in view of the
adjustment required for hypermultiplet gaugings
(\ref{eq:delta-g-Omega-2}), the real auxiliary fields $Y_{ij}{}^M$
are replaced by the complex fields
\begin{eqnarray}
\Y_{ij}{}^M = Y_{ij}{}^M - i g Z^{M , \ra} l_{ij \ra}\ .
\end{eqnarray}
One thus gets
\begin{eqnarray}\label{emvmg}
\delta X^M & = & \bar{\epsilon}^i \Om_i{}^M \ ,\nonumber\\
  \delta A_{\mu}{}^M& = & \varepsilon^{ij} \bar{\epsilon}_i
  \gamma_{\mu} \Om_j{}^M + \varepsilon_{ij}
  \bar{\epsilon}^i \gamma_{\mu} \Om^{j M}\ ,\nonumber\\
  \delta \Om_i{}^M & = & 2 \spa X^M \epsilon_i
  + \half \gamma^{\mu \nu} \H^-_{\mu\nu}{}^M
  \varepsilon_{ij} \epsilon^j + \Y_{ij}{}^M \ep^j \ ,\nn\\
  \delta Y_{ij}{}^M & = & 2 \bep_{(i} \spa \Om_{j)}{}^M + 2
  \vep_{ik} \vep_{jl} \bep^{(k} \spa \Om^{l) M}\ ,
\end{eqnarray}
and the gauge transformations (\ref{gabl}) now also act on the
vector multiplets,
\begin{eqnarray}\label{gaay}
\de A_\mu{}^M = - g Z^{M , \ra} \Xi_{\mu \ra}\ ,\quad \de
Y_{ij}{}^M = i g Z^{M \ra} \ga_{ij \ra}\ .
\end{eqnarray}

It can be shown that (\ref{emvmg}) are still off-shell multiplets.
The supersymmetry algebra closes as on (\ref{emvm}), up to
additional gauge transformations of the type (\ref{gaay}), with
parameters
\begin{eqnarray}
\Xi'_{\mu \sf{a}} & = & 4 \bep_{i [1} \ga^{\sigma} \ep_{2]}^i
B_{\sigma \mu \sf{a}} + 2i \vep^{ij} \bep_{j[1} \ga_{\mu}
\ep_{2]}^k l_{ik \sf{a}} - 2i \vep_{ij} \bep^j_{[1} \ga_{\mu}
\ep_{2]k} l^{ik}{}_{\sf{a}}\ ,\nn\\
\ga'_{ij \sf{a}} & = & 4(\bep_{(i[1} \ga^\rho \ep_{2]}^k \p_\rho
l_{j)k \sf{a}} - \vep_{ik} \vep_{jl} \bep_{[1}^{(k} \ga^\rho
\ep_{2]p} \p_\rho l^{l)p}{}_{\sf{a}})\nn\\
& & - 4 i \vep_{k(i} \bep_{[1}^k \ga^\rho \ep_{2]j)}\vep_{\rho
\alpha \beta \gamma} \p^\alpha B^{\beta \gamma}{}_{\sf{a}}\ .
\end{eqnarray}
Here $\ep_1$ and $\ep_2$ are the parameters of the supersymmetry
transformations, as appearing in the commutator $[\de_Q (\ep_1),
\de_Q (\ep_2)]$.

\vspace{3mm}

Then we construct supersymmetric Lagrangians in terms of the
off-shell multiplets (\ref{tens}) and (\ref{emvmg}).

Since just the $A_\mu{}^\Lambda$ are meant to play the role of
electric gauge fields, we introduce a Lagrangian of the form
(\ref{vecsh}) for the vector multiplets with upper indices
$\Lambda$ only. We replace $F_{\mu \nu}{}^\Lambda$ by $\H_{\mu
\nu}{}^\Lambda$ and $Y_{ij}{}^\Lambda$ by $\Y_{ij}{}^\Lambda$ and,
using the results we obtained before, add the topological term
\begin{eqnarray}\label{topterm}
\L_{\mathrm{top}} = \frac{1}{8} i g\,
\varepsilon^{\mu\nu\rho\sigma}\,
  \Theta^{\Lambda {\sf a}}\,B_{\mu\nu {\sf a}} \,
  \Big(2\,\partial_{\rho} A_{\sigma\,\Lambda}  -\vier
  g\Theta_{\Lambda}{}^{\sf b}B_{\rho\sigma {\sf b}}\Big)\ .
\end{eqnarray}
Recall that the equations of motion of the tensor fields $B_{\mu
\nu \ra}$ relate  the field strengths associated to the magnetic
vector gauge fields, $\H_{\mu \nu}{}_\Lambda$, to the dual field
strengths, $\G_{\mu \nu \Lambda}$,
\begin{eqnarray}\label{grelh}
- \frac{1}{8}i g \vep^{\mu \nu \rho \sigma} \Theta^{\Lambda {\sf
a}}(\mathcal{G}_{\rho \sigma} -\H_{\rho \sigma})_\Lambda & = & 0\
.
\end{eqnarray}

Since the $l_{ij \ra}$ enter the vector multiplets in a way
similar to $B_{\mu \nu \ra}$ we expect that the equations of
motion of $l_{ij \ra}$, when following from a supersymmetric
Lagrangian, relate $\Y_{ij}{}_\Lambda$ and the duals of
$\Y_{ij}{}^\Lambda$ in the same way as (\ref{grelh}) does with
$\H_{\mu \nu}{}_\Lambda$ and $\G_{\mu \nu \Lambda}$. To realize
this, the term
\begin{eqnarray}\label{topy}
\L_g^{'} = \frac{1}{8} g  \vep^{ik} \vep^{jl} \Theta^{\Lambda \ra}
l_{ij \ra} (Y_{kl \Lambda} + \vier i g \Theta_\Lambda{}^{\rb}
l_{kl \rb}) + h.c,
\end{eqnarray}
is added to the Lagrangian. As required, the equations of motion
of the $l_{ij \ra}$ then become
\begin{eqnarray}\label{yrelz}
- \frac{1}{8} \vep^{ik} \vep^{jl} \Theta^{\Lambda {\sf a}}(\Z_{ij}
- \Y_{ij})_\Lambda & = & 0\ ,
\end{eqnarray}
where
\begin{eqnarray}
\Z_{ij \Lambda} = F_{\Lambda \Sigma} \Y_{ij}{}^\Sigma - \half
F_{\Lambda \Sigma \Gamma} \bar{\Om}_i{}^\Sigma \Om_j{}^\Gamma\ ,
\end{eqnarray}
is the analog of $Z_{ij \Lambda}$ (which is defined in
(\ref{defiz})).

Similarly, the equations of motion of the tensor multiplet
fermions $\varphi^i{}_{\ra}$ and scalars $G_{\ra}$ should relate
the vector multiplet fermions $F_{\Lambda \Sigma} \Om_i{}^\Sigma$
and $\Om_{i \Lambda}$ and scalars $F_\Lambda$ and $X_\Lambda$.
This requires the inclusion of the terms
\begin{eqnarray}\label{lgom}
\L_g^{''} & = & \vier g \Theta^{\Lambda  \ra}
\bar{\varphi}^i{}_{\ra} (F_{\Lambda \Sigma} \Om_i{}^\Sigma -
\Om_{i \Lambda}) +
\mathrm{h.c.}\nn\\
& & + \vier g \Theta^{\Lambda \ra} G_{\ra} (F_\Lambda - X_\Lambda)
+ \mathrm{h.c.}\ ,
\end{eqnarray}
in the Lagrangian. It is straightforward to check that the model
thus obtained is indeed supersymmetric.

The abelian gauge group can be embedded in the rigid invariance
group associated with the hyper-K\"{a}hler space parameterized by
the hypermultiplet scalars. Since all $2n$ gauge fields come with
complete vector multiplets, gaugings of this type are similar to
gaugings with $2n$ electric charges. The corresponding
supersymmetric hypermultiplet Lagrangian therefore follows from
the results of section \ref{sec:hypermultiplets} in the case of
vanishing magnetic charges. It reads
\begin{eqnarray}\label{susl}
\L & = & \L_0 + 2g \bX^M t_M{}^{\ga}{}_{\al} \bOm_{\be \ga}
\bze^\al \ze^\be + \mathrm{h.c.}\nn\\
& & + 2g k^A{}_M \bga^i_{A \al} \vep_{ij} \bze^{\bal} \Om^{j M} +
\mathrm{h.c.}\nn\\
& & + g \mu^{ij}{}_M Y_{ij}{}^M\nn\\
& & - 2 g^2 k^A{}_M k^B{}_N g_{AB} X^M \bX^N\ ,
\end{eqnarray}
(note that, due to (\ref{eq:Z-Theta}), the terms in the
supersymmetry transformation of $\Om^M$ that are proportional to
$B_{\mu\nu \sf{a}}$ and $l_{ij \sf{a}}$ play no role), while the
order-$g$ variations of the hypermultiplet spinors are
\begin{equation}\label{ordgvar}
  \delta\zeta^\alpha =
  2g X^M \,k^A{}_M V^\alpha_{Ai}\,\varepsilon^{ij}\epsilon_j\ ,\qquad
  \delta\zeta^{\bar \alpha}=  2g{\bX}^M\,k^A{}_M  {\bar V}^{{\bar
  \alpha}i}_{A}\,\varepsilon_{ij}\epsilon^j\ .
\end{equation}
The complete Lagrangian is then given by (\ref{susl}) supplemented
with a Lagrangian of the form (\ref{vecsh}) for the electric
vector multiplets and the terms (\ref{topterm}), (\ref{topy}) and
(\ref{lgom}).

Eliminating $Y_{ij}{}_\Lambda$, $\Om_i{}_\Lambda$ and $X_\Lambda$
(and absorbing the real part of $- \half i g \Theta^{\Lambda \ra}
l_{ij \ra}$ in $Y_{ij}{}^\Lambda$) reproduces the results of
(\ref{sec:hypermultiplets}) for the subclass of abelian gaugings.

\renewcommand{\chaptermark}[1]{\markboth{\thechapter\ #1}{}}
\fancyhead{} \fancyhead[LE, RO]{\thepage}
\fancyhead[CO]{\slshape\leftmark}
\fancyhead[CE]{\slshape\leftmark}
\appendix
\chapter{Notation and conventions}\label{notcon}

We use $\mu, \nu, \cdots$ ($m,n, \cdots$) for four- (three-)
dimensional spacetime indices. $a, b, \cdots$ are four-dimensional
Lorentz indices. In the context of $N=2$ supersymmetry, $i, j,
\cdots$ are usually $SU(2)_R$ indices.

Our conventions for (anti-)symmetrization are
\begin{eqnarray}
[ab] = \half [ab - ba]\ ,\quad (ab) = \half (ab + ba)\ .
\end{eqnarray}

Gamma matrices we take, such that
\begin{eqnarray}
\gamma_a \gamma_b = \eta_{ab} + \gamma_{ab}\ ,\quad \gamma_5 = i
\gamma_0 \gamma_1 \gamma_2 \gamma_3\ ,
\end{eqnarray}
where we use $\eta_{ab} = (- + + +)$. A charge conjugation matrix
$C$ is defined, such that
\begin{eqnarray}
- \ga_{\mu}^T = C \ga_{\mu} C^{-1}\ , \quad \ga_5^T = C \ga_5
C^{-1}\ ,\quad C^T = -C\ .
\end{eqnarray}

In four dimensions the fully antisymmetric tensor reads
\begin{eqnarray}
\varepsilon^{a b c d} = e^{-1} \varepsilon^{\mu  \nu \lambda
\sigma} e^a_{\mu} e^b_{\nu} e^c_{\lambda} e^d_{\sigma}\ ,\quad
\varepsilon^{0123} = i\ ,\quad \varepsilon^{1234} = 1\ ,
\end{eqnarray}
where $\varepsilon^{1234}$ is a Euclidean component. The fully
antisymmetric tensor in three dimensions is defined analogously.

The dual of an antisymmetric tensor field $F_{ab}$ (in Minkowski
space) is defined by
\begin{eqnarray}
\tilde{F}_{ab} = \half \varepsilon_{a b c d} F^{cd}\ ,
\end{eqnarray}
such that its (anti-)selfdual part is given by
\begin{eqnarray}
F_{ab}^{\pm} = \half (F_{ab} \pm \tilde{F}_{ab})\ .
\end{eqnarray}

Under hermitian conjugation selfdual becomes antiselfdual and vice
versa. In the context of $N=2$, our conventions are such that
$SU(2)_R$ indices change place under complex conjugation.

The Dirac conjugate $\bar{\psi}$ of a Dirac spinor $\psi$ is
defined by
\begin{eqnarray}
\bar{\psi} = \psi^{\dag} \ga_0\ .
\end{eqnarray}
The (pseudo) reality condition
\begin{eqnarray}
\bar{\psi} = \psi^T C\ ,
\end{eqnarray}
defines a Majorana spinor.

Under complex conjugation there are the following identities,
\begin{eqnarray}\lb{idesp}
\bar{\psi} \ga_a \phi = - \bar{\phi} \ga_a \psi\ ,\quad \psi \phi
= \bar{\phi} \psi\ ,
\end{eqnarray}
for any two spinors $\psi$, $\phi$. From (\ref{idesp}) similar
identities for other bilinears can be derived. Furthermore,
\emph{Majorana} spinors $\psi$ and $\phi$ satisfy
\begin{eqnarray}
\bar{\psi} \ga_a \phi = - \bar{\phi} \ga_a \psi\ ,\quad \bar{\psi}
\phi = \bar{\phi} \psi\ .
\end{eqnarray}

If two spinors $\psi$ and $\phi$ do not form a bilinear, their
product can be decomposed on a basis of four-by-four matrices by
means of a Fierz rearrangement,
\begin{eqnarray}
\phi \bar{\psi} & = & - \vier (\bar{\psi} \phi) \mathbb{I} - \vier
(\bps \ga^a \phi) \ga_a - \vier (\bps \ga_5 \phi) \ga_5 + \vier
(\bps \ga^a \ga_5 \phi) \ga_a \ga_5\nn\\
& & + \frac{1}{8} (\bps \ga^{ab} \phi) \ga_{ab}\ .
\end{eqnarray}

Finally, we note the identities,
\begin{eqnarray}
\begin{array}{rcl} \ga_{ab} & = & - \half \vep_{abcd} \ga^{cd}
\ga_5\ ,\\
\ga^{ab} \ga_{ab} & = & -12\ ,\\
\ga^c \ga_{ab} \ga_c & = & 0\ ,\\
\[ \ga^c, \ga_{ab}\]  & = & 4 \de_{[a}{}^c \ga_{b]}\ ,\\
\[ \ga_{ab}, \ga^{cd} \] & = & - 8 \de_{[a}{}^{c]} \ga_{[b}{}^{d]}\
, \end{array} \quad
\begin{array}{rcl} \ga^b \ga_a \ga_b & = & - 2 \ga_a\ ,\\ \ga^{cd}
\ga_{ab} \ga_{cd} & = & 4 \ga_{ab}\ ,\\ \ga^{bc} \ga_a \ga_{bc} &
= & 0\ ,\\  \{ \ga^c, \ga_{ab} \} & = & 2 \vep_{ab}{}^{cd} \ga_5
\ga_d\ ,\\  \{ \ga_{ab}, \ga^{cd} \} & = & - 4 \de_{[a}{}^c
\de_{b]}{}^d +  2 \vep_{ab}{}^{cd} \ga_5\ . \end{array}
\end{eqnarray}

\vspace{10mm}

\textbf{Wick rotation}

\vspace{5mm}

The standard Wick rotation
\begin{equation}
t  =  - i \tau\ ,
\end{equation}
defines Euclidean Lagrangians
\begin{equation}
{\cal L}_{m}  =  i {\cal L}_{e}\ .
\end{equation}

The Wick rotation on tensors is
\begin{equation}
B_{tm}\rightarrow i B_{\tau m}\ ,\qquad B_{mn}\rightarrow B_{mn}\
,
\end{equation}
and similarly for vectors.

\vspace{5mm}

\chapter{Superconformal multiplets}\label{supcon}

 \textbf{The Weyl multiplet}

\vspace{5mm}

The \emph{dependent} fields of the Weyl multiplet:
\begin{eqnarray}
\omega_\mu{}^{ab} & = & - 2 e^{\nu[a} \p_{[\mu} e_{\nu]}{}^{b]} -
e^{\nu[a} e^{b] \sigma} e_{\mu c} \p_\sigma e_\nu{}^c - 2
e_\mu{}^{[a} e^{b] \nu} b_\nu \nn\\
& & - \vier (2 \bps_\mu{}^i \ga^{[a} \psi^{b]}{}_i +\bps^{ai}
\ga_\mu \psi^b{}_i + \mathrm{h.c.})\ ,\nn\\
\phi_\mu{}^i & = & (\half \ga^{\rho \si} \ga_\mu - \frac{1}{6}
\ga_\mu \ga^{\rho \si})(\mathbb{D}_\rho \psi_{\si i} -
\frac{1}{16} \ga^{\eta \lambda} T^{ij}_{\eta \lambda} \ga_\rho
\psi_{\si j} +
\vier \ga_{\rho \si} \chi^i)\ ,\nn\\
f_\mu{}^\mu & = & \frac{1}{6} R - D - (\frac{1}{12} e^{-1}
\vep^{\mu \nu \rho \sigma} \bps_\mu{}^i \ga_\nu \mathbb{D}_\rho
\psi_{\si i}\nn\\
& &  - \frac{1}{12} \bps_\mu{}^i \psi_\nu{}^j T_{ij}^{\mu \nu} -
\vier \bps_\mu{}^i \ga^\mu \chi_i + \mathrm{h.c.})\ .
\end{eqnarray}

\vspace{3mm}

The $Q$-supersymmetry, special supersymmetry and special conformal
transformation rules of the \emph{independent} fields of the Weyl
multiplet:
\begin{eqnarray}\lb{weyli}
\delta e_{\mu}{}^a & = & \bep^i \ga^a \psi_{\mu i} + \mathrm{h.c.}\ ,\nn\\
\delta \psi_{\mu}{}^i & = & 2 \mathbb{D}_\mu \ep^i - \acht
\ga^{\rho \sigma}
T^{ij}_{\rho \sigma} \ga_\mu \ep_j - \ga_\mu \eta^i\ ,\nn\\
\delta b_{\mu} & = & \half \bep^i \phi_{\mu i} - \frac{3}{4}
\bep^i \ga_\mu
\chi_i - \half \bar{\eta}^i \psi_{\mu i} + \mathrm{h.c.} + \Lambda^a_K e_{\mu a}\ ,\nn\\
\delta W_{\mu} & = & \half i \bep^i \phi_{\mu i} + \frac{3}{4} i
\bep^i \ga_\mu
\chi_i + \half i \bar{\eta}^i \psi_{\mu i} + \mathrm{h.c.}\ ,\nn\\
\delta \V_\mu{}^i{}_j & = & 2 \bep_j \phi_\mu^i - 3 \bep_j \ga_\mu
\chi^i
+ 2 \bar{\eta}_j \psi_\mu{}^i - (\mathrm{h.c.}; \mathrm{traceless})\ ,\nn\\
\delta T_{ab}^{ij} & = & 8 \bep^{[i} \hat{R}_{ab}(Q)^{j]}\ ,\nn\\
\delta \chi^i & = & - \frac{1}{12} \ga^{ab} \SD T_{ab}^{ij} \ep_j
+ \frac{1}{6} \hat{R}_{\mu \nu}(SU(2))^i{}_j \ga^{\mu \nu} \ep^j -
\frac{1}{3} i \hat{R}_{\mu \nu}(U(1)) \ga^{\mu \nu} \ep^i\nn\\
& &  + D \ep^i + \frac{1}{12} \ga^{\mu \nu} T^{ij}_{\mu \nu} \eta_j\ ,\nn\\
\delta D & = & \bep^i \SD \chi_i + \mathrm{h.c.}\ ,
\end{eqnarray}
where $\Lambda^a_K$ is the parameter associated with special
conformal transformations.


\vspace{3mm}

The derivatives $D_\mu$ appearing above are covariantized with
respect to all superconformal transformations, while
$\mathbb{D}_\mu$ denotes a derivative covariantized with respect
to Lorentz transformations, dilatations and $U(2)_R$
transformations only (see below).

Explicit expressions for the curvatures $\hat{R} (SU(2))$,
$\hat{R}(U(1))$ and $\hat{R}(Q)$ will also be given below.

\vspace{10mm}

\textbf{Vector multiplets}

\vspace{5mm}

The $Q$-supersymmetry and special supersymmetry transformation
rules of abelian vector multiplets:
\begin{eqnarray}\label{susyl}
\delta X^\Lambda & = & \bar{\epsilon}^i \Omega^\Lambda_i\
,\nonumber\\
\delta A^\Lambda_{\mu}& = & \varepsilon^{ij} \bar{\epsilon}_i
\gamma_{\mu} \Omega^\Lambda_j + \varepsilon_{ij} \bar{\epsilon}^i
\gamma_{\mu} \Omega^{j \Lambda} + 2 \bX^\Lambda \vep_{ij} \bep^i
\psi_{\mu}^j
+ 2 X^\Lambda \vep^{ij} \bep_i \psi_{j \mu}\ ,\nonumber\\
\delta \Omega_i{}^\Lambda & = & 2 \SD X^\Lambda \ep_i + \half
\gamma_{\mu \nu} \F^{- \mu \nu \Lambda} \varepsilon_{ij}
\epsilon^j + Y_{ij}{}^\Lambda
\epsilon^j + 2 X^\Lambda \eta_i\ ,\nonumber\\
\delta Y_{ij}{}^\Lambda & = & 2 \bar{\epsilon}_{(i} \SD
\Omega_{j)}{}^\Lambda + 2 \varepsilon_{ik} \varepsilon_{jl}
\bar{\epsilon}^{(k} \SD \Omega^{l) \Lambda}\ .
\end{eqnarray}
Here
\begin{eqnarray}
\F_{\mu \nu}{}^\Lambda & = & 2 \p_{[\mu} A_{\nu]}{}^\Lambda -
(\vep_{ij} \bar{\psi}_{[\mu}{}^i \ga_{\nu]} \Om^{j \Lambda} +
\vep_{ij} \bX^\Lambda \bar{\psi}_\mu{}^i \psi_\nu{}^j + \vier
\vep_{ij}
\bX^\Lambda T^{ij}_{\mu \nu} + \mathrm{h.c.} )\ ,\nn\\
\end{eqnarray}
such that
\begin{eqnarray}
\de \F_{\mu \nu}{}^\Lambda & = & 2 \vep^{ij} \bep_i \ga_{[\mu}
D_{\nu]} \Om_j{}^\Lambda + \mathrm{h.c.}\ .
\end{eqnarray}

\vspace{10mm}

\textbf{Hypermultiplets}

\vspace{5mm}

The $Q$-supersymmetry and special supersymmetry transformation
rules of hypermultiplets (without a coupling to vector
multiplets):
\begin{eqnarray}
\de \phi^A & = & 2 (\ga_{i \bal}^A \bep^i \ze^{\bal} +
\bga_{\bal}^{A i} \bep_i \ze^{\al})\ ,\nn\\
\de \ze^\al & = & \SDm A_i{}^\al \ep^i - \de_Q \phi^B
\Gamma_B{}^{\al}{}_{\be} \ze^\be + A_i{}^\al \eta^i\ ,
\end{eqnarray}
where
\begin{eqnarray}
A_i{}^{\al} = \chi^B V^{\al}_{Bi}\ ,\quad \D_\mu A_i{}^\al & = &
D_\mu A_i{}^\al + \p_\mu \phi^A \Gamma_{A \ \be}^{\ \al}
A_i{}^\be\ ,
\end{eqnarray}
with $D_\mu$ the
derivative covariantized with respect to all superconformal
transformations.

\newpage

\vspace{5mm}

\textbf{Weights and chirality}

\vspace{20mm}

\begin{tabular}{c||cccccccc|ccc}
\multicolumn{1}{c||}{}& \multicolumn{11}{c}{Weyl multiplet} \\
\hline \hline field & $e_\mu{}^a$ & $\psi_\mu{}^i$ & $b_\mu$ &
$W_\mu$ & $\V_\mu{}^i{}_j$ & $T_{ab}^{ij}$ & $\chi^i$ & $D$ &
$\omega_\mu{}^{ab}$ & $f_\mu{}^a$ & $\phi_\mu{}^i$\\
\hline $w$ & $-1$ & $- \half$ & $0$ & $0$ & $0$ & $1$ &
$\frac{3}{2}$ & $2$ & $0$ & $1$ & $\half$
\\
\hline $c$ & $0$ & $- \half$ & $0$ & $0$ & $0$ & $-1$ & $- \half$
& $0$ & $0$ & $0$ & $- \half$\\
\hline \multicolumn{1}{c||}{$\ga_5$} & \multicolumn{1}{c}{} &
\multicolumn{1}{c}{$+$} & \multicolumn{4}{c}{} &
\multicolumn{1}{c}{$+$} & \multicolumn{1}{c|}{} &
\multicolumn{2}{c}{} & \multicolumn{1}{c}{$-$} \\
\end{tabular}

\vspace{5mm}

Table B.I: Dilatational and $U(1)_R$ weights ($w$ and $c$,
respectively) and fermion chirality $(\ga_5)$ of the Weyl
multiplet component fields.

\vspace{20mm}

\begin{tabular}{c||cccc||cc||cc}
\multicolumn{1}{c||}{}& \multicolumn{4}{c||}{vector multiplet} &
\multicolumn{2}{c||}{hypermultiplet} &
\multicolumn{2}{c}{parameters}\\
\hline \hline field & $X^\Lambda$ & $\Om_i{}^\Lambda$ &
$A_\mu{}^\Lambda$ & $Y_{ij}{}^\Lambda$ & $\phi^A$ & $\ze^\al$  &
$\ep^i$ & $\eta^i$ \\
\hline $w$ & $1$ & $\frac{3}{2}$ & $0$ & $2$ & $1$ & $\frac{3}{2}$
& $- \half$ & $\half$
\\
\hline $c$ & $-1$ & $- \half$ & $0$ & $0$ & $0$ & $- \half$ & $- \half$ & $- \half$\\
\hline \multicolumn{1}{c||}{$\ga_5$} & \multicolumn{1}{c}{} &
\multicolumn{1}{c}{$+$} & \multicolumn{2}{c||}{} &
\multicolumn{1}{c}{} & \multicolumn{1}{c||}{$-$} &
\multicolumn{1}{c}{$+$} & \multicolumn{1}{c}{$-$} \\
\end{tabular}

\vspace{5mm}

Table B.II: Dilatational and $U(1)_R$ weights ($w$ and $c$,
respectively) and fermion chirality $(\ga_5)$ of the vector and
hypermultiplet component fields and of the supersymmetry
transformation parameters.

\vspace{15mm}

\emph{Covariant derivatives}

\vspace{5mm}

When a derivative is covariantized with respect to a set of
transformations $\delta_A$, it is given by
\begin{eqnarray}
D_\mu = \p_\mu - \sum_A \de_A (h_\mu (A))\ ,
\end{eqnarray}
where $h_\mu (A)$ is the gauge field associated with $\de_A$. For
the superconformal transformations, the gauge fields are
normalized like in \cite{deWit:1984px},
\begin{eqnarray}
h_\mu{}^{ab} (M) & = & \omega_\mu{}^{ab}\ ,\nn\\
h_\mu (D) & = & b_\mu\ ,\nn\\
h_\mu (U(1)_R) & = & W_\mu\ ,\nn\\
h_\mu{}^i{}_j (SU(2)_R) & = & - \half \V_\mu{}^i{}_j\ ,\nn\\
h_\mu{}^i (Q) & = & \half \psi_\mu{}^i\ ,\nn\\
h_\mu{}^i (S) & = & \half \phi_\mu{}^i\ ,\nn\\
h_\mu{}^a (K) & = & f_\mu{}^a\ .
\end{eqnarray}

\vspace{5mm}

\emph{Supercovariant curvatures}

\vspace{5mm}

\begin{eqnarray}
\hat{R}_{\mu \nu} (Q)^i & = & 2 \mathbb{D}_{[\mu} \psi_{\nu]}{}^i
- \ga_{[\mu} \phi_{\nu]}{}^i - \frac{1}{8} \ga^{\eta \lambda}
T^{ij}_{\eta \lambda} \ga_{[\mu} \psi_{\nu] j}\ ,\nn\\
\hat{R}_{\mu \nu} (U(1)) & = & 2 \p_{[\mu} A_{\nu]} - i (\half
\bps_{[\mu}{}^i \phi_{\nu] i} + \frac{3}{4} \bps_{[\mu}{}^i
\ga_{\nu]} \chi_i - \mathrm{h.c.})\ ,\nn\\
\hat{R}_{\mu \nu} (SU(2))^i{}_j & = & 2 \p_{[\mu} \V_{\nu]}{}^i{}_j
 + \V_{[\mu}{}^i{}_k \V_{\nu]}{}^k{}_j\nn\\
& & + (2 \bps_{[\mu}{}^i \phi_{\nu] j} - 3 \bps_{[\mu}{}^i
\ga_{\nu]} \chi_j - (\mathrm{h.c.}; \mathrm{traceless}))\ .
\end{eqnarray}

\backmatter
\renewcommand{\chaptermark}[1]{\markboth{ #1}{}}
\fancyhead{} \fancyhead[LE, RO]{\thepage}
\fancyhead[CO]{\slshape\nouppercase{\leftmark}}
\fancyhead[CE]{\slshape\nouppercase{\leftmark}}

\makeatletter
\def\thickhrulefill{\leavevmode \leaders \hrule height 1ex \hfill \kern \z@}
\def\@makechapterhead#1{%
  \vspace*{10\p@}%
  {\parindent \z@ \centering \reset@font
        \Huge \bfseries #1\par\nobreak
        \par
        \vspace*{10\p@}%
    \vskip 60\p@
  }}
\def\@makeschapterhead#1{%
  \vspace*{10\p@}%
  {\parindent \z@ \centering \reset@font
        \Huge \bfseries #1\par\nobreak
        \par
        \vspace*{10\p@}%
    \vskip 60\p@
  }}

\addcontentsline{toc}{chapter}{Bibliography}
\bibliographystyle{eigenutcaps}
\fontsize{11}{13pt}\selectfont
\bibliography{bib1}

\pagestyle{empty}

\end{document}